# Cedalion Tutorial: A Python-based framework for comprehensive analysis of multimodal fNIRS & DOT from the lab to the everyday world

E. Middell[1,2,†], L. Carlton[3], S. Moradi[1,2], T. Codina[1,2], T. Fischer[1,2], J. Cutler[1,2], S. Kelley[3], J. Behrendt[1,2], T. Dissanayake[1,2], N. Harmening[1,2], M. A. Yücel[3], D. A. Boas[3], A. von Lühmann[1,2,3,†,*]

[1] Intelligent Biomedical Sensing (IBS) Lab, Technische Universität Berlin, 10587 Berlin, Germany
[2] BIFOLD – Berlin Institute for the Foundations of Learning and Data, 10587 Berlin, Germany
[3] Neurophotonics Center, Biomedical Engineering, Boston University, Boston, MA 02215, USA

**Abstract**

**Significance:** Functional near-infrared spectroscopy (fNIRS) and diffuse optical tomography (DOT) are rapidly evolving toward wearable, multimodal, and data-driven, AI-supported neuroimaging in the everyday world. However, current analytical tools are fragmented across platforms, limiting reproducibility, interoperability, and integration with modern machine learning (ML) workflows.

**Aim:** *Cedalion* is a Python-based open-source framework designed to unify advanced model-based and data-driven analysis of multimodal fNIRS and DOT data within a reproducible, extensible, and community-driven environment.

**Approach:** Cedalion integrates forward modelling, photogrammetric optode co-registration, signal processing, GLM Analysis, DOT image reconstruction, and ML-based data-driven methods within a single standardized architecture based on the Python ecosystem. It adheres to SNIRF and BIDS standards, supports cloud-executable Jupyter notebooks, and provides containerized workflows for scalable, fully reproducible analysis pipelines that can be provided alongside original research publications.

**Results:** Cedalion connects established optical-neuroimaging pipelines with ML frameworks such as scikit-learn and PyTorch, enabling seamless multimodal fusion with EEG, MEG, and physiological data. It implements validated algorithms for signal-quality assessment, motion correction, GLM modelling, and DOT reconstruction, complemented by modules for simulation, data augmentation, and multimodal physiology analysis. Automated documentation links each method to its source publication, and continuous-integration testing ensures robustness. This tutorial paper provides seven fully executable notebooks that demonstrate core features.

**Conclusions:** Cedalion offers an open, transparent, and community extensible foundation that supports reproducible, scalable, cloud- and ML-ready fNIRS/DOT workflows for laboratory-based and real-world neuroimaging.

**Keywords**: fNIRS, DOT, multimodal, machine learning, data driven, physiology, everyday neuroscience.

† Co-First Authors, * correspondence: vonluehmann@tu-berlin.de

## 1 Introduction and Motivation

### 1.1 Towards data-driven fNIRS/DOT-based Neuroimaging in the Everyday World

Human neuroscience and neurotechnology are in the process of significant transformation, moving from conventional laboratory settings towards embracing the complexity of natural environments [1], [2], [3], [4], [5]. This shift to an ecologically valid depiction and decoding of human brain function will provide new scientific insights and breakthroughs in our understanding of neuronal development, health and ageing, and drive translation in medicine, psychiatry and brain-computer interfacing. However, substantial interdisciplinary challenges still need to be overcome. Naturalistic settings are difficult to study with conventional methods, sensors, and parametric experimental paradigms currently available. To better understand and deal with the complex interactions between the brain, the body, and the environment, we need unobtrusive and integrative multimodal platforms that can continuously monitor the embodied brain [6], [7] in order to discover and model the intertwined relationships between physiology, behavior, and cognition in everyday settings.

Thanks to the technological progress of the last years [8], [9], wearable functional near-infrared spectroscopy (fNIRS) and high-density diffuse optical tomography (HD-DOT) offer a pathway to continuous everyday brain monitoring. Recent HD-DOT studies have achieved the same spatial resolution of functional activity on the surface of the brain (cortex) as the latest fMRI-based Human Connectome Parcellation scheme [10], [11], enabling connectivity analysis or decoding performance comparable with fMRI [12], [13], [14], [15], [16]. However, natural behavior and motion create unique methodological challenges in fNIRS and DOT. It is non-trivial to distinguish evoked neuronal activity from complex systemic physiological activity [7], [17], [18], [19] and motion artifacts, or to perform multimodal co-modulation analysis with EEG, MEG, OPM-MEG, or other multivariate signals of physiology or behavior.

One way of addressing these challenges is to leverage recent advances in classical and deep learning (ML/AI) to better isolate neuronal components from confounding physiological variance, reduce artifacts and build generalizable and context-sensitive methods for brain imaging and decoding. ML/AI can explain variance and improve contrast by uncovering complex patterns and relationships between multimodal signals from brain, physiology, and behavior, for which no straightforward analytical models exist. While ML has transformed numerous scientific fields, its impact on fNIRS – let alone DOT – has only just begun [20]. Consequently, there is both a growing need and great promise for analytical tools that converge the transformative potential of wearable fNIRS / DOT combined with multimodal ML-driven signal processing methods. See recent reviews on deep learning for fNIRS [20], [21] and ML for multimodal fNIRS-EEG sensor fusion [22], [23] for an overview of recent progress and remaining limitations. In this tutorial we introduce the **Cedalion toolbox**, a comprehensive suite for reproducible, state-of-the-art fNIRS analysis in channel and cortical image space (i.e. DOT) that supports multimodal signal analysis and facilitates the development and application of ML and AI tools.

## 1.2 Moving beyond the State of the Art

Current analytical toolkits for fNIRS and DOT—such as HOMER2/3 [24], AtlasViewer [25], Brain AnalyzIR [26], NeuroDOT [27], and NIRStorm [28] – have mastered conventional signal processing of data from structured paradigms. While powerful, these platforms are predominantly MATLAB-based and offer limited capabilities for multimodal neuroimaging with advanced ML. In recent years, there has been a clear trend towards Python-based development for ML applications [29], [30], [31]. Python is open source and presents unique advantages for toolbox extensibility and integration with multimodal and ML-driven analytical workflows. Cedalion capitalizes on the Python ecosystem to provide (see also **Figure 1**):

- **Direct ML integration** with widely used libraries such as scikit-learn and PyTorch, enabling

cutting-edge fNIRS/DOT processing and Machine Learning within a unified pipeline.

- **Compliance with standards such as SNIRF** [32] **and BIDS** [33], ensuring compatibility and interoperability with other neuroimaging modalities and community tools, including the existing fNIRS/DOT MATLAB toolboxes.

- **Direct integration with established toolkits for neuroimaging and physiological signals**, including MNE-Python (EEG/MEG) [31], NeuroKit2 (EDA, ECG, EMG) [34], and Nipype (f/MRI workflows) [35], facilitating seamless multimodal analysis.

- **Robust, intuitive data structures** built on Xarray and Pandas, using multidimensional labeled arrays with physical units for easy data handling and minimizing indexing errors.

- **Advanced visualization** for direct inspection of results in brain- or parcel-space, improving (neuro)physiological interpretability of high-dimensional data.

- **Integration of Jupyter Notebooks** for combining code, documentation, and results in a single interactive environment. Hosted execution environments (e.g., Google Colab) allow entire analyses to run in the cloud without local installation to lower barriers to adoption, and to greatly improve reproducibility of results by the scientific community.

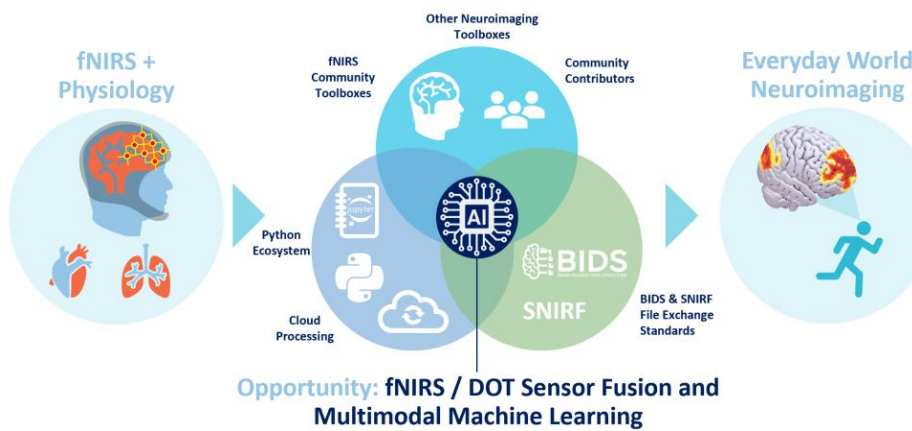

**Figure 1**: Key architectural considerations and motivation of the Cedalion toolbox. Cedalion brings together community contributions, file exchange standards, HPC and cloud processing within a python framework for advanced data-driven multimodal fNIRS / DOT analysis towards neuroimaging in the everyday world.

Cedalion combines these general architectural considerations with established core fNIRS/DOT functionality from Homer3/Atlas Viewer, and significantly expanded features:

1) **Anatomical head models and image reconstruction:** Support for individual anatomies and template-based head models, functional atlases/parcellations, accurate photogrammetric sensor co-registration, forward-model photon simulation, and several variants of regularized DOT image reconstruction (see Sections 2.1, 2.2 and 2.5).

2) **Signal processing and GLM analysis**: State-of-the-art preprocessing methods for fNIRS and DOT data, including channel quality metrics, artifact rejection, filtering, detrending, resampling, modified Beer-Lambert law (mBBL) conversion and fully customizable General Linear Model (GLM) analysis (see Sections 2.3 and 2.4).

3) **ML and data-driven analysis:** Functionality for multimodal data fusion, signal decomposition, and single-trial decoding. Containerized execution on scientific computing clusters for performance-intense model training and inference. Realistic data augmentation with synthetic HRFs and motion artifacts for method validation and model training (see Sections 2.6 and **Error! Reference source not found.**).

Cedalion is developed with a **community-centric philosophy** that encourages broad participation and ensures contributor recognition by clear attribution mechanisms. Code contributions are credited in the function docstrings and on the documentation website. Implemented methods are automatically linked to their source publications in a searchable bibliography— making it easy for researchers to properly cite all methods that they employed for their analyses with the toolbox, instead of citing only the toolbox itself. This way, Cedalion aims to work as a shared framework that amplifies both visibility and accessibility of individual contributions.

Reproducibility and quality are maintained via version control and continuous integration pipelines that automatically test code for breaking changes and validate documentation builds. The toolbox is designed with a strong focus on reproducibility, open science principles and reusability via a permissive MIT license. It comes with comprehensive documentation in form of autogenerated API descriptions, detailed examples in form of Jupyter notebooks and downloadable example datasets. To **maximize transparency and replicability as well as to lower the barrier to adoption**, all examples and any user-published notebooks using Cedalion can be fully executed in the cloud (e.g., via Google Colab), enabling anyone to reproduce analyses without local setup. This provides the research community with a tool for publishing papers alongside fully replicable analysis pipelines run on the original data by anyone.

In this manuscript we introduce the Cedalion toolbox and provide a comprehensive tutorial with seven detailed supplementary Jupyter notebook examples. Cedalion is freely available via www.cedalion.tools. We invite the community to contribute to the project and to make it a lever for the convergence of wearable fNIRS/DOT and multimodal machine learning, towards everyday world neuroscience and neurotechnology.

## 2 Toolbox

This section introduces the Cedalion toolbox, its architecture, data structures, and core functional packages and modules, organized into eight subsections. Each subsection is accompanied by a comprehensive tutorial that explains key elements and design choices through rendered Jupyter notebook examples (see Appendix S: Supplemental Material). These notebooks can be explored in a static, rendered form, or executed interactively either locally, on a computing cluster, or in the cloud [DOI-REGISTERED CODE OCEAN LINK PROVIDED HERE UPON PUBLICATION]. By providing both code, automatic example-data fetching and visualizations, the notebooks support hands-on testing and understanding of the toolbox. **Figure 2** offers a graphical overview, while full API documentation and additional example notebooks are available at www.cedalion.tools/docs.

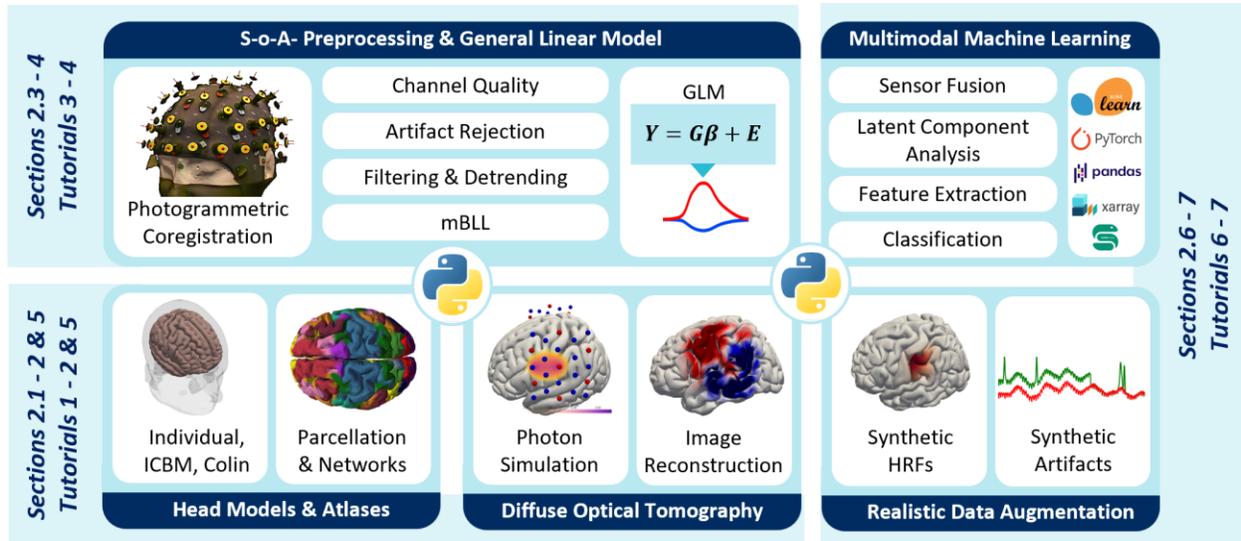

**Figure 2:** Toolbox overview and graphical guide to sections and tutorials.

## 2.0 Architecture and Data Structures

The toolbox primarily processes sampled multivariate time series, i.e. data with multiple values per time point. Examples include wavelength-dependent NIRS measurements, chromophore concentrations across brain regions or 3D accelerometer data.

Such data fit naturally into NumPy's multidimensional arrays [36], which allow organizing data along dimensions such as time, channel, or signal component (e.g., wavelength, chromophore). Combined with array programming techniques like indexing, broadcasting, and reductions, this layout provides powerful means to manipulate and analyze time series data.

Each array position is described by additional metadata, such as dimension names, timestamps for samples, channel labels ("S1D1"), optode labels ("S1", "D3"), and chromophores. These act as coordinates and can be queried for data selection (e.g., extract HbO concentration changes in channel "S1D1" for the first 10 seconds). The Xarray Python package [37] provides labeled multidimensional arrays called `DataArray` that support coordinate-based indexing and keep coordinates aligned with the time series data during shape-changing array operations. Cedalion stores time series and corresponding metadata in Xarray's data structures as shown in **Figure 3**.

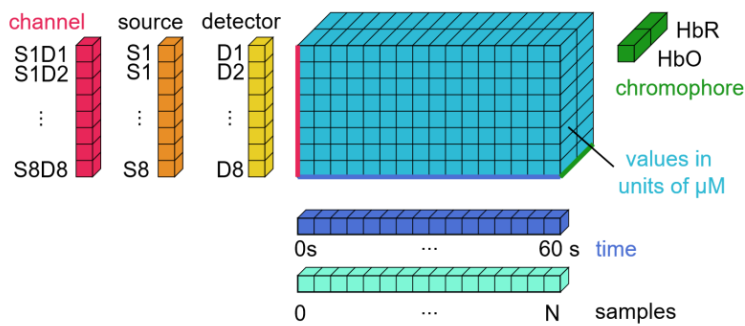

**Figure 3:** Visualization of array layout of multidimensional time series with corresponding coordinates and units. Example has three dimensions with multiple coordinate axes per dimension (i.e. time and sample coordinates across time dimension). Values carry physical units (i.e. μM - micro-Molar) that allow automatic consistency checks and conversion.

Compared to position-based indexing, label-based indexing improves the readability of the code. In the following example `x_numpy` and `x_xarray` denote two arrays storing amplitude time series. Time varies along the second dimension, and each line calculates the time-averaged amplitude:

```
avg_amplitude = x_numpy.mean(1)
avg_amplitude = x_xarray.mean("time")
```

`DataArrays` implement the array interface protocol, ensuring compatibility with NumPy functions and other libraries that accept NumPy arrays. Common ML libraries (e.g., scikit-learn, PyTorch) operate on multidimensional array inputs, making integration straightforward.

The multidimensional layout with labeled dimensions aligns well with typical fNIRS transformations that modify some dimensions while leaving others intact. For instance, applying the Beer-Lambert law replaces the wavelength dimension with the chromophore, while image reconstruction swaps the channel dimension for a voxel, vertex, or parcel dimension. Dimensions and coordinates use canonical names such as *time*, *channel*, and *chromo*, while allowing custom labels when practical. Additional flexibility comes from making toolbox functions dimension-agnostic. For instance, a bandpass filter operating along the time axis works equally well on a single-channel amplitude series or a multi-channel, multi-chromophore concentration series.

Cedalion also employs labeled multidimensional arrays for non-time series data. For example, digitized 3D optode positions are stored alongside coordinates that specify optode labels and types. Information about different coordinate reference systems is encoded in dimension names to prevent errors related to mismatched reference systems (see **S1 – Tutorial Notebook 1: Head Models and Forward Modeling**). Additionally, Xarray's support for physical units helps avoid common mistakes due to unit inconsistencies.

In the example below the amplitudes array contains an amplitude time series and the array `geo3d` stores optode coordinates in centimeters. The function `cedalion.nirs.channel_distances` infers from `amplitudes` which optode pairs form channels and uses their coordinates from the `geo3D` array to calculate their distances also in centimeters. Element-wise comparisons with a quantity in different units (here millimeters) produce the boolean array *is_short*, which is then used to select only short-distance channels from the amplitudes time series. Unit conversion (mm $\leftrightarrow$ cm) is automatically taken care of.

```
distances = cedalion.nirs.channel_distances(amplitudes, geo3d)
is_short = distances <= 15 * cedalion.units.mm
short_channel_amplitudes = amplitudes.sel(channel=is_short)
```

In Cedalion, different categories of data (e.g., time series and coordinates) are represented as labeled multidimensional arrays, making them indistinguishable to Python's type system. To manage this, Cedalion uses type annotations and aliases to define array schemas that specify expected dimension names and coordinates and validates these schemas at runtime.

Xarray allows attaching accessor objects to extend the `DataArray` type with category-specific functionality. Cedalion uses this feature selectively: for example, the `.points` accessor provides methods such as `geo3d.points.apply_transform` that applies transformations between different coordinate systems. However, the overall architecture of the toolbox primarily follows a functional design. Most functionality is implemented as standalone functions organized into topical subpackages (see Tables 2-6). These functions take `DataArray` objects as input, with type hints indicating the category of data expected. Parameters that carry physical units must be quantified (i.e. include explicit units) to avoid ambiguities. Comprehensive docstrings accompany all functions to document their expected inputs, outputs, behavior and scientific references, and form one pillar of the toolbox's documentation.

Processing and analyzing fNIRS data produce multiple analysis objects from different categories. To manage these, Cedalion provides the `Recording` container, which carries time series and related objects through the program in ordered dictionaries much like a segmented tray carrying items in separate compartments. The structure of the Recording container closely mirrors the SNIRF file format [32]. When reading a SNIRF file, Cedalion populates a `Recording` instance with data in the following attributes: `.timeseries`, `.geo3d`, `.stim`, `.aux_ts`, and `.meta_data`. Conversely, `cedalion.io.write_snirf` exports a `Recording` container back into a SNIRF file. Continuous wave (CW), frequency domain (FD) and time domain (TD) fNIRS data are all supported according to SNIRF specification. **Table 1** provides an overview of common array layouts and their corresponding fields in the `Recording` container.

Table 1: Common array layouts and coordinates used in Cedalion. For brevity not all coordinates are always shown. Note that multiple coordinates per dimension are possible. Highlighted rows expand the snirf specification.

| Data Type | Layout | Coordinates | Units e.g. |
|---|---|---|---|
| raw amplitudes or optical densities → rec["amp"] → rec["od"] | time x channel x wavelength | time: [0s, 0.1s, 0.2s,...] channel: [S1D1, S1D2, …] source (dim: channel): [S1, S2,...] detector (dim: channel): [D1, D2,...] wavelength: [760nm, 850nm] | "V" or "unitless" |
| concentration changes in channel-space → rec["conc"] | time x channel x chromo | time: [0s, 0.1s, 0.2s,...] channel: [S1D1, S1D2, …] chromo: [HbO, HbR] | "µM" |
| stimuli → rec.stim | pandas.DataFrame | | |
| coordinates → rec.geo3D | label x 3D pos | label: [S1, D1, Nz, Cz,..] | "mm" |
| accelerometer → rec.aux_ts["accel"] | time x axis | time: [-10s, …, 0s, …, 30s] | "V" or "m/s²" |
| concentration changes in image-space → rec["img"] | time x (vertex\|voxel) x chromo | time: [0s, 0.1s, 0.2s,...] vertex: [0,1,2,3,...] chromo: [HbO, HbR] | "µM" |
| concentration changes in parcel space → rec["parcel"] | time x parcel x chromo | time: [0s, 0.1s, 0.2s,...] parcel: [Somatomotor A, Visual B, …] chromo: [HbO, HbR] | "µM" |
| epoched concentration changes → rec["epochs"] | trial x reltime x (channel\|vertex\|parcel) x chromo | reltime: [-10s, …, 0s, …, 30s] channel: [S1D1, S1D2, …] chromo: [HbO, HbR] trial: [FingerTapping/Left, FingerTapping/Right, …] subject (dim: trial): [S001, S002] | "µM" |
| boolean masks, e.g. SNR or motion → rec.masks["snr"] | | | "unitless" (boolean) |
| headmodel →rec.headmodel | cedalion.dot.TwoSurfaceHeadModel | | |

By centering its data structures around the SNIRF and BIDS standards as common interfaces, Cedalion maintains compatibility with other toolboxes and platforms across different stages of channel-space analysis. Users can process their data in a modular way, importing and exporting to SNIRF files in between steps, if needed. However, SNIRF was primarily designed for raw fNIRS time-series data in channel space with only a few provisions for derived data. Certain fields in `Recording`, such as `.head_model`, `.masks`, `._aux_obj` or time series in image space, have no equivalent in the SNIRF specification. We are working with the SNIRF maintainers and the SfNIRS standardization committee on solutions to improve standardized exchange of processed data in the future.

**Table 2** shows a general overview of data-structure and I/O-related packages in Cedalion. The `cedalion.dataclasses` package contains classes and functions for creating time series and geometric data. The `cedalion.data` package provides seamless access to example datasets for fNIRS and HD-DOT, along with head models and precomputed forward model results. These datasets are hosted online. They are versioned, get cached after download and updated when the dataset changes. To the user, they are loadable with a single command, such as:

```
rec = cedalion.data.get_fingertappingDOT()
```

Simple access to the required data lets users easily run example notebooks locally or in the cloud and quickly experiment with new implementations or analyses.

**Table 2:** Core Functions for Data Structures and I/O. Not all subpackages are listed for brevity.

| Path | Description | Ref. |
|---|---|---|
| cedalion | Top-level toolbox | |
| └─.**dataclasses** | Data classes used throughout Cedalion. | |
| └─.**typing** | Type aliases, in particular for arrays with data schemas. | |
| └─.**physunits** | Physical units, built on pint_xarray's unit registry. | |
| └─.**xrutils** | Utility functions for Xarray objects. | |
| └─.**data** | Cedalion data, example datasets and utility functions. | |
| └─.**io** | **Modules for data I/O** | |
|    └─.snirf | Reading and writing SNIRF files | [32] |
|    └─.bids | Reading BIDS data. | [33] |
|    └─.anatomy | Reading and processing anatomical data. | |
|    └─.forward_model | Reading and writing forward model computation results | |
|    └─.photogrammetry | Reading photogrammetry output file formats. | |
|    └─.probe_geometry | Reading and writing probe geometry files | |

**Pipelines.** There is substantial diversity in fNIRS-analysis pipelines [38]. With increasing complexity of a project, an analysis may involve multiple processing steps, draw on methods from several toolboxes in different scripting languages, and require substantial computing resources. One of our development goals is to help users in constructing such heterogeneous processing graphs. To accomplish this, we are integrating with Snakemake [39], an established workflow-management system. Snakemake provides means to specify complex workflows as directed acyclic graphs of dependent processing steps. Specified in this way, the complete processing pipeline of an analysis can be executed from a single command. Because each processing step may have distinct runtime requirements, it is practical to run steps as individual processes and exchange data through intermediate, preferably standardized, files. Snakemake offers dependency and runtime-environment management and supports workflow execution across local machines as well as computing clusters (see **Figure 4**).

For Cedalion, this integration entails bundling existing functionality into larger, reusable pipeline building blocks, which encapsulate complexity where possible while still providing the functional diversity users need. Relevant parameters can be modified by the user via text-based YAML configuration files. For example, in a preprocessing block, users need flexibility in selecting which artifact-correction methods to apply, in what orders, and with which parameter settings. By encapsulating complexity and offering a configuration-file interface, that is simpler than Python code, these building blocks make it easy to express and adjust such choices for otherwise fixed pipelines and for users who are new to Python and its ecosystem. At the same time, advanced users can leverage Snakemake's flexibility to define workflows that combine these building blocks with other tools, enabling them to design the complex processing schemes their projects require.

Finally, we believe that making pipelines easy to package and publish will lower the barrier for researchers to share complete analysis workflows, making it more likely that pipelines are reproduced, adapted and extended by others. This, in turn, will help improve overall reproducibility and transparency in fNIRS research.

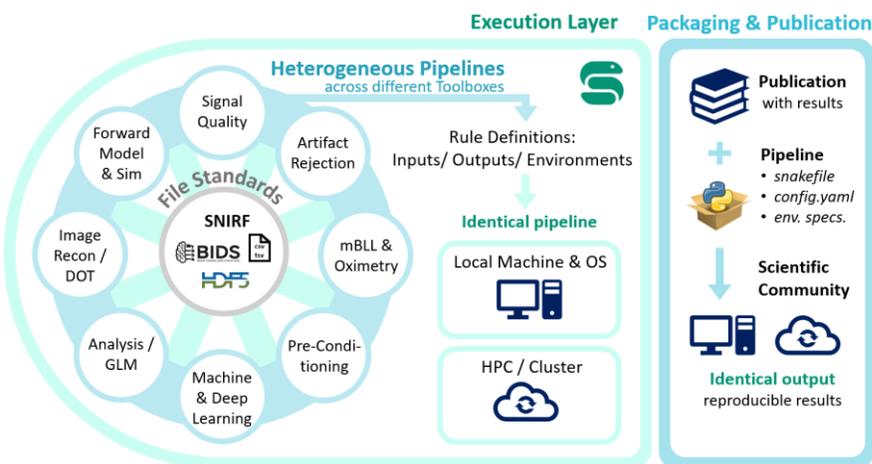

**Figure 4:** Pipelines in Cedalion. Through integration of environment specification with Snakemake and SNIRF / BIDS standardized file-exchange, heterogeneous processing steps across community toolboxes can be executed identically on local machines and clusters via rule-based pipeline configuration and definition (snakefile, config.yaml).

## 2.1 Head Models and Forward Modelling

### Overview

Cedalion incorporates and expands established methods from AtlasViewer [25] for head model generation, forward modelling and image reconstruction / diffuse optical tomography. The main interface to this functionality is provided by classes in the `cedalion.dot` package, whereas assisting geometric calculations are provided in `cedalion.geometry`.

The `cedalion.dot.TwoSurfaceHeadmodel` class supports the handling of segmented MRI head scans, the construction of triangulated brain and scalp surfaces, the derivation of voxel-vertex mappings as well as coordinate transformations between voxel and scanner spaces.

Cedalion can process individual MRIs if available, but also provides head anatomies based on the ICBM-152 [40] and Colin27 [41] atlases, including a brain parcellation scheme based on the Schaefer atlas [42] (see **Figure 5**).

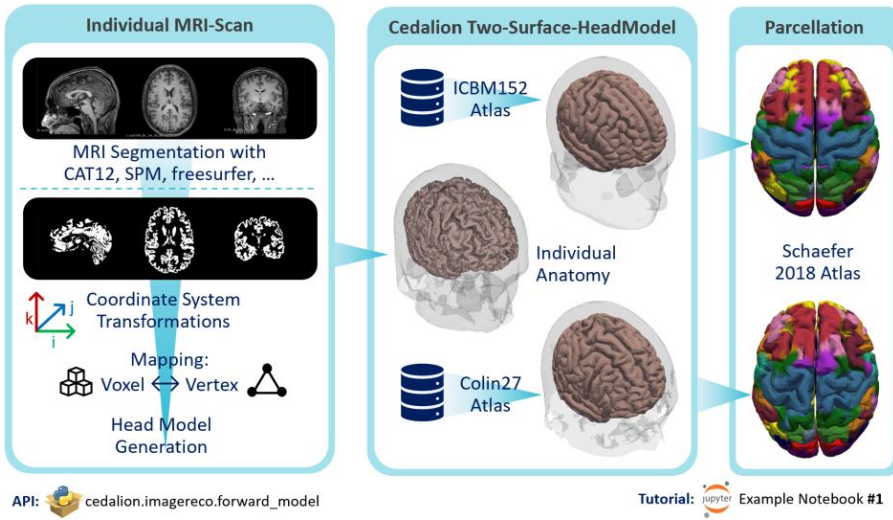

**Figure 5:** Head models in Cedalion. Segmentation masks from individual MRI scans as well as existing Atlases (ICBM152 and Colin27) can be used for (two surface) head-model generation, coordinate transformation, voxel-vertex mapping and parcellation in Cedalion.

Based on common fiducial points, the head model allows the registration of labeled points (i.e. existing probe geometries that comprise optodes, electrodes or landmarks) onto the scalp surface. The reduction of the voxelized head anatomy to two surfaces simplifies the inverse problem of image reconstruction by reducing the number of unknowns and limiting reconstructed absorption changes to areas where hemodynamic changes are expected and measurable (see Section 2.5 for more details).

The `cedalion.dot.ForwardModel` class provides methods for modelling photon migration in the voxelized and tissue-segmented head model both based on Monte-Carlo simulations with MCX / MCX-CL [43] and the finite-element method based on NIRFASTer [44]. From the calculated photon fluences it can then compute the sensitivity matrix $A$ for forward modeling that maps absorption or corresponding concentration changes $\Delta C$ in brain/scalp to optical signal changes in channel space $Y$ (see **Figure 6**).

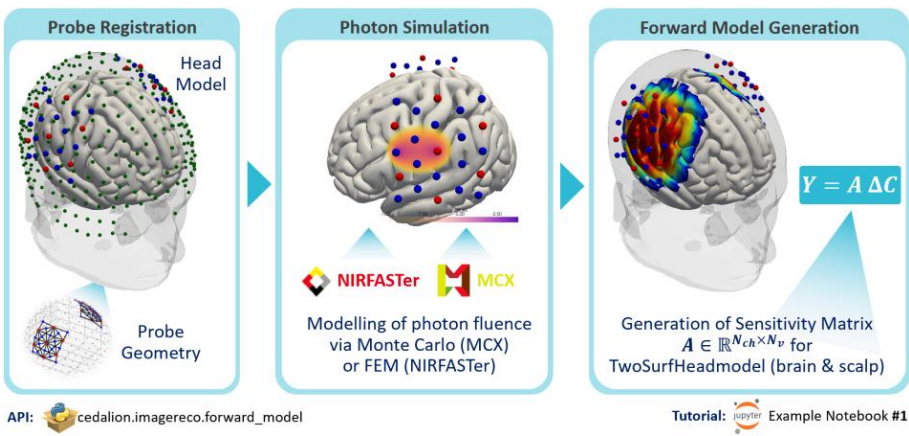

**Figure 6:** Forward Modeling in Cedalion. Integration Cedalion's head models and SNIRF probe geometries with MCX/MCX-CL and NIRFASTer enable both Monte-Carlo and FEM-based photon propagation modeling for Forward Model generation.

**Table 3** provides an overview of relevant packages and modules for head and forward model.

**Table 3:** Core Functions for Head Models and Forward Modelling. Not all subpackages are shown for brevity.

| Path | Description | Ref. |
|---|---|---|
| cedalion | Top-level toolbox | |
| └─**.geometry** | **Modules for geometric calculations.** | |
|   ├─.landmarks | Constructing the 10-10-system on the scalp surface. | [45] |
|   ├─.registration | Registration of optodes to scalp surfaces. | |
|   ├─.photogrammetry | Modules for photogrammetric sensor registration. | |
|   └─.segmentation | Functionality to work with segmented MRI scans. | |
| └─**.dot** | **Modules for DOT image reconstruction** | |
|   ├─.TwoSurfaceHeadModel | Class to represent a segmented head. | [25] |
|   ├─.get_standard_headmodel() | Access to Colin27 and ICBM-152 models. | [41], [46] |
|   ├─.ForwardModel | Class for simulating light transport in tissue. | [43], [44] |
|   └─.tissue_properties | Optical properties for light transport simulation. | [25] |

*Tutorial 1*

Tutorial notebook 1 (**S1 – Tutorial Notebook 1: Head Models and Forward Modeling**) introduces how Cedalion handles head models and forward modelling for diffuse optical tomography. At the beginning, configuration variables allow the user to choose between processing standard or custom head anatomies, the photon-propagation method, and whether to use precomputed forward-model results. Optode coordinates from an example finger-tapping dataset get loaded, and the notebook discusses how 3D probe geometry and channel definitions are represented in `xarray.DataArrays`.

When handling geometric data, Cedalion strictly tracks coordinate reference systems to avoid accidental confusion. For segmented MRI scans and probe geometries, the user must distinguish between the 'digitized' optode space, voxel space ('ijk'), and MRI scanner space ('ras'), as well as the affine transformations that translate between them.

Afterwards, different ways of constructing a `TwoSurfaceHeadModel` object, either from segmentations of the Colin27 or ICBM152 atlases or from a custom MRI scan, are shown. In the former case, Cedalion provides the function `cedalion.dot.get_standard_headmodel`. In the latter, the user can either choose Cedalion's built-in functions for deriving and processing brain and scalp surfaces (`TwoSurfaceHeadModel.from_segmentation`) or load high-resolution cortical surfaces produced by tools such as Freesurfer or CAT12 (`TwoSurfaceHeadModel.from_surfaces`).

The notebook demonstrates several strategies for registering an existing probe geometry to a head model. The user can scale the head model to match measurements of the subject's head or digitized landmarks. Alternatively, the probe geometry can be transformed and snapped onto the scalp surface of the unscaled head model, adjusting channel distances accordingly.

Finally, the notebook builds a `cedalion.dot.ForwardModel` object. For computational efficiency, the forward model is calculated in voxel space. Fluence distributions and the sensitivity matrix get computed. A 3D visualization of the resulting sensitivity matrix concludes the first tutorial.

## 2.2 Photogrammetric Optode Co-Registration

### Overview

The `cedalion.geometry` package also incorporates functionality for processing photogrammetric scans from a 3D scanner or smartphone for automatic optode digitization, labeling, and co-registration (see **Figure 7**). This method has increasingly been established as a replacement for conventional manual or electromagnetic-field-based optode registration techniques. Photogrammetry provides accurate 3D positions of optodes on the subject's scalp, which increases the spatial accuracy of fNIRS and DOT and reduces inter-subject variance [47], [48], [49].

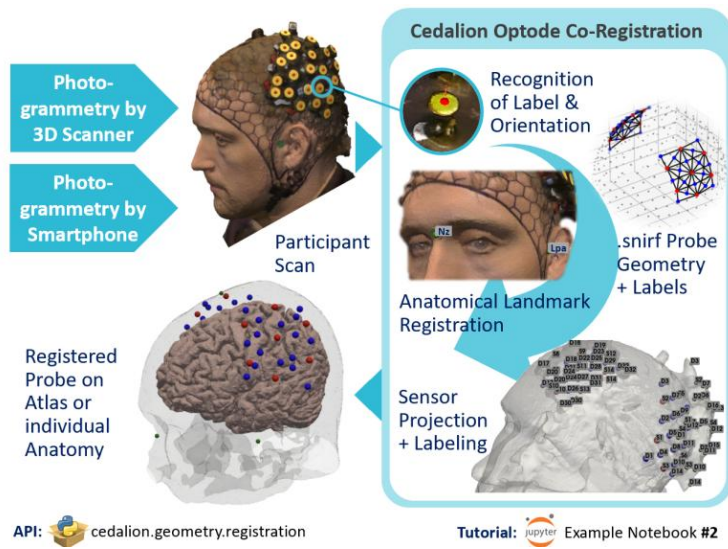

**Figure 7**: Photogrammetric Optode Co-Registration in Cedalion. Photogrammetric scans of visually labeled optodes, either from a smartphone or a 3D scanner, can be automatically processed for optode detection, registration of orientation and position in 3D space relative to anatomical landmarks, and mapping pre-designed probe geometries with Source and Detector labels from SNIRF files. Enables precise co-registration of individual probe positions onto atlases or individual head geometries to reduce localization errors to anatomical brain regions within and across subjects.

### Tutorial 2

Tutorial notebook 2 (**S2 – Tutorial Notebook 2: Photogrammetric Optode Co-Registration**) demonstrates a workflow for deriving optode coordinates from a photogrammetric head scan and mapping them onto a predefined montage. Circular colored stickers were attached to the spring tops of NIRx NIRSport2 optodes. A populated cap with such prepared optodes was scanned with a Structured-Light Photogrammetric 3D scanner (EINSTAR 3D Scanner, Shining 3D, Hangzhou, China), producing a textured triangle mesh. After loading the textured mesh, the pipeline identifies optode stickers by color and computes their surface normals. The sticker coordinates are shifted inward along the normal direction by the known optode length to find the points where the optodes touch the scalp surface. Finally, by comparing the found positions to known montage coordinates, optode labels are assigned.

The `ColoredStickerProcessor` class implements sticker detection by selecting vertices whose colors fall within configurable regions of the HSV color space. Furthermore, the class provides diagnostic visualizations to inspect and adjust the classification process. If automatic detection misses or misclassifies optodes, users can interactively add or remove points using the

`OptodeSelector` class.

Furthermore, the user must select the position of five anatomical landmarks (Nz, Iz, Cz, LPA, RPA) on the mesh. If these are difficult to identify, the user may identify three optodes instead.

Based on these fiducial points, the known montage is aligned. An iterative closest point (ICP) algorithm matches optode positions from the scan with those from the known montage, thereby assigning labels to the optodes.

### 2.3 Signal Processing

*Overview*

A great number of signal processing methods exist in the fNIRS community for signal quality assessment, artifact detection/correction and signal modeling. While the choice of methods and pipelines is diverse (see [38] for a recent large study investigating the choice of pipelines across 35 international teams), some approaches are increasingly adopted within the community as they have shown particular robustness and reliability across studies.

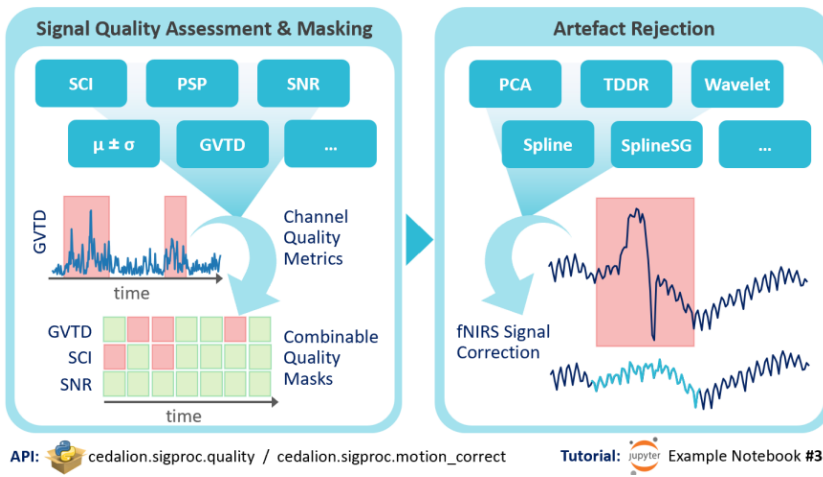

**Figure 8:** Quality assessment and artifact rejection. Established quality metrics such as SCI, PSP, SNR etc., can be used for the generation of multidimensional (Xarray-based) quality masks that can be flexibly combined or carried along. Widely used artefact rejection methods can clean data in an unsupervised way (e.g., TDDR, Wavelet, GVTD) or via supervision from quality masks (e.g. Spline, SplineSG, PCA, ...).

Expanding Homer2/3's functionality, Cedalion implements a range of these widely used methods (see **Figure 8** and **Table 4**) for:

1) Signal quality assessment in `cedalion.sigproc.quality`, including Global Variance of the Time Derivative (GVTD) [50], SNR, Scalp-Coupling Index and Peak Spectral Power [51],
2) Motion correction in `cedalion.sigproc.motion`, including Temporal Derivative Distribution Repair (TDDR) [52], Wavelet [53], spline [54], splineSG [55]. recursive PCA [25], and more),
3) General Linear Model analysis (see next subsection).

Standard functionality for frequency filtering, epoching or correcting data is complemented by a mechanism of masking bad segments in time series (such as timepoints or whole channels): Any quality or motion detection metric can be used to generate boolean masks with the same array dimensions as the analyzed time series. These can be logically combined and collapsed across metrics or dimensions (e.g., time or wavelength) and carried along through the analysis pipeline in `rec.masks`. The following Tutorial 3 provides more details on this concept.

**Table 4:** Core Functions for Signal Processing & Analysis

| Path | Description | Ref. |
|---|---|---|
| cedalion | Top-level toolbox | |
| ├─**.sigproc** | **Module for processing and analyzing fNIRS signals** | |
|   ├─**.nirs** | **Module for fNIRS conversion & preprocessing**<br>e.g. *beer_lambert(), int2od(), od2conc()* | [24] |
|   ├─**.quality** | **Module for signal quality metrics & channel pruning** | |
|     ├─.snr() | Signal-to-noise ratio for each channel. | [24] |
|     ├─.gvtd() | Global Variance of the Temporal Derivative | [50] |
|     ├─.id_motion() | Identify motion artifacts in an fNIRS input data array. | [24] |
|     ├─.sci() | Scalp-Coupling Index (SCI). | [51] |
|     ├─.psp() | Peak Spectral Power (PSP). | [51] |
|     ├─.prune_ch() | Prune channels from data using quality masks. | |
|     └─…. | … and more | |
|   ├─**.motion** | **Module for motion correction of fNIRS data.** | |
|     ├─.PCA_recurse() | … using recursive PCA method | [24] |
|     ├─.spline() | … using spline method | [54] |
|     ├─.splineSG() | … using splineSG method | [55] |
|     ├─.wavelet() | Wavelet-based motion correction | [53] |
|     ├─.tddr() | Temporal Derivative Distribution Repair-based correction. | [52] |
|     └─…. | … and more | |
|   ├─**.physio** | Methods for systemic physiology in fNIRS | |
|     ├─.ampd() | Automatic Multiscale Peak Detection | [56] |
|     └─…. | … and more | |
|   ├─**.epochs** | **Module for Epoching of time series based on events**<br>e.g, *to_epochs()* | |
|   ├─**.frequency** | **Module for Frequency-related signal processing**<br>*e.g., freq_filter(), sampling_rate()* | |
|   ├─**.time** | **Module for Time-related signal processing** | |
| ├─**.models.glm** | **Module with General Linear Models for fNIRS data.** | |
|   ├─.basis_functions | Temporal basis functions for the GLM, including Gamma, GammaDeriv, Dirac, Gaussian, and more | [57] |
|   ├─.design_matrix | Functions to create the design matrix for the GLM. | |
|   └─.solve | GLM solver (including OLS, AR-IRLS, …) | [24], [58] |
| ├─**.math.ar_model** | **Module for Autoregressive Modeling** | |
|   ├─.ar_filter() | Computes and applies an AR filter on a time series | |
|   └─…. | … and more | |

*Tutorial 3*

Tutorial notebook 3 (**S3 – Tutorial Notebook 3: Signal Processing**) walks through the workflow for loading, inspecting, and preprocessing a continuous-wave fNIRS recording of two motor tasks. It begins by examining the recording container in more detail, illustrating how it organizes fNIRS and auxiliary time series, probe geometry, and stimulus events.

Stimulus events are stored as a `pandas.DataFrame` to which additional functionality is attached through the custom accessor, `.cd`. The function `rec.stim.cd.rename_events` translates integer event codes into descriptive names. Users are free to choose any labels, but using a consistent naming scheme, such as "BallSqueezing/Left" and "BallSqueezing/Right", makes later selection easier, because related events can be retrieved by matching shared parts of the label (e.g. select labels starting with "BallSqueezing" in this case).

Next, the notebook computes two channel-wise signal-quality metrics: the Scalp Coupling Index (SCI) and the Peak Spectral Power ratio (PSP), both implemented in `cedalion.sigproc.quality`. These metrics are calculated in each channel over sliding, non-overlapping windows. For each channel and window, both the metric value and a Boolean indicating whether it surpasses a configurable threshold are computed. These boolean masks represent periods when the signal quality, based on a single criterion, is deemed insufficient. More advanced signal quality selection rules can be formulated by using logical combinations of these masks or by applying logical reduction operations across the array dimensions. A combined mask is generated, and the percentage of clean time per channel is calculated and plotted on the scalp to identify consistently low-quality channels.

The notebook then computes optical densities from raw amplitudes and applies two methods to correct for motion artifacts: Temporal Derivative Distribution Repair (TDDR) and wavelet-based correction. The notebook concludes by demonstrating the effectiveness of these methods using the Global Variance of Temporal Derivatives (GVTD) metric and by plotting individual time traces.

*2.4 Model-Driven (GLM) Analysis*

*Overview*

The General Linear Model (GLM) [59], [60] for fNIRS explains the time series (Y) for each channel and chromophore individually as a linear superposition of regressors, some representing the hemodynamic response and others addressing nuisance effects such as superficial or physiological components and signal drifts. The regressors form the design matrix (G) and the fit determines the unknown coefficients (β) for each regressor in the presence of unmodeled noise (E).

Regressors can vary across channels, such as when using the nearest short channel for short-channel regression techniques. Therefore, regressors are managed by the `DesignMatrix` class, which distinguishes between common and channel-wise regressors, building each channel's design matrix for the model fit. Multiple functions for creating `DesignMatrix` objects are available. Available nuisance regressors range from different polynomials, to Fourier (cosine) basis regressors, to short-channel regressors, to auxiliary data-driven physiology regressors, such as those derived via tCCA [61]. Hemodynamic response regressors can be generated from paradigm-encoded basis functions, for which Cedalion offers several options including Gamma functions (with or without derivatives) as well as Gaussian basis functions. By combining `DesignMatrix` objects, users can build models of any complexity needed to describe their data. To fit the GLM, the *statsmodels* library [62] is used, giving users numerous options for robust coefficient estimation, uncertainty assessment, and statistical testing. The AR-IRLS algorithm is also available to address serially correlated noise (see **Figure 9** and **Table 4).**

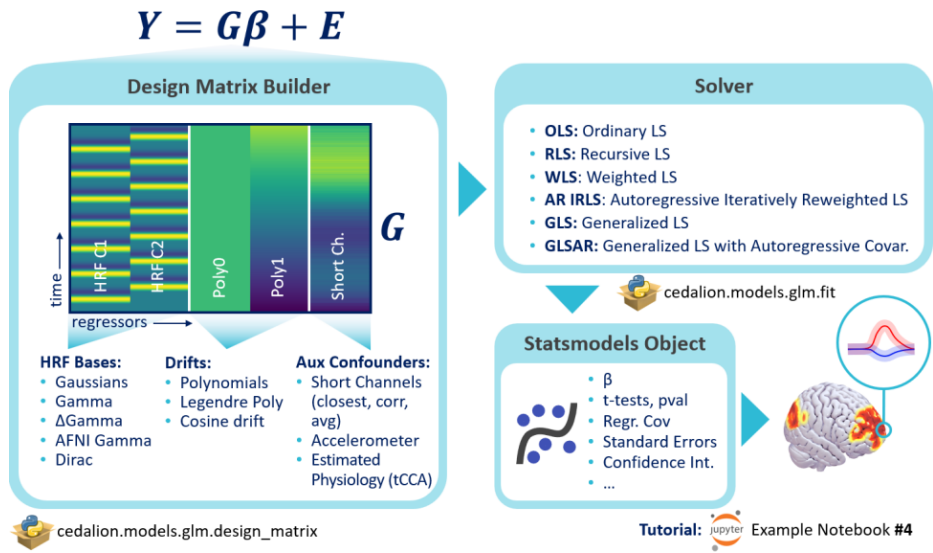

**Figure 9:** General Linear Model (GLM) in Cedalion. The Design Matrix Builder permits flexible combination of HRF kernels, drift and physiology regressors. A variety of solvers (from OLS to AR IRLS) can be used to estimate the fits, provided as a *statsmodels* object, leveraging the *statsmodels* Python package for advanced statistical analysis.

*Tutorial 4*

Tutorial notebook 4 (**S4 – Tutorial Notebook 4: Model-Driven (GLM) Analysis**) illustrates a workflow for modeling the motor task dataset using a GLM. Building on the preprocessing steps from the previous notebook, optical densities are converted to concentration changes via the modified Beer-Lambert law, and a low-pass filter removes the cardiac component. Source-detector distances are computed to separate long from short channels. Only long channels are modeled, while short channels are averaged to form a nuisance regressor. The hemodynamic response is modeled as a combination of Gaussian kernels, and the model is fitted using the AR-IRLS method. The notebook illustrates how string labels for regressors and coefficients help to identify and select specific components of the fitted model. After illustrating goodness-of-fit metrics, a hypothesis test for the HRF coefficients is performed, and channels with significant activations are plotted on the scalp. The notebook concludes by extracting and visualizing the HRF along with its uncertainty.

## 2.5 DOT - Image Reconstruction

*Overview*

Cedalion implements AtlasViewer's surface–based image reconstruction functionality for diffuse optical tomography (DOT) and extends it by incorporating additional regularization strategies and spatial bases [63]. DOT using a Two-Surface Head Model reduces the dimensionality of the image space from 100–500 k voxels to approximately 10–50 k surface vertices, thereby simplifying the inversion of the forward model. The approach maps volumetric tissue sensitivity to vertices of two representative surfaces—scalp and cortex—by attributing contributions from the scalp to the superficial surface, and by pooling contributions from gray and white matter to the cortical surface. This two-surface representation is a simplification justified by the depth-dependent photon sensitivity in diffuse optical tomography: the optical sensitivity decays exponentially with depth, and volumetric contributions within superficial and cortical compartments are highly correlated. The two-surface reduction is well matched to CW and typical FD HD-DOT settings with limited depth discrimination, whereas TD measurements can retain more depth information and may warrant time-resolved/volumetric reconstruction rather than a two-surface model. As shown by Boas et al. [64], the limited depth resolution of DOT causes partial-volume effects that can be mitigated by incorporating anatomical priors. The Two-Surface Head Model used in AtlasViewer [25] capitalizes on this property by effectively representing the main tissue classes that contribute to the measured signal while maintaining computational efficiency and anatomical interpretability. **Table 5** provides an overview of relevant packages and modules.

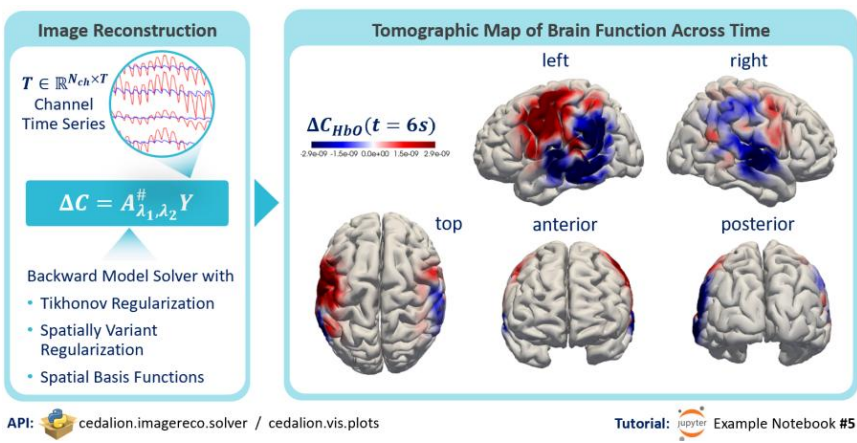

**Figure 10**: Image Reconstruction in Cedalion. Forward model inversion includes options for Tikhonov (measurement noise) regularization and Spatially Variant Regularization, and use of spatial basis functions such as Gaussian kernels. Results can be visualized in still images and as GIFs across time. Color bar unit is micro-Molar.

**Table 5:** Core Functions for DOT image reconstruction.

| Path | Description | Ref. |
|---|---|---|
| cedalion | Top-level toolbox | |
| └─.dot | **Modules for DOT image reconstruction** | |
|   ├─.TwoSurfaceHeadModel | Class to represents a segmented head. | [25] |
|   ├─.get_standard_headmodel() | Access to Colin27 and ICBM-152 models. | [41], [46] |
|   ├─.ForwardModel | Class for simulating light transport in tissue. | [43], [44] |
|   ├─.ImageRecon | Image reconstruction methods. | [25] |
|   ├─.GaussianSpatialBasisFunctions | Basis functions for regularization and dimensionality reduction | [63] |
|   ├─.tissue_properties | Optical properties for light transport simulation. | [25] |
|   └─.utils | Utility functions for image reconstruction. | |

*Tutorial 5*

Tutorial notebook 5 (**S5 – Tutorial Notebook 5: Image Reconstruction**) demonstrates Cedalion's image-reconstruction capabilities to explain measured optical density changes through hemoglobin concentration changes in the brain and scalp, thereby shifting the analysis from channel space to image space and then parcel space.

In the motor task dataset used in the previous tutorials, the finger tapping trials are epoched and block-averaged, demonstrating the flexibility of the time series data structures and yielding an HRF estimate to be reconstructed. The head model is set up, and a precomputed sensitivity matrix is loaded (computing the sensitivity is described in tutorial notebook 1).

The image reconstruction functionality is implemented in the class `cedalion.dot.ImageRecon`, which during construction takes the sensitivity matrix as well as measurement and spatial regularization parameters. With these options, the user can control the balance between image noise and spatial resolution and rescale sensitivities to suppress scalp-reconstructed activity [63]. Instead of solving the inverse problem for each vertex, activations can be modeled as a combination of Gaussian spatial basis functions defined on the brain and scalp surfaces by passing an instance of `cedalion.dot.GaussianSpatialBasisFunctions` to `ImageRecon`.

Furthermore, the user can choose how to reconstruct concentration changes from optical densities: either directly, by incorporating extinction coefficients into the sensitivity matrix, or indirectly, by first reconstructing absorption changes for each wavelength and then converting these to concentration changes. The indirect approach is preferred, as it reduces cross-talk between chromophores.

The function `ImageRecon.reconstruct` solves the inverse problem and several ways of visualizing the results are demonstrated. The transition from channel space to image space is reflected in the returned time series array, where the channel dimension is replaced by a vertex dimension. When using a standard head model, the vertex dimension has parcel coordinates which map each brain-surface vertex to its corresponding label in the Schaefer 2018 atlas with 17 networks and 600 regions of interest. Grouping vertices by shared labels makes it straightforward to average activations within each parcel, yielding time series array in which the vertex dimension is replaced by a parcel dimension. This demonstration of shifting the analysis to parcel space concludes the notebook.

## 2.6 Data-Driven (ML) Analysis

### Overview

Being built within the Python ecosystem, Cedalion is designed for integration of advanced fNIRS/DOT analysis with state-of-the-art machine learning and deep learning libraries such as scikit-learn [65] and PyTorch [66]. The Xarray-based data structures map directly onto the input and output tensors used by these libraries, ensuring seamless interoperability. This enables preprocessing and feature extraction within Cedalion before passing the prepared data to machine-learning libraries for model training and inference. Cedalion can extract a range of features from fNIRS data, including statistical features, GLM-derived coefficients, and frequency-domain features based on Fourier and wavelet transforms [67], [68].

Beyond classical ML pipelines, the Xarray representation integrates also into PyTorch-based deep learning workflows, where for example parcel/channel-informed normalization supports model development.

Looking ahead, the rapid emergence of deep-learning architectures for tasks such as artifact correction and latent-component analysis motivates the development of convenient APIs for deep-learning model inference in future Cedalion releases. For example, the community will be able to leverage deep latent-component models to obtain data-driven features for network analysis or classical ML pipelines, as well as generalized artifact-correction/detection models similar to [69].

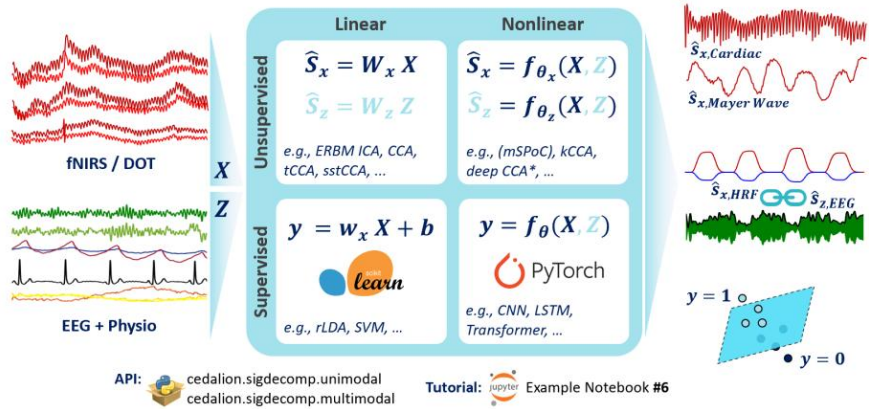

Beyond interfaces to established ML libraries, Cedalion contains dedicated methods for data-driven analysis of multivariate and multimodal fNIRS/DOT data (see **Figure 11** and **Table 6**). The package `cedalion.sigdecomp` contains several families of unimodal, and multimodal methods in the domain of sensor fusion-based signal decomposition for unsupervised fNIRS/DOT analysis, see [23] for a recent review. Among them are independent component analysis (ICA)-based methods such as EBM and ERBM ICA [70], [71] and their constrained variants, canonical correlation analysis (CCA) with a range of regularization options, priors, and temporal embedding, e.g., for tCCA GLM [61], and (multimodal) "source power comodulation" (SPoC) methods [72], [73] optimized for fNIRS-EEG band power analysis. See Tutorial 6 for more details.

**Table 6:** Core Functions for Multimodal Fusion, Data Driven Analysis and ML

| Path | Description | Ref. |
|---|---|---|
| cedalion | Top-level toolbox | |
| └─.sigdecomp | **Package for signal/source decomposition** | |
|  └─.unimodal | **Module for Unimodal Signal Decomposition** | |
|   └─.ICA_EBM() | ICA by Entropy Bound Minimization | [74] |
|   └─.ICA_ERBM() | ICA by Entropy Bound Rate Minimization (ICA-ERBM) | [70] |
|   └─.spoc | Source Power Co-modulation (SPoC) algorithm. | [73] |
|  └─.multimodal | **Module for Multimodal Signal Decomposition** | |
|   └─.cca | Canonical Correlation Analysis Variants (CCA, ssCCA, ElasticNetCCA, PLS) | [23], [75] |
|   └─.tcca | CCA using temporal embedding (tCCA, tssCCA, tElasticNetCCA), | [23], [75] |
|   └─.mspoc | Multimodal Source Power Co-Modulation Analysis | [72] |
| └─.mlutils | **Package with Machine Learning Utility Functions** | |
|  └─.cv | Cross-validation techniques | [76] |
|  └─.features | Functionality to extract features from fNIRS time series. | |

By interfacing these packages with DOT image reconstruction to a brain parcellation atlas, Cedalion enables machine learning from a structured image space based on shared coordinates across heterogeneous datasets from various subjects and studies [77], [78].

*Tutorial 6*

Tutorial notebook 6 (**S6 – Tutorial Notebook 6: Data-Driven (ML) Analysis**) first demonstrates how Cedalion's multimodal source-decomposition methods recover shared neural sources from simulated fNIRS–EEG data in `sim.datasets.synthetic_fnirs_eeg`. A bimodal toy dataset with controllable signal-to-noise ratio is generated, standardized, and split into train/test sets. The tutorial applies Canonical Correlation Analysis (CCA) to maximize correlation of latent components from EEG bandpower and fNIRS signals to enable, for instance, neurovascular coupling (NVC) analysis. It then demonstrates regularized variants - ElasticNet, Sparse, and Ridge CCA - and shows how hyperparameter choices affect performance. Temporally embedded CCA (tCCA) is then introduced to model realistic time lags caused by neurovascular coupling, followed by the multimodal Source Power Co-Modulation (mSPoC) algorithm, which computes EEG bandpower after inversion of the generative model to further improve latent component analysis performance. Correlations with known simulated sources and visual comparisons are used to evaluate how well each method recovers the underlying multimodal brain signals. Please see [23] for more details on methods and multimodal data simulation.

The second part illustrates Cedalion's implementation of Independent Component Analysis by Entropy Rate Bound Minimization (ICA-ERBM) for extracting physiological components from fNIRS recordings. After preprocessing a segment of the motor-task dataset, the `cedalion.sigdecomp.unimodal.ICA_ERBM` function decomposes 30 channels into the same number of independent sources. Ranking these sources by spectral power around 1 Hz and 0.1 Hz reveals cardiac and Mayer-wave components, demonstrating ICA-ERBM's ability to isolate meaningful physiological signals across different bands of the physiological power spectrum in an unsupervised way.

The third part demonstrates how Cedalion can be interfaced with scikit-learn to train a single-trial classifier. A cross-validation scheme is used and combined with short-channel regression to account for physiological noise [76]. Here, special care must be taken to ensure that the GLM is not fitted on any data from the respective test set. After subtracting nuisance components, the time series is split into epochs from which features are extracted. The notebook demonstrates how the layout of the data changes from Cedalion's representation of time series to samples expected by scikit-learn. In this conversion coordinate data is preserved which allows to trace each feature back to its origin. At the end of the notebook an LDA classifier is trained and output performance and feature importances are assessed.

*2.7  Data Augmentation*

  *Overview*

Development of conventional and data-driven processing methods often requires ground truth activation or artifacts for methods validation, or to increase the diversity of data for training of larger ML/DL models. The `cedalion.sim` simulation package (see **Table 7**) provides structured functionality to add an expandable range of synthetic motion artifacts and synthetic hemodynamic responses to real fNIRS/DOT data. Using the forward model, spatio-temporal hemodynamic response functions can be injected at user-defined brain coordinates and projected to channel space where the activation can then be added to real fNIRS recordings to test a method's reconstruction or classification performance with known ground truth. See Tutorial 7 and [78] for more details.

**Table 7:** Core Functions for Data Simulation and Augmentation.

| Path | Description | Ref. |
|---|---|---|
| cedalion | Top-level toolbox | |
| └ **.sim** | **Package for data augmentation with synthetic ground truth** | |
| ├ .synthetic_artifact | Functions for generating synthetic artifacts in fNIRS data. | [78] |
| ├ .synthetic_hrf | Functions for augmenting data with synthetic hemodyn. responses. | |
| └ .datasets | Functions for generating entirely synthetic datasets. | |

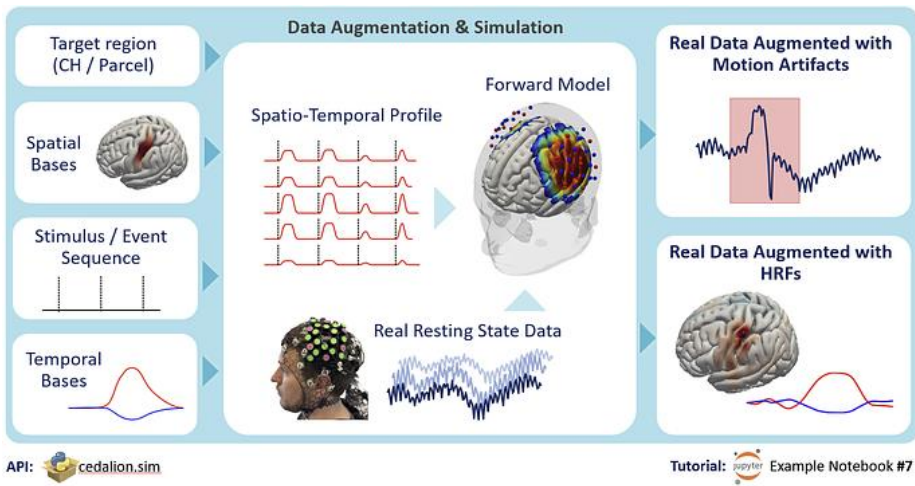

**Figure 12**: Modules for Synthetic Data-Augmentation. Realistic hemodynamic response functions on the cortex can be generated from customizable spatio-temporal bases and event sequences. Forward modeling enables projection to and addition of HRFs to real resting state background data in channel space.

Tutorial 7

Tutorial notebook 7 (**S7 – Tutorial Notebook 7: Data Augmentation**) explains how fNIRS datasets may be systematically augmented with synthetic activations or artifacts using pre-defined basis functions.

The package `cedalion.sim.synthetic_hrf` provides functionality for creating simulated hemodynamic responses. The notebook processes a resting state dataset recorded with a NinjaNIRS 2022 device covering the whole head from [63]. The Colin27 head model is set up and precomputed sensitivities are loaded. Using the `build_spatial_activation` function, concentration changes with a Gaussian spatial profile are generated. The scale and intensity parameters determine the concentration change at the center and how rapidly the activation decays with geodesic distance along the convoluted brain surface. Using the forward model results, the activation in image space is translated into channel space. The `build_stim_df` function generates stimulus events, with random onset times and durations. Combined with a chosen hemodynamic response function (using the same options available for the GLM), this defines the temporal profile of the activation. The produced time series contains for each channel the change in optical density caused by the synthetic activation. This synthetic component is then added to the resting-state dataset. By reconstructing the augmented time series and comparing the reconstruction result to the ground truth, the effectiveness of the image reconstruction can be evaluated.

To assess the impact of motion artifacts on an analysis or to evaluate the effectiveness of artifact-removal algorithms, functions from `cedalion.sim.synthetic_hrf` augment time series with simulated artifacts. Baseline-shift and spike artifacts are inserted at random time points and channels, after which the TDDR + wavelet artifact-removal procedure from Tutorial 3 is applied. Comparing the original, augmented, and cleaned time series illustrates both the effects of the simulated artifacts and the performance of the cleaning method.

## 3    Discussion and Outlook

The Cedalion toolbox is a community effort towards a solution that bridges model-based and data-driven analysis for functional near-infrared spectroscopy (fNIRS) and diffuse optical tomography (DOT), with a particular emphasis on FAIR (Findable, Accessible, Interoperable, Reusable) principles [79], multimodal integration, and machine learning (ML) readiness. By uniting core optical neuroimaging functionality with the flexibility of the Python ecosystem, Cedalion complements and extends existing MATLAB-based toolboxes such as *AtlasViewer, Brain AnalyzIR, Homer2/3, and NeuroDOT.* It addresses the need for interoperable, transparent, and scalable workflows capable of supporting both conventional neuroimaging analyses and emerging ML-based paradigms.

### 3.1  Advancing the fNIRS/DOT and Multimodal Ecosystem

Cedalion's architecture facilitates seamless integration with multimodal and physiological data sources, providing a coherent analysis framework for fNIRS and DOT together with EEG, MEG, OPM-MEG, and other complementary biosignals such as Blood Pressure, PPG, EDA, ECG or EMG. This interoperability opens avenues for combined analyses of neural and systemic signals that have historically been treated in isolation. Further, Cedalion's SNIRF-centered, modality-agnostic architecture is not limited to conventional fNIRS and DOT but can support plugin-style extensions to other optical neuroimaging methods that can be represented within the same data abstractions. As an example, a dedicated speckle contrast optical spectroscopy (SCOS) package builds on Cedalion's core infrastructure to support recent SCOS and high-density SCOS approaches [80], [81] that process and infer blood flow index time series, demonstrating how emerging modalities can be integrated into the core framework.

By adhering to community standards (SNIRF, BIDS) and incorporating cloud-executable environments (Jupyter/Colab), the toolbox directly responds to the challenges in reproducibility and transparency highlighted by the FRESH study [38], where analytical variability across fNIRS pipelines was shown to critically affect outcomes. Integrated helper functionality that can be run both locally, and on the cloud, like Cedalion's BIDS conversion tool for the creation of BIDS-compliant datasets from an arbitrary collection of snirf files, aims to help lower the barrier to adoption of data standards in the community. Cedalion promotes consistency and reproducibility through structured data models, containerized workflows, and open documentation—key enablers for cross-laboratory comparability and collective scientific advancement.

Key architectural design choices provide a natural interface with ML frameworks such as *scikit-*

*learn* and *PyTorch* to facilitate data-driven exploration of latent neural and physiological structures in the data. This way, the toolbox provides an infrastructure to translate methodological insights from the emerging fields of deep learning for fNIRS [20], [21] and multimodal fusion [23], [61], [82], [83], [84] into practical, community-accessible research tools.

## 3.2 Limitations and Critical Considerations

Cedalion's scope and architecture also come with limitations that merit transparent discussion. First, while Python facilitates open and reproducible science, it introduces a learning curve for researchers accustomed to MATLAB-based toolchains. To mitigate this, the toolbox documentation is primarily built on extensive tutorials based on Jupyter Notebooks that can be executed with only marginal Python proficiency.

Second, certain features of Cedalion's advanced functionality, particularly for photon simulation, DOT reconstruction, and ML training, may exceed the resources available to individuals or smaller labs. While this is a general challenge not specific to the toolbox, we provide mitigation strategies that offer alternative computational pathways. For instance, both GPU- and CPU-based methods are supported for calculating the DOT forward model using Monte Carlo (MCX) or FEM-based (NIRFASTer) photon transport simulations. If local infrastructure cannot accommodate computationally intensive operations (e.g., deep learning model training), Cedalion's containerized design enables seamless migration to scientific computing clusters or cloud-based environments.

Third, as machine learning becomes increasingly accessible for fNIRS analysis, essential methodological standards from computer science are easily overlooked. While Cedalion enables the implementation and development of advanced models and architectures, the responsibility for maintaining methodological rigor rests with the researcher. Achieving generalizable and interpretable data-driven solutions requires strict attention to model validation, prevention of overfitting, and preservation of interpretability.

Fourth, the diversity of devices, file formats, and acquisition paradigms remains a major obstacle for standardization. Although Cedalion supports SNIRF and BIDS, continuous alignment with evolving standards and active collaboration with community initiatives will be essential to achieve the envisioned full interoperability across modalities and toolboxes. One major remaining gap is the definition of a standard beyond the SNIRF specification to enable exchanging analysis results, including DOT image-space data, corresponding head-models, sensitivity profiles or voxel/vertex/parcel-mapping.

Finally, systematic benchmarking of community pipelines, including those built with Cedalion, against each other on standardized datasets remains an open and necessary effort, echoing the reproducibility initiatives exemplified by the FRESH consortium.

## 3.3 Outlook and Future Directions

With its third major release, Cedalion has reached a stable foundation for multimodal and ma-chine-learning–ready fNIRS and DOT analysis. The focus now shifts from consolidation to expan-sion. Looking forward, Cedalion aims to support the ongoing transformation of neuroimaging from fragmented analyses toward integrated, continuous, and interpretable workflows that span laboratory and real-world settings.

A key focus of upcoming development lies in pipeline usability. While the architectural foundations for modular processing chains are implemented, ease of use must approach that of established MATLAB toolboxes such as Homer, where complete pipelines can be executed from compact configuration files. Using Snakemake, Cedalion's next releases will extend this accessibility to both developers and end users by providing high-level interfaces, preconfigured template (YAML) files, and executable examples for common workflows.

Machine learning integration remains an area of active expansion. Current functionality emphasizes interfacing with established frameworks and includes decomposition methods such as ICA, CCA, and SPoC. Future work will focus on providing the means to integrate lightweight pretrained models for generalizable data-driven applications such as signal quality assessment, artifact rejection, or physiology-aware analysis.

Standardization and data exchange are central to Cedalion's long-term vision both to enable the training of more complex machine learning models with larger amounts of data, and to aid the enhancement of general reproducibility in fNIRS/DOT-based neuroscience. Building on its SNIRF- and BIDS-based data structures, ongoing work addresses the previously mentioned absence of standardized formats for analysis results and DOT reconstructions. These efforts align with community-driven standardization initiatives toward fully interoperable data and tool ecosystems.

Ultimately, Cedalion is conceived as a toolbox from the community for the community, standing on the shoulders of the many who built the foundations of optical neuroimaging. In Greek myth, Cedalion stood upon the shoulders of the blinded giant Orion to guide him toward the rising sun, where his sight was restored. Likewise, the toolbox exists to bundle and extend—not replace— the collective expertise on which it is built. Consequently, as its strength derives from that shared foundation, a major future focus will lie on continued education and integration of community contribution to aid the advancement of global fNIRS and DOT research.

### Author Contributions

E.M., D.A.B. and A.v.L. conceptualized the toolbox architecture, implemented, and validated core algorithms of the toolbox. L.C., S.M., T.C, T.F., J.C., S.K., J.B., T.D., and N.H., implemented and validated core algorithms of the toolbox. E.M. compiled the tutorial Jupyter notebooks. M.A.Y. and D.A.B. provided user feedback, validated algorithms and reviewed and edited the manuscript. A.v.L. and E.M. wrote the original draft, reviewed and edited the manuscript.

### Disclosures

The authors declare that there are no financial interests, commercial affiliations, or other potential conflicts of interest that could have influenced the objectivity of this research or the writing of this paper

### Code, Data, and Materials

The Cedalion toolbox presented in this article, alongside documentation, example data and Jupyter notebooks are publicly available on [www.cedalion.tools](www.cedalion.tools). The seven tutorials for this manuscript are available in rendered form in **Appendix S: Supplemental Material** and provided in a versioned executable container on [DOI-REGISTERED CODE OCEAN LINK PROVIDED HERE UPON PUBLICATION].

*Acknowledgments*

The authors would like to thank Masha Iudina, Sung Anh, Theodore Huppert, Qianqian Fang and Jiaming Ciao for their support, contributions and feedback during toolbox development. The IBS Lab under AvL gratefully acknowledges support from BMFTR BIFOLD25B and from the European Research Council (ERC) under the European Union's Horizon Europe research and innovation program (ERC Starting Grant, Grant Agreement No. 101163363). DAB gratefully acknowledges support from NIH U01-EB029856 and NIH UG3-EB034710. This paper was edited using feedback from ChatGPT-5.2 (OpenAI).

### References

[1]   S. Sonkusare, M. Breakspear, and C. Guo, "Naturalistic Stimuli in Neuroscience: Critically Acclaimed," *Trends Cogn. Sci.*, vol. 23, no. 8, pp. 699–714, Aug. 2019, doi: 10.1016/j.tics.2019.05.004.

[2]   A. von Lühmann *et al.*, "Toward Neuroscience of the Everyday World (NEW) using functional near-infrared spectroscopy," *Curr. Opin. Biomed. Eng.*, vol. 18, p. 100272, June 2021, doi: 10.1016/j.cobme.2021.100272.

[3]   M. A. Schmuckler, "What Is Ecological Validity? A Dimensional Analysis," *Infancy*, vol. 2, no. 4, pp. 419–436, Oct. 2001, doi: 10.1207/S15327078IN0204_02.

[4]   P. Pinti *et al.*, "The present and future use of functional near-infrared spectroscopy (fNIRS) for cognitive neuroscience," *Ann. N. Y. Acad. Sci.*, vol. 1464, no. 1, pp. 5–29, Mar. 2020, doi: 10.1111/nyas.13948.

[5]   J. L. Park, P. A. Dudchenko, and D. I. Donaldson, "Navigation in Real-World Environments: New Opportunities Afforded by Advances in Mobile Brain Imaging," *Front. Hum. Neurosci.*, vol. 12, p. 361, Sept. 2018, doi: 10.3389/fnhum.2018.00361.

[6]   J. Kiverstein and M. Miller, "The embodied brain: towards a radical embodied cognitive neuroscience," *Front. Hum. Neurosci.*, vol. 9, May 2015, doi: 10.3389/fnhum.2015.00237.

[7]   F. Scholkmann, I. Tachtsidis, M. Wolf, and U. Wolf, "Systemic physiology augmented functional near-infrared spectroscopy: a powerful approach to study the embodied human brain," *Neurophotonics*, vol. 9, no. 03, July 2022, doi: 10.1117/1.NPh.9.3.030801.

[8]   E. E. Vidal-Rosas, A. von Lühmann, P. Pinti, and R. J. Cooper, "Wearable, high-density fNIRS and diffuse optical tomography technologies: a perspective," *Neurophotonics*, vol. 10, no. 2, 2023, doi: 10.1117/1.NPh.10.2.023513.

[9]   H. Ayaz *et al.*, "Optical imaging and spectroscopy for the study of the human brain: status report," *Neurophotonics*, vol. 9, no. S2, Aug. 2022, doi: 10.1117/1.NPh.9.S2.S24001.

[10]  C. Habermehl *et al.*, "Somatosensory activation of two fingers can be discriminated with ultrahigh-density diffuse optical tomography," *NeuroImage*, vol. 59, no. 4, pp. 3201–3211, Feb. 2012, doi: 10.1016/j.neuroimage.2011.11.062.

[11] B. R. White and J. P. Culver, "Quantitative evaluation of high-density diffuse optical tomography: in vivo resolution and mapping performance," *J. Biomed. Opt.*, vol. 15, no. 02, p. 1, Mar. 2010, doi: 10.1117/1.3368999.

[12] A. T. Eggebrecht *et al.*, "Mapping distributed brain function and networks with diffuse optical tomography," *Nat. Photonics*, vol. 8, no. 6, pp. 448–454, June 2014, doi: 10.1038/nphoton.2014.107.

[13] S. L. Ferradal *et al.*, "Functional Imaging of the Developing Brain at the Bedside Using Diffuse Optical Tomography," *Cereb. Cortex*, vol. 26, no. 4, pp. 1558–1568, Apr. 2016, doi: 10.1093/cercor/bhu320.

[14] J. Tang, A. LeBel, S. Jain, and A. G. Huth, "Semantic reconstruction of continuous language from non-invasive brain recordings," *Nat. Neurosci.*, vol. 26, no. 5, pp. 858–866, May 2023, doi: 10.1038/s41593-023-01304-9.

[15] M. Fogarty *et al.*, "Functional brain mapping using whole-head very-high-density diffuse optical tomography," *Imaging Neurosci.*, June 2025, doi: 10.1162/IMAG.a.54.

[16] J. C. Pang *et al.*, "Geometric constraints on human brain function," *Nature*, vol. 618, no. 7965, pp. 566–574, June 2023, doi: 10.1038/s41586-023-06098-1.

[17] I. Tachtsidis and F. Scholkmann, "False positives and false negatives in functional near-infrared spectroscopy: issues, challenges, and the way forward," *Neurophotonics*, vol. 3, no. 3, p. 031405, Mar. 2016, doi: 10.1117/1.NPh.3.3.031405.

[18] T. J. Huppert, "Commentary on the statistical properties of noise and its implication on general linear models in functional near-infrared spectroscopy," *Neurophotonics*, vol. 3, no. 1, p. 010401, Mar. 2016, doi: 10.1117/1.NPh.3.1.010401.

[19] M. A. Yücel *et al.*, "Short separation regression improves statistical significance and better localizes the hemodynamic response obtained by near-infrared spectroscopy for tasks with differing autonomic responses," *Neurophotonics*, vol. 2, no. 3, p. 035005, Sept. 2015, doi: 10.1117/1.NPh.2.3.035005.

[20] C. Eastmond, A. Subedi, S. De, and X. Intes, "Deep learning in fNIRS: a review," *Neurophotonics*, vol. 9, no. 04, July 2022, doi: 10.1117/1.NPh.9.4.041411.

[21] T. Dissanayake, K.-R. Müller, and A. Von Lühmann, "Deep Learning from Diffuse Optical Oximetry Time-Series: An fNIRS-Focused Review of Recent Advancements and Future Directions," *IEEE Rev. Biomed. Eng.*, vol. 14, pp. 1–22, 2025, doi: 10.1109/RBME.2025.3617858.

[22] R. Li, D. Yang, F. Fang, K.-S. Hong, A. L. Reiss, and Y. Zhang, "Concurrent fNIRS and EEG for Brain Function Investigation: A Systematic, Methodology-Focused Review," *Sensors*, vol. 22, no. 15, p. 5865, Aug. 2022, doi: 10.3390/s22155865.

[23] T. Codina, B. Blankertz, and A. Von Lühmann, "Multimodal fNIRS-EEG Sensor Fusion: Review of Data-Driven Methods and Perspective for Naturalistic Brain Imaging," *Imaging Neurosci.*, vol. 3, no. 1, p. IMAG.a.974, 2025, doi: 10.1162/IMAG.a.974.

[24] T. J. Huppert, S. G. Diamond, M. A. Franceschini, and D. A. Boas, "HomER: a review of time-series analysis methods for near-infrared spectroscopy of the brain," *Appl. Opt.*, vol. 48, no. 10, pp. 280–298, 2009.

[25] C. M. Aasted *et al.*, "Anatomical guidance for functional near-infrared spectroscopy: AtlasViewer tutorial," *Neurophotonics*, vol. 2, no. 2, p. 020801, May 2015, doi: 10.1117/1.NPh.2.2.020801.

[26] H. Santosa, X. Zhai, F. Fishburn, and T. Huppert, "The NIRS Brain AnalyzIR Toolbox," *Algorithms*, vol. 11, no. 5, p. 73, May 2018, doi: 10.3390/a11050073.

[27] A. T. Eggebrecht and J. P. Culver, "NeuroDOT: an extensible Matlab toolbox for streamlined optical functional mapping," in *Diffuse Optical Spectroscopy and Imaging VII*, H. Dehghani and H. Wabnitz, Eds., Munich, Germany: SPIE, July 2019, p. 26. doi: 10.1117/12.2527164.

[28] É. Delaire *et al.*, "NIRSTORM: a Brainstorm extension dedicated to functional near-infrared spectroscopy data analysis, advanced 3D reconstructions, and optimal probe design," *Neurophotonics*, vol. 12, no. 02, May 2025, doi: 10.1117/1.NPh.12.2.025011.

[29] E. Speh, A. Segel, Y. Thacker, D. Marcus, M. D. Wheelock, and A. T. Eggebrecht, "NeuroDOTpy: A Python Neuroimaging Toolbox for DOT," in *Biophotonics Congress: Optics in the Life Sciences 2023 (OMA, NTM, BODA, OMP, BRAIN)*, Vancouver, British Columbia: Optica Publishing Group, 2023, p. JTu4B.24. doi: 10.1364/BODA.2023.JTu4B.24.

[30] J. Benerradi, J. Clos, A. Landowska, M. F. Valstar, and M. L. Wilson, "Benchmarking framework for machine learning classification from fNIRS data," *Front. Neuroergonomics*, vol. 4, p. 994969, Mar. 2023, doi: 10.3389/fnrgo.2023.994969.

[31] A. Gramfort *et al.*, "MNE software for processing MEG and EEG data," *NeuroImage*, vol. 86, pp. 446–460, Feb. 2014, doi: 10.1016/j.neuroimage.2013.10.027.

[32] S. Tucker *et al.*, "Introduction to the shared near infrared spectroscopy format," *Neurophotonics*, vol. 10, no. 01, 2023, doi: 10.1117/1.NPh.10.1.013507.

[33] R. Luke *et al.*, "NIRS-BIDS: Brain Imaging Data Structure Extended to Near-Infrared Spectroscopy," *Sci. Data*, vol. 12, no. 1, p. 159, Jan. 2025, doi: 10.1038/s41597-024-04136-9.

[34] D. Makowski *et al.*, "NeuroKit2: A Python toolbox for neurophysiological signal processing," *Behav. Res. Methods*, vol. 53, no. 4, pp. 1689–1696, Aug. 2021, doi: 10.3758/s13428-020-01516-y.

[35] K. Gorgolewski *et al.*, "Nipype: A Flexible, Lightweight and Extensible Neuroimaging Data Processing Framework in Python," *Front. Neuroinformatics*, vol. 5, 2011, doi: 10.3389/fninf.2011.00013.

[36] S. Van Der Walt, S. C. Colbert, and G. Varoquaux, "The NumPy Array: A Structure for Efficient Numerical Computation," *Comput. Sci. Eng.*, vol. 13, no. 2, pp. 22–30, Mar. 2011, doi: 10.1109/MCSE.2011.37.

[37] S. Hoyer and J. Hamman, "xarray: N-D labeled Arrays and Datasets in Python," *J. Open Res. Softw.*, vol. 5, no. 1, p. 10, Apr. 2017, doi: 10.5334/jors.148.

[38] M. A. Yücel *et al.*, "fNIRS reproducibility varies with data quality, analysis pipelines, and researcher experience," *Commun. Biol.*, vol. 8, no. 1, p. 1149, Aug. 2025, doi: 10.1038/s42003-025-08412-1.

[39] F. Mölder *et al.*, "Sustainable data analysis with Snakemake," *F1000Research*, vol. 10, p. 33, Jan. 2021, doi: 10.12688/f1000research.29032.1.

[40] V. Fonov, A. C. Evans, K. Botteron, C. R. Almli, R. C. McKinstry, and D. L. Collins, "Unbiased average age-appropriate atlases for pediatric studies," *NeuroImage*, vol. 54, no. 1, pp. 313–327, Jan. 2011, doi: 10.1016/j.neuroimage.2010.07.033.

[41] C. J. Holmes, R. Hoge, L. Collins, R. Woods, A. W. Toga, and A. C. Evans, "Enhancement of MR Images Using Registration for Signal Averaging," *J. Comput. Assist. Tomogr.*, vol. 22,

no. 2, 1998, [Online]. Available: https://journals.lww.com/jcat/fulltext/1998/03000/enhancement_of_mr_images_using_registration_for.32.aspx

[42] A. Schaefer *et al.*, "Local-Global Parcellation of the Human Cerebral Cortex from Intrinsic Functional Connectivity MRI," *Cereb. Cortex*, vol. 28, no. 9, pp. 3095–3114, Sept. 2018, doi: 10.1093/cercor/bhx179.

[43] Q. Fang and D. A. Boas, "Monte Carlo Simulation of Photon Migration in 3D Turbid Media Accelerated by Graphics Processing Units," *Opt. Express*, vol. 17, no. 22, p. 20178, Oct. 2009, doi: 10.1364/OE.17.020178.

[44] H. Dehghani *et al.*, "Near infrared optical tomography using NIRFAST: Algorithm for numerical model and image reconstruction," *Commun. Numer. Methods Eng.*, vol. 25, no. 6, pp. 711–732, June 2009, doi: 10.1002/cnm.1162.

[45] R. Oostenveld and P. Praamstra, "The five percent electrode system for high-resolution EEG and ERP measurements," *Clin. Neurophysiol.*, vol. 112, no. 4, pp. 713–719, Apr. 2001, doi: 10.1016/S1388-2457(00)00527-7.

[46] J. Mazziotta *et al.*, "A probabilistic atlas and reference system for the human brain: International Consortium for Brain Mapping (ICBM)," *Philos. Trans. R. Soc. Lond. B. Biol. Sci.*, vol. 356, no. 1412, pp. 1293–1322, Aug. 2001, doi: 10.1098/rstb.2001.0915.

[47] S. Homölle and R. Oostenveld, "Using a structured-light 3D scanner to improve EEG source modeling with more accurate electrode positions," *J. Neurosci. Methods*, vol. 326, p. 108378, Oct. 2019, doi: 10.1016/j.jneumeth.2019.108378.

[48] S. Srinivasan, D. Acharya, E. Butters, L. Collins-Jones, F. Mancini, and G. Bale, "Subject-specific information enhances spatial accuracy of high-density diffuse optical tomography," *Front. Neuroergonomics*, vol. 5, p. 1283290, Feb. 2024, doi: 10.3389/fnrgo.2024.1283290.

[49] I. Mazzonetto, M. Castellaro, R. J. Cooper, and S. Brigadoi, "Smartphone-based photogrammetry provides improved localization and registration of scalp-mounted neuroimaging sensors," *Sci. Rep.*, vol. 12, no. 1, p. 10862, June 2022, doi: 10.1038/s41598-022-14458-6.

[50] A. Sherafati *et al.*, "Global motion detection and censoring in high-density diffuse optical tomography," *Hum. Brain Mapp.*, vol. 41, no. 14, pp. 4093–4112, Oct. 2020, doi: 10.1002/hbm.25111.

[51] L. Pollonini, H. Bortfeld, and J. S. Oghalai, "PHOEBE: a method for real time mapping of optodes-scalp coupling in functional near-infrared spectroscopy," *Biomed. Opt. Express*, vol. 7, no. 12, p. 5104, Dec. 2016, doi: 10.1364/BOE.7.005104.

[52] F. A. Fishburn, R. S. Ludlum, C. J. Vaidya, and A. V. Medvedev, "Temporal Derivative Distribution Repair (TDDR): A motion correction method for fNIRS," *NeuroImage*, vol. 184, pp. 171–179, Jan. 2019, doi: 10.1016/j.neuroimage.2018.09.025.

[53] B. Molavi and G. A. Dumont, "Wavelet-based motion artifact removal for functional near-infrared spectroscopy," *Physiol. Meas.*, vol. 33, no. 2, p. 259, 2012.

[54] F. Scholkmann, S. Spichtig, T. Muehlemann, and M. Wolf, "How to detect and reduce movement artifacts in near-infrared imaging using moving standard deviation and spline interpolation," *Physiol. Meas.*, vol. 31, no. 5, pp. 649–662, May 2010, doi: 10.1088/0967-3334/31/5/004.

[55] S. Jahani, S. K. Setarehdan, D. A. Boas, and M. A. Yücel, "Motion artifact detection and correction in functional near-infrared spectroscopy: a new hybrid method based on spline interpolation method and Savitzky–Golay filtering," *Neurophotonics*, vol. 5, no. 01, p. 1, Feb. 2018, doi: 10.1117/1.NPh.5.1.015003.

[56] F. Scholkmann, J. Boss, and M. Wolf, "An Efficient Algorithm for Automatic Peak Detection in Noisy Periodic and Quasi-Periodic Signals," *Algorithms*, vol. 5, no. 4, pp. 588–603, Nov. 2012, doi: 10.3390/a5040588.

[57] K. J. Friston, *Statistical parametric mapping: the analysis of functional brain images*, 1st ed. Amsterdam Boston: Elsevier / Academic Press, 2007.

[58] J. W. Barker, A. Aarabi, and T. J. Huppert, "Autoregressive model based algorithm for correcting motion and serially correlated errors in fNIRS," *Biomed. Opt. Express*, vol. 4, no. 8, p. 1366, Aug. 2013, doi: 10.1364/BOE.4.001366.

[59] S. G. Diamond *et al.*, "Dynamic physiological modeling for functional diffuse optical tomography," *NeuroImage*, vol. 30, no. 1, pp. 88–101, Mar. 2006, doi: 10.1016/j.neuroimage.2005.09.016.

[60] E. Kirlilna, N. Yu, A. Jelzow, H. Wabnitz, A. M. Jacobs, and I. Tachtsidis, "Identifying and quantifying main components of physiological noise in functional near infrared spectroscopy on the prefrontal cortex," *Front. Hum. Neurosci.*, vol. 7, 2013, doi: 10.3389/fnhum.2013.00864.

[61] A. von Lühmann, X. Li, K.-R. Müller, D. A. Boas, and M. A. Yücel, "Improved physiological noise regression in fNIRS: A multimodal extension of the General Linear Model using temporally embedded Canonical Correlation Analysis," *NeuroImage*, vol. 208, p. 116472, Mar. 2020, doi: 10.1016/j.neuroimage.2019.116472.

[62] S. Seabold and J. Perktold, "Statsmodels: Econometric and Statistical Modeling with Python," in *Proc. of the 9th Python in Science Conference*, in 1, vol. 7. 2010, pp. 92–96.

[63] L. Carlton *et al.*, "Surface-Based Image Reconstruction Optimization for High-Density Functional Near Infrared Spectroscopy," *Neurophotonics*, vol. (in review), 2026.

[64] D. A. Boas, A. M. Dale, and M. A. Franceschini, "Diffuse optical imaging of brain activation: approaches to optimizing image sensitivity, resolution, and accuracy," *NeuroImage*, vol. 23, pp. S275–S288, Jan. 2004, doi: 10.1016/j.neuroimage.2004.07.011.

[65] F. Pedregosa *et al.*, "Scikit-learn: Machine learning in Python," *J. Mach. Learn. Res.*, vol. 12, pp. 2825–2830, 2011.

[66] A. Paszke *et al.*, "PyTorch: An Imperative Style, High-Performance Deep Learning Library," in *Advances in Neural Information Processing Systems*, H. Wallach, H. Larochelle, A. Beygelzimer, F. d'Alché-Buc, E. Fox, and R. Garnett, Eds., Curran Associates, Inc., 2019. [Online]. Available: https://proceedings.neurips.cc/paper_files/paper/2019/file/bdbca288fee7f92f2bfa9f7012727740-Paper.pdf

[67] R. Fernandez Rojas, X. Huang, and K.-L. Ou, "A Machine Learning Approach for the Identification of a Biomarker of Human Pain using fNIRS," *Sci. Rep.*, vol. 9, no. 1, p. 5645, Apr. 2019, doi: 10.1038/s41598-019-42098-w.

[68] Y. Zhu *et al.*, "Classifying Major Depressive Disorder Using fNIRS During Motor Rehabilitation," *IEEE Trans. Neural Syst. Rehabil. Eng.*, vol. 28, no. 4, pp. 961–969, Apr. 2020, doi: 10.1109/TNSRE.2020.2972270.

[69] A. Bizzego, M. Neoh, G. Gabrieli, and G. Esposito, "A Machine Learning Perspective on fNIRS Signal Quality Control Approaches," *IEEE Trans. Neural Syst. Rehabil. Eng.*, vol. 30, pp. 2292–2300, 2022, doi: 10.1109/TNSRE.2022.3198110.

[70] G.-S. Fu, R. Phlypo, M. Anderson, X.-L. Li, and T. Adali, "Blind Source Separation by Entropy Rate Minimization," *IEEE Trans. Signal Process.*, vol. 62, no. 16, pp. 4245–4255, Aug. 2014, doi: 10.1109/TSP.2014.2333563.

[71] A. von Lühmann, Z. Boukouvalas, K.-R. Müller, and T. Adalı, "A new blind source separation framework for signal analysis and artifact rejection in functional Near-Infrared Spectroscopy," *NeuroImage*, vol. 200, pp. 72–88, Oct. 2019, doi: 10.1016/j.neuroimage.2019.06.021.

[72] S. Dähne, F. Biessmann, F. C. Meinecke, J. Mehnert, S. Fazli, and K.-R. Müller, "Integration of Multivariate Data Streams With Bandpower Signals," *IEEE Trans. Multimed.*, vol. 15, no. 5, pp. 1001–1013, Aug. 2013, doi: 10.1109/TMM.2013.2250267.

[73] S. Dähne *et al.*, "SPoC: a novel framework for relating the amplitude of neuronal oscillations to behaviorally relevant parameters," *neuroimage*, vol. 86, no. 0, pp. 111–122, 2014, doi: http://dx.doi.org/10.1016/j.neuroimage.2013.07.079.

[74] X.-L. Li and T. Adali, "Independent Component Analysis by Entropy Bound Minimization," *IEEE Trans. Signal Process.*, vol. 58, no. 10, pp. 5151–5164, Oct. 2010, doi: 10.1109/TSP.2010.2055859.

[75] X. Zhuang, Z. Yang, and D. Cordes, "A technical review of canonical correlation analysis for neuroscience applications," *Hum. Brain Mapp.*, vol. 41, no. 13, pp. 3807–3833, Sept. 2020, doi: 10.1002/hbm.25090.

[76] A. von Lühmann, A. Ortega-Martinez, D. A. Boas, and M. A. Yücel, "Using the General Linear Model to Improve Performance in fNIRS Single Trial Analysis and Classification: A Perspective," *Front. Hum. Neurosci.*, vol. 14, p. 30, Feb. 2020, doi: 10.3389/fnhum.2020.00030.

[77] S. Moradi, T. Dissanayake, N. Harmening, E. Middell, and A. von Lühmann, "Cross-Modal Learning from fMRI to DOT: Leveraging Large-Scale Neuroimaging for fNIRS Decoding," *IEEE J. Biomed. Health Inform.*, vol. (in review), 2026.

[78] T. Fischer, E. Middell, S. Moradi, and A. von Lühmann, "fNIRS Single-Trial Decoding Improves Systematically with Higher Optode Density, Model-Based Noise Regression, and Image Reconstruction," *Neurophotonics*, vol. (in review), 2026.

[79] M. D. Wilkinson *et al.*, "The FAIR Guiding Principles for scientific data management and stewardship," *Sci. Data*, vol. 3, no. 1, p. 160018, Mar. 2016, doi: 10.1038/sdata.2016.18.

[80] B. Kenny. Kim *et al.*, "Mapping human cerebral blood flow with high-density, multi-channel speckle contrast optical spectroscopy," *Commun. Biol.*, vol. 8, no. 1, p. 1553, Nov. 2025, doi: 10.1038/s42003-025-08915-x.

[81] A. C. Howard *et al.*, "Validation of the Linearity in Image Reconstruction Methods for Speckle Contrast Optical Tomography," *IEEE J. Sel. Top. Quantum Electron.*, vol. 31, no. 4: Adv. in Neurophoton. for Non, pp. 1–8, July 2025, doi: 10.1109/JSTQE.2025.3581407.

[82] H. Dashtestani *et al.*, "Structured sparse multiset canonical correlation analysis of simultaneous fNIRS and EEG provides new insights into the human action-observation network," *Sci. Rep.*, vol. 12, no. 1, p. 6878, Apr. 2022, doi: 10.1038/s41598-022-10942-1.

[83] J. Cao, T. J. Huppert, P. Grover, and J. M. Kainerstorfer, "Enhanced spatiotemporal

resolution imaging of neuronal activity using joint electroencephalography and diffuse optical tomography," *Neurophotonics*, vol. 8, no. 01, Jan. 2021, doi: 10.1117/1.NPh.8.1.015002.

[84] Y. Gao, B. Jia, M. Houston, and Y. Zhang, "Hybrid EEG-fNIRS Brain Computer Interface Based on Common Spatial Pattern by Using EEG-Informed General Linear Model," *IEEE Trans. Instrum. Meas.*, vol. 72, pp. 1–10, 2023, doi: 10.1109/TIM.2023.3276509.

**Appendix S: Supplemental Material**

*S1 – Tutorial Notebook 1: Head Models and Forward Modeling*

Please see the supplementary PDF "Cedalion_Tutorial_Supplement.pdf", Section 1, for the rendered notebook, as well as the executable version on [DOI-REGISTERED CODE OCEAN LINK PROVIDED HERE UPON PUBLICATION]

*S2 – Tutorial Notebook 2: Photogrammetric Optode Co-Registration*

Please see the supplementary PDF "Cedalion_Tutorial_Supplement.pdf", Section 2, for the rendered notebook, as well as the executable version on [DOI-REGISTERED CODE OCEAN LINK PROVIDED HERE UPON PUBLICATION]

*S3 – Tutorial Notebook 3: Signal Processing*

Please see the supplementary PDF "Cedalion_Tutorial_Supplement.pdf", Section 3, for the rendered notebook, as well as the executable version on [DOI-REGISTERED CODE OCEAN LINK PROVIDED HERE UPON PUBLICATION]

*S4 – Tutorial Notebook 4: Model-Driven (GLM) Analysis*

Please see the supplementary PDF "Cedalion_Tutorial_Supplement.pdf", Section 4, for the rendered notebook, as well as the executable version on [DOI-REGISTERED CODE OCEAN LINK PROVIDED HERE UPON PUBLICATION]

*S5 – Tutorial Notebook 5: Image Reconstruction*

Please see the supplementary PDF "Cedalion_Tutorial_Supplement", Section 5, for the rendered notebook, as well as the executable version on [DOI-REGISTERED CODE OCEAN LINK PROVIDED HERE UPON PUBLICATION]

*S6 – Tutorial Notebook 6: Data-Driven (ML) Analysis*

Please see the supplementary PDF "Cedalion_Tutorial_Supplement", Section 6, for the rendered notebook, as well as the executable version on <mark>[DOI-REGISTERED CODE OCEAN LINK PROVIDED HERE UPON PUBLICATION]</mark>

*S7 – Tutorial Notebook 7: Data Augmentation*

Please see the supplementary PDF "<mark>Cedalion_Tutorial_Supplement</mark>", Section 7, for the rendered notebook, as well as the executable version on <mark>[DOI-REGISTERED CODE OCEAN LINK PROVIDED HERE UPON PUBLICATION]</mark>

## **Caption List**

**Fig. 1** Key architectural considerations and motivation of the Cedalion toolbox. Cedalion brings together community contributions, file exchange standards, HPC and cloud processing within a python framework for advanced data-driven multimodal fNIRS / DOT analysis towards neuroimaging in the everyday world.

**Fig. 2** Toolbox overview and graphical guide to sections and tutorials.

**Fig. 3** Visualization of array layout of multidimensional time series with corresponding coordinates and units. Example has three dimensions with multiple coordinate axes per dimension (i.e. time and sample coordinates across time dimension). Values carry physical units (i.e. µM - micro-Molar) that allow automatic consistency checks and conversion.

**Fig. 4** Pipelines in Cedalion. Through integration of environment specification with Snakemake and SNIRF / BIDS standardized file-exchange, heterogeneous processing steps across community toolboxes can be executed identically on local machines and clusters via rule-based pipeline configuration and definition (snakefile, config.yaml).

**Fig. 5** Head models in Cedalion. Segmentation masks from individual MRI scans as well as existing Atlases (ICBM152 and Colin27) can be used for (two surface) head-model generation, coordinate transformation, voxel-vertex mapping and parcellation in Cedalion.

**Fig. 6** Forward Modeling in Cedalion. Integration Cedalion's head models and SNIRF probe geometries with MCX/MCX-CL and NIRFASTer enable both Monte-Carlo and FEM-based photon propagation modeling for Forward Model generation.

**Fig. 7** Photogrammetric Optode Co-Registration in Cedalion. Photogrammetric scans of visually labeled optodes, either from a smartphone or a 3D scanner, can be automatically processed for optode detection, registration of orientation and position in 3D space relative to anatomical landmarks, and mapping pre-designed probe geometries with Source and Detector labels from SNIRF files. Enables precise co-registration of individual probe positions onto atlases or individual head geometries to reduce localization errors to anatomical brain regions within and across subjects.

**Fig. 8** Quality assessment and artifact rejection. Established quality metrics such as SCI, PSP, SNR etc., can be used for the generation of multidimensional (Xarray-based) quality masks that can be flexibly combined or carried along. Widely used artefact rejection methods can clean data in an unsupervised way (e.g., TDDR, Wavelet, GVTD) or via supervision from quality masks (e.g. Spline, SplineSG, PCA, …).

**Fig. 9** General Linear Model (GLM) in Cedalion. The Design Matrix Builder permits flexible combination of HRF kernels, drift and physiology regressors. A variety of solvers (from OLS to AR IRLS) can be used to estimate the fits, provided as a statsmodels object, leveraging the statsmodels Python package for advanced statistical analysis.

**Fig. 10** Image Reconstruction in Cedalion. Forward model inversion includes options for Tikhonov (measurement noise) regularization and Spatially Variant Regularization, and use of spatial basis functions such as Gaussian kernels. Results can be visualized in still images and as GIFs across time. Color bar unit is micro-Molar.

**Fig. 11** Data Driven Analysis. Aside from direct interfacing with sklearn and pytorch libraries for classical machine learning and deep learning, Cedalion provides tailored implementations for unimodal and multimodal linear and nonlinear unsupervised signal decomposition methods, including the CCA family, and (constrained) ICA.

**Fig. 12** Modules for Synthetic Data-Augmentation. Realistic hemodynamic response functions on the cortex can be generated from customizable spatio-temporal bases and event sequences. Forward modeling enables projection to and addition of HRFs to real resting state background data in channel space.

**Table 1** Common array layouts and coordinates used in Cedalion. For brevity not all coordinates are always shown. Note that multiple coordinates per dimension are possible. Highlighted rows expand the snirf specification.

**Table 2** Core Functions for Data Structures and I/O. Not all subpackages are listed for brevity.

**Table 3** Core Functions for Head Models and Forward Modelling. Not all subpackages are shown for brevity.

**Table 4** Core Functions for Signal Processing & Analysis

**Table 5** Core Functions for DOT image reconstruction.

**Table 6** Core Functions for Multimodal Fusion, Data Driven Analysis and ML

**Table 7** Core Functions for Data Simulation and Augmentation

# Supplementary Material

The paper is accompanied by 7 Jupyter notebooks which are provided in rendered form in this PDF and which can be interactively explored on Google Colab by following these links:

- S1: Head models and Forward Modelling
- S2: Photogrammetric Optode Co-Registration
- S3: Signal Processing
- S4: Model-driven (GLM)
- S5: DOT - Image Reconstruction
- S6: Data-Driven (ML) Analysis
- S7: Data Augmentation

Please use the PDF's document outline to navigate.

The notebooks will eventually be hosted on Code Ocean and GitHub.

- **Code Ocean**: *Link and DOI pending*
- **GitHub**: https://github.com/ibs-lab/cedalion/tree/dev/examples/tutorial

To cite any of these notebooks please use the following:

`FIXME DOI issued by Code Ocean`

The notebooks are also part of the Cedalion's official documentation, available at

http://cedalion.tools/docs

# S1: Head models and Forward Modelling

This notebook introduces how Cedalion handles head models and forward modelling for diffuse optical tomography.

```python
[1]: # This cells setups the environment when executed in Google Colab.
     try:
         import google.colab
         !curl -s https://raw.githubusercontent.com/ibs-lab/cedalion/dev/scripts/
         ↪colab_setup.py -o colab_setup.py
         # Select branch with --branch "branch name" (default is "dev")
         %run colab_setup.py
     except ImportError:
         pass
```

```python
[2]: from pathlib import Path
     from tempfile import TemporaryDirectory

     import cedalion
     import cedalion.data
     import cedalion.dataclasses as cdc
     import cedalion.dot
     import cedalion.io
     import cedalion.nirs
     import cedalion.vis.anatomy
     import cedalion.vis.blocks as vbx
     import matplotlib.pyplot as p
     import numpy as np
     import pyvista as pv
     import xarray as xr
     from cedalion import units
     from matplotlib.colors import ListedColormap

     # set pyvista's jupyter backend to 'server' for interactive 3D plots
     pv.set_jupyter_backend("static")
     # pv.set_jupyter_backend('server')

     xr.set_options(display_expand_data=False);
```

## Configuration

The constants in the next cell control the behavior of the notebook.

```python
[3]: # set this to either 'colin27', 'icbm152', 'custom_from_segmentation',
     ↪'custom_from_surfaces'
     HEAD_MODEL = "colin27"

     # set this to either 'mcx' or 'nirfaster'
     FORWARD_MODEL = "mcx"

     # set this to False to actually run the forward model
     PRECOMPUTED_SENSITIVITY = True
```

Create a temporary directory as a working directory.

```
[4]:  working_directory = TemporaryDirectory()
      tmp_dir_path = Path(working_directory.name)
```

## Loading and inspecting an example dataset

This notebook loads one of Cedalion's example datasets.

Using a function from the `cedalion.data` package, the following cell will download a snirf file with a recording of a fingertapping experiment.

The content of the snirf file is returned as a `cedalion.dataclasses.Recording` object.

```
[5]:  rec = cedalion.data.get_fingertappingDOT()
```

The Recording container carries time series and related objects through the program in ordered dictionaries much like a segmented tray carrying items in separate compartments.

This notebook mainly needs the probe geometry and channel definitions, detailed below.

### Probe Geometry

The snirf file contains the 3D probe geometry. When loaded in Cedalion, the probe geometry is represented in a `xarray.DataArray` with shape (number of labeled points x 3).

It is available as an attribute of the `Recording` container:

```
[6]:  rec.geo3d
```

```
[6]:  <xarray.DataArray (label: 346, digitized: 3)> Size: 8kB
      [mm] -77.82 15.68 23.17 -61.91 21.23 56.49 … 14.23 -38.28 81.95 -0.678 -37.03
      Coordinates:
          type      (label) object 3kB PointType.SOURCE … PointType.LANDMARK
        * label     (label) <U6 8kB 'S1' 'S2' 'S3' 'S4' … 'FFT10h' 'FT10h' 'FTT10h'
      Dimensions without coordinates: digitized
```

This array stores 3D coordinates for 346 points. The first dimension, named `'label'`, indexes the points, while the second dimension holds the three coordinate values for each point. In Cedalion, the name of the second dimension serves as the label for the coordinate's reference system, which in this case is set to `'digitized'`.

Two coordinate arrays are linked to `'label'` dimension. The `'label'` coordinate assigns a string name to each 3D point (for example, `'S1'` for the first source or `'Nz'` for the nasion). Using values of `cedalion.dataclasses.PointType`, the `'type'` coordinate is used to distinguish between sources, detectors and landmarks.

The values stored in the geo3d array have physical units associated with them. In this case, the coordinates are given in millimeters.

This particular schema of representing geometry information in a `xarray.DataArray` is used throughout Cedalion. Arrays following this schema are called `cedalion.typing.LabeledPoints`.

### Channel Definitions

This dataset contains one continuous-wave raw amplitudes time series. It is stored in the `Recording` container within the `timeseries` dictionary under the key `'amp'`.

```
[7]: rec.timeseries.keys()
```

```
[7]: odict_keys(['amp'])
```

Channels are defined by source-detector pairs. Cedalion uses two ways of representing channel definitions: measurement lists and time series coordinates.

In the SNIRF file, channels are specified in measurement lists, a tabular data structure that lists for each channel source, detector, wavelength and data type.

For each loaded time series, the measurement list is stored as a `pandas.DataFrame` in `rec._measurement_lists` under the same key.

```
[8]: meas_list = rec._measurement_lists["amp"]
     display(meas_list)
```

|     | sourceIndex | detectorIndex | wavelengthIndex | wavelengthActual | \ |
|-----|-------------|---------------|-----------------|------------------|---|
| 0   | 1           | 1             | 1               | None             |   |
| 1   | 1           | 2             | 1               | None             |   |
| 2   | 1           | 4             | 1               | None             |   |
| 3   | 1           | 5             | 1               | None             |   |
| 4   | 1           | 6             | 1               | None             |   |
| ..  | ...         | ...           | ...             | ...              |   |
| 195 | 14          | 27            | 2               | None             |   |
| 196 | 14          | 28            | 2               | None             |   |
| 197 | 14          | 29            | 2               | None             |   |
| 198 | 14          | 31            | 2               | None             |   |
| 199 | 14          | 32            | 2               | None             |   |

|     | wavelengthEmissionActual | dataType | dataUnit | dataTypeLabel | dataTypeIndex | \ |
|-----|--------------------------|----------|----------|---------------|---------------|---|
| 0   | None                     | 1        | None     | raw-DC        | 1             |   |
| 1   | None                     | 1        | None     | raw-DC        | 1             |   |
| 2   | None                     | 1        | None     | raw-DC        | 1             |   |
| 3   | None                     | 1        | None     | raw-DC        | 1             |   |
| 4   | None                     | 1        | None     | raw-DC        | 1             |   |
| ..  | ...                      | ...      | ...      | ...           | ...           |   |
| 195 | None                     | 1        | None     | raw-DC        | 1             |   |
| 196 | None                     | 1        | None     | raw-DC        | 1             |   |
| 197 | None                     | 1        | None     | raw-DC        | 1             |   |
| 198 | None                     | 1        | None     | raw-DC        | 1             |   |
| 199 | None                     | 1        | None     | raw-DC        | 1             |   |

|     | sourcePower | detectorGain | moduleIndex | sourceModuleIndex | \ |
|-----|-------------|--------------|-------------|-------------------|---|
| 0   | None        | None         | None        | None              |   |
| 1   | None        | None         | None        | None              |   |
| 2   | None        | None         | None        | None              |   |
| 3   | None        | None         | None        | None              |   |
| 4   | None        | None         | None        | None              |   |
| ..  | ...         | ...          | ...         | ...               |   |
| 195 | None        | None         | None        | None              |   |
| 196 | None        | None         | None        | None              |   |
| 197 | None        | None         | None        | None              |   |
| 198 | None        | None         | None        | None              |   |
| 199 | None        | None         | None        | None              |   |

```
     detectorModuleIndex channel source detector  wavelength chromo
0                    None    S1D1     S1       D1       760.0   None
1                    None    S1D2     S1       D2       760.0   None
2                    None    S1D4     S1       D4       760.0   None
3                    None    S1D5     S1       D5       760.0   None
4                    None    S1D6     S1       D6       760.0   None
..                    ...     ...    ...      ...         ...    ...
195                  None  S14D27    S14      D27       850.0   None
196                  None  S14D28    S14      D28       850.0   None
197                  None  S14D29    S14      D29       850.0   None
198                  None  S14D31    S14      D31       850.0   None
199                  None  S14D32    S14      D32       850.0   None

[200 rows x 19 columns]
```

The `'amp'` time series is represented as a `xarray.DataArray` with dimensions `'channel'`, `'wavelength'` and `'time'`. Three coordinate arrays are linked to the `'channel'` dimension, specifying for each channel a string label as well as the string label and the corresponding source and detector labels (e.g. the first channel `'S1D1'` is between source `'S1'` and detector `'D1'`).

`[9]:` `rec.timeseries["amp"]`

`[9]:`
```
<xarray.DataArray (channel: 100, wavelength: 2, time: 8794)> Size: 14MB
[V] 0.0874 0.08735 0.08819 0.08887 0.0879 … 0.09108 0.09037 0.09043 0.08999
Coordinates:
  * time        (time) float64 70kB 0.0 0.2294 0.4588 … 2.017e+03 2.017e+03
    samples     (time) int64 70kB 0 1 2 3 4 5 … 8788 8789 8790 8791 8792 8793
  * channel     (channel) object 800B 'S1D1' 'S1D2' 'S1D4' … 'S14D31' 'S14D32'
    source      (channel) object 800B 'S1' 'S1' 'S1' 'S1' … 'S14' 'S14' 'S14'
    detector    (channel) object 800B 'D1' 'D2' 'D4' 'D5' … 'D29' 'D31' 'D32'
  * wavelength  (wavelength) float64 16B 760.0 850.0
Attributes:
    data_type_group:  unprocessed raw
```

The function `cedalion.vis.anatomy` visualizes the probe geometry. Optode locations are obtained from the geo3d array, while channel definitions are derived from the time series.

Sources, detectors and landmarks are represented as red, blue and grey dots, respectively. The nasion is highlighted in yellow.

`[10]:` `cedalion.vis.anatomy.plot_montage3D(rec["amp"], rec.geo3d)`

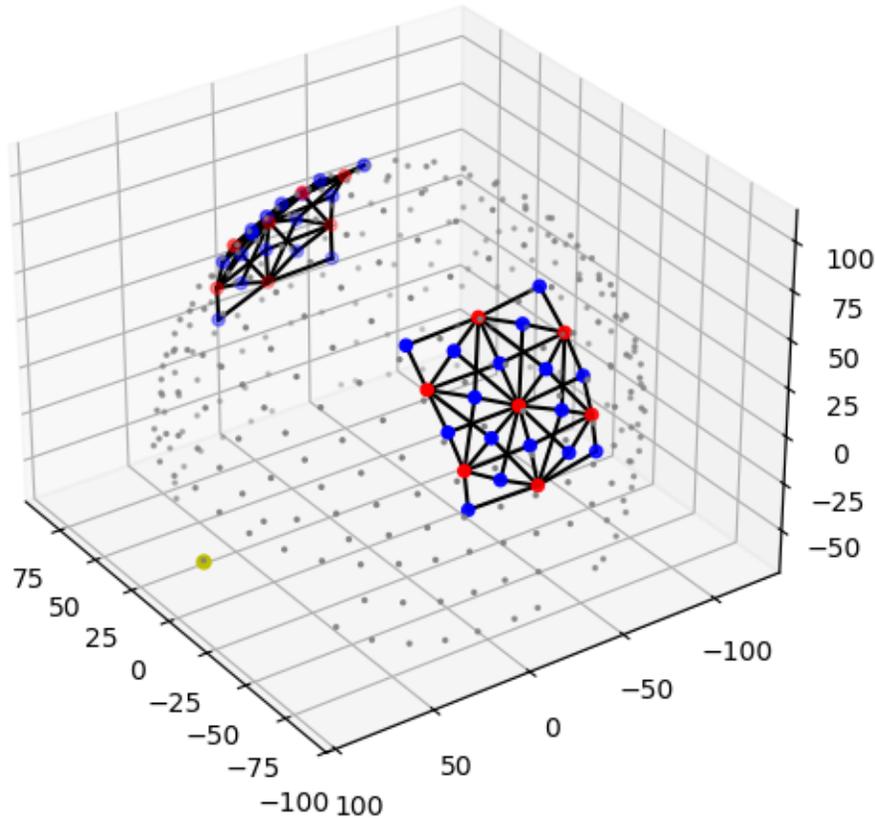

## Coordinate systems

When working with probe and head model geometries, it is important to distinguish between coordinates from three different coordinate reference systems (CRS):

- **Digitization space** (`CRS='digitized'`)

  This is the coordinate space used by the digitization or montage planning tool. In this example, optode positions are given in millimeters with the origin located at the head center and the y-axis pointing in the superior direction. The CRS has the user-configurable name `'digitized'`. Other digitization tools may use different names, units or coordinate systems.

- **Voxel space** (`CRS='ijk'`)

  Tissue masks from segmented MRI scans are in voxel space, where coordinate axes are aligned with voxel edges. Independent of the MRI scanner resolution or the physical voxel size, voxel space is defined such that each voxel has unit edge length and dimensionless coordinates.

  Axis-aligned grids are computationally efficient, which is why the photon propagation code MCX uses this coordinate system.

- **Scanner space / RAS space** (`CRS='ras'`)

  Scanner space differs from voxel space in that it uses physical units and its coordinate axes point in the **R**ight, **A**nterior, and **S**uperior directions. Therefore, scanner space is also referred to as RAS space. Both voxel and scanner spaces are related through affine transformations which are stored in the head model as `t_ijk2ras` and `t_ras2ijk`.

The scanner space of the standard Colin27 and ICBM-152 heads is widely used as a common coordinate reference system to share results. To highlight this particularly important CRS the head models returned by `get_standard_headmodel` use the label `'mni'` instead of `'ras'` for the scanner space.

To avoid confusion between different coordinate systems, Cedalion is explicit about the CRS associated with each point cloud or surface.

As shown above a dimension name of the `geo3d` array is used to track the CRS. Another way to inspect the CRS is by using the `.points` accessor:

```
[11]: print(f"The coordinate reference system of the probe geometry is '{rec.geo3d.points.
      ↪crs}'")
```

```
The coordinate reference system of the probe geometry is 'digitized'
```

### The TwoSurfaceHeadModel

The forward model simulates photon propagation through different tissue types. Starting from an anatomical MRI scan, tissue segmentation assigns each voxel to a specific tissue type and corresponding optical properties.

The image reconstruction method implemented in Cedalion does not estimate absorption changes for all voxels. Instead, the inverse problem is simplified by defining two surfaces representing the scalp and the brain and by restricting the reconstruction of absorption changes to voxels in their vicinitiy.

The class `cedalion.dot.TwoSurfaceHeadModel` groups together the segmentation mask, landmark positions and affine transformations as well as the scalp and brain surfaces.

### Constructing the TwoSurfaceHeadModel

There are three ways how to construct a TwoSurfaceHeadModel: - using the function `cedalion.dot.get_standard_headmodel` - from tissue segmentation masks. Cedalion derives scalp and brain surfaces. - from tissue segmentation masks and triangulated surface meshes generated by an external tool

**FIXME** explain better

- The segmentation masks are in individual niftii files.
- The dict `mask_files` maps mask filenames relative to `SEG_DATADIR` to short labels. These labels describe the tissue type of the mask.
- The variable `landmarks_file` holds the path to a file containing landmark positions in scanner space (RAS). One way to create this file is by using Slicer3D.

```
[12]: if HEAD_MODEL in ["colin27", "icbm152"]:
          # use a factory method for the standard head models
          head_ijk = cedalion.dot.get_standard_headmodel(HEAD_MODEL)

      elif HEAD_MODEL == "custom_from_segmentation":
          # point these to a directory with segmentation masks
          segm_datadir = Path("/path/to/dir/with/segmentation_masks")
          mask_files = {
              "csf": "mask_csf.nii",
              "gm": "mask_gray.nii",
              "scalp": "mask_skin.nii",
              "skull": "mask_bone.nii",
              "wm": "mask_white.nii",
```

```python
    }
    # the landmarks must be in scanner (RAS) space.
    # For example Slicer3D can be used to pick the landmarks.
    landmarks_file = Path("/path/to/landmarks.mrk.json")

    # if available provide a mapping between vertices and labels
    parcel_file = None

    # Construct a head model from segmentation mask.
    head_ijk = cedalion.dot.TwoSurfaceHeadModel.from_segmentation(
        segmentation_dir=segm_datadir,
        mask_files=mask_files,
        landmarks_ras_file=landmarks_file,
        parcel_file=parcel_file,
        # adjust these to control mesh parameters
        brain_face_count=None,
        scalp_face_count=None,
        smoothing=0.0,
    )
elif HEAD_MODEL == "custom_from_surfaces":
    # point these to a directory with segmentation masks
    segm_datadir = Path("/path/to/dir/with/segmentation_masks")
    mask_files = {
        "csf": "mask_csf.nii",
        "gm": "mask_gray.nii",
        "scalp": "mask_skin.nii",
        "skull": "mask_bone.nii",
        "wm": "mask_white.nii",
    }
    # the landmarks must be in scanner (RAS) space.
    # For example Slicer3D can be used to pick the landmarks.
    landmarks_file = Path("/path/to/landmarks.mrk.json")

    # if available provide a mapping between vertices and labels
    parcel_file = None

    # Likely, better brain and scalp surfaces are achievable from
    # specialized segmentation tools.
    head_ijk = cedalion.dot.TwoSurfaceHeadModel.from_surfaces(
        segmentation_dir=segm_datadir,
        mask_files=mask_files,
        landmarks_ras_file=landmarks_file,
        parcel_file=parcel_file,
        brain_surface_file=segm_datadir / "mask_brain.obj",
        scalp_surface_file=segm_datadir / "mask_scalp.obj",
    )
else:
    raise ValueError("unknown head model")
```

**Inspecting the Head Model**

The created head model is in voxel space (CRS='ijk'). The string representation shows its different components: the segmentation masks for different tissue types, the brain and scalp surfaces as well as the landmarks.

```
[13]: head_ijk
```

```
[13]: TwoSurfaceHeadModel(
          crs: ijk
          tissue_types: csf, gm, scalp, skull, wm
          brain faces: 29988 vertices: 15002 units: dimensionless
          scalp faces: 20096 vertices: 10050 units: dimensionless
          landmarks: 73
      )
```

The segmentation masks are 3D image cubes (dimensions i,j,k) that are stacked along a fourth dimension 'segmentation_type'. Each tissue mask has a unique integer assigned which are used to mark voxels that belong to that tissue type. The string label of each tissue type (e.g. 'gm', 'csf') map to tabulated optical properties in `cedalion.dot.tissue_properties.TISSUE_LABELS`.

```
[14]: head_ijk.segmentation_masks
```

```
[14]: <xarray.DataArray (segmentation_type: 5, i: 181, j: 217, k: 181)> Size: 36MB
      0 0 0 0 0 0 0 0 0 0 0 0 0 0 0 0 0 0 … 0 0 0 0 0 0 0 0 0 0 0 0 0 0 0 0 0 0
      Coordinates:
        * segmentation_type  (segmentation_type) <U5 100B 'csf' 'gm' … 'skull' 'wm'
      Dimensions without coordinates: i, j, k
```

Visualize segmentation masks

```
[15]: collapsed_mask = head_ijk.segmentation_masks.sum("segmentation_type")

      cmap = ListedColormap(["black", "cyan", "lightgray", "pink", "gray", "white"])
      p.pcolormesh(collapsed_mask[:,:, collapsed_mask.sizes["k"]//2], cmap=cmap, vmin=-0.5,
      ↪vmax=5.5)
      cb = p.colorbar()
      cb.set_ticks([0,1,2,3,4,5])
      cb.set_ticklabels(["air", "CSF", "gray matter", "skin", "skull", "white matter"])
      p.xlabel("i")
      p.ylabel("j");
```

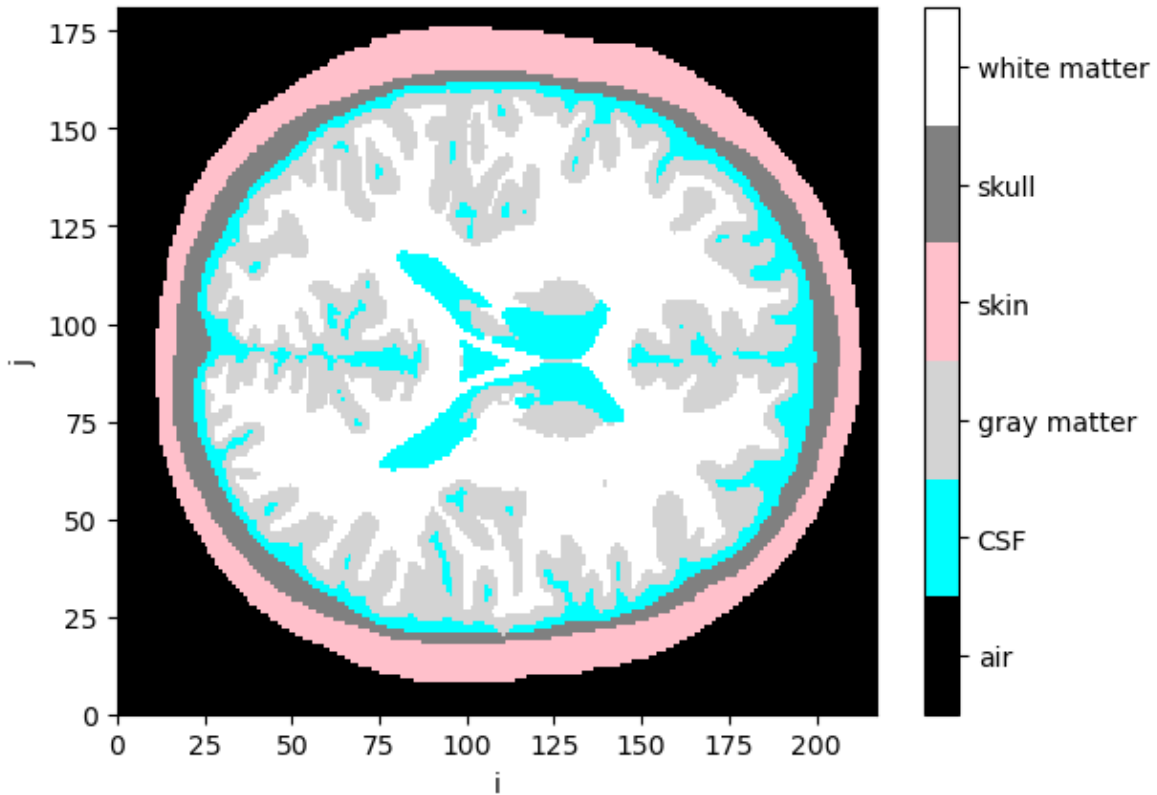

Landmarks are stored as `LabeledPoints`.

```
[16]: head_ijk.landmarks
```

```
[16]: <xarray.DataArray (label: 73, ijk: 3)> Size: 2kB
      [] 90.0 7.435 48.92 3.924 106.0 24.01 … 127.4 25.63 109.0 139.8 26.66 93.47
      Coordinates:
        * label    (label) <U3 876B 'Iz' 'LPA' 'RPA' 'Nz' … 'PO1' 'PO2' 'PO4' 'PO6'
          type     (label) object 584B PointType.LANDMARK … PointType.LANDMARK
      Dimensions without coordinates: ijk
```

The brain and scalp surfaces are represented by triangulated meshes, which are represented by instances of `cedalion.dataclasses.TrimeshSurface`.

```
[17]: head_ijk.brain
```

```
[17]: TrimeshSurface(faces: 29988 vertices: 15002 crs: ijk units: dimensionless vertex_coords:
      ['parcel'])
```

```
[18]: head_ijk.scalp
```

```
[18]: TrimeshSurface(faces: 20096 vertices: 10050 crs: ijk units: dimensionless vertex_coords:
      [])
```

The attribute `.crs` indicates that the head model and all its components is in voxel space.

```
[19]: head_ijk.crs
```

```
[19]:  'ijk'
```

**Transformations**

The affine transformations between different coordinate reference systems are represented as 4x4 matrices which Cedalion also stores in `DataArrays`.

The dimension names are chosen such that applying the transformation through matrix multiplication transforms `LabeledPoints` from one CRS to the other.

The transformation matrix to translate from voxel to scanner space is available as:

```
[20]:  head_ijk.t_ijk2ras
```

```
[20]:  <xarray.DataArray (mni: 4, ijk: 4)> Size: 128B
       [mm] 1.0 0.0 0.0 -90.0 0.0 1.0 0.0 -126.0 0.0 0.0 1.0 -72.0 0.0 0.0 0.0 1.0
       Dimensions without coordinates: mni, ijk
```

And the inverse transformation matrix to translate from scanner to voxel space is available as:

```
[21]:  head_ijk.t_ras2ijk
```

```
[21]:  <xarray.DataArray (ijk: 4, mni: 4)> Size: 128B
       [1/mm] 1.0 -1.616e-14 -8.192e-15 90.0 … 3.253e-16 -1.74e-16 -8.828e-17 1.0
       Dimensions without coordinates: ijk, mni
```

The head model provides the function `.apply_transform` to change the coordinate systems for all components:

```
[22]:  head_ras = head_ijk.apply_transform(head_ijk.t_ijk2ras)
       display(head_ras)
```

```
TwoSurfaceHeadModel(
  crs: mni
  tissue_types: csf, gm, scalp, skull, wm
  brain faces: 29988 vertices: 15002 units: millimeter
  scalp faces: 20096 vertices: 10050 units: millimeter
  landmarks: 73
)
```

**Scaling head models or registering optode positions**

- cedalion has functionality to scale the head model to given head sizes or probe geometries
- alternatively, the probe geometry can be scaled to match the head size, changing channel distances in doing so
- the scaling affects how large a head model is in scanner space. the result of the scaling is represented in the trafos that connect voxel and scanner space

```
[23]:  def compare_trafos(t_orig, t_scaled):
           """Compare the length of a unit vector after applying the transform."""
           t_orig = t_orig.pint.dequantify().values
           t_scaled = t_scaled.pint.dequantify().values

           for i, dim in enumerate("xyz"):
               unit_vector = np.zeros(4)
               unit_vector[i] = 1.0
```

```
        n1 = np.linalg.norm(t_orig @ unit_vector)
        n2 = np.linalg.norm(t_scaled @ unit_vector)

        print(
            f"After scaling, a unit vector along the {dim}-axis is {n2 / n1 * 100.0:.
↪1f}% "
            "of the original length."
        )
```

**Scale head model using geodesic distances**

Transform the head model to the subject's head size as defined by three measurements: - the head circumference - the distance from Nz through Cz to Iz - the distance from LPA through Cz to RPA

These distances define a ellipsoidal scalp surface to which the head model is fit.

The resulting head will be in CRS='ellipsoid'.

```
[24]: head_scaled = head_ijk.scale_to_headsize(
          circumference=51*units.cm,
          nz_cz_iz=33*units.cm,
          lpa_cz_rpa=33*units.cm
      )
      display(head_scaled)
```

```
TwoSurfaceHeadModel(
  crs: ellipsoid
  tissue_types: csf, gm, scalp, skull, wm
  brain faces: 29988 vertices: 15002 units: millimeter
  scalp faces: 20096 vertices: 10050 units: millimeter
  landmarks: 73
)
```

Visualize original (green) and scaled (red) scalp surfaces and landmarks.

```
[25]: plt = pv.Plotter()
      vbx.plot_surface(plt, head_ras.scalp, color="#4fce64", opacity=.1)
      vbx.plot_labeled_points(plt, head_ras.landmarks, color="g")
      vbx.plot_surface(plt, head_scaled.scalp, color="#ce5c4f")
      vbx.plot_labeled_points(plt, head_scaled.landmarks, color="r")
      plt.show()
```

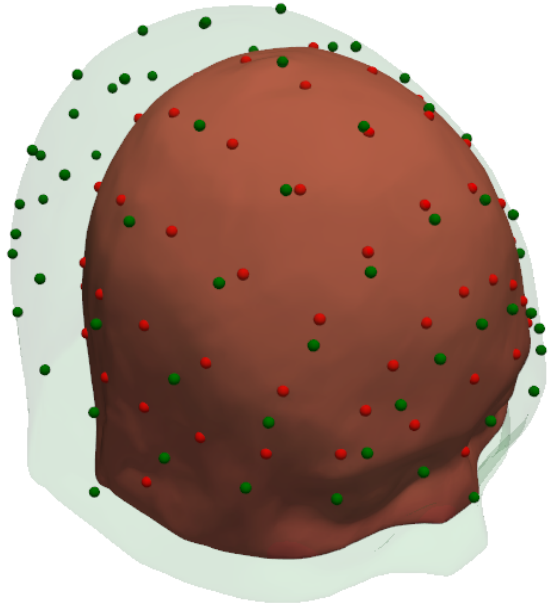

```
[26]: compare_trafos(head_ijk.t_ijk2ras, head_scaled.t_ijk2ras)
```

After scaling, a unit vector along the x-axis is 89.8% of the original length.
After scaling, a unit vector along the y-axis is 76.4% of the original length.
After scaling, a unit vector along the z-axis is 87.3% of the original length.

**Scale head model using digitized landmark positions**

Transform the head model to the subject's head size by fitting landmarks of the head model to digitized landmarks.

The resulting head will be in the same CRS as the digitized landmarks. This also affects axis orientations.

```
[27]: # Construct LabeledPoints with landmark positions.
      # Use the digitized landmarks from 41_photogrammetric_optode_coregistration.ipynb
      geo3d = cdc.build_labeled_points(
          [
              [14.00420712, -7.84856869, 449.77840004],
              [99.09920059, 29.72154755, 620.73876117],
              [161.63815139, -48.49738938, 494.91210993],
              [82.8771277, 79.79500128, 498.3338802],
              [15.17214095, -60.56186128, 563.29621021],
          ],
          crs="digitized",
```

```
        labels=["Nz", "Iz", "Cz", "LPA", "RPA"],
        types = [cdc.PointType.LANDMARK] * 5,
        units="mm"
    )

    # shift the center of the landmarks to plot original and
    # scaled versions next to each other
    geo3d = geo3d - geo3d.mean("label") + (0,200,0)*cedalion.units.mm

    head_scaled = head_ijk.scale_to_landmarks(geo3d)
```

Visualize original (green) and scaled (red) scalp surfaces and landmarks.

The digitized landmarks are plotted in yellow.

```
[28]: plt = pv.Plotter()
      vbx.plot_surface(plt, head_ras.scalp, color="#4fce64", opacity=.1)
      vbx.plot_labeled_points(plt, head_ras.landmarks, color="g")
      vbx.plot_surface(plt, head_scaled.scalp, color="#ce5c4f", opacity=.8)
      vbx.plot_labeled_points(plt, head_scaled.landmarks, color="r")
      vbx.plot_labeled_points(plt, geo3d, color="y", )
      vbx.camera_at_cog(plt, head_scaled.scalp, (600,0,0), fit_scene=True)
      plt.show()
```

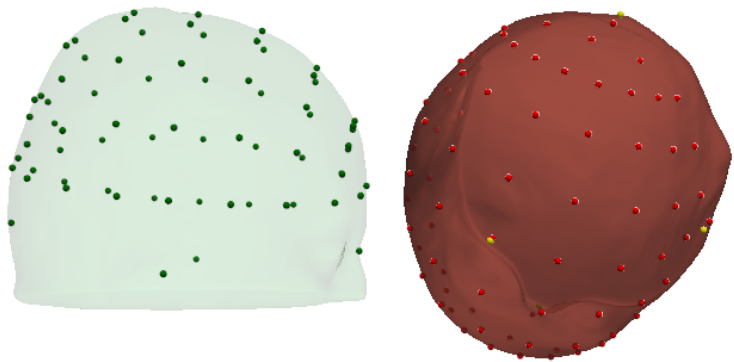

```
[29]: compare_trafos(head_ijk.t_ijk2ras, head_scaled.t_ijk2ras)
```

```
After scaling, a unit vector along the x-axis is 98.2% of the original length.
After scaling, a unit vector along the y-axis is 95.9% of the original length.
After scaling, a unit vector along the z-axis is 93.0% of the original length.
```

**Scale digitized optode positions to head model**

Without changing the head model size, transform the optode positions to fit the head model's landmarks.

This function calculates an unconstrained affine transformation to bring landmarks as best as possible into alignment. Any remaining distance between transformed optode positions from the scalp surface is removed by snapping each transformed optode to its nearest scalp vertex.

To derive the unconstrained transformation the digitized coordinates and head model landmarks must contain at least the positions of "Nz", "Iz", "Cz", "LPA", "RPA".

If less landmarks are available `align_and_snap_to_scalp` can be called with `mode='trans_rot_isoscale`, which restricts the affine to translations, rotations and isotropic scaling, which have less degrees of freedom.

```
[30]: geo3d_snapped_ras = head_ras.align_and_snap_to_scalp(rec.geo3d)
```

The transformed digitized optode and landmark positions are plotted in red (sources), blue (detectors) and green (landmarks).

The fiducial landmarks of the head model are plotted in yellow.

```
[31]: plt = pv.Plotter()
vbx.plot_surface(plt, head_ras.scalp, color="#4fce64", opacity=.1)
vbx.plot_labeled_points(plt, geo3d_snapped_ras)
vbx.plot_labeled_points(
    plt, head_ras.landmarks.sel(label=["Nz", "Iz", "Cz", "LPA", "RPA"]), color="y"
)
plt.show()
```

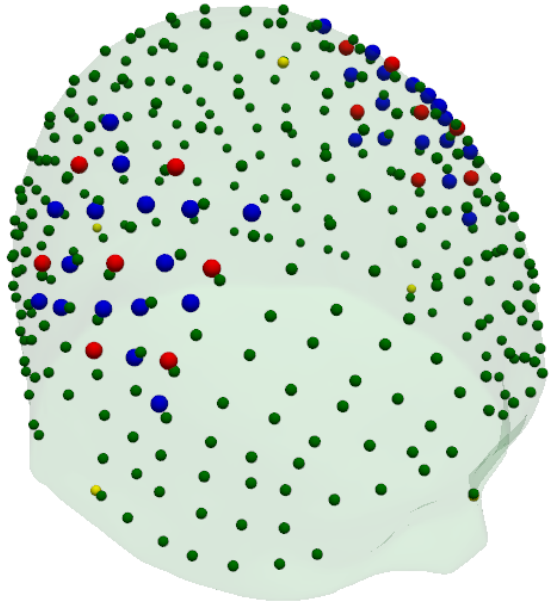

```
[32]: bins = np.arange(0, 40.5, 1)
      p.hist(cedalion.nirs.channel_distances(rec["amp"], rec.geo3d), bins, label="before
      ↪scaling", fc="g", alpha=.5)
      p.hist(cedalion.nirs.channel_distances(rec["amp"], geo3d_snapped_ras), bins,
      ↪label="after scaling", fc="r", alpha=.5);
      p.legend()
      p.xlabel("channel distance / mm")
      p.ylabel("# channel")
      p.xlim(0,40);
```

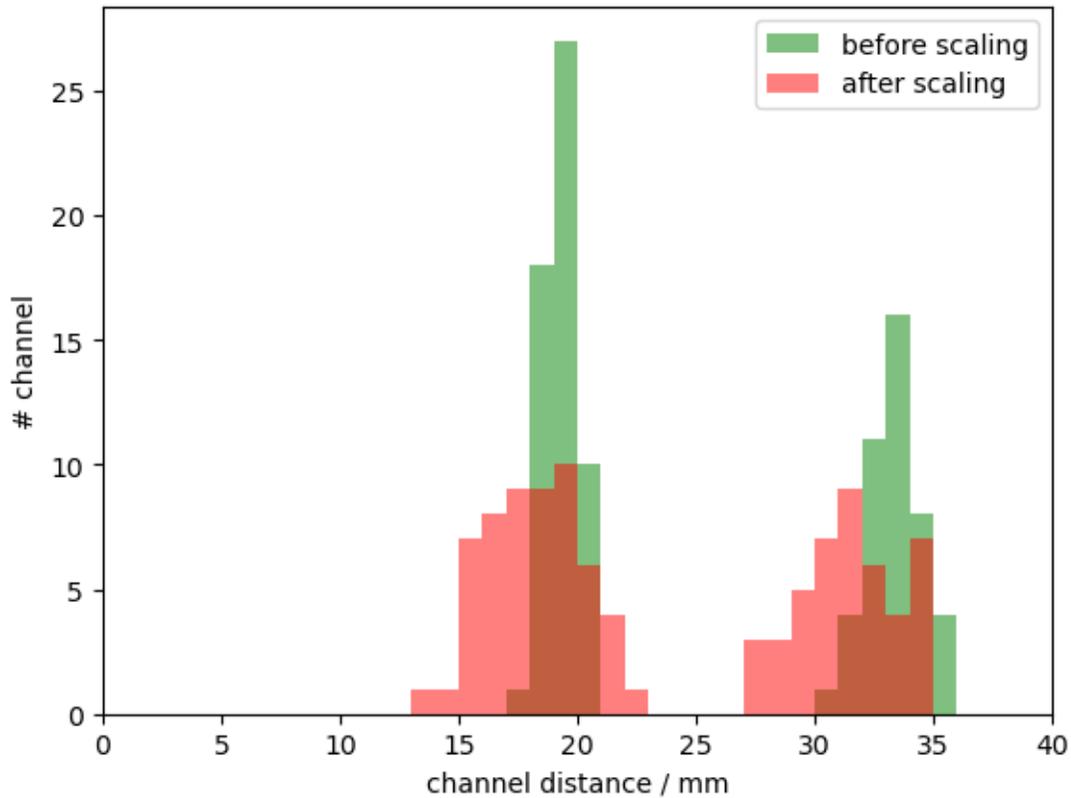

## Simulate light propagation in tissue

`cedalion.dot.ForwardModel` is a wrapper around pmcx and NIRFASTER. It transforms the data in the head model to the input format expected by these photon propagators and offers functionality to calculate the sensitivity matrix.

The forward model operates in voxel space, so we transform the aligned optodes into this CRS.

```
[33]: geo3d_snapped_ijk = geo3d_snapped_ras.points.apply_transform(head_ijk.t_ras2ijk)
```

```
[34]: fwm = cedalion.dot.ForwardModel(head_ijk, geo3d_snapped_ijk, meas_list)
```

## Run the simulation

The `compute_fluence_mcx` and `compute_fluence_nirfaster` methods simulate a light source at each optode position and calculate the fluence in each voxel. By setting `FORWARD_MODEL`, you can choose between the pmcx or NIRFASTer package to perform this simulation.

```
[35]: if PRECOMPUTED_SENSITIVITY:
          Adot = cedalion.data.get_precomputed_sensitivity("fingertappingDOT", HEAD_MODEL)
      else:
          fluence_fname = working_directory / "fluence.h5"
          sensitivity_fname = working_directory / "sensitivity.h5"

          # calculate fluence into fluence file
          if FORWARD_MODEL == "MCX":
```

```
        fwm.compute_fluence_mcx(fluence_fname)
    elif FORWARD_MODEL == "NIRFASTER":
        fwm.compute_fluence_nirfaster(fluence_fname)

    # calculate sensitivity from fluence into sensitivity file
    fwm.compute_sensitivity(fluence_fname, sensitivity_fname)

    # load sensitivity
    Adot = cedalion.io.load_Adot(sensitivity_fname)
```

The result is a sensitivity matrix for each wavelength that describes how changes in absorption at a given brain vertex affect the optical density in a given channel.

[36]: `Adot`

```
[36]: <xarray.DataArray (channel: 100, vertex: 25052, wavelength: 2)> Size: 40MB
      1.212e-17 1.212e-17 3.962e-20 3.962e-20 … 1.96e-18 3.815e-16 3.815e-16
      Coordinates:
          parcel      (vertex) object 200kB 'VisCent_ExStr_8_LH' … 'scalp'
          is_brain    (vertex) bool 25kB True True True True … False False False
        * channel     (channel) object 800B 'S1D1' 'S1D2' 'S1D4' … 'S14D31' 'S14D32'
          source      (channel) object 800B 'S1' 'S1' 'S1' 'S1' … 'S14' 'S14' 'S14'
          detector    (channel) object 800B 'D1' 'D2' 'D4' 'D5' … 'D29' 'D31' 'D32'
        * wavelength  (wavelength) float64 16B 760.0 850.0
      Dimensions without coordinates: vertex
      Attributes:
          units:    mm
```

Visualize the sensitivity

```
[37]: # select only a subset of labeled points to plot
      # Here we select sources and detectors via their labelsthat start with S or D
      geo3d_plot_ijk = geo3d_snapped_ijk.sel(
          label=geo3d_snapped_ijk.label.str.contains("S|D")
      )

      plotter = cedalion.vis.anatomy.sensitivity_matrix.Main(
          sensitivity=Adot,
          brain_surface=head_ijk.brain,
          head_surface=head_ijk.scalp,
          labeled_points= geo3d_plot_ijk,
      )
      plotter.plot(high_th=0, low_th=-3)
      plotter.plt.show()
```

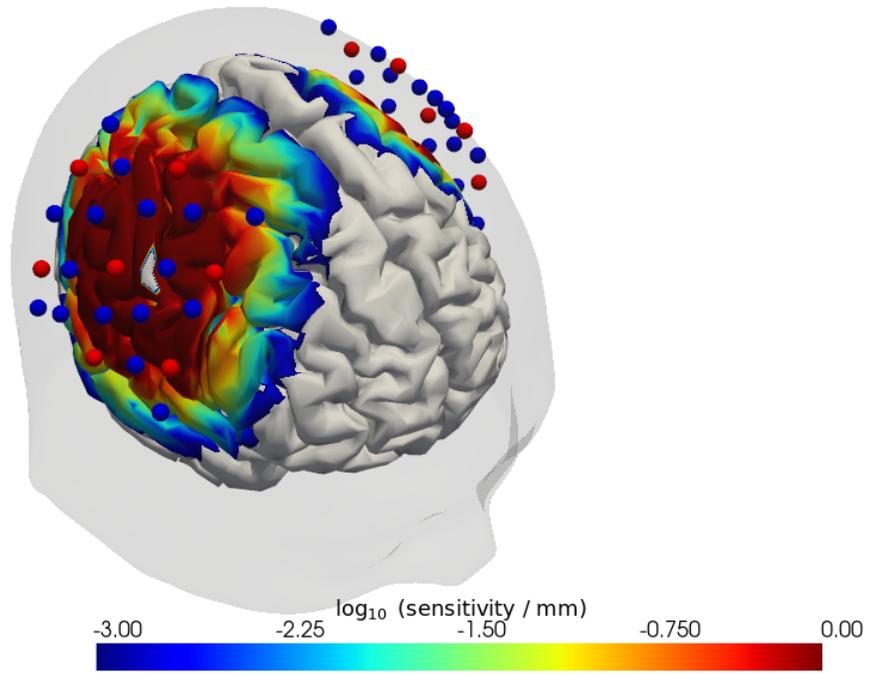

log₁₀ (sensitivity / mm)

# S2: Photogrammetric Optode Co-Registration

Photogrammetry offers a possibility to get subject-specific optode coordinates. This notebook illustrates the individual steps to obtain these coordinates from a textured triangle mesh and a predefined montage.

```python
[1]: # This cells setups the environment when executed in Google Colab.
try:
    import google.colab
    !curl -s https://raw.githubusercontent.com/ibs-lab/cedalion/dev/scripts/
    ↪colab_setup.py -o colab_setup.py
    # Select branch with --branch "branch name" (default is "dev")
    %run colab_setup.py
except ImportError:
    pass
```

```python
[2]: import logging

import matplotlib.image as mpimg
import matplotlib.pyplot as plt

import numpy as np
import pyvista as pv
import xarray as xr

import cedalion
import cedalion.dataclasses as cdc
import cedalion.data
import cedalion.geometry.registration
import cedalion.io
import cedalion.vis
import cedalion.vis.blocks as vbx
from cedalion.vis.anatomy import OptodeSelector
from cedalion.geometry.photogrammetry.processors import (
    ColoredStickerProcessor,
    geo3d_from_scan,
)
from cedalion.geometry.registration import find_spread_points

xr.set_options(display_expand_data=False)
```

```python
[2]: <xarray.core.options.set_options at 0x7c590c075610>
```

## 0. Choose between interactive and static mode

This example notebook provides two modes, controlled by the constant `INTERACTIVE`: - a static mode intended for rendering the documentation - an interactive mode, in which the 3D visualizations react to user input. The camera position can be changed. More importantly, the optode and landmark picking needs these interactive plots.

```python
[3]: INTERACTIVE = False
```

```python
if INTERACTIVE:
    # option 1: render in the browser
    # pv.set_jupyter_backend("client")
    # option 2: offload rendering to a server process using trame
    pv.set_jupyter_backend("server")
else:
    pv.set_jupyter_backend("static")  # static rendering (for documentation page)
```

## 1. Loading the triangulated surface mesh

Use `cedalion.io.read_einstar_obj` to read the textured triangle mesh produced by the Einstar scanner. By default we use an example dataset. By setting the **fname_** variables the notebook can operate on another scan.

```python
# insert here your own files if you do not want to use the example
fname_scan = ""   # path to .obj scan file
fname_snirf = ""  # path to .snirf file for montage information
fname_montage_img = ""  # path to an image file of the montage

if not fname_scan:
    fname_scan, fname_snirf, fname_montage_img = (
        cedalion.data.get_photogrammetry_example_scan()
    )

surface_mesh = cedalion.io.read_einstar_obj(fname_scan)
display(surface_mesh)
```

```
TrimeshSurface(faces: 869980 vertices: 486695 crs: digitized units: millimeter␣
↪vertex_coords: [])
```

## 2. Identifying sticker vertices

Processors are meant to analyze the textured mesh and extract positions. The `ColoredStickerProcessor` searches for colored vertices that form circular areas. We operate in HSV color space and the colors must be specified by their ranges in hue and value. These can be found by usig a color pipette tool on the texture file.

Multiple classes with different colors can be specified. In the following only yellow stickers for class "O(ptode)" are searched. But it could be extended to search also for differently colored sticker. (e.g. "L(andmark)").

For each sticker the center and the normal is derived. Labels are generated from the class name and a counter, e.g. "O-01, O-02, …"

```python
processor = ColoredStickerProcessor(
    colors={
        "O" : ((0.11, 0.21, 0.7, 1)),  # (hue_min, hue_max, value_min, value_max)
        #"L" : ((0.25, 0.37, 0.35, 0.6))
    }
)

sticker_centers, normals, details = processor.process(surface_mesh, details=True)
display(sticker_centers)
```

```
[[ 89.200134    6.045319 647.938599]
 [ 89.200134    5.245316 647.979248]
 [ 91.200134    8.845306 647.925781]
 …
 [124.331436   49.604759 453.649017]
 [122.423584   49.688946 453.390503]
 [124.424706   48.602425 456.144684]]
[[ 91.200134    8.845306 647.925781]
 [ 89.600143    3.965302 647.806641]
 [ 90.320129    3.245316 647.687866]
 …
 [ 34.000137  -95.936989 536.323303]
 [ 34.80014   -96.269691 536.323303]
 [ 34.400146  -96.094498 536.323303]]
O (0.11, 0.21, 0.7, 1)

[0.51869679 0.52571218 0.09720632 0.09024364 0.0031426  0.26716014
 0.15729909 0.15603078 0.18434642 0.21127269 0.6031898  0.15173879
 0.13589549 0.5367809  0.08970815 0.43860341 0.24223317 0.22831593
 0.11666005 0.38538707 0.06208547 0.26157969 0.21756175 0.33913646
 0.18674489 0.1367986  0.19878102 0.00108876 0.41054799 0.15589406
 0.11664845 0.2340569  0.15650441 0.19405528 0.25735362 0.276244
 0.02862721 0.35069713 0.51928721 0.2630901  0.0427068  0.27555675
 0.20007741 0.29776479 0.09225413 0.21428844 0.05078502 0.29102101
 0.35556447 0.81571727 0.09213511 0.31188405 0.34762441 0.0583965
 0.97653422 0.14968739 0.19396413 0.22924693 0.59414052 0.22320201
 0.79269725]
6.004194440998015
surface.crs digitized

<xarray.DataArray (label: 61, digitized: 3)> Size: 1kB
[mm] 168.2 -1.85 604.7 72.76 -29.73 641.0 … -89.01 501.7 203.5 -9.854 514.2
Coordinates:
  * label    (label) <U4 976B 'O-28' 'O-05' 'O-37' … 'O-61' 'O-50' 'O-55'
    type     (label) object 488B PointType.UNKNOWN … PointType.UNKNOWN
    group    (label) <U1 244B 'O' 'O' 'O' 'O' 'O' … 'O' 'O' 'O' 'O' 'O'
Dimensions without coordinates: digitized
```

Visualize the surface and extraced results.

```python
camera_pos = sticker_centers.mean("label").pint.dequantify() - np.array([-500,0,0])
camera_focal_point = sticker_centers.mean("label").pint.dequantify()
camera_up = (0., 0. ,1.)

pvplt = pv.Plotter()
vbx.plot_surface(pvplt, surface_mesh, opacity=1.0)
vbx.plot_labeled_points(pvplt, sticker_centers, color="r")
vbx.plot_vector_field(pvplt, sticker_centers, normals)

pvplt.camera.position = camera_pos
pvplt.camera.focal_point = camera_focal_point
pvplt.camera.up = camera_up
```

```
pvplt.show()
```

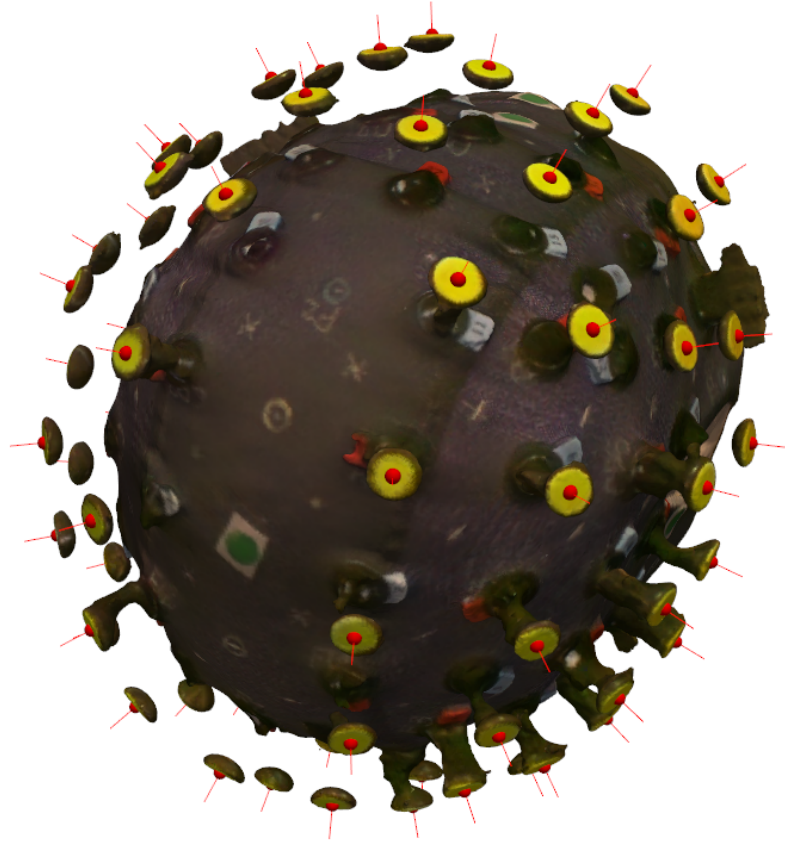

The details object is a container for debuging information. It also provides plotting functionality.

The following scatter plot shows the vertex colors in the hue-value plane in which the vertex classification operates.

The black rectangle illustrates the classification criterion.

```
[7]: details.plot_vertex_colors()
```

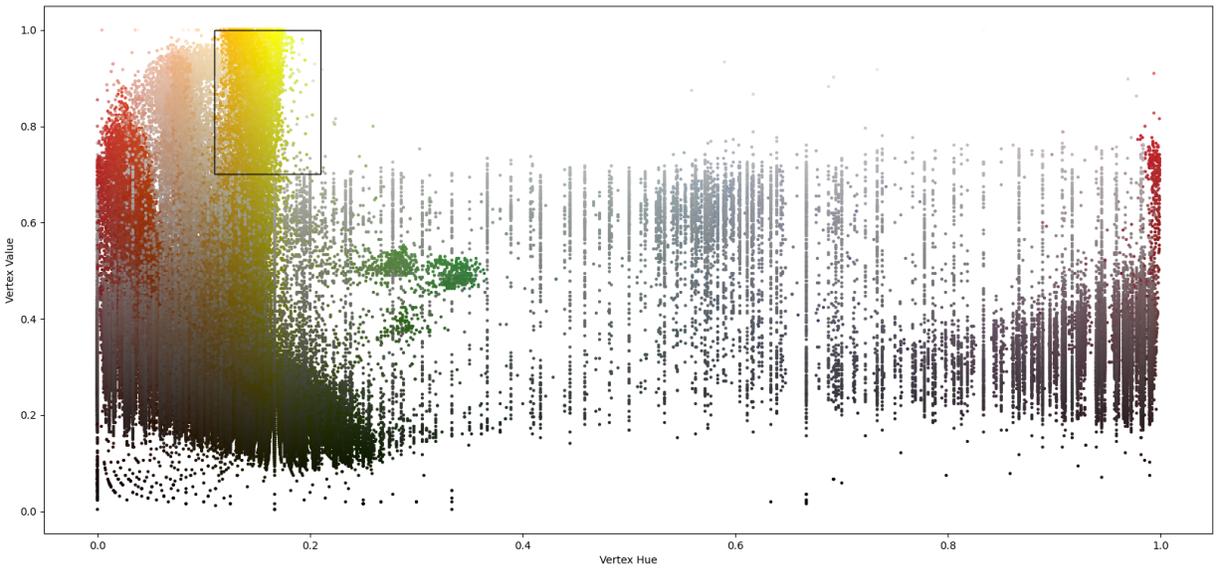

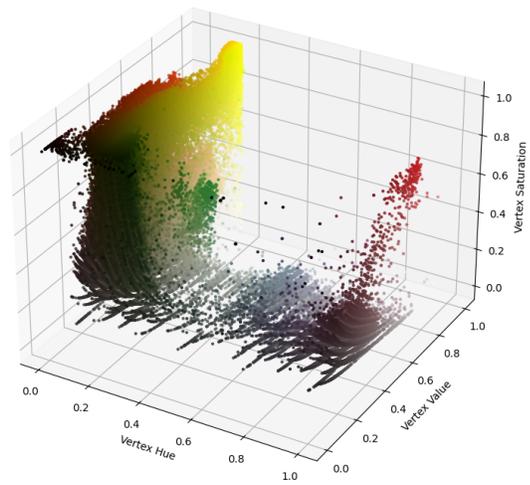

The following plots show for each cluster (tentative group of sticker vertices) The vertex positions perpendicular to the sticker normal as well as the minimum enclosing circle which is used to find the sticker's center.

```
[8]: details.plot_cluster_circles()
```

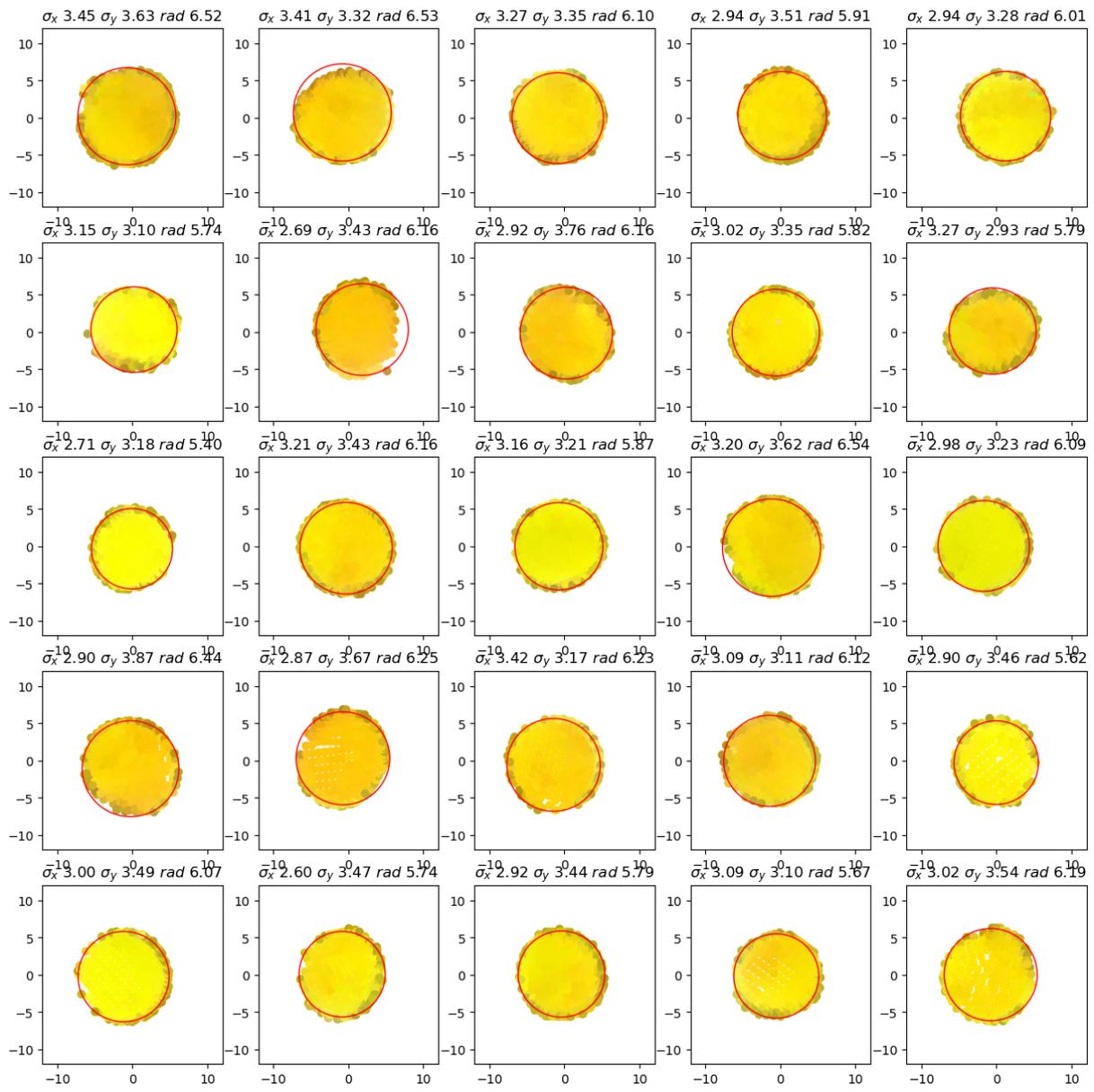

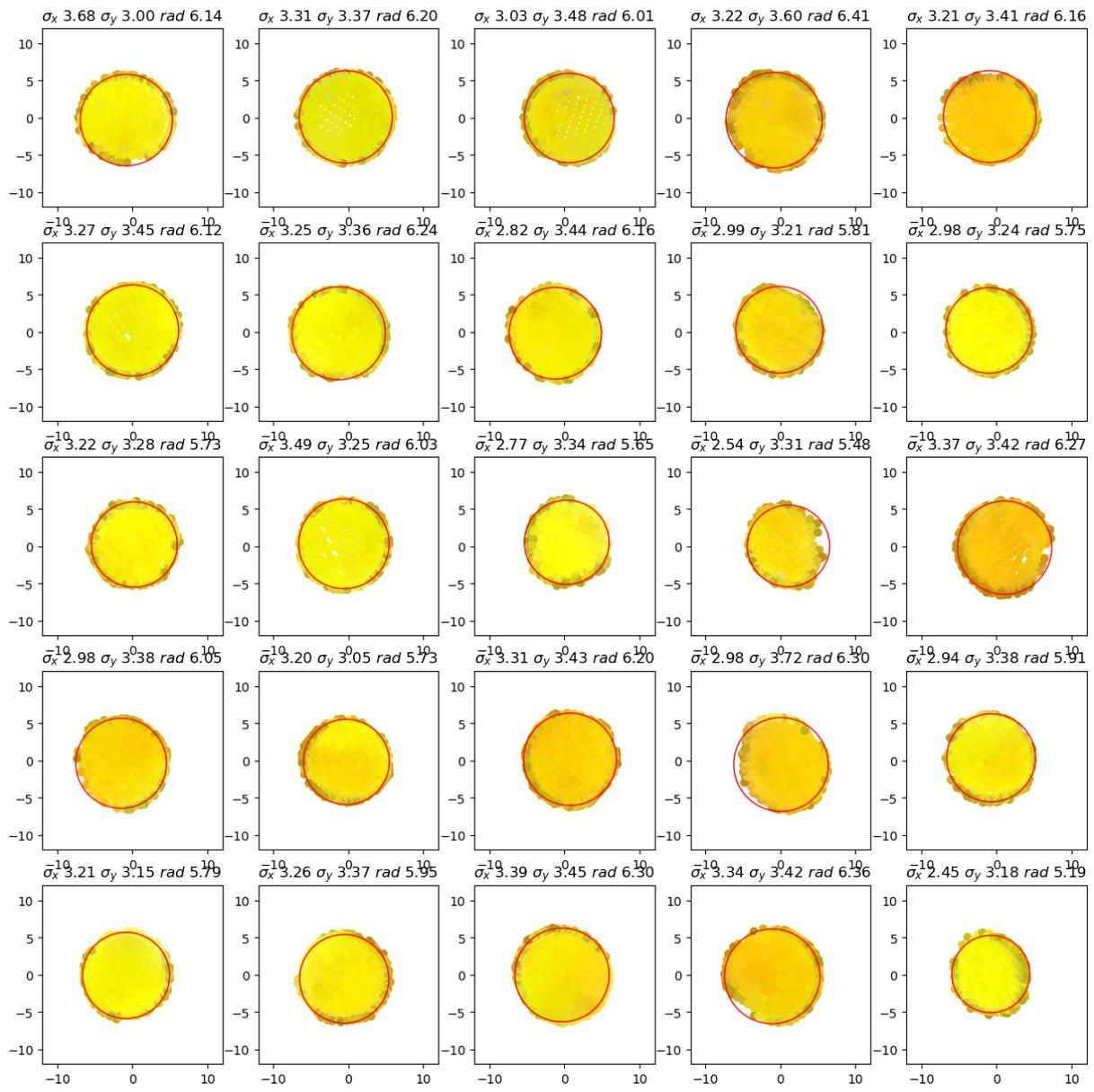

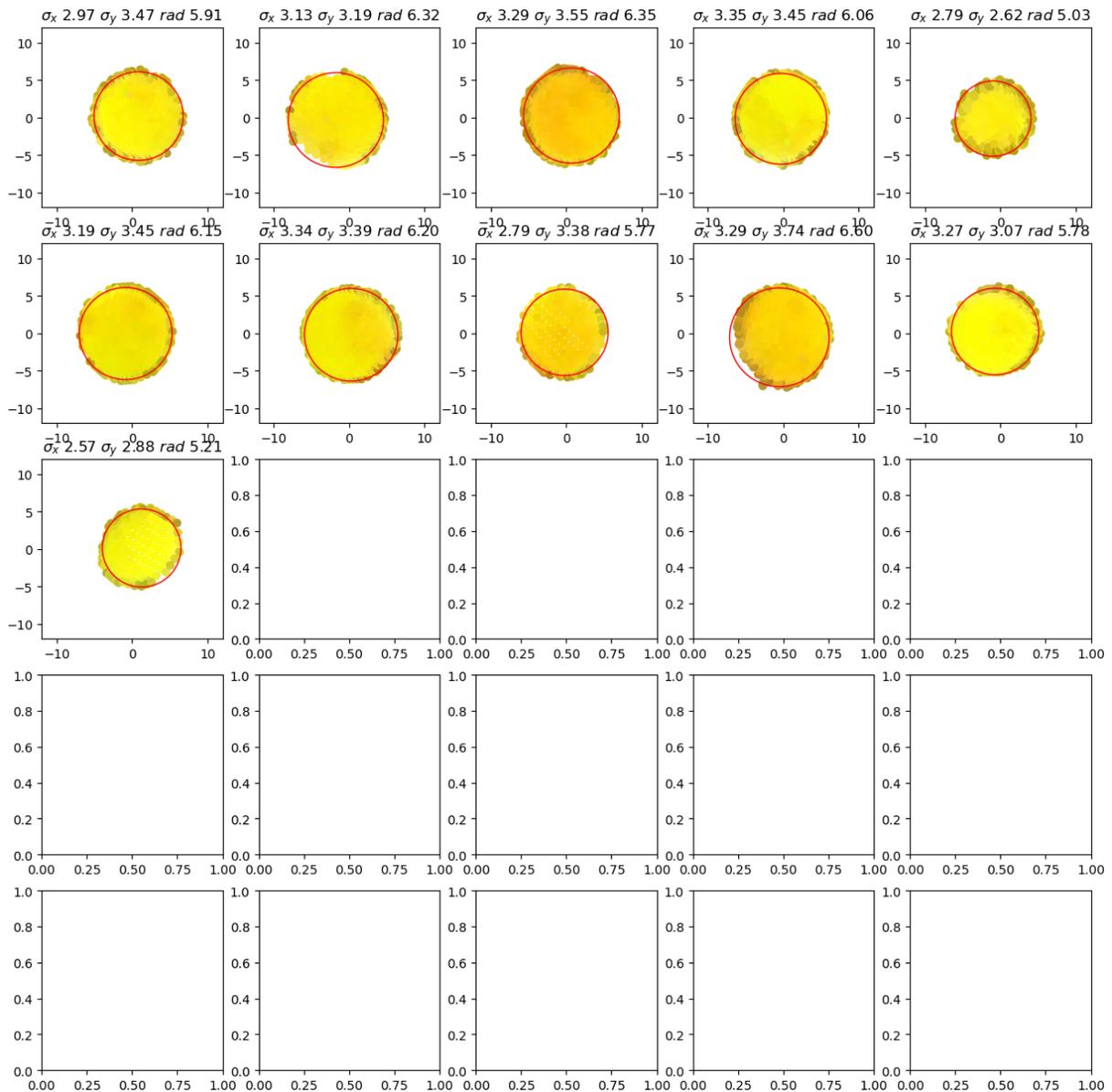

### 3. Manual corrections of sticker detection

If not all optodes were found automatically, there's way to remove or add them manually.

The `OptodeSelect` class provides an interactive visualization of the head scan and the detected stickers (red spheres):

By clicking with the right mouse button on: - a sphere, a misidentified sticker can be removed. - somewhere on the surface, a new sticker position can be added.

```
[9]: optode_selector = OptodeSelector(surface_mesh, sticker_centers, normals)
     optode_selector.plot()
     optode_selector.enable_picking()
     vbx.plot_surface(optode_selector.plotter, surface_mesh, opacity=1.0)

     optode_selector.plotter.show()
```

Right click to place or remove optode

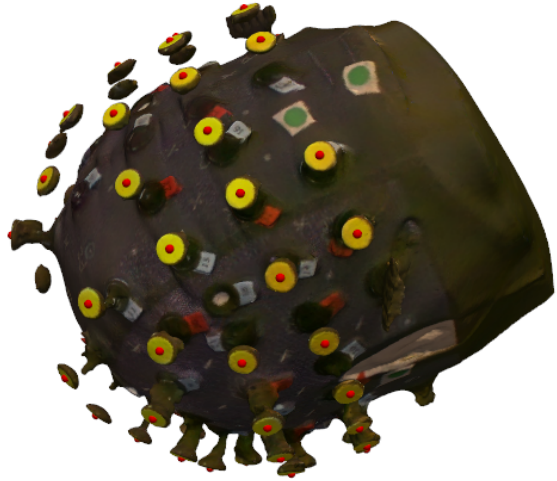

Interactions modify the `optode_selector.points` and `optode_selector.normals`. After selecting all optodes, update `sticker_centers` and `normals`:

```
[10]:  sticker_centers = optode_selector.points
       normals = optode_selector.normals
       display(sticker_centers)
```

```
<xarray.DataArray (label: 61, digitized: 3)> Size: 1kB
[mm] 168.2 -1.85 604.7 72.76 -29.73 641.0 … -89.01 501.7 203.5 -9.854 514.2
Coordinates:
  * label    (label) <U4 976B 'O-28' 'O-05' 'O-37' … 'O-61' 'O-50' 'O-55'
    type     (label) object 488B PointType.UNKNOWN … PointType.UNKNOWN
    group    (label) <U1 244B 'O' 'O' 'O' 'O' 'O' 'O' … 'O' 'O' 'O' 'O' 'O'
Dimensions without coordinates: digitized
```

## 4. Project from sticker to scalp surface

Finally, to get from the sticker centers to the scalp coordinates we have to subtract the known lenght of the optodes in the direction of the normals:

```
[11]:  optode_length = 22.6 * cedalion.units.mm

       scalp_coords = sticker_centers.copy()
       mask_optodes = sticker_centers.group == "O"
```

```
scalp_coords[mask_optodes] = (
    sticker_centers[mask_optodes] - optode_length * normals[mask_optodes]
)
# we make a copy of this raw set of scalp coordinates to use later in the 2nd case of
# the coregistration example that showcases an alternative route if landmark-based
# coregistration fails
scalp_coords_altcase = scalp_coords.copy()

display(scalp_coords)
```

```
<xarray.DataArray (label: 61, digitized: 3)> Size: 1kB
[mm] 152.5 -3.156 588.5 70.89 -19.87 620.8 … -70.83 507.3 181.5 -9.023 519.2
Coordinates:
  * label    (label) <U4 976B 'O-28' 'O-05' 'O-37' … 'O-61' 'O-50' 'O-55'
    type     (label) object 488B PointType.UNKNOWN … PointType.UNKNOWN
    group    (label) <U1 244B 'O' 'O' 'O' 'O' 'O' 'O' … 'O' 'O' 'O' 'O' 'O'
Dimensions without coordinates: digitized
```

Visualize sticker centers (red) and scalp coordinates (green).

```
[12]: pvplt = pv.Plotter()
      vbx.plot_surface(pvplt, surface_mesh, opacity=0.3)
      vbx.plot_labeled_points(pvplt, sticker_centers, color="r")
      vbx.plot_labeled_points(pvplt, scalp_coords, color="g")
      vbx.plot_vector_field(pvplt, sticker_centers, normals)
      pvplt.show()
```

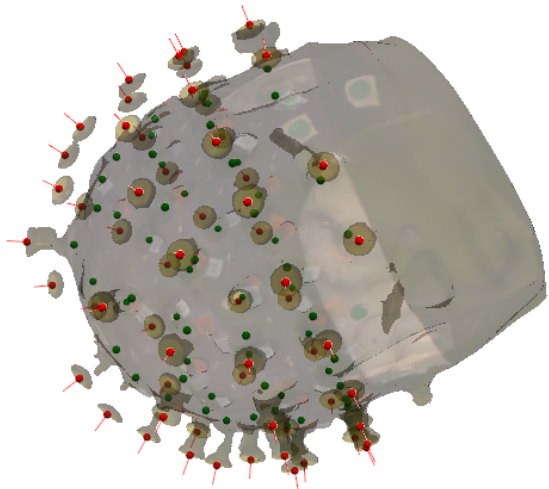

## 5. Specify landmarks on scanned head surface

### 5.1. Pick positions in interactive plot

When using the `plot_surface` function with parameter `pick_landmarks` set to *True*, the plot becomes interactive and allows to pick the positions of 5 landmarks. These are "Nz", "Iz", "Cz", "Lpa", "RpA".

After clicking on the mesh, a green sphere marks the picked location. The sphere has a label attached. If this label is not visible, try to zoom further into the plot (mouse wheel). By clicking again with right mouse button on the sphere one can cycle through the different labels or remove a misplaced landmark.

It halps to add colored markers at the landmark positions when preparing the subject. Here green stickers where used.

```
[13]:   pvplt = pv.Plotter()
        get_landmarks = vbx.plot_surface(pvplt, surface_mesh, opacity=1.0, pick_landmarks=True)
        pvplt.show()
```

Right click to place or change the landmark label

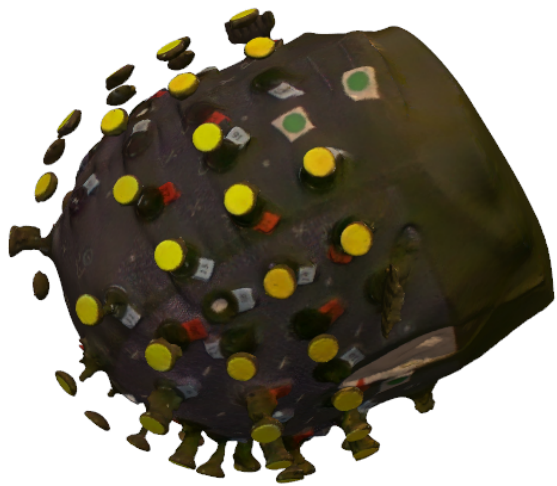

### 5.2. Retrieve picked positions from interactive plot

The `plot_surface` function returns a function `get_landmarks`. Call this function to obtain: * 1st value - coordinates of picked landmarks * 2nd - labels of corresponding landmarks

```
[14]:  if INTERACTIVE:
           landmark_coordinates, landmark_labels = get_landmarks()
       else:
           # For documentation purposes and to enable automatically rendered example notebooks
           # we provide the hand-picked coordinates here, too.
           landmark_labels = ["Nz", "Iz", "Cz", "Lpa", "Rpa"]
           landmark_coordinates = np.asarray(
               [
                   [14.00420712, -7.84856869, 449.77840004],
                   [99.09920059, 29.72154755, 620.73876117],
                   [161.63815139, -48.49738938, 494.91210993],
                   [82.8771277, 79.79500128, 498.3338802],
                   [15.17214095, -60.56186128, 563.29621021],
               ]
           )

       display(landmark_labels)
       display(landmark_coordinates)

       assert len(set(landmark_labels)) == 5, "please select 5 landmarks"
```

```
['Nz', 'Iz', 'Cz', 'Lpa', 'Rpa']

array([[ 14.00420712,  -7.84856869, 449.77840004],
       [ 99.09920059,  29.72154755, 620.73876117],
       [161.63815139, -48.49738938, 494.91210993],
       [ 82.8771277 ,  79.79500128, 498.3338802 ],
       [ 15.17214095, -60.56186128, 563.29621021]])
```

### 5.3 Wrap landmark positions and labels in a xarray.DataArray structure

- insert *landmark_coordinates* and *landmark_labels*

```
[15]:  coordinates = landmark_coordinates
       labels = landmark_labels

       types = [cdc.PointType.LANDMARK] * 5
       groups = ["L"] * 5

       landmarks = xr.DataArray(
           np.vstack(coordinates),
           dims=["label", "digitized"],
           coords={
               "label": ("label", labels),
               "type": ("label", types),
               "group": ("label", groups),
           },
       ).pint.quantify("mm")

       display(landmarks)
```

```
<xarray.DataArray (label: 5, digitized: 3)> Size: 120B
[mm] 14.0 -7.849 449.8 99.1 29.72 620.7 … 82.88 79.8 498.3 15.17 -60.56 563.3
```

```
Coordinates:
  * label      (label) <U3 60B 'Nz' 'Iz' 'Cz' 'Lpa' 'Rpa'
    type       (label) object 40B PointType.LANDMARK … PointType.LANDMARK
    group      (label) <U1 20B 'L' 'L' 'L' 'L' 'L'
Dimensions without coordinates: digitized
```

## 6. Mapping the scanned optode positions to a predefined montage.

So far the optode positions found in the photogrammetric head scan carry only generic labels. In oder to identify them, they must be matched with a definition of the actual montage.

Snirf files store next to the actual time series data also the probe geometry, i.e. 3D coordinates of each source and detector. To label the optodes found in the photogrammetric scan, we map each optode to its counterpart in the snirf file.

The snirf coordinates are written during the data acquisition and are typically obtained by arranging the montage on a template head like ICBM-152 or colin27. So despite their similarity, the probe geometries in the snirf file and those from the head scan have differences because of different head geometries aand different coordinate systems.

### 6.1 Load the montage information from .snirf file

```python
[16]: # read the example snirf file. Specify a name for the coordinate reference system.
      rec = cedalion.io.read_snirf(fname_snirf, crs="aligned")[0]

      # read 3D coordinates of the optodes
      montage_elements = rec.geo3d

      # landmark labels must match exactly. Adjust case where they don't match.
      montage_elements = montage_elements.points.rename({"LPA": "Lpa", "RPA": "Rpa"})
```

### 6.2 Find a transformation to align selected landmarks to montage coordinates

The coordinates in the snirf file and from the photogrammetric scan use different coordinate reference systems (CRS). In Cedalion the user needs to explicitly name different CRSs. Here the labels 'digitized' and 'aligned' were used.

The following plot shows the probe geometry from the snirf file and the landmarks from the head scan. Two black lines Nz-Iz and Lpa-Rpa are added to guide the eye.

```python
[17]: f = plt.figure(figsize=(12,5))
      ax1 = f.add_subplot(1,2,1, projection="3d")
      ax2 = f.add_subplot(1,2,2, projection="3d")
      colors = {cdc.PointType.SOURCE: "r", cdc.PointType.DETECTOR: "b"}
      sizes = {cdc.PointType.SOURCE: 20, cdc.PointType.DETECTOR: 20}

      for i, (type, x) in enumerate(montage_elements.groupby("type")):
          x = x.pint.to("mm").pint.dequantify()
          ax1.scatter(x[:, 0], x[:, 1], x[:, 2], c=colors.get(type, "g"), s=sizes.get(type,
      ↪2))

      for i, (type, x) in enumerate(landmarks.groupby("type")):
          x = x.pint.to("mm").pint.dequantify()
          ax2.scatter(x[:, 0], x[:, 1], x[:, 2], c=colors.get(type, "g"), s=20)
```

```
for ax, points in [(ax1, montage_elements), (ax2, landmarks)]:
    points = points.pint.to("mm").pint.dequantify()
    ax.plot([points.loc["Nz",0], points.loc["Iz",0]],
            [points.loc["Nz",1], points.loc["Iz",1]],
            [points.loc["Nz",2], points.loc["Iz",2]],
            c="k"
            )
    ax.plot([points.loc["Lpa",0], points.loc["Rpa",0]],
            [points.loc["Lpa",1], points.loc["Rpa",1]],
            [points.loc["Lpa",2], points.loc["Rpa",2]],
            c="k"
            )

ax1.set_title(f"from snirf | crs: {montage_elements.points.crs}")
ax2.set_title(f"from scan | crs: {landmarks.points.crs}");
```

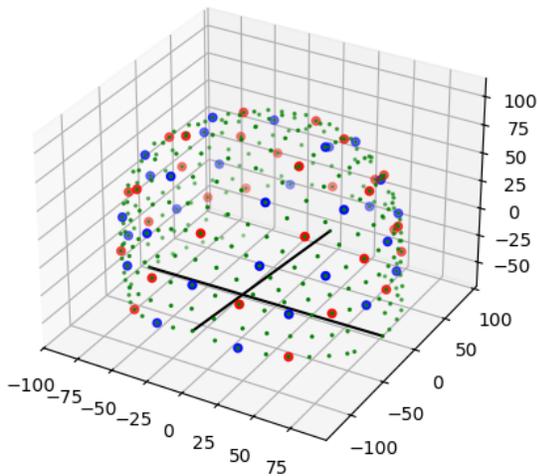
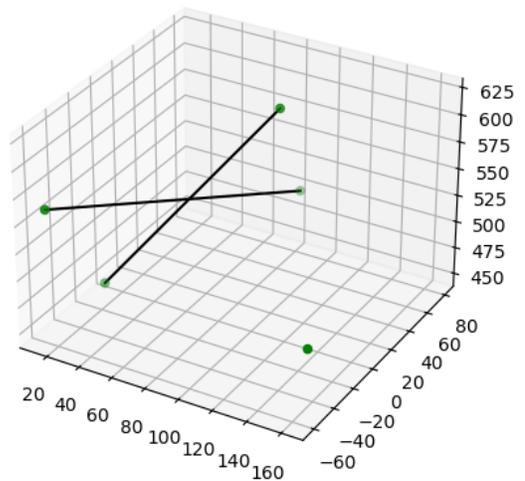

Subsequently, to bring the coordinates into the same space, from the landmarks a transformation (translations and rotations) is derived. This transforms the coordinates from the snirf file to the CRS of the photogramettric scan.

The following plot illustrates the transformed coordinates of sources (red) and detectors (blue). Deviations between these coordinates and the head surface are expected, since the optode positions where specified on a different head geometry.

[18]:
```
trafo = cedalion.geometry.registration.register_trans_rot(landmarks, montage_elements)

filtered_montage_elements = montage_elements.where(
    (montage_elements.type == cdc.PointType.SOURCE)
    | (montage_elements.type == cdc.PointType.DETECTOR),
    drop=True,
)
filtered_montage_elements_t = filtered_montage_elements.points.apply_transform(trafo)
```

```
pvplt = pv.Plotter()
vbx.plot_surface(pvplt, surface_mesh, color="w", opacity=.2)
vbx.plot_labeled_points(pvplt, filtered_montage_elements_t)
pvplt.show()
```

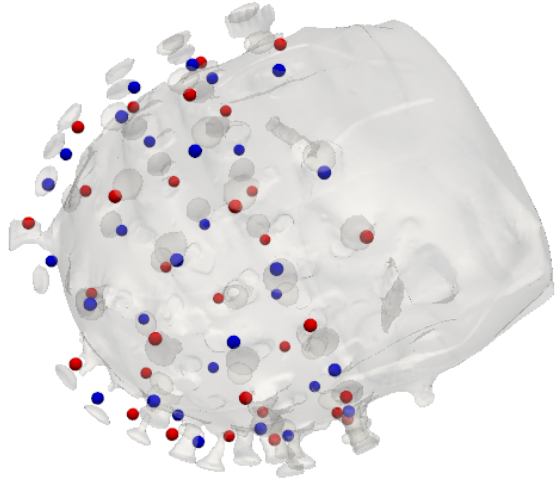

### 6.3 Iterative closest point algorithm to find labels for detected optode centers

Finally, the mapping is derived by iteratively trying to find a transformation that yilds the best match between the snirf and the scanned coordinates.

The following plot visualizes the result: * Green points represent optode centers * Next to them there shall be labels assumed by ICP algorithm (*show_labels = True*)

```
[19]: # iterative closest point registration
idx = cedalion.geometry.registration.icp_with_full_transform(
    scalp_coords, filtered_montage_elements_t, max_iterations=100
)

# extract labels for detected optodes
label_dict = {}
for i, label in enumerate(filtered_montage_elements.coords["label"].values):
    label_dict[i] = label
```

```python
labels = [label_dict[index] for index in idx]

# write labels to scalp_coords
scalp_coords = scalp_coords.assign_coords(label=labels)

# add landmarks
geo3Dscan = geo3d_from_scan(scalp_coords, landmarks)

display(geo3Dscan)
```

```
<xarray.DataArray (label: 66, digitized: 3)> Size: 2kB
[mm] 152.5 -3.156 588.5 70.89 -19.87 620.8 … 79.8 498.3 15.17 -60.56 563.3
Coordinates:
    type     (label) object 528B PointType.SOURCE … PointType.LANDMARK
    group    (label) <U1 264B 'O' 'O' 'O' 'O' 'O' 'O' … 'L' 'L' 'L' 'L' 'L'
  * label    (label) <U3 792B 'S21' 'S23' 'D13' 'D8' … 'Iz' 'Cz' 'Lpa' 'Rpa'
Dimensions without coordinates: digitized
Attributes:
    units:    mm
```

```python
[20]: f,ax = plt.subplots(1,2, figsize=(12,6))
cedalion.vis.anatomy.scalp_plot(
    rec["amp"],
    montage_elements,
    cedalion.nirs.channel_distances(rec["amp"], montage_elements),
    ax=ax[0],
    optode_labels=True,
    cb_label="channel dist. / mm",
    cmap="plasma",
    vmin=25,
    vmax=42,
)
ax[0].set_title("montage from snirf file")
cedalion.vis.anatomy.scalp_plot(
    rec["amp"],
    geo3Dscan,
    cedalion.nirs.channel_distances(rec["amp"], geo3Dscan),
    ax=ax[1],
    optode_labels=True,
    cb_label="channel dist. / mm",
    cmap="plasma",
    vmin=25,
    vmax=42,
)
ax[1].set_title("montage from photogrammetric scan")
plt.tight_layout()
```

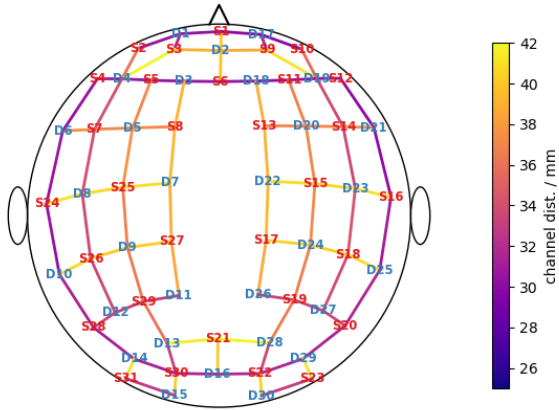 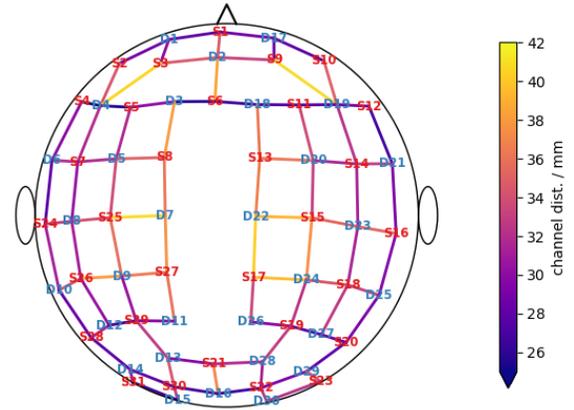

Visualization of successfull assignment *(show_labels = True)*

```
[21]: pvplt = pv.Plotter()
      cedalion.vis.anatomy.plot_brain_and_scalp(None, surface_mesh.mesh, None, None,
       ↪plotter=pvplt)
      vbx.plot_labeled_points(pvplt, geo3Dscan, show_labels=True)
      pvplt.show()
```

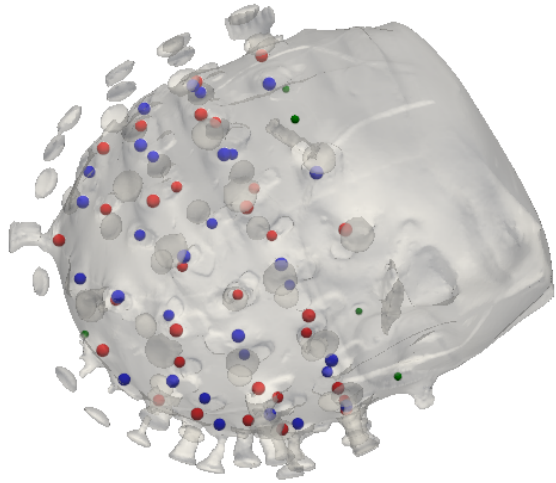

### 6.4 Alternative approach without landmarks

Mapping the optode labels can fail for example because of a bad landmark selection.

In such cases it is possible to find a new transformation by manually labeling three optodes. This is done by selecting them in a given order. For that it helps to have a visualization of the montage of your experiment.

```python
[22]: if fname_montage_img:
          # Load and display the image
          img = mpimg.imread(fname_montage_img)
          plt.figure(figsize=(12, 10))
          plt.imshow(img)
          plt.axis("off")   # Turn off axis labels and ticks
          plt.show()
      else:
          print("No montage image specified.")
```

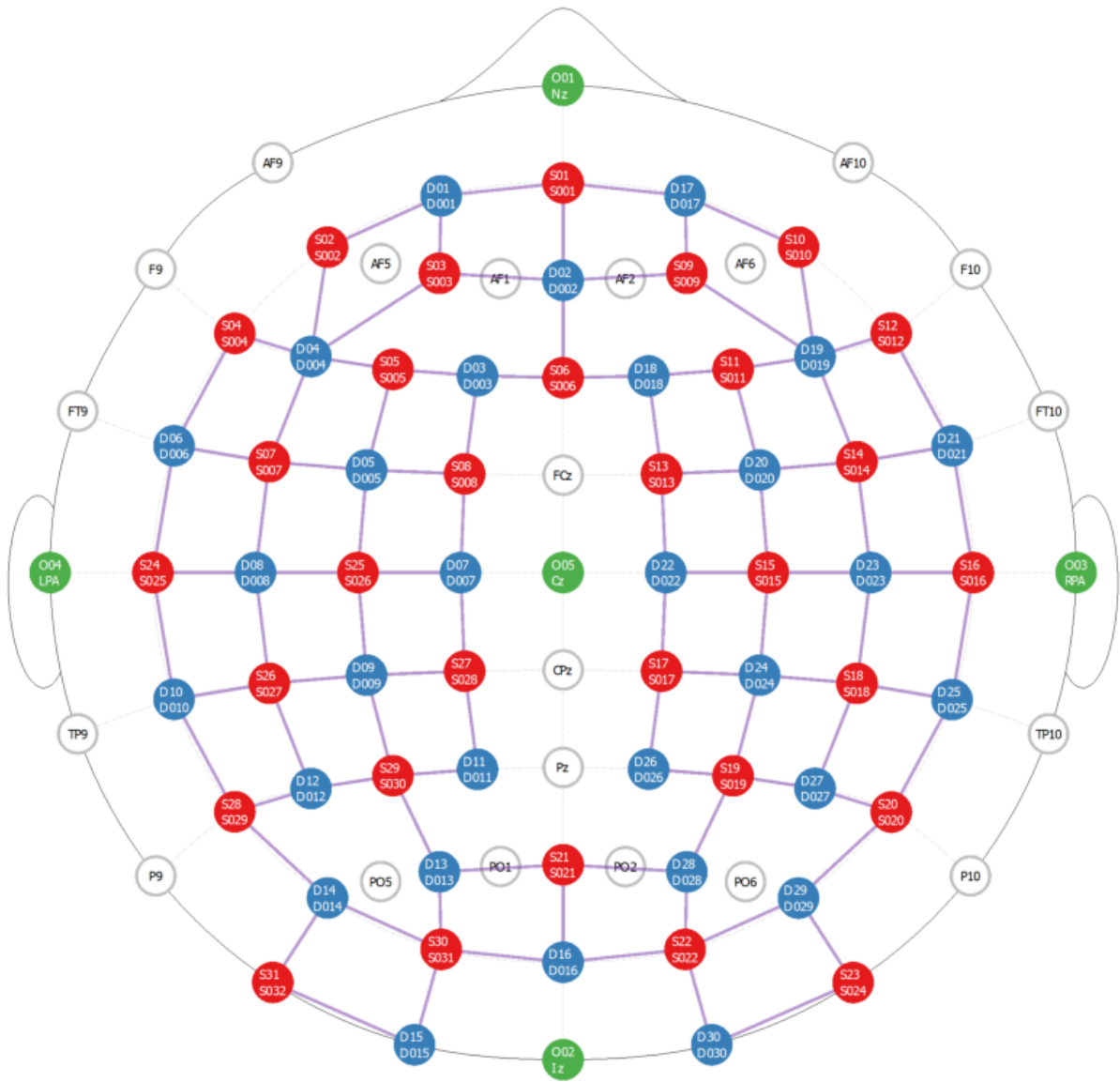

Search for three optodes that are evenly spreaded across the head surface. Afterwards prompt the uer to right click on each of them.

```
[23]: spread_point_labels = find_spread_points(filtered_montage_elements)
      print("Select those points")
      print(spread_point_labels)

      points = []
      pvplt = pv.Plotter()
      vbx.plot_surface(pvplt, surface_mesh, opacity=1.0)
      vbx.plot_labeled_points(pvplt, sticker_centers, color="r", ppoints = points)
      pvplt.show()
```

/opt/miniconda3/envs/cedalion_250922/lib/python3.11/site-
packages/xarray/core/variable.py:315: UnitStrippedWarning: The unit of the quantity is

```
stripped when downcasting to ndarray.
  data = np.asarray(data)

Select those points
['S1' 'D30' 'S31']
```

Right-click or press P to pick under the mouse

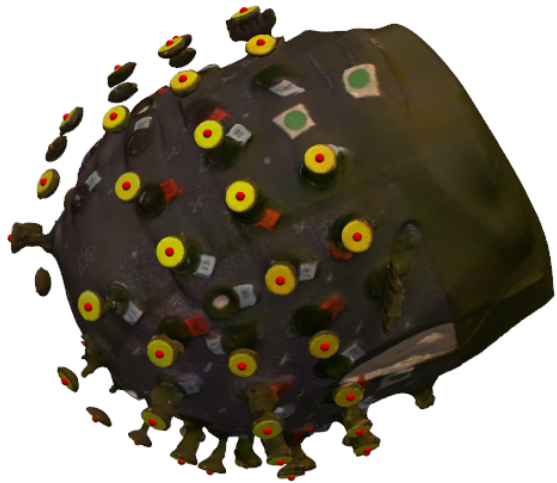

Retrieve picked positions:

```
[24]: if INTERACTIVE:
          labeled_points = points
      else:
          # For documentation purposes and to enable automatically rendered example notebooks
          # we provide the hand-picked coordinates here, too.
          labeled_points = [19, 52, 50]

      labeled_points
```

```
[24]: [19, 52, 50]
```

Write the selected labels to the corresponding points of **xarray.DataArray** scalp_coords:

```
[25]: new_labels = scalp_coords_altcase.label.values.copy()
      for i, idx in enumerate(labeled_points):
          new_labels[idx] = spread_point_labels[i]
```

```python
scalp_coords_altcase = scalp_coords_altcase.assign_coords(label=new_labels)
scalp_coords_altcase
```

[25]: <xarray.DataArray (label: 61, digitized: 3)> Size: 1kB
      [mm] 152.5 -3.156 588.5 70.89 -19.87 620.8 … -70.83 507.3 181.5 -9.023 519.2
      Coordinates:
          type     (label) object 488B PointType.UNKNOWN … PointType.UNKNOWN
          group    (label) <U1 244B '0' '0' '0' '0' '0' '0' … '0' '0' '0' '0' '0'
        * label    (label) <U4 976B 'O-28' 'O-05' 'O-37' … 'O-61' 'O-50' 'O-55'
      Dimensions without coordinates: digitized

Find the affine transformation for the newly labeled points and apply it to the montage optodes

```python
trafo2 = cedalion.geometry.registration.register_trans_rot(
    scalp_coords_altcase, montage_elements
)

filtered_montage_elements = montage_elements.where(
    (montage_elements.type == cdc.PointType.SOURCE)
    | (montage_elements.type == cdc.PointType.DETECTOR),
    drop=True,
)
filtered_montage_elements_t = filtered_montage_elements.points.apply_transform(trafo2)
```

and run ICP algorithm for label assignment once again, extract labels for detected optodes and plot the results

```python
# iterative closest point registration
idx = cedalion.geometry.registration.icp_with_full_transform(
    scalp_coords_altcase, filtered_montage_elements_t, max_iterations=100
)
# extract labels for detected optodes
label_dict = {}
for i, label in enumerate(filtered_montage_elements.coords["label"].values):
    label_dict[i] = label
labels = [label_dict[index] for index in idx]

# write labels to scalp_coords
scalp_coords_altcase = scalp_coords_altcase.assign_coords(label=labels)

# add landmarks
geo3Dscan_alt = geo3d_from_scan(scalp_coords_altcase, landmarks)
```

```python
f,ax = plt.subplots(1,2, figsize=(12,6))
cedalion.vis.anatomy.scalp_plot(
    rec["amp"],
    montage_elements,
    cedalion.nirs.channel_distances(rec["amp"], montage_elements),
    ax=ax[0],
    optode_labels=True,
    cb_label="channel dist. / mm",
    cmap="plasma",
```

```
    vmin=25,
    vmax=42,
)

ax[0].set_title("montage from snirf file")
cedalion.vis.anatomy.scalp_plot(
    rec["amp"],
    geo3Dscan_alt,
    cedalion.nirs.channel_distances(rec["amp"], geo3Dscan_alt),
    ax=ax[1],
    optode_labels=True,
    cb_label="channel dist. / mm",
    cmap="plasma",
    vmin=25,
    vmax=42,
)
ax[1].set_title("montage from photogrammetric scan")
plt.tight_layout()
```

montage from snirf file

montage from photogrammetric scan

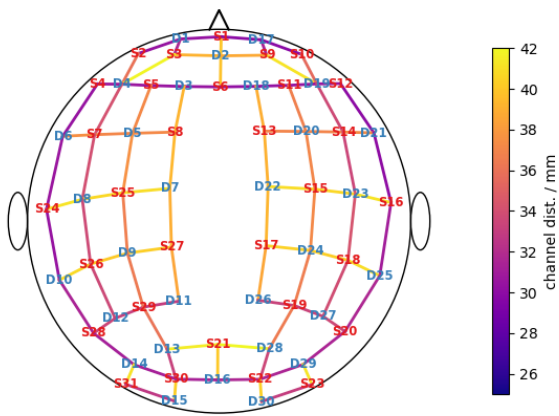
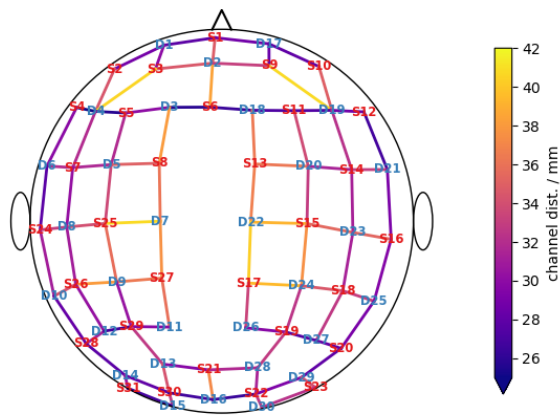

# S3: Signal Processing

This notebook demonstrates Cedalion's capabilities to assess signal quality and correct motion artifacts.

Several signal quality metrics are implemented in the package `cedalion.sigproc.quality`. From these metrics boolean masks are created which indicate whether the quality of a given time point or segment is acceptable or not. By combining these boolean masks, complex selection criteria can be formulated.

```
[1]: # This cells setups the environment when executed in Google Colab.
     try:
         import google.colab
         !curl -s https://raw.githubusercontent.com/ibs-lab/cedalion/dev/scripts/
     ↪colab_setup.py -o colab_setup.py
         # Select branch with --branch "branch name" (default is "dev")
         %run colab_setup.py
     except ImportError:
         pass
```

```
[2]: import cedalion
     import cedalion.data
     import cedalion.sigproc.motion as motion
     import cedalion.sigproc.quality as quality
     import cedalion.vis.anatomy
     import cedalion.vis.blocks as vbx
     import cedalion.vis.colors as colors
     import matplotlib.pyplot as p
     import numpy as np
     import pandas as pd
     import xarray as xr
     from cedalion import units
     from cedalion.vis.quality import plot_quality_mask

     xr.set_options(display_expand_data=False)
```

```
[2]: <xarray.core.options.set_options at 0x77e30fbde310>
```

## Load Data

The example starts by loading an example datasets via `cedalion.data`:

```
[3]: rec = cedalion.data.get_fingertappingDOT()
```

## Recording Container

The `Recording` container carries time series and related objects through the program in ordered dictionaries.

The dataset contains a single time series of fNIRS raw amplitudes. This is stored in the attribute `rec.timeseries` with key `'amp'`:

```
[4]: rec.timeseries.keys()
```

```
[4]: odict_keys(['amp'])
```

Among the data stored in the `Recording`container, the time series are access most frequently. Therefore, a shortcut is provided. The user can access items in `.timeseries` directly on the `Recording` container:

```
[5]: rec["amp"] is rec.timeseries["amp"]
```

```
[5]: True
```

Next to the fNIRS time series, the dataset contains also time series data from auxiliary sensors which are stored in `.aux_ts`:

```
[6]: rec.aux_ts.keys()
```

```
[6]: odict_keys(['ACCEL_X_1', 'ACCEL_Y_1', 'ACCEL_Z_1', 'GYRO_X_1', 'GYRO_Y_1', 'GYRO_Z_1',
     'ExGa1', 'ExGa2', 'ExGa3', 'ExGa4', 'ECG', 'Respiration', 'PPG', 'SpO2', 'Heartrate',
     'GSR', 'Temperature'])
```

Information about stimulus events is stored in a `pandas.DataFrame` under `.stim`. For each stimulus event the onset and duration is stored in seconds. Each event also includes a value indicating stimulus strength (e.g., the loudness of an auditory stimulus), which can be used to scale the amplitude of the modeled hemodynamic response. Finally, the trial type string label allows distinguishing different event types.

```
[7]: rec.stim
```

```
[7]:          onset  duration  value  trial_type
     0     23.855104      10.0    1.0           1
     1     54.132736      10.0    1.0           1
     2     84.410368      10.0    1.0           1
     3    114.688000      10.0    1.0           1
     4    146.112512      10.0    1.0           1
     ..          ...       ...    ...         ...
     125  1431.535616      10.0    1.0           5
     126  1526.038528      10.0    1.0           5
     127  1650.819072      10.0    1.0           5
     128  1805.418496      10.0    1.0           5
     129  1931.116544      10.0    1.0           5

     [130 rows x 4 columns]
```

When loading SNIRF files, trial types are derived from numeric stimulus markers. Cedalion registers the accessor `.cd` on DataFrames through which the user can modify stimulus dataframes. In the following the function `.cd.rename_events` is used to translate the numeric stimulus markers into descriptive names.

Users are free to choose any labels, but a consistent naming scheme is recommended because it makes selecting events easier. The dataset contains two motor tasks: finger tapping and ball squeezing executed both with the left and right hand:

```
[8]: rec.stim.cd.rename_events(
         {
             "1": "Rest",
             "2": "FTapping/Left",
             "3": "FTapping/Right",
             "4": "BallSqueezing/Left",
             "5": "BallSqueezing/Right",
         }
     )

     rec.stim
```

```
[8]:          onset  duration  value           trial_type
     0      23.855104      10.0    1.0                 Rest
     1      54.132736      10.0    1.0                 Rest
     2      84.410368      10.0    1.0                 Rest
     3     114.688000      10.0    1.0                 Rest
     4     146.112512      10.0    1.0                 Rest
     ..           ...       ...    ...                  ...
     125  1431.535616      10.0    1.0    BallSqueezing/Right
     126  1526.038528      10.0    1.0    BallSqueezing/Right
     127  1650.819072      10.0    1.0    BallSqueezing/Right
     128  1805.418496      10.0    1.0    BallSqueezing/Right
     129  1931.116544      10.0    1.0    BallSqueezing/Right

     [130 rows x 4 columns]
```

Selecting all BallSqueezing tasks:

```
[9]: with pd.option_context("display.max_rows", 5):
         display(rec.stim[rec.stim.trial_type.str.startswith("BallSqueezing")])
```

```
          onset  duration  value           trial_type
     97    8.486912      10.0    1.0    BallSqueezing/Left
     98  161.021952      10.0    1.0    BallSqueezing/Left
     ..          ...       ...    ...                  ...
     128  1805.418496      10.0    1.0    BallSqueezing/Right
     129  1931.116544      10.0    1.0    BallSqueezing/Right

     [33 rows x 4 columns]
```

Selecting all motor tasks with the left hand:

```
[10]: with pd.option_context("display.max_rows", 5):
          display(rec.stim[rec.stim.trial_type.str.endswith("Left")])
```

```
          onset  duration  value           trial_type
     65    99.549184      10.0    1.0          FTapping/Left
     66   129.597440      10.0    1.0          FTapping/Left
     ..          ...       ...    ...                  ...
     112  1962.541056      10.0    1.0    BallSqueezing/Left
     113  1994.194944      10.0    1.0    BallSqueezing/Left

     [33 rows x 4 columns]
```

## Time Series

The `'amp'` time series is represented as a `xarray.DataArray` with dimensions `'channel'`, `'wavelength'` and `'time'`. Three coordinate arrays are linked to the `'channel'` dimension, specifying for each channel a string label as well as the string label and the corresponding source and detector labels (e.g. the first channel `'S1D1'` is between source `'S1'` and detector `'D1'`). Coordinates of the `'wavelength'` dimension indicate that this CW-fNIRS measurement was done at 760 and 850 nm. The `'time'` dimensions has timestamps and an absolute sample counter as coordinates.

The `DataArray` is quantified in units of Volts.

```
[11]: rec["amp"]
```

```
[11]: <xarray.DataArray (channel: 100, wavelength: 2, time: 8794)> Size: 14MB
      [V] 0.0874 0.08735 0.08819 0.08887 0.0879 … 0.09108 0.09037 0.09043 0.08999
      Coordinates:
        * time        (time) float64 70kB 0.0 0.2294 0.4588 … 2.017e+03 2.017e+03
          samples     (time) int64 70kB 0 1 2 3 4 5 … 8788 8789 8790 8791 8792 8793
        * channel     (channel) object 800B 'S1D1' 'S1D2' 'S1D4' … 'S14D31' 'S14D32'
          source      (channel) object 800B 'S1' 'S1' 'S1' 'S1' … 'S14' 'S14' 'S14'
          detector    (channel) object 800B 'D1' 'D2' 'D4' 'D5' … 'D29' 'D31' 'D32'
        * wavelength  (wavelength) float64 16B 760.0 850.0
      Attributes:
          data_type_group:  unprocessed raw
```

### Montage

As described in the first tutorial, the probe geometry is stored in `rec.geo3d` as a `DataArray` of type `cedalion.typing.LabeledPoints`.

```
[12]: rec.geo3d
```

```
[12]: <xarray.DataArray (label: 346, digitized: 3)> Size: 8kB
      [mm] -77.82 15.68 23.17 -61.91 21.23 56.49 … 14.23 -38.28 81.95 -0.678 -37.03
      Coordinates:
          type     (label) object 3kB PointType.SOURCE … PointType.LANDMARK
        * label    (label) <U6 8kB 'S1' 'S2' 'S3' 'S4' … 'FFT10h' 'FT10h' 'FTT10h'
      Dimensions without coordinates: digitized
```

```
[13]: cedalion.vis.anatomy.plot_montage3D(rec["amp"], rec.geo3d)
```

Using functions of `matplotlib` time trace for a single channel can be plotted. Note, how the time series `amp` is indexed by label to select channel and wavelengths.

The package `cedalion.vis.blocks` (imported as `vbx`) provides visualizations building blocks, such as adding stimulus markers to a plot.

```
[14]: # example time trace
      amp = rec["amp"]
      ch = "S12D25"
      f, ax = p.subplots(1, 1, figsize=(12, 3))
      ax.set_prop_cycle("color", cedalion.vis.colors.COLORBREWER_Q8)
      ax.plot(amp.time, amp.sel(channel=ch, wavelength=760), label="amp. 760 nm")
      ax.plot(amp.time, amp.sel(channel=ch, wavelength=850), label="amp. 850 nm")
      vbx.plot_stim_markers(ax, rec.stim, y=1)
      ax.set_xlabel("time / s")
      ax.set_ylabel("amplitude / V")
      ax.set_xlim(0, 150)
      ax.legend(loc="upper right")
      ax.set_title(ch);
```

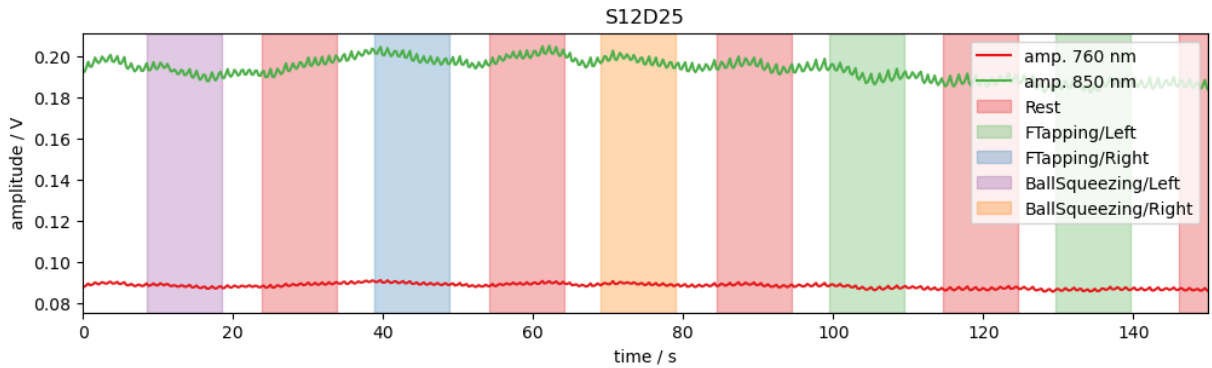

## Quality Metrics : SCI & PSP

To assess the signal quality the Scalp Coupling Index (SCI) and Peak Spectral Power (PSP) are calculated with functions from `cedalion.sigproc.quality`. Both metrics are computed in 10-second sliding windows. Each metric a configurable threshold. Values exceeding the threshold indicate good signal quality. The functions `sci` and `psp` each return two `DataArrays`: one with the metric values, and one with a boolean mask indicating where the threshold is exceeded.

Since the metrics are calculated for each time window the time axis changed.

```
[15]: sci_threshold = 0.75
      window_length = 10 * units.s
      sci, sci_mask = quality.sci(rec["amp"], window_length, sci_threshold)

      psp_threshold = 0.03
      psp, psp_mask = quality.psp(rec["amp"], window_length, psp_threshold)

      display(sci.rename("sci"))
      display(sci_mask.rename("sci_mask"))
```

```
<xarray.DataArray 'sci' (channel: 100, time: 200)> Size: 160kB
0.9949 0.9949 0.9775 0.9958 0.9914 0.9941 … 0.9369 0.9498 0.9628 0.9718 0.9545
Coordinates:
  * time       (time) float64 2kB 0.0 10.09 20.19 … 1.998e+03 2.008e+03
    samples    (time) int64 2kB 0 44 88 132 176 220 … 8580 8624 8668 8712 8756
  * channel    (channel) object 800B 'S1D1' 'S1D2' 'S1D4' … 'S14D31' 'S14D32'
    source     (channel) object 800B 'S1' 'S1' 'S1' 'S1' … 'S14' 'S14' 'S14'
    detector   (channel) object 800B 'D1' 'D2' 'D4' 'D5' … 'D29' 'D31' 'D32'

<xarray.DataArray 'sci_mask' (channel: 100, time: 200)> Size: 20kB
True True True True True True True True … True True True True True True True
Coordinates:
  * time       (time) float64 2kB 0.0 10.09 20.19 … 1.998e+03 2.008e+03
    samples    (time) int64 2kB 0 44 88 132 176 220 … 8580 8624 8668 8712 8756
  * channel    (channel) object 800B 'S1D1' 'S1D2' 'S1D4' … 'S14D31' 'S14D32'
    source     (channel) object 800B 'S1' 'S1' 'S1' 'S1' … 'S14' 'S14' 'S14'
    detector   (channel) object 800B 'D1' 'D2' 'D4' 'D5' … 'D29' 'D31' 'D32'
```

Visualize the metrics and quality masks:

```
[16]:  # define three colomaps: redish below a threshold, blueish above
       sci_norm, sci_cmap = colors.threshold_cmap("sci_cmap", 0.0, 1.0, sci_threshold)
       psp_norm, psp_cmap = colors.threshold_cmap("psp_cmap", 0.0, 0.30, psp_threshold)

       def plot_sci(sci):
           # plot the heatmap
           f, ax = p.subplots(1, 1, figsize=(12, 10))

           m = ax.pcolormesh(sci.time, np.arange(len(sci.channel)), sci, shading="nearest",␣
       ↪cmap=sci_cmap, norm=sci_norm)
           cb = p.colorbar(m, ax=ax)
           cb.set_label("SCI")
           ax.set_xlabel("time / s")
           p.tight_layout()
           ax.yaxis.set_ticks(np.arange(len(sci.channel)))
           ax.yaxis.set_ticklabels(sci.channel.values, fontsize=7)

       def plot_psp(psp):
           f, ax = p.subplots(1, 1, figsize=(12, 10))

           m = ax.pcolormesh(psp.time, np.arange(len(psp.channel)), psp, shading="nearest",␣
       ↪cmap=psp_cmap, norm=psp_norm)
           cb = p.colorbar(m, ax=ax)
           cb.set_label("PSP")
           ax.set_xlabel("time / s")
           p.tight_layout()
           ax.yaxis.set_ticks(np.arange(len(psp.channel)))
           ax.yaxis.set_ticklabels(psp.channel.values, fontsize=7)
```

```
[17]:  plot_sci(sci)
       plot_quality_mask(sci > sci_threshold, f"SCI > {sci_threshold}")
       plot_psp(psp)
       plot_quality_mask(psp > psp_threshold, f"PSP > {psp_threshold}")
```

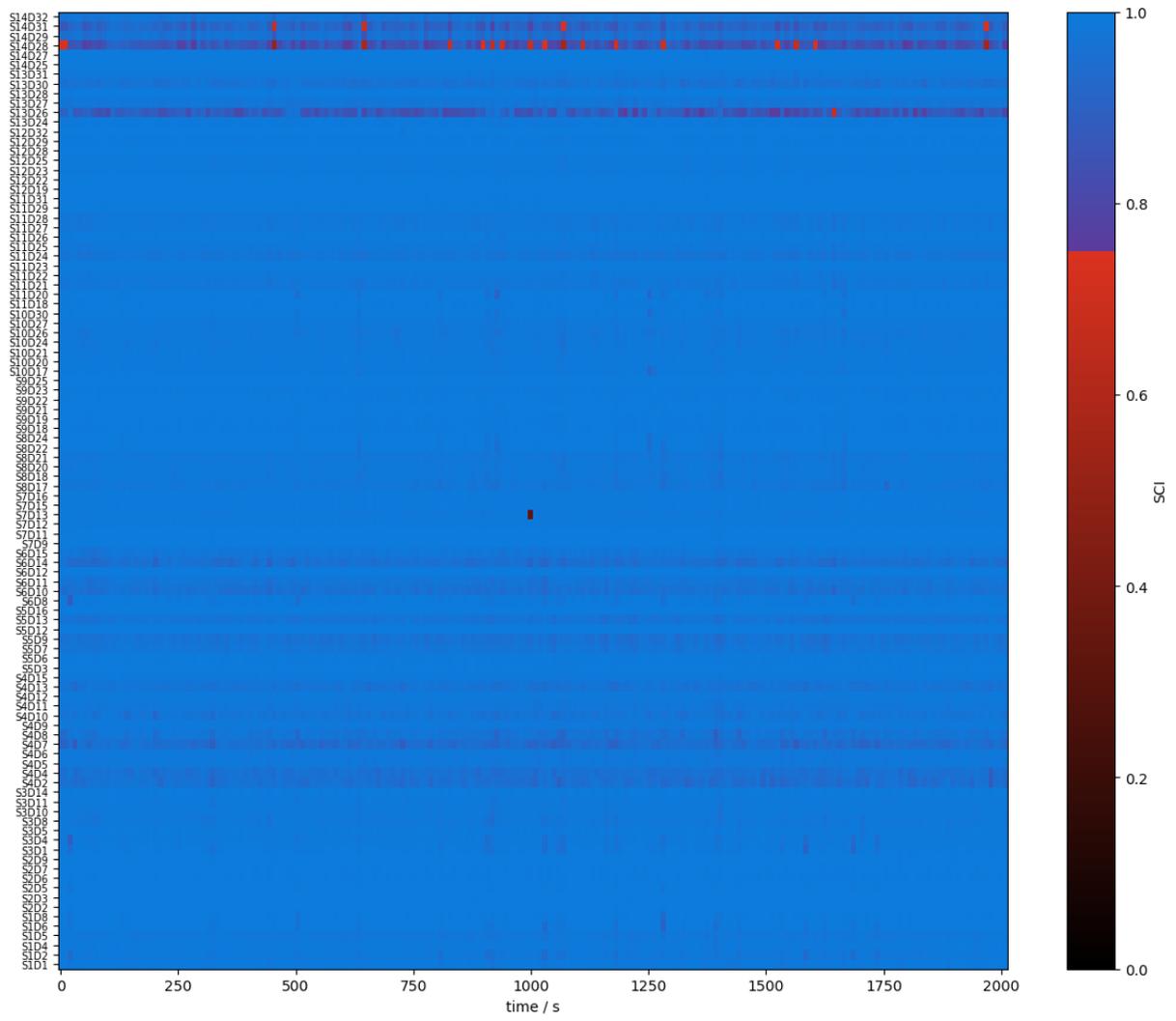

time / s

SCI > 0.75

CLEAN

TAINTED

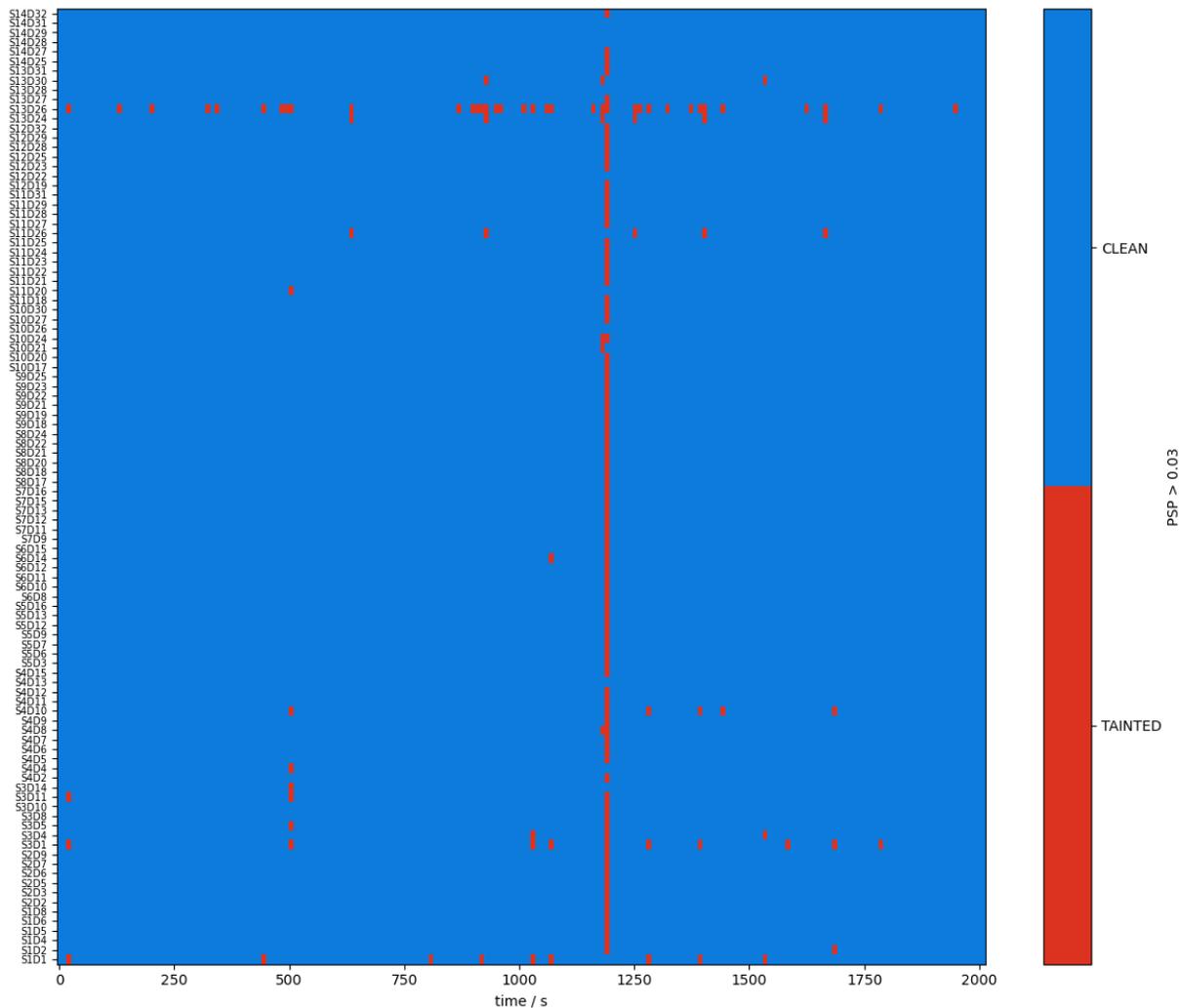

A window shall be considered clean only if both SCI and PSP exceed their respective thresholds. The user can combine the two boolean masks with a logical **and**, then compute the fraction of time windows where both conditions are true.

```
[18]: combined_mask = sci_mask & psp_mask

display(combined_mask)
plot_quality_mask(combined_mask, "combined_mask")
```

```
<xarray.DataArray (channel: 100, time: 200)> Size: 20kB
True True False True True True True True … True True True True True True True
Coordinates:
  * time      (time) float64 2kB 0.0 10.09 20.19 … 1.998e+03 2.008e+03
    samples   (time) int64 2kB 0 44 88 132 176 220 … 8580 8624 8668 8712 8756
  * channel   (channel) object 800B 'S1D1' 'S1D2' 'S1D4' … 'S14D31' 'S14D32'
    source    (channel) object 800B 'S1' 'S1' 'S1' 'S1' … 'S14' 'S14' 'S14'
    detector  (channel) object 800B 'D1' 'D2' 'D4' 'D5' … 'D29' 'D31' 'D32'
```

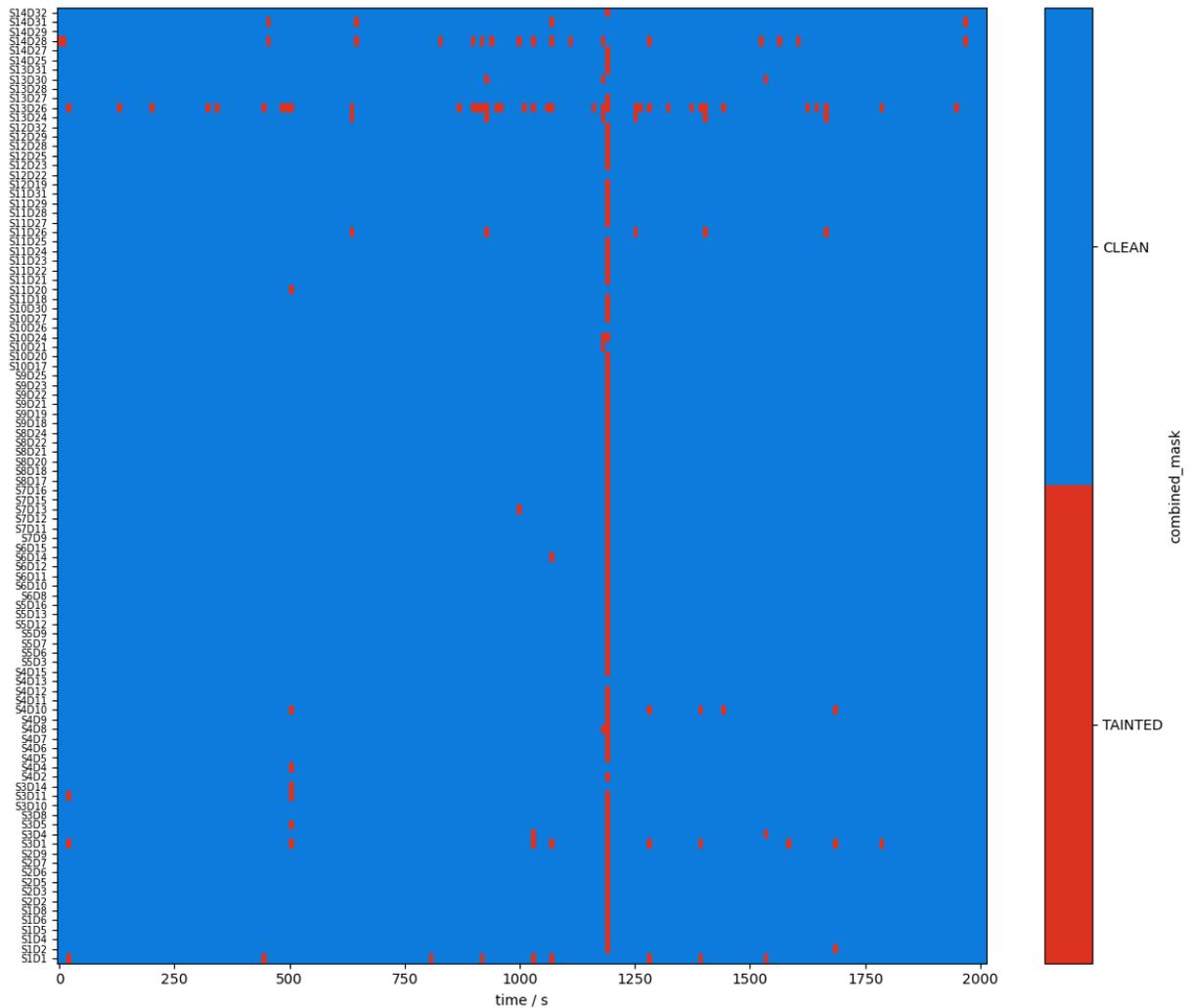

The next cell calculates the percentage of clean time windows per channel. Afterwards, two channels are identified that are clean in fewer than 95% of the time windows.

```
[19]: perc_time_clean = combined_mask.sum(dim="time") / len(sci.time)

      display(perc_time_clean)

      print("Channels clean less than 95% of the recording:")
      display(perc_time_clean[perc_time_clean < 0.95])
```

```
<xarray.DataArray (channel: 100)> Size: 800B
0.955 0.99 0.995 0.995 0.995 0.995 0.995 … 0.995 0.995 0.91 1.0 0.98 0.995
Coordinates:
  * channel    (channel) object 800B 'S1D1' 'S1D2' 'S1D4' … 'S14D31' 'S14D32'
    source     (channel) object 800B 'S1' 'S1' 'S1' 'S1' … 'S14' 'S14' 'S14'
    detector   (channel) object 800B 'D1' 'D2' 'D4' 'D5' … 'D29' 'D31' 'D32'

Channels clean less than 95% of the recording:

<xarray.DataArray (channel: 2)> Size: 16B
0.815 0.91
```

```
Coordinates:
  * channel    (channel) object 16B 'S13D26' 'S14D28'
    source     (channel) object 16B 'S13' 'S14'
    detector   (channel) object 16B 'D26' 'D28'
```

Visualize the percentage of clean times with a scalp plot:

```python
[20]: f, ax = p.subplots(1,1,figsize=(6.5,6.5))

cedalion.vis.anatomy.scalp_plot(
    rec["amp"],
    rec.geo3d,
    perc_time_clean,
    ax,
    cmap="RdYlGn",
    vmin=0.7,
    vmax=1,
    title=None,
    cb_label="Percentage of clean time",
    channel_lw=2,
    optode_labels=True
)
f.tight_layout()
```

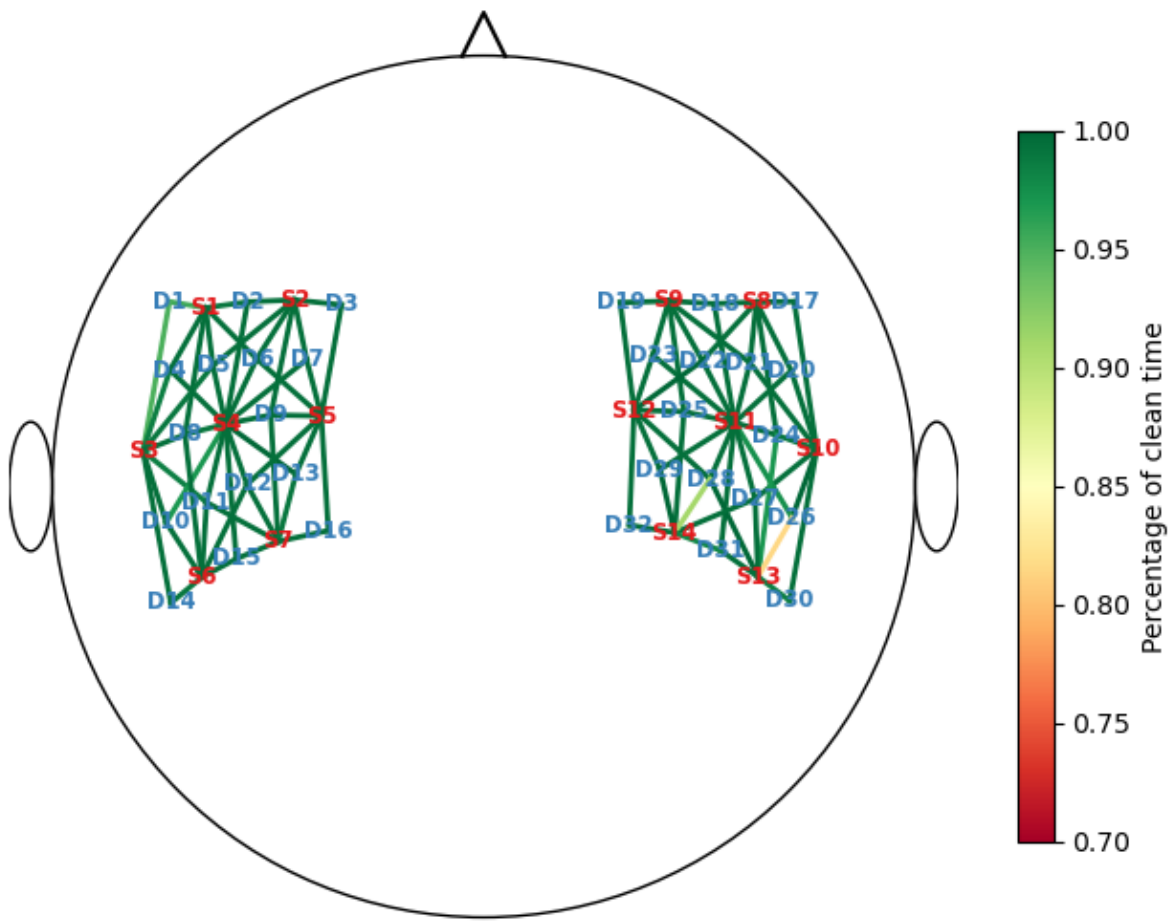

### Correct Motion Artifacts

Using the function `cedalion.nirs.cw.int2od` the raw amplitudes are converted to optical densities. Two motion-artifact correction methods are then applied.

- First, Temporal Derivative Distribution Repair (TDDR) is used to repair unusually large jumps in the time series.
- Second, a wavelet-based motion correction is applied.

The correction algorithms operate on optical densities. After correction, the corresponding corrected amplitudes are derived.

The modified time series are stored under different names in the `Recording` container.

```
[21]: rec["od"] = cedalion.nirs.cw.int2od(rec["amp"])
      rec["od_tddr"] = motion.tddr(rec["od"])
```

```python
rec["od_wavelet"] = motion.wavelet(rec["od_tddr"])
rec["amp_corrected"] = cedalion.nirs.cw.od2int(
    rec["od_wavelet"], rec["amp"].mean("time")
)
```

Recalculate the SCI and PSP metrics on the corrected amplitudes and visualize the combined masks before and after correction.

```python
[22]: # recalculate sci & psp on cleaned data
sci_corr, sci_corr_mask = quality.sci(
    rec["amp_corrected"], window_length, sci_threshold
)
psp_corr, psp_corr_mask = quality.psp(
    rec["amp_corrected"], window_length, psp_threshold
)
combined_corr_mask = sci_corr_mask & psp_corr_mask
```

```python
[23]: plot_quality_mask(combined_mask, "combined mask")
plot_quality_mask(combined_corr_mask, "combined corrected mask")
```

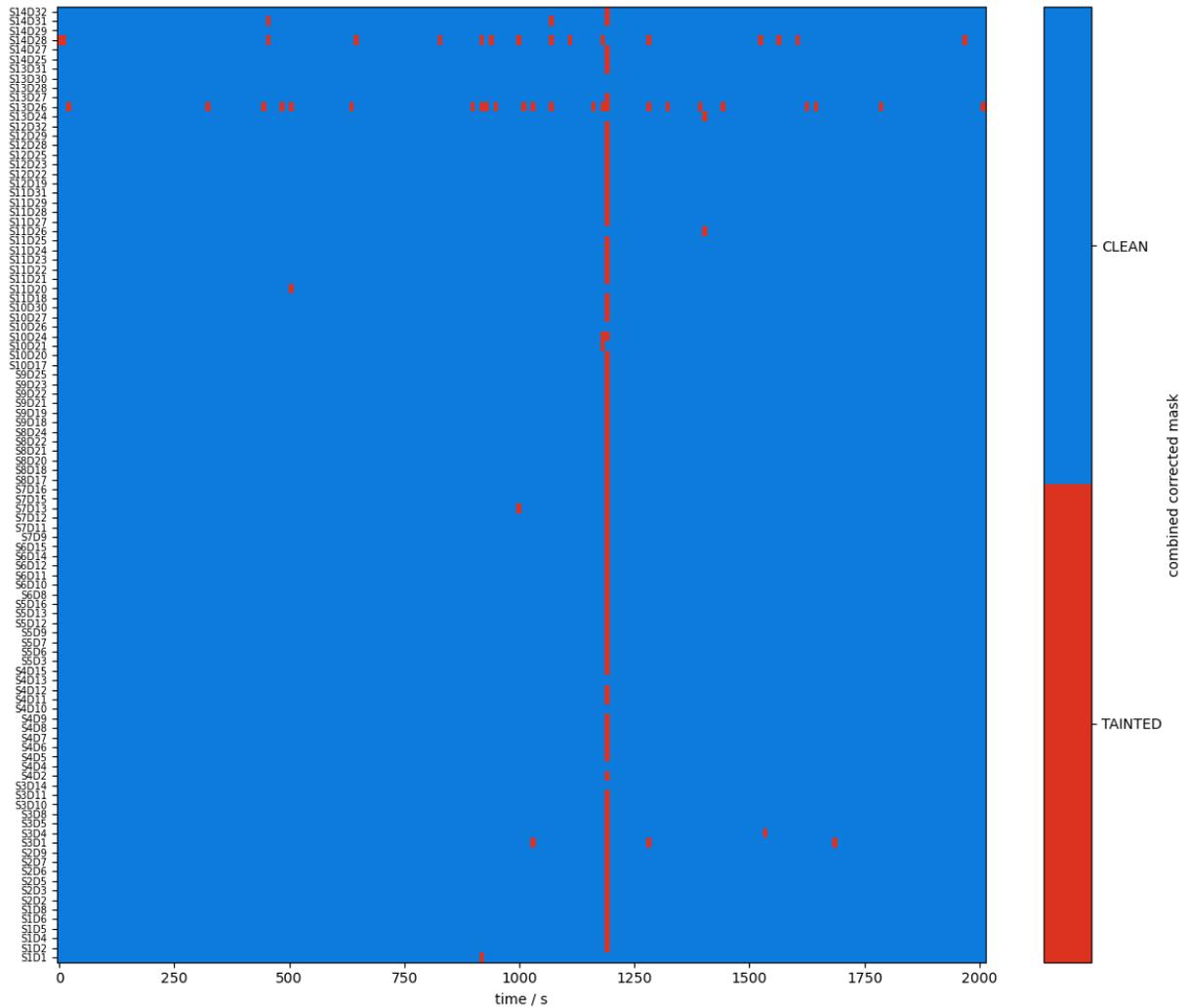

## Compare masks before and after motion artifact correction

Using logical operations on the quality masks, the next cell identifies the time windows affected by the correction algorithms.

```
[24]: changed_windows = (combined_mask == quality.TAINTED) & (combined_corr_mask == quality.
      ↪CLEAN)

      plot_quality_mask(
          changed_windows,
          "mask of time windows cleaned by motion correction",
          bool_labels=["unchanged", "improved"],
      )

      changed_windows = (combined_mask == quality.CLEAN) & (combined_corr_mask == quality.
      ↪TAINTED)

      plot_quality_mask(
          changed_windows,
```

```
        "mask of time windows corrupted by motion correction",
        bool_labels=["unchanged", "worsened"],
)
```

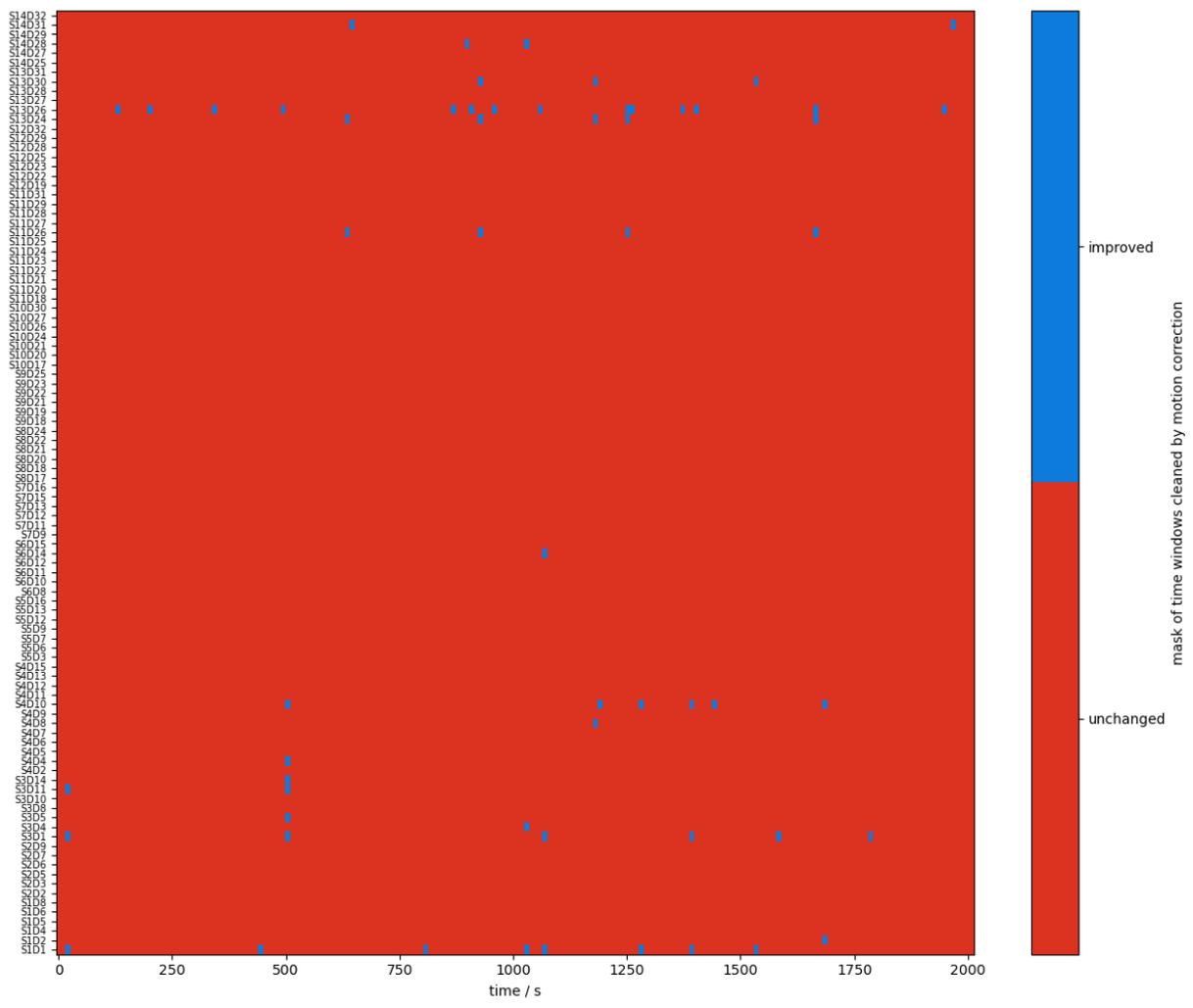

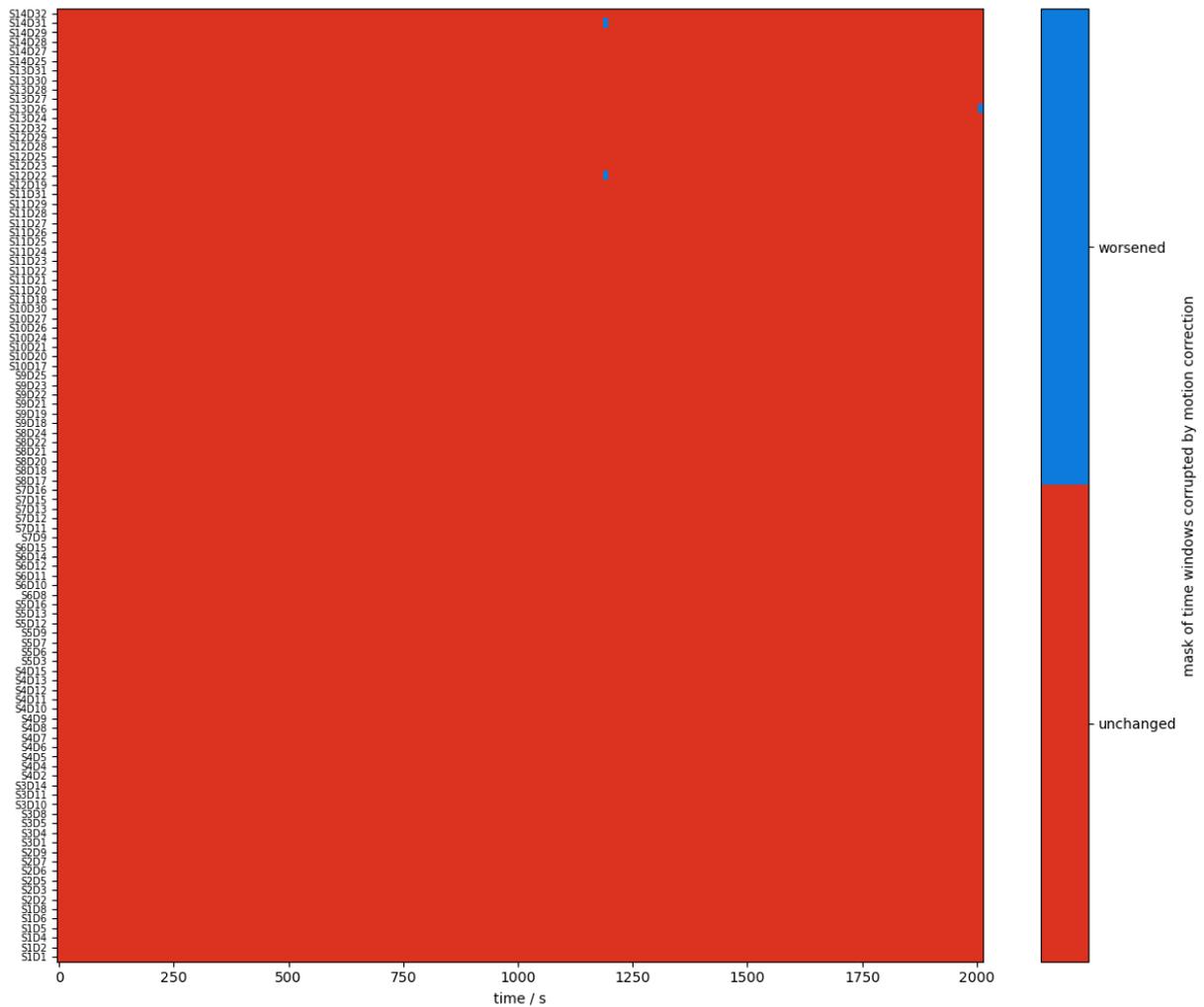

Recalculate and visualize the percentage of clean time:

```
[25]: perc_time_clean_corr = combined_corr_mask.sum(dim="time") / len(sci.time)

f, ax = p.subplots(1,2,figsize=(14,6.5))

cedalion.vis.anatomy.scalp_plot(
    rec["amp"],
    rec.geo3d,
    perc_time_clean,
    ax[0],
    cmap="RdYlGn",
    vmin=0.80,
    vmax=1,
    title="before correction",
    cb_label="Percentage of clean time",
    channel_lw=2,
    optode_labels=True
)
```

```
cedalion.vis.anatomy.scalp_plot(
    rec["amp"],
    rec.geo3d,
    perc_time_clean_corr,
    ax[1],
    cmap="RdYlGn",
    vmin=0.80,
    vmax=1,
    title="after correction",
    cb_label="Percentage of clean time",
    channel_lw=2,
    optode_labels=True
)
f.tight_layout()
```

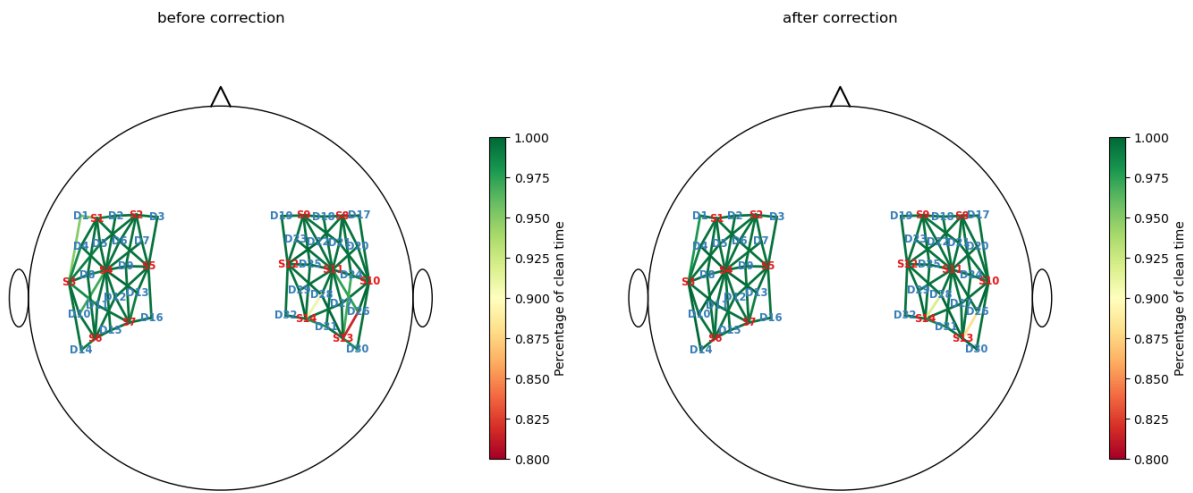

## Global Variance of the Temporal Derivative (GVTD)

The GVTD metric allows identifying global bad time segments. Here, it is calculated for the original and corrected amplitudes.

```
[26]: gvtd, gvtd_mask = quality.gvtd(rec["amp"])
      gvtd_corr, gvtd_prr_mask = quality.gvtd(rec["amp_corrected"])
```

The 10 segments with highest GVTD values are selected:

```
[27]: top10_bad_segments = sorted(
          [seg for seg in quality.mask_to_segments(combined_mask.all("channel"))],
          key=lambda t: gvtd.sel(time=slice(t[0], t[1])).max(),
          reverse=True,
      )[:10]
```

The following plot shows the GVTD metric before and after corrections. The 10 selected segments are highlighted in red. The combined masks are further reduced to indicate, if all channels at a given time are clean.

The impact of the correction methods is evident in the reduced number of spikes in the GVTD trace and the improvements seen in the quality mask. Time windows that previously did not satisfy the all-channels-clean criteria meet it after correction.

```
[28]: f,ax = p.subplots(4,1,figsize=(16,6), sharex=True)
      ax[0].plot(gvtd.time, gvtd)
      ax[1].plot(combined_mask.time, combined_mask.all("channel"))
      ax[2].plot(gvtd_corr.time, gvtd_corr)
      ax[3].plot(combined_corr_mask.time, combined_corr_mask.all("channel"))
      ax[0].set_ylim(0, 0.02)
      ax[2].set_ylim(0, 0.02)
      ax[0].set_ylabel("GVTD")
      ax[2].set_ylabel("GVTD")
      ax[1].set_ylabel("all channels clean\n (before)")
      ax[3].set_ylabel("all channels clean\n (after)")
      ax[3].set_xlabel("time / s")

      for i in range(4):
          vbx.plot_segments(ax[i], top10_bad_segments)
```

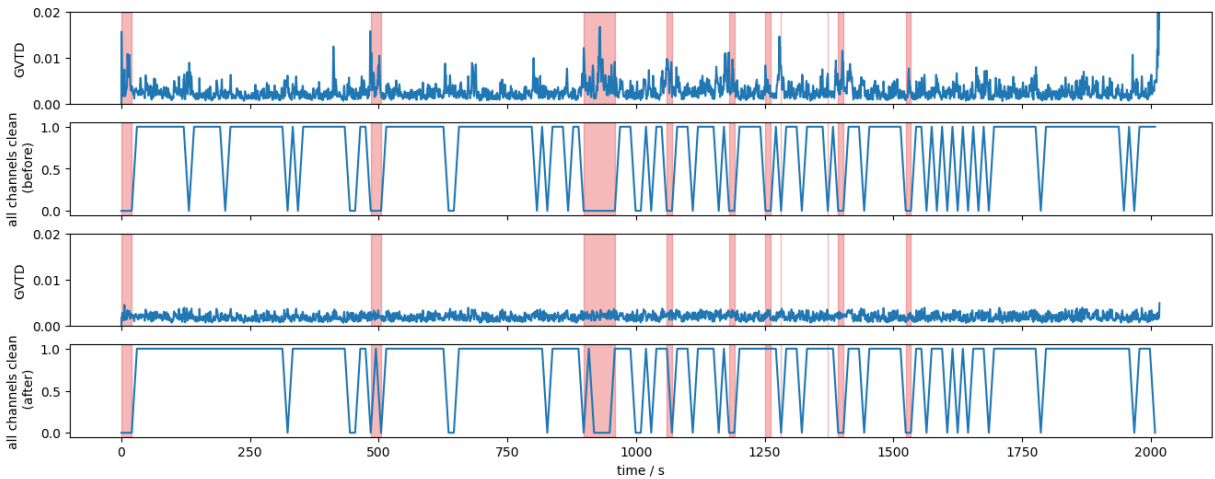

### Visualize motion correction in selected segments

To illustrate the effect of TDDR and wavelet-based motion correction on the time series, the following cell plots the amplitudes for the ten selected segments in two channels before and after correction.

```
[29]: example_channels = ["S4D10", "S13D26"]

      f, ax = p.subplots(5,4, figsize=(16,16), sharex=False)
      ax = ax.T.flatten()
      padding = 15
      i = 0
      for ch in example_channels:
          for (start, end) in top10_bad_segments:
              ax[i].set_prop_cycle(color=["#e41a1c", "#ff7f00", "#377eb8", "#984ea3"])
              for wl in rec["od"].wavelength.values:
```

```
        sel = rec["od"].sel(time=slice(start-padding, end+padding), channel=ch,
    ↪wavelength=wl)
        ax[i].plot(sel.time, sel, label=f"{wl:.0f} nm orig")
        sel = rec["od_wavelet"].sel(time=slice(start-padding, end+padding),
    ↪channel=ch, wavelength=wl)
        ax[i].plot(sel.time, sel, label=f"{wl:.0f} nm corr")
        ax[i].set_title(ch)
    ax[i].legend(ncol=2, loc="upper center")
    ylim = ax[i].get_ylim()
    ax[i].set_ylim(ylim[0], ylim[1]+0.25*(ylim[1]-ylim[0])) # make space for legend

    i += 1

p.tight_layout()
```

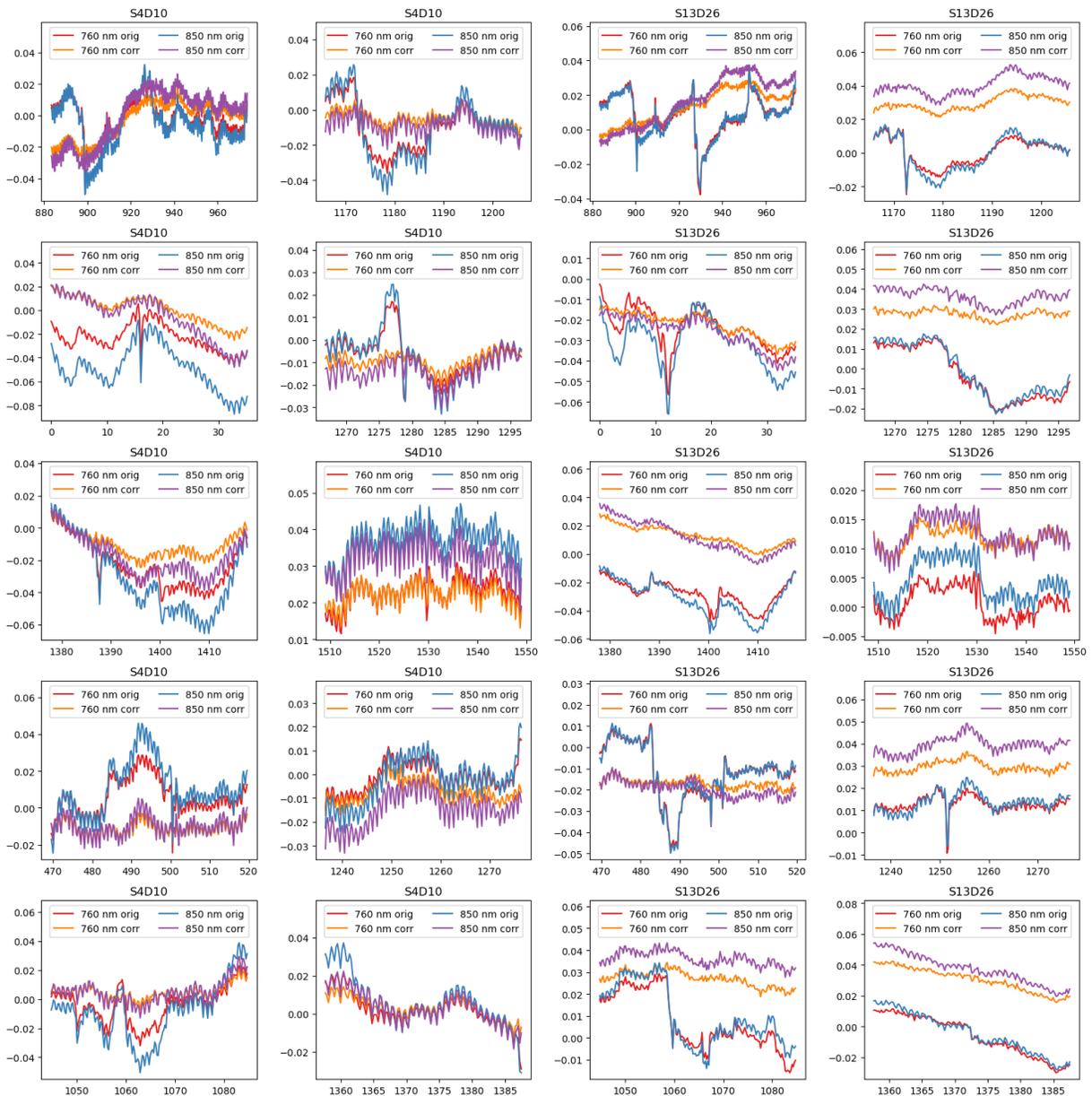

## Final channel selection

Before correction, the channels "S13D26" and "S14D28" had the most unclean time windows. Although the correction methods recovered some of these windows, the fraction of clean windows for both channels still remains below 95%.

```
[30]: perc_time_clean_corr[perc_time_clean_corr < 0.95]
```

```
[30]: <xarray.DataArray (channel: 2)> Size: 16B
      0.88 0.92
      Coordinates:
        * channel    (channel) object 16B 'S13D26' 'S14D28'
          source     (channel) object 16B 'S13' 'S14'
          detector   (channel) object 16B 'D26' 'D28'
```

To prune bad channels from time series the function `cedalion.sigproc.quality.prun_ch` is available. It takes a list of quality masks, combines them and discards channels from the time series.

```
[31]: signal_quality_selection_masks = [perc_time_clean >= .95]

      rec["amp_pruned"], pruned_channels = quality.prune_ch(
          rec["amp"], signal_quality_selection_masks, "all"
      )
      display(rec["amp_pruned"])
      display(pruned_channels)
```

```
<xarray.DataArray (channel: 98, wavelength: 2, time: 8794)> Size: 14MB
[V] 0.0874 0.08735 0.08819 0.08887 0.0879 … 0.09108 0.09037 0.09043 0.08999
Coordinates:
  * time        (time) float64 70kB 0.0 0.2294 0.4588 … 2.017e+03 2.017e+03
    samples     (time) int64 70kB 0 1 2 3 4 5 … 8788 8789 8790 8791 8792 8793
  * channel     (channel) object 784B 'S1D1' 'S1D2' 'S1D4' … 'S14D31' 'S14D32'
    source      (channel) object 784B 'S1' 'S1' 'S1' 'S1' … 'S14' 'S14' 'S14'
    detector    (channel) object 784B 'D1' 'D2' 'D4' 'D5' … 'D29' 'D31' 'D32'
  * wavelength  (wavelength) float64 16B 760.0 850.0
Attributes:
    data_type_group:  unprocessed raw

array(['S13D26', 'S14D28'], dtype=object)
```

# S4: Model-driven (GLM) Analysis

This tutorial demonstrates how to use a General Linear Model (GLM) to model the recorded time series as a superposition of hemodynamic responses and nuisance effects.

```python
[1]:  # This cells setups the environment when executed in Google Colab.
      try:
          import google.colab
          !curl -s https://raw.githubusercontent.com/ibs-lab/cedalion/dev/scripts/
          ↪colab_setup.py -o colab_setup.py
          # Select branch with --branch "branch name" (default is "dev")
          %run colab_setup.py
      except ImportError:
          pass
```

```python
[2]:  import cedalion
      import cedalion.sigproc.quality as quality
      import cedalion.sigproc.motion as motion
      import cedalion.sigproc.physio as physio

      from cedalion import units
      import cedalion.xrutils as xrutils
      import cedalion.models.glm as glm
      import cedalion.data

      import cedalion.vis.blocks as vbx
      from cedalion.vis.anatomy import scalp_plot, plot_montage3D
      from cedalion.vis.colors import p_values_cmap

      import numpy as np
      import xarray as xr

      import matplotlib.pyplot as p
      from statsmodels.stats.multitest import multipletests

      xr.set_options(display_expand_data=False)
      xrutils.unit_stripping_is_error()
```

## Load Data

The dataset is loaded and events are renamed. Afterwards, the stimulus data frame is summarized.

```python
[3]:  rec = cedalion.data.get_fingertappingDOT()

      # assign better trial_type labels
      rec.stim.cd.rename_events(
          {
              "1": "Control",
              "2": "FTapping/Left",
              "3": "FTapping/Right",
              "4": "BallSqueezing/Left",
```

```
                "5": "BallSqueezing/Right",
        }
)
```

```
[4]:   # count trials
       rec.stim.groupby("trial_type")[["trial_type"]].count().rename(
           {"trial_type": "# trials"},
           axis=1,
       )
```

```
[4]:                      # trials
       trial_type
       BallSqueezing/Left        17
       BallSqueezing/Right       16
       Control                   65
       FTapping/Left             16
       FTapping/Right            16
```

**Trim dataset**

Trim the dataset to roughly 5 minutes to keep computation times low while still preserving three trials per condition. Also exclude the 'BallSqueezing' conditions.

```
[5]:   tmin = 5
       tmax = 315

       rec.stim = rec.stim[
           (tmin <= rec.stim.onset)
           & (rec.stim.onset <= tmax)
           & (rec.stim.trial_type.isin(["Control", "FTapping/Left", "FTapping/Right"]))
       ]
       # use a slice to select a time range
       rec["amp_cropped"] = rec["amp"].sel(time=slice(tmin,tmax))

       # count remaining trials
       rec.stim.groupby("trial_type")[["trial_type"]].count().rename(
           {"trial_type": "# trials"},
           axis=1,
       )
```

```
[5]:                      # trials
       trial_type
       Control                   10
       FTapping/Left              3
       FTapping/Right             3
```

**Preprocessing**

The same motion-correction methods as described in the third tutorial notebook are applied, and the bad channels identified there are removed. The cleaned optical-density time series are then converted to concentration changes. Finally, a low-pass filter with a 0.5 Hz cutoff is applied to remove the cardiac component.

```
[6]: rec["od"] = cedalion.nirs.cw.int2od(rec["amp_cropped"])
     rec["od_tddr"] = motion.tddr(rec["od"])
     rec["od_wavelet"] = motion.wavelet(rec["od_tddr"])

     bad_channels = ['S13D26', 'S14D28']

     rec["od_clean"] = rec["od_wavelet"].sel(channel=~rec["od"].channel.isin(bad_channels))

     # differential pathlength factors
     dpf = xr.DataArray(
         [6, 6],
         dims="wavelength",
         coords={"wavelength": rec["amp"].wavelength},
     )

     rec["conc"] = cedalion.nirs.cw.od2conc(rec["od_clean"], rec.geo3d, dpf, ␣
     ↪spectrum="prahl")

     # Here we use a lowpass-filter to remove the cardiac component.
     # Drift will be modeled in the design matrix.

     fmin = 0 * units.Hz
     fmax = 0.5 * units.Hz

     rec["conc_filtered"] = cedalion.sigproc.frequency.freq_filter(rec["conc"], fmin, fmax)

     TS_NAME = "conc_filtered"
```

```
[7]: display(rec[TS_NAME])
```

```
<xarray.DataArray 'concentration' (chromo: 2, channel: 98, time: 1352)> Size: 2MB
[μM] 0.8276 0.7548 0.7043 0.6868 0.6978 … -0.08036 -0.07753 -0.07426 -0.07086
Coordinates:
  * chromo     (chromo) <U3 24B 'HbO' 'HbR'
  * time       (time) float64 11kB 5.046 5.276 5.505 5.734 … 314.5 314.7 314.9
    samples    (time) int64 11kB 22 23 24 25 26 27 … 1369 1370 1371 1372 1373
  * channel    (channel) object 784B 'S1D1' 'S1D2' 'S1D4' … 'S14D31' 'S14D32'
    source     (channel) object 784B 'S1' 'S1' 'S1' 'S1' … 'S14' 'S14' 'S14'
    detector   (channel) object 784B 'D1' 'D2' 'D4' 'D5' … 'D29' 'D31' 'D32'
```

## Montage and Channel Distances

The hexagonal montage has longer (3-3.5cm) and shorter (~1.7-2.2cm) channels. We define a cut-off at 22.5 mm to separate long and short channels.

```
[8]: plot_montage3D(rec["amp"], rec.geo3d)
```

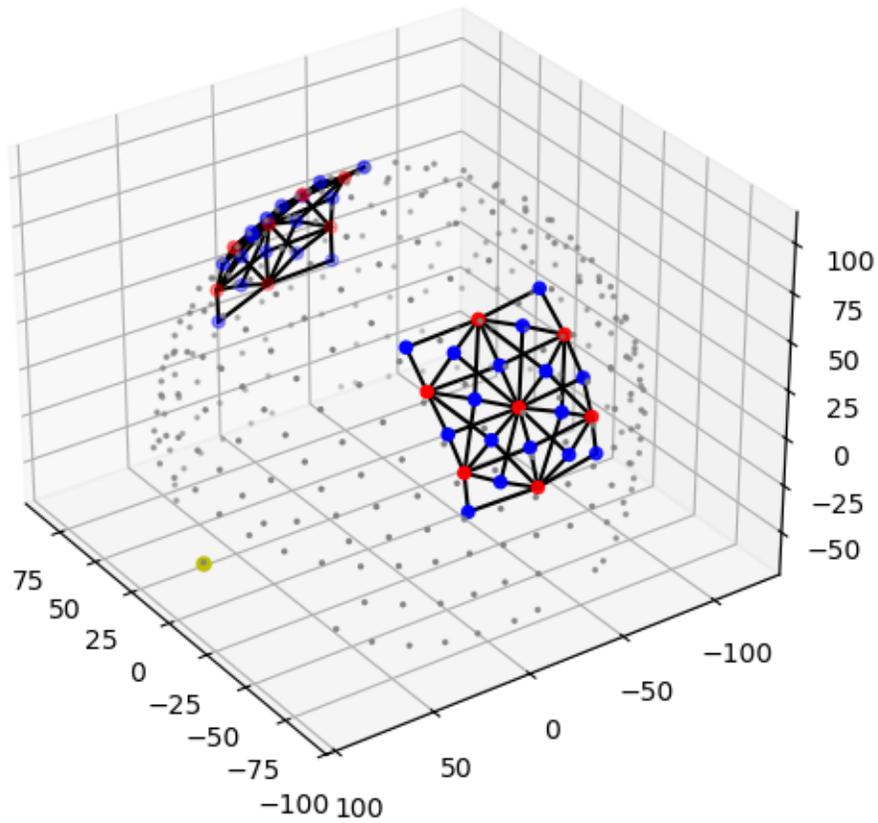

Calculate and histogram channel distances:

```
[9]: distances = cedalion.nirs.channel_distances(rec["amp"], rec.geo3d)

p.figure(figsize=(8,4))
p.hist(distances, 40)
p.axvline(22.5, c="r", ls="--")
p.xlabel("channel distance / mm")
p.ylabel("channel count");
```

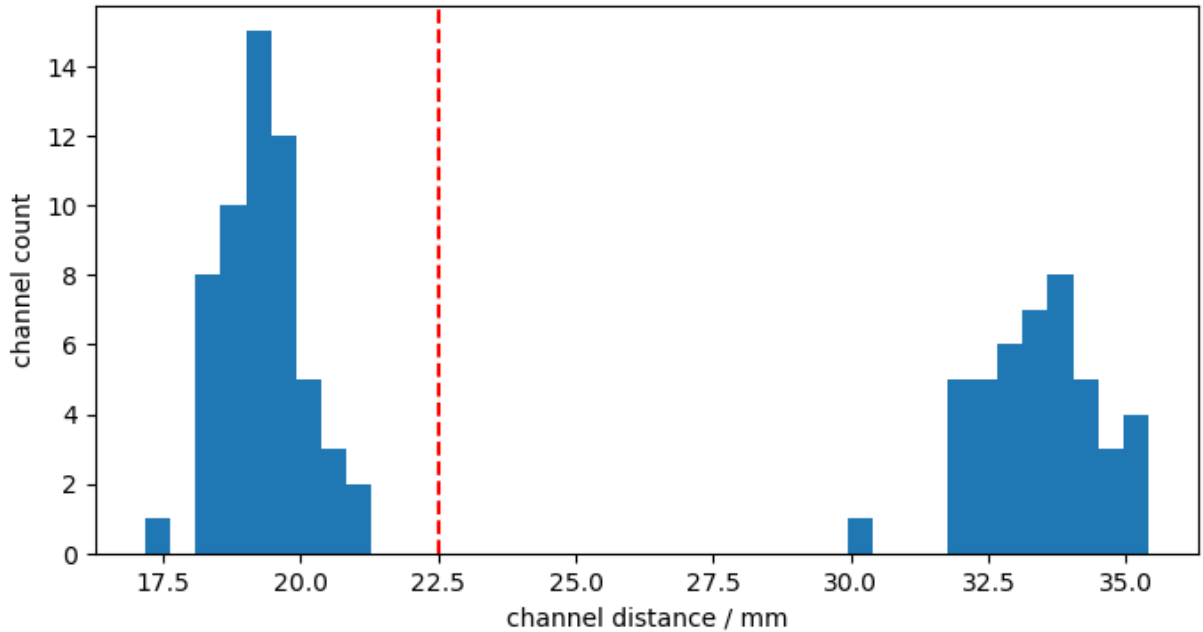

Split long and short channels

```
[10]: rec["ts_long"], rec["ts_short"] = cedalion.nirs.split_long_short_channels(
          rec[TS_NAME], rec.geo3d, distance_threshold=22.5 * units.mm
      )

      # This renaming is for display purposes only.
      display(rec["ts_long"].rename("long channels"))
      display(rec["ts_short"].rename("short channels"))
```

```
<xarray.DataArray 'long channels' (chromo: 2, channel: 44, time: 1352)> Size: 952kB
[µM] 0.3715 0.3284 0.2975 0.2849 0.2892 … 0.006362 0.008495 0.01122 0.01438
Coordinates:
  * chromo     (chromo) <U3 24B 'HbO' 'HbR'
  * time       (time) float64 11kB 5.046 5.276 5.505 5.734 … 314.5 314.7 314.9
    samples    (time) int64 11kB 22 23 24 25 26 27 … 1369 1370 1371 1372 1373
  * channel    (channel) object 352B 'S1D6' 'S1D8' 'S2D5' … 'S14D25' 'S14D27'
    source     (channel) object 352B 'S1' 'S1' 'S2' 'S2' … 'S13' 'S14' 'S14'
    detector   (channel) object 352B 'D6' 'D8' 'D5' 'D9' … 'D28' 'D25' 'D27'

<xarray.DataArray 'short channels' (chromo: 2, channel: 54, time: 1352)> Size: 1MB
[µM] 0.8276 0.7548 0.7043 0.6868 0.6978 … -0.08036 -0.07753 -0.07426 -0.07086
Coordinates:
  * chromo     (chromo) <U3 24B 'HbO' 'HbR'
  * time       (time) float64 11kB 5.046 5.276 5.505 5.734 … 314.5 314.7 314.9
    samples    (time) int64 11kB 22 23 24 25 26 27 … 1369 1370 1371 1372 1373
  * channel    (channel) object 432B 'S1D1' 'S1D2' 'S1D4' … 'S14D31' 'S14D32'
    source     (channel) object 432B 'S1' 'S1' 'S1' 'S1' … 'S14' 'S14' 'S14'
    detector   (channel) object 432B 'D1' 'D2' 'D4' 'D5' … 'D29' 'D31' 'D32'
```

## Fitting a General Linear Model

The GLM for fNIRS explains the time series (Y) for each channel and chromophore individually as a linear superposition of regressors, some representing the hemodynamic response and others addressing nuisance effects such as superficial or physiological components and signal drifts. The regressors form the design matrix (G) and the fit determines the unknown coefficients ( ) for each regressor in the presence of unmodeled noise (E).

$$Y = G\beta + E$$

## Building the Design Matrix

The design matrix comprises the hemodynamic response, short channel and drift regressors.

We model the hemodynamic response function (HRF) using a set of Gaussian basis kernels. The basis is defined by the durations of pre- and post-stimulus periods, the temporal spacing between successive kernels, and the standard deviation (width) of each Gaussian.

Using an instance of `cedalion.models.glm.basis_functions.GaussianKernels`, a basis for the long-channel time series is created and visualized.

```
[11]: gaussian_kernels = glm.basis_functions.GaussianKernels(
          t_pre=2 * units.s,
          t_post=15 * units.s,
          t_delta=1.5 * units.s,
          t_std=2 * units.s,
      )

      hrf_basis = gaussian_kernels(rec["ts_long"])

      for i in range(hrf_basis.sizes["component"]):
          p.plot(hrf_basis.time, hrf_basis[:,i])
          p.xlabel("relative time / s")
      p.axvline(0, c="k")
```

```
[11]: <matplotlib.lines.Line2D at 0x7612569e1ed0>
```

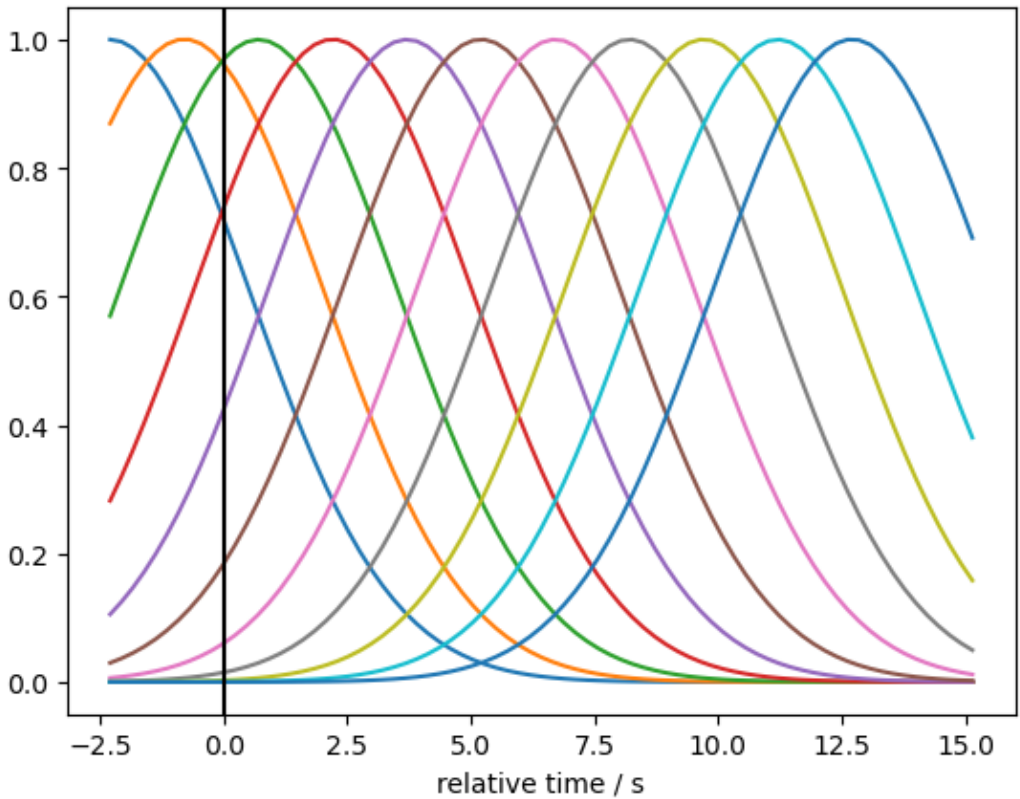

In Cedalion the class `glm.DesignMatrix` manages the regressors.

Functions in the package `glm.design_matrix` create regressors and return them as instances of `DesignMatrix`. These can be combined with the `&`-operator.

In this example, all short channels are averaged to model a global (mostly) superficial component.

Any remaining signal drift is modeled with cosine basis functions up tp 0.02 Hz.

```
[12]: dms = (
          glm.design_matrix.hrf_regressors(
              rec["ts_long"],
              rec.stim,
              gaussian_kernels,
          )
          & glm.design_matrix.drift_cosine_regressors(rec[TS_NAME], fmax=0.02 * units.Hz)
          & glm.design_matrix.average_short_channel_regressor(rec["ts_short"])
      )
```

When fitting a GLM to fNIRS data, some regressors are shared across all channels, wheras others are channel-specific. For example, if superficial physiology is modeled using the nearest short-separation channel for each measurement channel, the resulting short-channel regressor is channel-specific and therefore differs across channels.

Consequently, the `DesignMatrix` distinguishes between common regressors and channel-wise regressors. In this example the design matrix has only common regressors.

Each regressor has a string label for clarity. The convention used by Cedalion is to use labels of the form `'HRF <trial_typ> <number>'` for the HRF regressors and `'Drift <type> <number>'` for the drift components.

As with the trial-type labels, adopting this naming scheme makes it straightforward to select regressors. In this instance, the stimulus 'BallSqueezing/Left' is modeled with multiple regressors. They can be easily distinguished from other HRF or drift terms by selecting labels that begin with 'HRF BallSqueezing/Left'.

```
[13]: display(dms)
      display(dms.common)
      display(dms.channel_wise)
```

```
DesignMatrix(common=['HRF Control 00','HRF Control 01','HRF Control 02','HRF Control
↪03','HRF Control 04','HRF Control 05','HRF Control 06','HRF Control 07','HRF Control
↪08','HRF Control 09','HRF Control 10','HRF FTapping/Left 00','HRF FTapping/Left
↪01','HRF FTapping/Left 02','HRF FTapping/Left 03','HRF FTapping/Left 04','HRF
↪FTapping/Left 05','HRF FTapping/Left 06','HRF FTapping/Left 07','HRF FTapping/Left
↪08','HRF FTapping/Left 09','HRF FTapping/Left 10','HRF FTapping/Right 00','HRF
↪FTapping/Right 01','HRF FTapping/Right 02','HRF FTapping/Right 03','HRF FTapping/
↪Right 04','HRF FTapping/Right 05','HRF FTapping/Right 06','HRF FTapping/Right
↪07','HRF FTapping/Right 08','HRF FTapping/Right 09','HRF FTapping/Right 10','Drift
↪Cos 0','Drift Cos 1','Drift Cos 2','Drift Cos 3','Drift Cos 4','Drift Cos 5','Drift
↪Cos 6','Drift Cos 7','Drift Cos 8','Drift Cos 9','Drift Cos 10','Drift Cos
↪11','short'], channel_wise=[])
```

```
<xarray.DataArray (time: 1352, regressor: 46, chromo: 2)> Size: 995kB
0.0 0.0 0.0 0.0 0.0 0.0 0.0 0.0 … 0.9999 0.9999 -0.9999 -0.9999 0.4137 0.03177
Coordinates:
  * time       (time) float64 11kB 5.046 5.276 5.505 5.734 … 314.5 314.7 314.9
  * regressor  (regressor) <U21 4kB 'HRF Control 00' … 'short'
  * chromo     (chromo) <U3 24B 'HbO' 'HbR'
    samples    (time) int64 11kB 22 23 24 25 26 27 … 1369 1370 1371 1372 1373
```

```
[]
```

**Visualize the design matrix**

```
[14]: # select common regressors
      dm = dms.common
      display(dm)

      # using xr.DataArray.plot
      f, ax = p.subplots(1,1,figsize=(12,10))
      dm.sel(chromo="HbO", time=dm.time < 600).T.plot()
      p.title("Shared Regressors")
      p.show()
```

```
<xarray.DataArray (time: 1352, regressor: 46, chromo: 2)> Size: 995kB
0.0 0.0 0.0 0.0 0.0 0.0 0.0 0.0 … 0.9999 0.9999 -0.9999 -0.9999 0.4137 0.03177
Coordinates:
  * time       (time) float64 11kB 5.046 5.276 5.505 5.734 … 314.5 314.7 314.9
  * regressor  (regressor) <U21 4kB 'HRF Control 00' … 'short'
  * chromo     (chromo) <U3 24B 'HbO' 'HbR'
    samples    (time) int64 11kB 22 23 24 25 26 27 … 1369 1370 1371 1372 1373
```

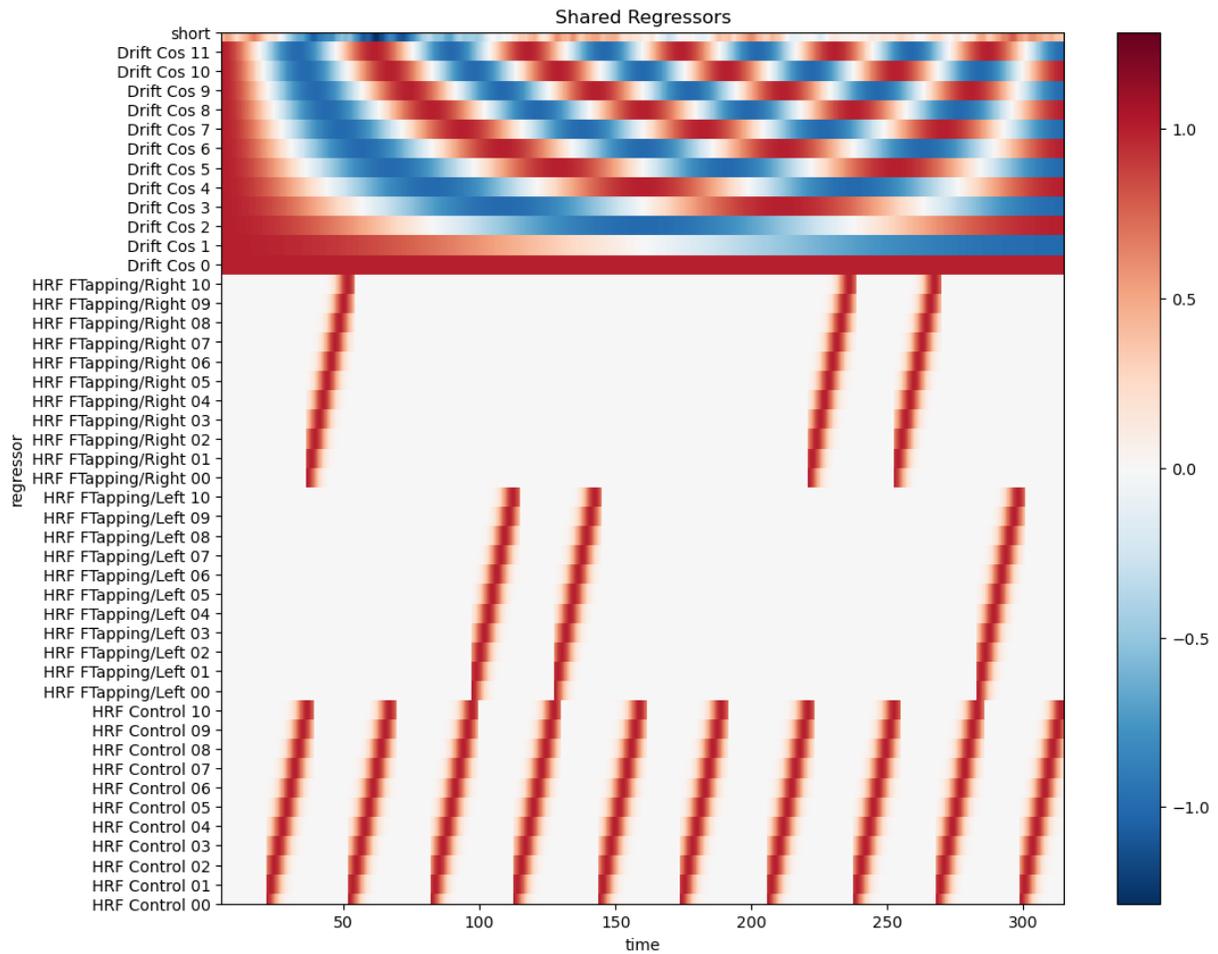

```
[15]:  # line plots of all regressors
       f, ax = p.subplots(2,1,sharex=True, figsize=(12,6))

       for i, chromo in enumerate(["HbO", "HbR"]):
           for reg in dm.regressor.values:
               label = reg if i == 0 else None
               ax[i].plot(dm.time, dm.sel(chromo=chromo, regressor=reg), label=label)

           vbx.plot_stim_markers(ax[i], rec.stim, y=1)
           ax[i].grid()
           ax[i].set_title(chromo)
           ax[i].set_ylim(-1.5,1.5)
       f.suptitle("All Common Regressors")

       f.legend(ncol=5, loc="upper center", bbox_to_anchor=(0.5, 0))
       #ax[0].set_xlim(0,240);
       ax[1].set_xlabel("time / s");
       f.set_tight_layout(True)
```

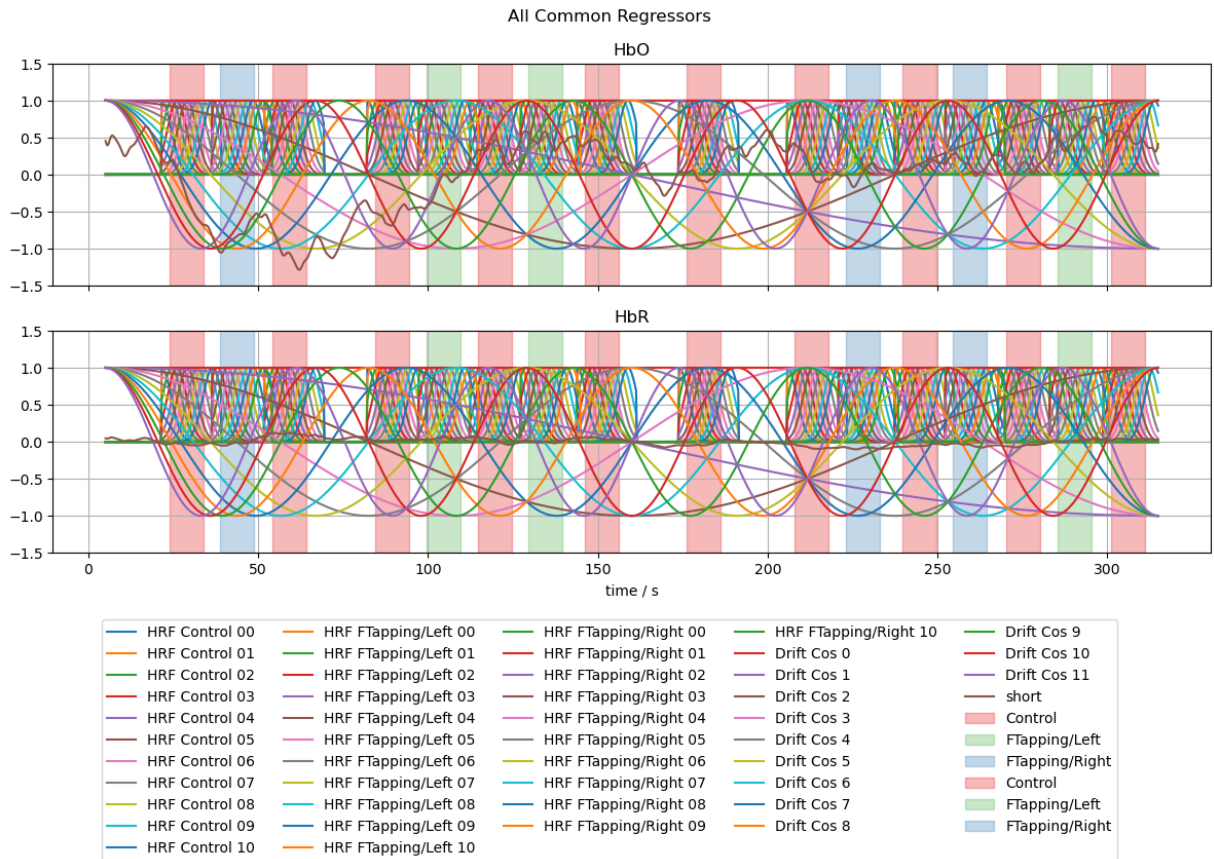

**Fitting the Model**

To fit the GLM, use `glm.fit`, which requires the time series and the design matrix as inputs. Optionally, you can choose a noise model from the following currently supported options:

- ols (default): ordinary least squares
- rls: recursive least squares
- wls: weighted least squares
- ar_irls: autoregressive iteratively reweighted least squares
- gls: generalized least squares
- glsar: generalized least squares with autoregressive covariance structure

Model fitting is implemented using the `statsmodels` library. Because each channel-chromophore pair is fitted independently, the output consists of one `statsmodels.RegressionResults` object per channel and chromophore. These are arranged in a `DataArray` object with dimensions (`'channel'`, `'chromo'`).

The user can investigate fit results, assess parameter uncertainties and perform hypothesis tests through functions of the `RegressionResults` objects. To facilitate calling these functions on all result objects in a `DataArray`, Cedalion provides the `.sm` accessor. For example, model coefficients (betas) can be obtained with `result.sm.params`, which calls `params` on all result objects and returns a DataArray with the parameters. For the full list of available methods and attributes, see the [statsmodels documentation](#). .

```
[16]: #results = glm.fit(rec["ts_long"], dms, noise_model="ols", max_jobs=-1)
results = glm.fit(rec["ts_long"], dms, noise_model="ar_irls", max_jobs=-1)

display(results)
```

```
<xarray.DataArray (channel: 44, chromo: 2)> Size: 704B
<statsmodels.robust.robust_linear_model.RLMResultsWrapper object at 0x76124fe…
Coordinates:
  * chromo    (chromo) <U3 24B 'HbO' 'HbR'
  * channel   (channel) object 352B 'S1D6' 'S1D8' 'S2D5' … 'S14D25' 'S14D27'
    source    (channel) object 352B 'S1' 'S1' 'S2' 'S2' … 'S13' 'S14' 'S14'
    detector  (channel) object 352B 'D6' 'D8' 'D5' 'D9' … 'D28' 'D25' 'D27'
Attributes:
    description:  AR_IRLS
```

**Inspecting Fit Results**

The result of the fit is an array of statsmodels result objects. Each contains the fitted parameters and other functionality to inspect fit parameter uncertainty or to perform hypothesis tests.

```
[17]: results[0,0].item().params
```

```
[17]: HRF Control 00          0.276679
      HRF Control 01         -1.322791
      HRF Control 02          3.381821
      HRF Control 03         -5.959108
      HRF Control 04          7.910174
      HRF Control 05         -8.189405
      HRF Control 06          6.586288
      HRF Control 07         -4.007499
      HRF Control 08          1.722847
      HRF Control 09         -0.484760
      HRF Control 10          0.070155
      HRF FTapping/Left 00    0.234327
      HRF FTapping/Left 01   -0.918640
      HRF FTapping/Left 02    1.663890
      HRF FTapping/Left 03   -1.340175
      HRF FTapping/Left 04   -0.837867
      HRF FTapping/Left 05    3.703104
      HRF FTapping/Left 06   -4.886997
      HRF FTapping/Left 07    3.755480
      HRF FTapping/Left 08   -1.792997
      HRF FTapping/Left 09    0.528106
      HRF FTapping/Left 10   -0.077728
      HRF FTapping/Right 00  -0.858642
      HRF FTapping/Right 01   4.243697
      HRF FTapping/Right 02  -11.265726
      HRF FTapping/Right 03   20.680841
      HRF FTapping/Right 04  -28.569771
      HRF FTapping/Right 05   30.562608
      HRF FTapping/Right 06  -25.349188
      HRF FTapping/Right 07   15.793779
      HRF FTapping/Right 08   -6.882461
      HRF FTapping/Right 09    1.949848
      HRF FTapping/Right 10   -0.283047
      Drift Cos 0             0.019738
      Drift Cos 1            -0.062680
```

```
Drift Cos 2     -0.048399
Drift Cos 3     -0.004764
Drift Cos 4      0.098752
Drift Cos 5      0.038450
Drift Cos 6      0.049932
Drift Cos 7      0.015421
Drift Cos 8      0.003108
Drift Cos 9      0.050348
Drift Cos 10     0.032108
Drift Cos 11    -0.038858
short            0.654660
dtype: float64
```

Cedalion provides the `.sm` accessor on arrays of result objects to make accessing the statsmodels functionality easier.

```
[18]: display(results.sm.params)
```

```
<xarray.DataArray (channel: 44, chromo: 2, regressor: 46)> Size: 32kB
0.2767 -1.323 3.382 -5.959 7.91 … -0.008269 -0.0007833 0.01253 0.01076 0.2799
Coordinates:
  * regressor  (regressor) object 368B 'HRF Control 00' … 'short'
  * chromo     (chromo) <U3 24B 'HbO' 'HbR'
  * channel    (channel) object 352B 'S1D6' 'S1D8' 'S2D5' … 'S14D25' 'S14D27'
    source     (channel) object 352B 'S1' 'S1' 'S2' 'S2' … 'S13' 'S14' 'S14'
    detector   (channel) object 352B 'D6' 'D8' 'D5' 'D9' … 'D28' 'D25' 'D27'
Attributes:
    description:  AR_IRLS
```

The `glm.predict` function takes the original time series, the design matrix and the fitted parameters. Using these it predicts the sum of all regressors scaled with the best fitted parameters.

When only a subset of the fitted parameters is provided, only those regressors are considered. This allows to predict the signal or background components separately.

```
[19]: predicted = glm.predict(rec["ts_long"], results.sm.params, dms)

      predicted_background = glm.predict(
          rec["ts_long"],
          results.sm.params.sel(
              regressor=results.sm.params.regressor.str.startswith("Drift")
              | results.sm.params.regressor.str.startswith("short")
          ),
          dms,
      )
      predicted_signal = glm.predict(
          rec["ts_long"],
          results.sm.params.sel(regressor=results.sm.params.regressor.str.startswith("HRF")),
          dms,
      )

      meas_hrf_only = rec["ts_long"].pint.dequantify() - predicted_background
```

Compare the preprocessed data to the full model

```
[20]: channels = ["S3D1", "S12D22", "S9D25", "S11D31"]

      f, ax = p.subplots(len(channels),1, figsize=(16, 3*len(channels)), sharex=True)

      for i_ch, ch in enumerate(channels):
          ax[i_ch].plot(rec["ts_long"].time, rec["ts_long"].sel(channel=ch, chromo="HbO"),␣
      ↪"-", c="orange", label="Data HbO")
          ax[i_ch].plot(rec["ts_long"].time, rec["ts_long"].sel(channel=ch, chromo="HbR"),␣
      ↪"-", c="purple", label="Data HbR")
          ax[i_ch].plot(predicted_background.time, predicted.sel(channel=ch, chromo="HbO"),␣
      ↪"r-", label="Model HbO")
          ax[i_ch].plot(predicted_background.time, predicted.sel(channel=ch, chromo="HbR"),␣
      ↪"b-", label="Model HbR")
          vbx.plot_stim_markers(ax[i_ch], rec.stim, y=1)
          ax[i_ch].set_title(f"{ch} - {distances.sel(channel=ch).item().magnitude:.1f} mm")
          ax[i_ch].set_ylim(-1.75,1.75)
          ax[i_ch].legend(loc="upper left", ncol=8)
```

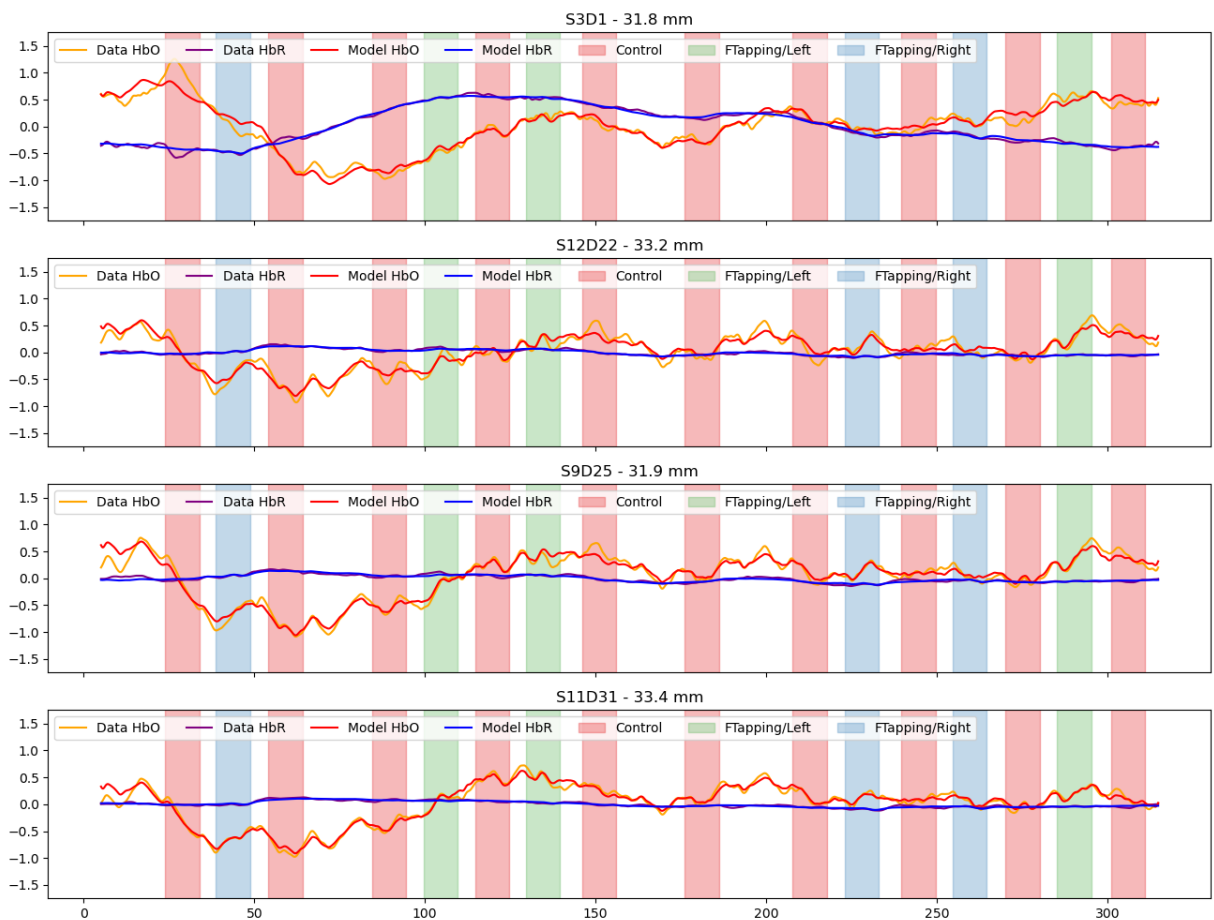

Subtract the components modelled by short and drift regressors and show the fitted activations

```
[21]: f, ax = p.subplots(len(channels),1, figsize=(16, 3*len(channels)), sharex=True)
```

```python
for i_ch, ch in enumerate(channels):
    ax[i_ch].plot(meas_hrf_only.time, meas_hrf_only.sel(channel=ch, chromo="HbO"),␣
↪"-", c="orange", label="Data - Model Drift HbO" )
    ax[i_ch].plot(meas_hrf_only.time, meas_hrf_only.sel(channel=ch, chromo="HbR"),␣
↪"-", c="purple", label="Data - Model Drift HbR")
    ax[i_ch].plot(predicted_signal.time, predicted_signal.sel(channel=ch,␣
↪chromo="HbO"), "r-", label="Model HRF HbO")
    ax[i_ch].plot(predicted_signal.time, predicted_signal.sel(channel=ch,␣
↪chromo="HbR"), "b-", label="Model HRF HbR")
    vbx.plot_stim_markers(ax[i_ch], rec.stim, y=1)

    ax[i_ch].legend(loc="upper left", ncol=8)
    ax[i_ch].set_title(ch)
    #ax[i_ch].set_ylim(-1.25, 1.25)
    ax[i_ch].set_ylim(-.5, .5)
```

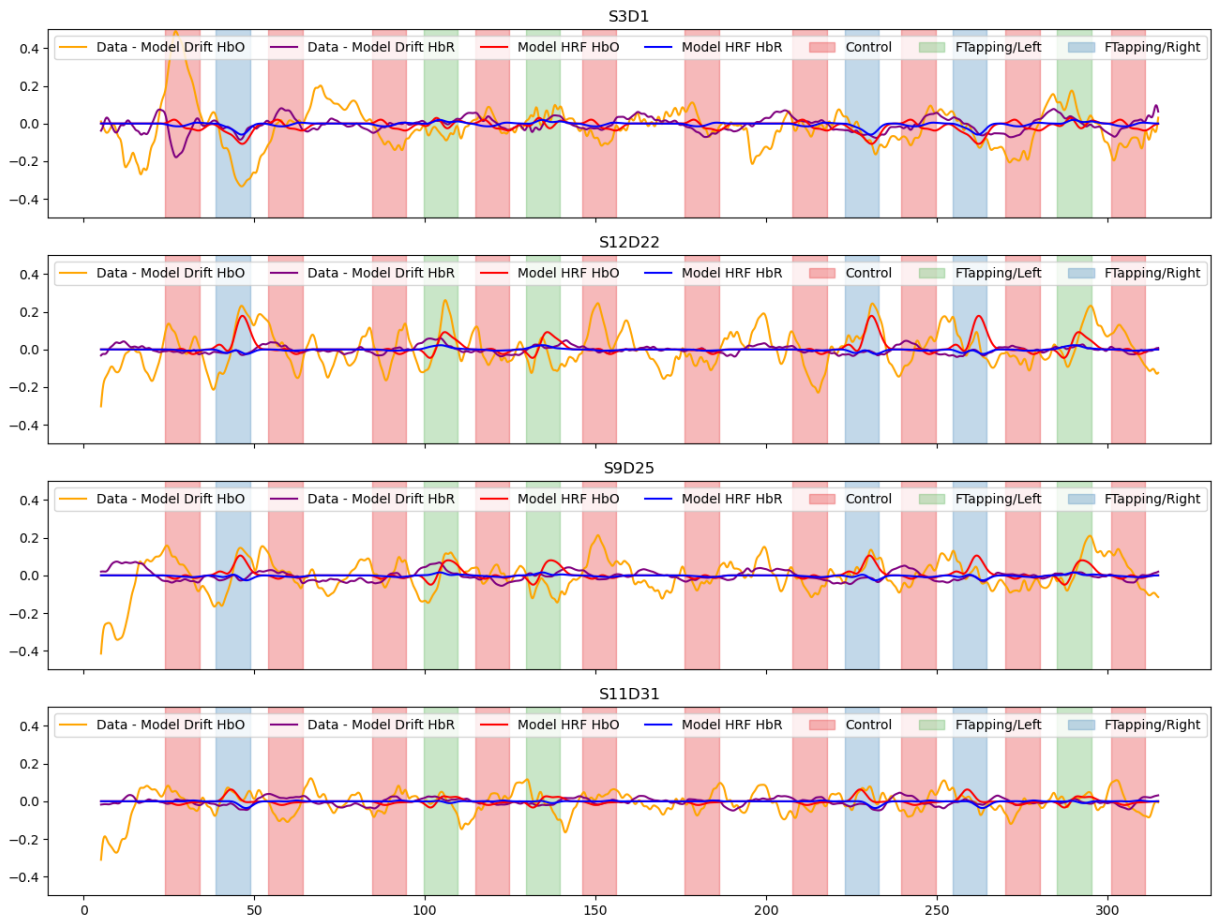

**Plot goodness of fit metrics**

To assess how well the fitted model describes the data, to metrics are calculated and visualized in scalp plots - the median absolute residuals (MAR) and - the coefficient of determination $R^2$

Note the scale difference between HbO and HbR for the MAR metric.

```
[22]: mar = np.abs(predicted - rec["ts_long"].pint.dequantify()).median("time")

      ss_res = ((rec["ts_long"].pint.dequantify() - predicted)**2).sum("time")
      ss_tot = ((rec["ts_long"] - rec["ts_long"].mean("time"))**2).pint.dequantify().
      ↪sum("time")
      r2 = (1 - ss_res / ss_tot)

      f,ax = p.subplots(2,2, figsize=(12,10))

      for i, chromo in enumerate(["HbO", "HbR"]):
          scalp_plot(
              rec["ts_long"],
              rec.geo3d,
              # results.sm.pvalues.sel(regressor=reg, chromo=chromo),
              mar.sel(chromo=chromo),
              ax[0,i],
              cmap="RdYlGn_r",
              vmin=0,
              vmax={"HbO" : 0.15, "HbR" : 0.05}[chromo],
              title=f"MAR - {chromo}",
              cb_label="MAR / µM",
              channel_lw=2,
              optode_labels=True,
              optode_size=24,
          )
          scalp_plot(
              rec["ts_long"],
              rec.geo3d,
              r2.sel(chromo=chromo),
              ax[1,i],
              cmap="RdYlGn",
              vmin=0,
              vmax=1,
              title=f"$R^2$ - {chromo}",
              cb_label="$R^2$",
              channel_lw=2,
              optode_labels=True,
              optode_size=24,
          )

      f.set_tight_layout(True)
```

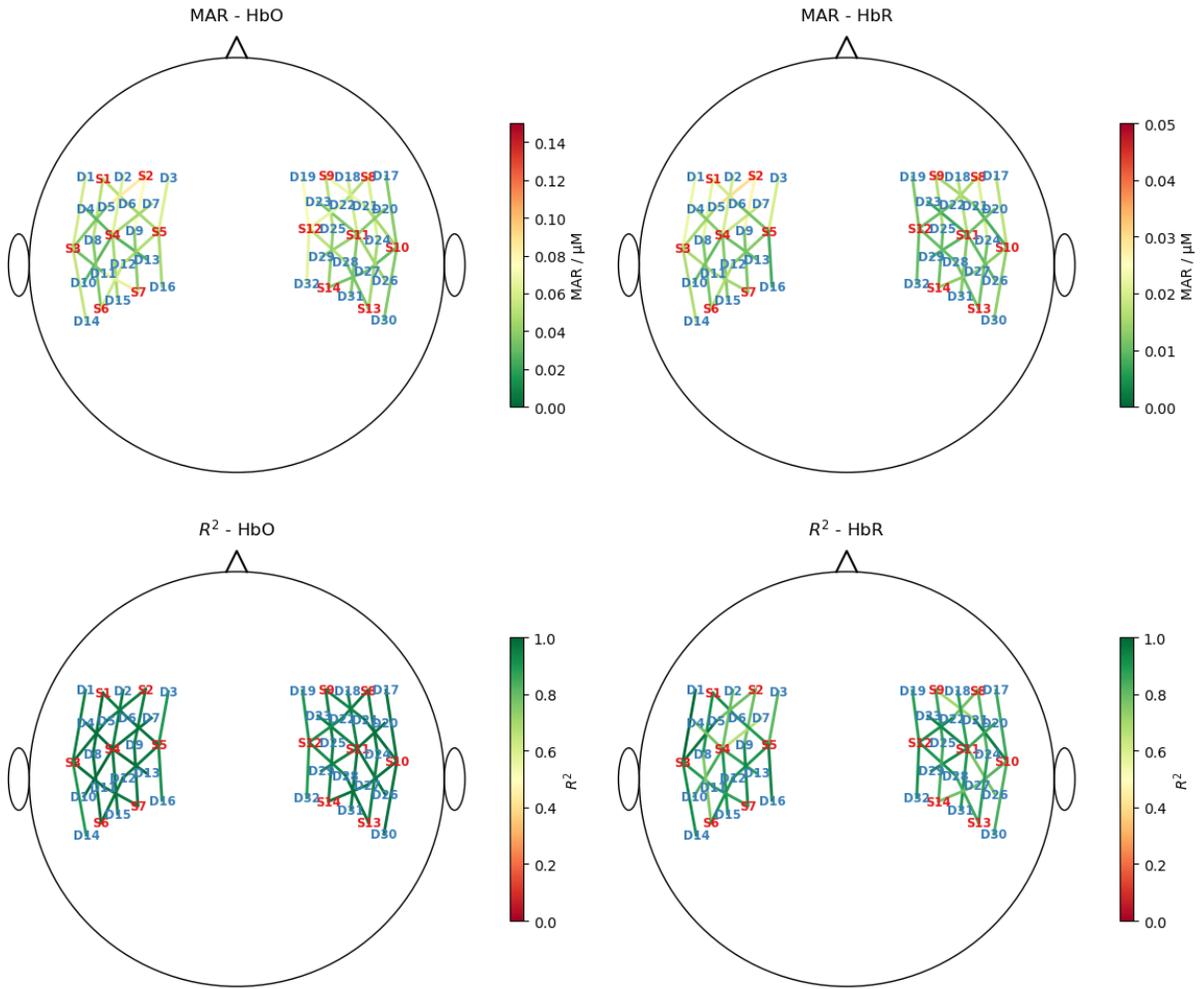

## Hypothesis tests

We would like to investigate which channels show activation during finger tapping. Answering this question with a hypothesis test requires specifying a null hypothesis. In this case, the null hypothesis states that the *size of the activation* during left- or right-hand finger tapping does not differ from the control condition.

For simple models with one regressor per condition, `statsmodels` provides a convenient interface for formulating such null hypotheses directly.

But in the present case, the HRF for each condition is modeled using multiple Gaussian kernels and the *size of the activation* needs to be formalized.

The null hypothesis H0 can be formulated as a linear combination of the fitted parameters $\beta_i$ and a contrast vector $c_i$ which should be zero if the null hypothesis is true. For example, in a fit with only three parameters, this would look like:

$$H0 : x = \begin{pmatrix} \beta_1 \\ \beta_2 \\ \beta_3 \end{pmatrix}^\top \cdot \begin{pmatrix} c_1 \\ c_2 \\ c_3 \end{pmatrix} = 0$$

Through the specification of different contrast vectors, different hypothesis can be formulated. For example:

$$c = \begin{pmatrix} 1 \\ 0 \\ -1 \end{pmatrix} \rightarrow H0 : \beta_1 = \beta_3$$

i.e the contrast vector encodes the hypothesis that there is no difference between $\beta_1$ and $\beta_3$.

Each such hypothesis is then tested, by checking if the linear combination significantly deviates from zero.

In the present case, we define the *size of the activation* as the area under the HRF. Because the model is linear, the area under the fitted HRF can be computed as a weighted sum of the Gaussian kernel areas $A_i$, with the weights given by the fitted parameters $\beta_i$:

$$A_{HRF} = \begin{pmatrix} \beta_1 \\ \beta_2 \\ \vdots \\ \beta_N \end{pmatrix}^{\top} \cdot \begin{pmatrix} A_1 \\ A_2 \\ \vdots \\ A_n \end{pmatrix}$$

Comparing with the formulas above, it becomes clear that by using the areas of the kernel basis functions as contrast weights and by assigning opposite signs to the conditions being compared, we can formulate the null hypothesis that the *size of the activation* (as quantified by HRF area) does not differ between two conditions.

Finally, we restrict the area computation to a time window in which the main hemodynamic response is expected, so that the calculated area reflects the activation of interest rather than unrelated fluctuations.

The following cell defines a function that computes contrast vectors from the HRF areas during 5 to 10 seconds post-stimulus then uses this function to formulate the two hypotheses: - "HRF FTapping/Left = HRF Control" and - "HRF FTapping/Right = HRF Control".

```python
[23]: def gaussian_kernel_timewindowed_auc_contrast(
          dms, df_stim, condition1: str, condition2: str, tmin: float, tmax: float
      ):
          """This function computes contrast vectors based on the time-windowed are of the
      ↪regressors."""

          time = dms.common.time

          # create two masks, that for each condition contains 1.0 only for
          # time samples between onset+tmin and onset+tmax. All other entries
          # zero
          mask_cond1 = np.zeros(len(time))
          mask_cond2 = np.zeros(len(time))

          for _, row in df_stim.iterrows():
              t1, t2 = row["onset"]+tmin, row["onset"]+tmax
              if row["trial_type"].startswith(condition1):
                  mask_cond1[(t1 <= time) & (time <= t2)] = 1.
              if row["trial_type"].startswith(condition2):
                  mask_cond2[(t1 <= time) & (time <= t2)] = 1.

          # each gaussian regressor is multiplied with the mask for its condition. This sets
          # all parts of the regressor outside the time window to zero. Through integration
      ↪the remaining
```

```python
    # area under the curve is calculated.

    nregressors = dms.common.sizes["regressor"]
    contrast = np.zeros(nregressors)
    for i in range(nregressors):
        if dms.common.regressor.values[i].startswith(f"HRF {condition1}"):
            contrast[i] = np.trapezoid(dms.common[:,i,0]*mask_cond1, dms.common.time)
        if dms.common.regressor.values[i].startswith(f"HRF {condition2}"):
            contrast[i] = - np.trapezoid(dms.common[:,i,0]*mask_cond2, dms.common.time)

    return contrast

hypothesis_labels = [
    "HRF FTapping/Left = HRF Control",
    "HRF FTapping/Right = HRF Control",
]

hypotheses = [
    gaussian_kernel_timewindowed_auc_contrast(dms, rec.stim, "FTapping/Left",
 ↪"Control", 5, 10),
    gaussian_kernel_timewindowed_auc_contrast(dms, rec.stim, "FTapping/Right",
 ↪"Control", 5, 10),
]

display(hypotheses)
```

```
[array([ -0.38536386,  -1.54846768,  -4.83084777, -11.81761093,
        -22.81664425, -35.12676303, -43.40391258, -43.25178978,
        -34.75902577, -22.41329998, -11.52180284,   0.11269669,
          0.45496975,   1.42568445,   3.50208166,   6.7877193 ,
         10.48759214,  13.0030286 ,  12.99994119,  10.48098719,
          6.78010497,   3.49667564,   0.        ,   0.        ,
          0.        ,   0.        ,   0.        ,   0.        ,
          0.        ,   0.        ,   0.        ,   0.        ,
          0.        ,   0.        ,   0.        ,   0.        ,
          0.        ,   0.        ,   0.        ,   0.        ,
          0.        ,   0.        ]),
 array([ -0.38536386,  -1.54846768,  -4.83084777, -11.81761093,
        -22.81664425, -35.12676303, -43.40391258, -43.25178978,
        -34.75902577, -22.41329998, -11.52180284,   0.        ,
          0.        ,   0.        ,   0.        ,   0.        ,
          0.        ,   0.        ,   0.12240493,   0.48687159,
          1.50425073,   3.64608706,   6.97863256,  10.65571472,
         13.06351251,  12.91859368,  10.303389  ,   6.59305506,
          3.36289301,   0.        ,   0.        ,   0.        ,
          0.        ,   0.        ,   0.        ,   0.        ,
          0.        ,   0.        ,   0.        ,   0.        ,
          0.        ,   0.        ])]
```

Using the calculated contrasts (i.e. hypotheses) t-tests are performed in each channel and chromophore. The accessor `.sm` is used to call the `t_test` function on all results, yielding a `DataArray` of `ContrastResults`:

```
[24]: contrast_results = results.sm.t_test(hypotheses)
      display(contrast_results)
```

```
<xarray.DataArray (channel: 44, chromo: 2)> Size: 704B
                          Test for Constraints                        …
Coordinates:
  * chromo     (chromo) <U3 24B 'HbO' 'HbR'
  * channel    (channel) object 352B 'S1D6' 'S1D8' 'S2D5' … 'S14D25' 'S14D27'
    source     (channel) object 352B 'S1' 'S1' 'S2' 'S2' … 'S13' 'S14' 'S14'
    detector   (channel) object 352B 'D6' 'D8' 'D5' 'D9' … 'D28' 'D25' 'D27'
Attributes:
    description:  AR_IRLS
```

Inspecting a single contrast result: the two rows labeled 'c0' and 'c1' correspond to the two hypotheses/contrasts "HRF FTapping/Left = HRF Control" and "HRF FTapping/Right = HRF Control".

```
[25]: display(contrast_results[0,0].item())
```

```
<class 'statsmodels.stats.contrast.ContrastResults'>
                     Test for Constraints
==============================================================================
              coef    std err          z      P>|z|      [0.025      0.975]
------------------------------------------------------------------------------
c0           1.0722      0.886      1.211      0.226      -0.664       2.808
c1           1.4259      0.885      1.611      0.107      -0.309       3.161
==============================================================================
```

**Apply FDR-control and visualize corrected p-values for all channels**

In the following, the p-values of all tests are visualized in a scalp plot. A Benjamini-Hochberg FDR correction is applied.

```
[26]: # create a colormap for p-values
      norm, cmap = p_values_cmap()

      nhypo = len(hypotheses)
      f, ax = p.subplots(2,len(hypotheses),figsize=(6 * nhypo,8))

      for i_row, chromo in enumerate(["HbO", "HbR"]):
          for i_hypo, hypo in enumerate(hypotheses):

              # get p_values for all channels and apply fdr correction
              # use Benjamini-Hochberg here
              p_values = contrast_results.sm.p_values().sel(chromo=chromo, hypothesis=i_hypo)
              _, q_values, _, _ = multipletests(p_values, alpha=0.05, method="fdr_bh")

              scalp_plot(
                  rec["ts_long"],
                  rec.geo3d,
                  np.log10(q_values),
                  ax[i_row][i_hypo],
                  cmap=cmap,
```

```
            norm=norm,
            title=f"'{hypothesis_labels[i_hypo]}'?\n{chromo}",
            cb_label="BH-corrected p-value",
            channel_lw=2,
            optode_labels=False,
            cb_ticks_labels=[(np.log10(i), str(i)) for i in [0.0001, 0.001, 0.005, 0.
    ↳01, 0.05, 1.]],
            optode_size=24,
        )

f.tight_layout()
```

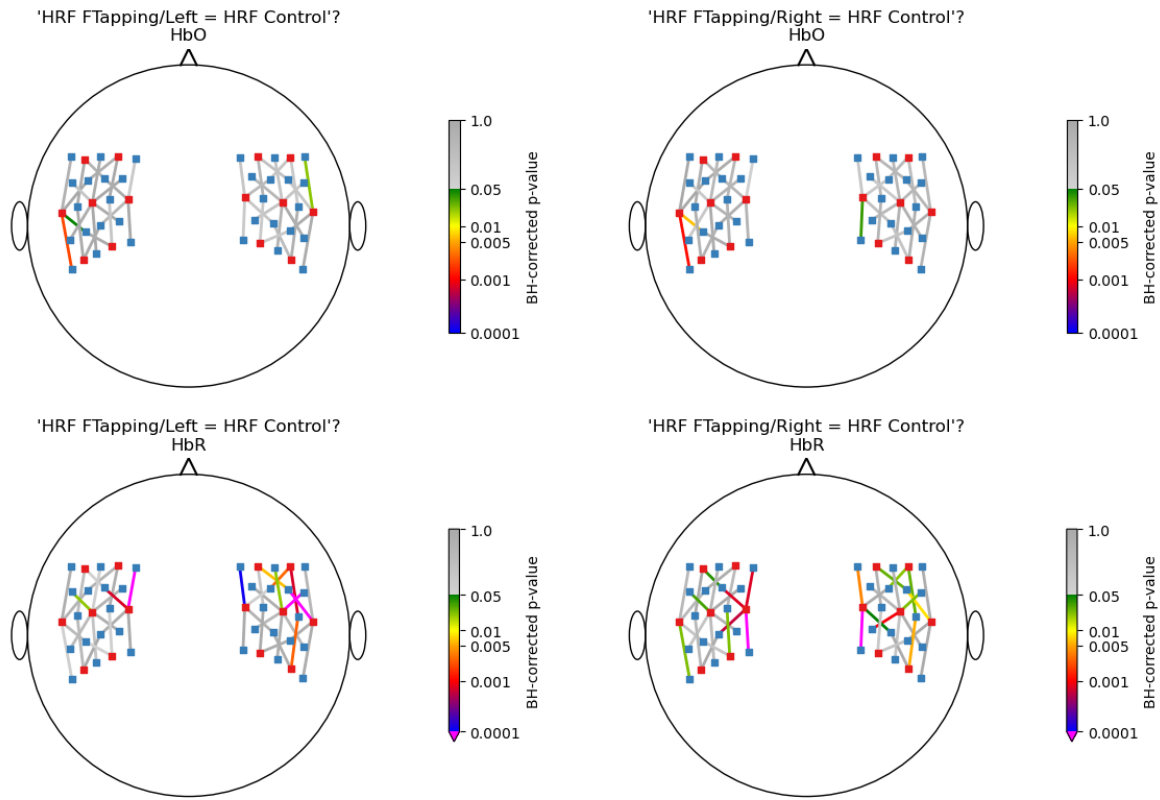

## Extract and plot HRFs with uncertainties

For extracting the HRF a special design matrix must be built. It spans only the duration of a single trial, contains only the HRF regressors and that HRF regressors are not convolved over the stimulus duration. The last point is irrelevant for Gaussian-kernel bases, since these kernels are not convolved in the first place.

```
[27]: dms_extract = glm.design_matrix.hrf_extract_regressors(rec["ts_long"], rec.stim,⎵
    ↳gaussian_kernels)
    display(dms_extract.common)
```

```
<xarray.DataArray (time: 77, regressor: 33, chromo: 2)> Size: 41kB
1.0 1.0 0.8694 0.8694 0.5698 0.5698 … 0.1581 0.3804 0.3804 0.6912 0.6912
Coordinates:
```

```
* time        (time) float64 616B -2.294 -2.064 -1.835 … 14.68 14.91 15.14
* regressor   (regressor) <U21 3kB 'HRF Control 00' … 'HRF FTapping/Right 10'
* chromo      (chromo) <U3 24B 'HbO' 'HbR'
```

Passing `dms_extract` to `glm.predict` would predict the hemodynamic response function for the parameters estimated in the fit.

In addition, we would like to assess how the uncertainties of the fitted parameters translate into uncertainties of the extracted HRFs.

Using the covariance matrix of the fitted parameters and a multivariate normal distribution, the function `glm.predict_with_uncertainty` samples sets of parameters around the best fit values. This ensemble of parameter sets, which are still compatible with the reported uncertainties of the fit, are used to calculte the mean HRF as well as an error band (standard deviation) around the mean response.

[28]:
```python
hrfs_control_mean, hrfs_control_std = glm.predict_with_uncertainty(
    rec["ts_long"],
    results,
    dms_extract,
    results.sm.params.regressor.str.startswith("HRF Control"),
)

for i_hypo, trial_type in enumerate(["HRF FTapping/Left", "HRF FTapping/Right"]):
    hrfs_mean, hrfs_std = glm.predict_with_uncertainty(
        rec["ts_long"],
        results,
        dms_extract,
        results.sm.params.regressor.str.startswith(trial_type),
    )

    p_values_hbo = contrast_results.sm.p_values().sel(chromo="HbO", hypothesis=i_hypo)
    p_values_hbr = contrast_results.sm.p_values().sel(chromo="HbR", hypothesis=i_hypo)

    # apply fdr correction (q-values are shown in captions)
    _, q_values_hbo, _, _ = multipletests(p_values_hbo, alpha=0.05, method="fdr_bh")
    _, q_values_hbr, _, _ = multipletests(p_values_hbr, alpha=0.05, method="fdr_bh")

    channels = hrfs_mean.channel.values

    f, ax = p.subplots(5, 9, figsize=(16, 9), sharex=True, sharey=True)
    ax = ax.flatten()

    for i_ch, ch in enumerate(channels[: len(ax)]):
        q_hbo = np.log10(q_values_hbo[i_ch])
        q_hbr = np.log10(q_values_hbr[i_ch])

        mm_hbo = hrfs_mean.sel(channel=ch, chromo="HbO")
        mm_hbr = hrfs_mean.sel(channel=ch, chromo="HbR")
        ss_hbo = hrfs_std.sel(channel=ch, chromo="HbO")
        ss_hbr = hrfs_std.sel(channel=ch, chromo="HbR")

        ax[i_ch].plot(mm_hbo.time, mm_hbo, "r-", label="HbO")
        ax[i_ch].fill_between(
```

```
        mm_hbo.time, mm_hbo - ss_hbo, mm_hbo + ss_hbo, fc="r", alpha=0.3
    )
    ax[i_ch].plot(mm_hbr.time, mm_hbr, "b-", label="HbR")
    ax[i_ch].fill_between(
        mm_hbr.time, mm_hbr - ss_hbr, mm_hbr + ss_hbr, fc="b", alpha=0.3
    )

    mm_hbo = hrfs_control_mean.sel(channel=ch, chromo="HbO")
    mm_hbr = hrfs_control_mean.sel(channel=ch, chromo="HbR")
    ss_hbo = hrfs_control_std.sel(channel=ch, chromo="HbO")
    ss_hbr = hrfs_control_std.sel(channel=ch, chromo="HbR")

    ax[i_ch].plot(mm_hbo.time, mm_hbo, "orange", label="HbO Control")
    ax[i_ch].fill_between(
        mm_hbo.time, mm_hbo - ss_hbo, mm_hbo + ss_hbo, fc="orange", alpha=0.3
    )
    ax[i_ch].plot(mm_hbr.time, mm_hbr, "magenta", label="HbR Control")
    ax[i_ch].fill_between(
        mm_hbr.time, mm_hbr - ss_hbr, mm_hbr + ss_hbr, fc="magenta", alpha=0.3
    )

    ax[i_ch].set_title(
        f"{ch}\nlogq: HbO:{q_hbo:.1f} HbR:{q_hbr:.1f}", fontdict={"fontsize": 8}
    )
    ax[i_ch].set_ylim(-0.3, 0.3)
    ax[i_ch].grid()
    ax[i_ch].axvline(0, c="k", lw=2)
    if i_ch == 0:
        ax[i_ch].legend(loc="upper left")
f.suptitle(trial_type)
f.set_tight_layout(True)
```

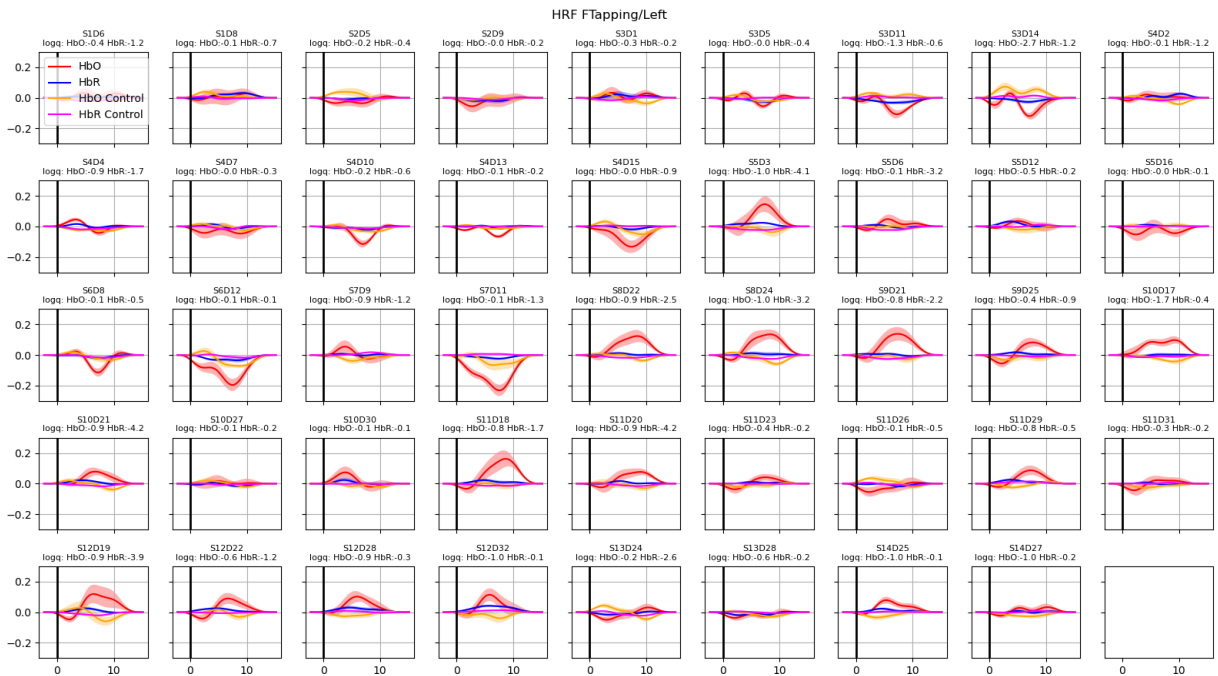

HRF FTapping/Right

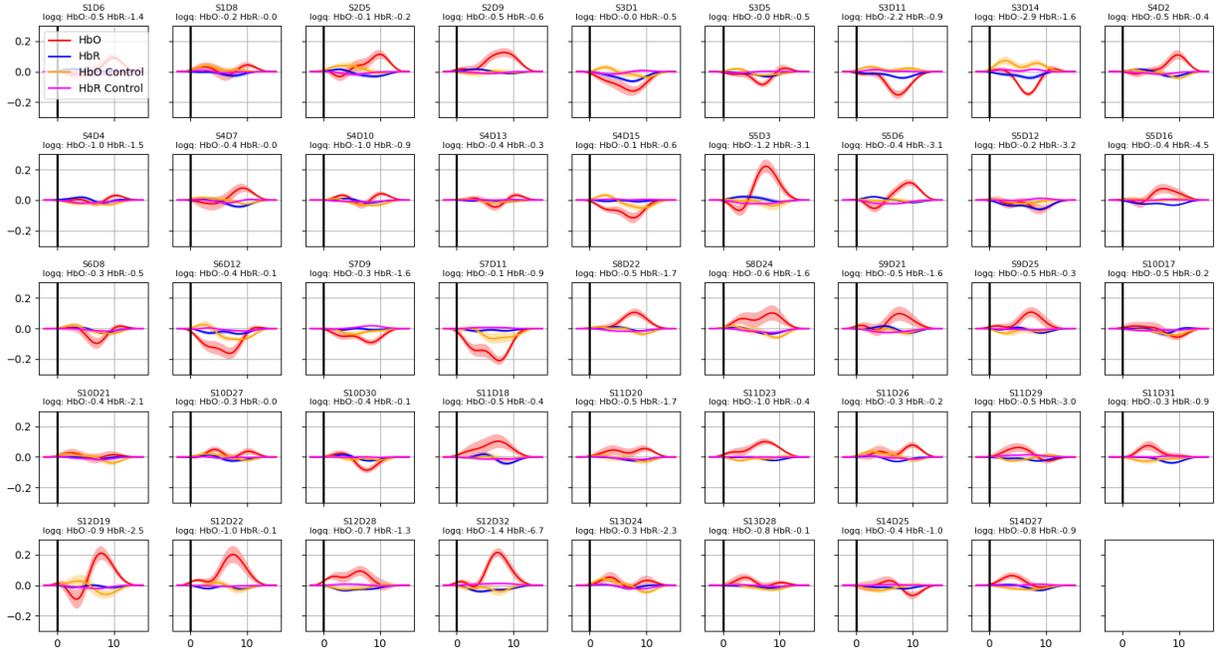

# S5: DOT - Image Reconstruction

```python
[1]:  # This cells setups the environment when executed in Google Colab.
      try:
          import google.colab
          !curl -s https://raw.githubusercontent.com/ibs-lab/cedalion/dev/scripts/
      ↪colab_setup.py -o colab_setup.py
          # Select branch with --branch "branch name" (default is "dev")
          %run colab_setup.py
      except ImportError:
          pass
```

## Notebook configuration

Decide for an example with a sparse probe or a high density probe for DOT. The notebook will load example data accordingly.

Also specify, if precomputed results of the photon propagation should be used and if the 3D visualizations should be interactive.

```python
[2]:  # the head model used in this example
      HEAD_MODEL = "colin27"

      # the dataset used in this example
      DATASET = "fingertappingDOT"

      # used precomputed forward model results
      PRECOMPUTED_SENSITIVITY = True

      FORWARD_MODEL = "MCX"

      # set this flag to True to enable interactive 3D plots
      INTERACTIVE_PLOTS = False
```

```python
[3]:  import pyvista as pv

      if INTERACTIVE_PLOTS:
          pv.set_jupyter_backend('server')
      else:
          pv.set_jupyter_backend('static')

      from pathlib import Path
      from tempfile import TemporaryDirectory

      import cedalion
      import cedalion.data
      import cedalion.dataclasses as cdc
      import cedalion.dot
      import cedalion.dot as dot
      import cedalion.io
      import cedalion.sigproc.motion as motion
      import cedalion.sigproc.physio as physio
```

```python
import cedalion.sigproc.quality as quality
import cedalion.vis.anatomy.sensitivity_matrix
import cedalion.vis.blocks as vbx
import matplotlib.pyplot as p
import numpy as np
import xarray as xr
from cedalion import units
from cedalion.io.forward_model import load_Adot
from IPython.display import Image

xr.set_options(display_expand_data=False);
```

```python
[4]: # helper function to display gifs in rendered notbooks
     def display_image(fname : str):
         display(Image(data=open(fname,'rb').read(), format='png'))
```

## Working Directory

In this notebook tHe output of the fluence and sensitivity calculations are stored in a temporary directory. This will be deleted when the notebook ends.

```python
[5]: temporary_directory = TemporaryDirectory()
     working_directory = Path(temporary_directory.name)
```

## Load a DOT finger-tapping dataset

- loading via `cedalion.data`
- give stimulus events descriptive string labels

```python
[6]: rec = cedalion.data.get_fingertappingDOT()

     rec.stim.cd.rename_events(
         {
             "1": "Control",
             "2": "FTapping/Left",
             "3": "FTapping/Right",
             "4": "BallSqueezing/Left",
             "5": "BallSqueezing/Right",
         }
     )
```

The location of the probes is obtained from the snirf metadata.

The units ('mm') are adopted and the coordinate system is named 'digitized'.

```python
[7]: display(rec.geo3d)
     print(f"CRS='{rec.geo3d.points.crs}'")
```

```
<xarray.DataArray (label: 346, digitized: 3)> Size: 8kB
[mm] -77.82 15.68 23.17 -61.91 21.23 56.49 ... 14.23 -38.28 81.95 -0.678 -37.03
Coordinates:
    type        (label) object 3kB PointType.SOURCE … PointType.LANDMARK
  * label       (label) <U6 8kB 'S1' 'S2' 'S3' 'S4' … 'FFT10h' 'FT10h' 'FTT10h'
Dimensions without coordinates: digitized
```

```
CRS='digitized'
```

```
[8]: cedalion.vis.anatomy.plot_montage3D(rec["amp"], rec.geo3d)
```

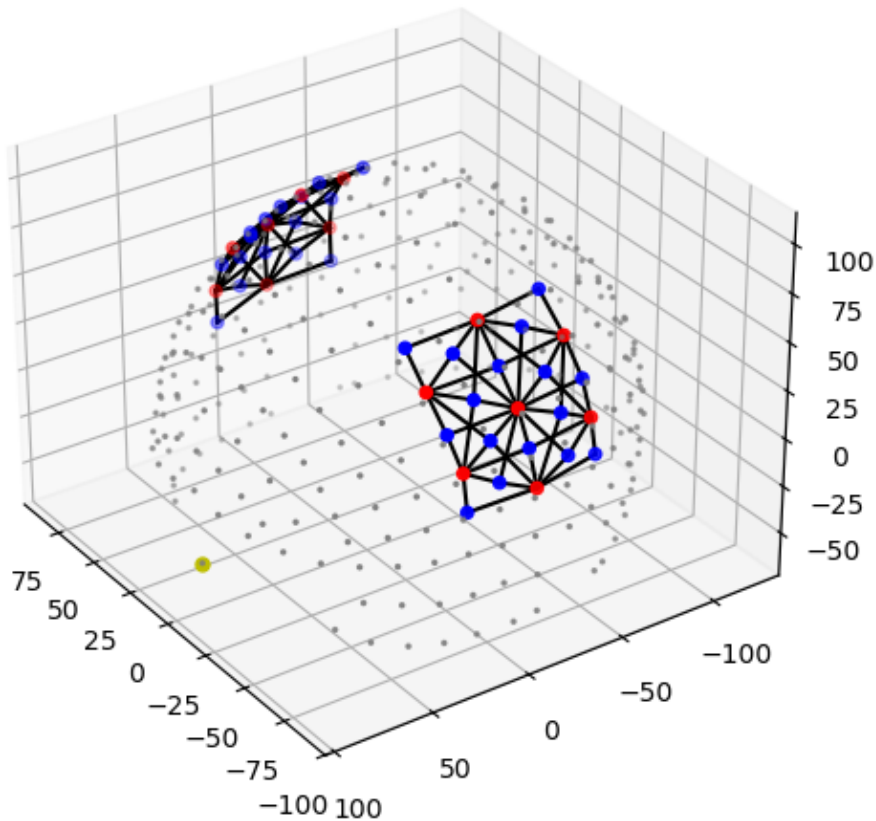

**Preprocessing**

```
[9]: rec["od"] = cedalion.nirs.cw.int2od(rec["amp"])
     rec["od_tddr"] = motion.tddr(rec["od"])
     rec["od_wavelet"] = motion.wavelet(rec["od_tddr"])

     # see 2_tutorial_preprocessing.ipynb for channel selection
     bad_channels = ['S13D26', 'S14D28']

     rec["od_clean"] = rec["od_wavelet"].sel(channel=~rec["od"].channel.isin(['S13D26',␣
     ↪'S14D28']))

     od_var = quality.measurement_variance(rec["od_clean"], calc_covariance=False)

     rec["od_mean_subtracted"], global_comp = physio.global_component_subtract(
         rec["od_clean"], ts_weights=1 / od_var, k=0
     )

     rec["od_freqfiltered"] = rec["od_mean_subtracted"].cd.freq_filter(
```

```
    fmin=0.01, fmax=0.5, butter_order=4
)
```

**Calculate block averages in optical density**

```
[10]:  # segment data into epochs
       epochs = rec["od_freqfiltered"].cd.to_epochs(
               rec.stim, # stimulus dataframe
               ["FTapping/Left", "FTapping/Right"],  # select events, discard the others
               before=5 * cedalion.units.s, # seconds before stimulus
               after=30 * cedalion.units.s, # seconds after stimulus
       )

       # calculate baseline
       baseline = epochs.sel(reltime=(epochs.reltime < 0)).mean("reltime")
       # subtract baseline
       epochs_blcorrected = epochs - baseline

       # group trials by trial_type. For each group individually average the epoch dimension
       blockaverage = epochs_blcorrected.groupby("trial_type").mean("epoch")

       # Plot block averages. Please ignore errors if the plot is too small in the HD case

       noPlts2 = int(np.ceil(np.sqrt(len(blockaverage.channel))))
       f,ax = p.subplots(noPlts2,noPlts2, figsize=(12,10))
       ax = ax.flatten()
       for i_ch, ch in enumerate(blockaverage.channel):
           for ls, trial_type in zip(["-", "--"], blockaverage.trial_type):
               ax[i_ch].plot(blockaverage.reltime, blockaverage.sel(wavelength=760,␣
       ↪trial_type=trial_type, channel=ch), "r", lw=2, ls=ls)
               ax[i_ch].plot(blockaverage.reltime, blockaverage.sel(wavelength=850,␣
       ↪trial_type=trial_type, channel=ch), "b", lw=2, ls=ls)

           ax[i_ch].grid(1)
           ax[i_ch].set_title(ch.values)
           ax[i_ch].set_ylim(-.01, .01)
           ax[i_ch].set_axis_off()
           ax[i_ch].axhline(0, c="k")
           ax[i_ch].axvline(0, c="k")

       p.suptitle("760nm: r | 850nm: b | left: - | right: --")
       p.tight_layout()
```

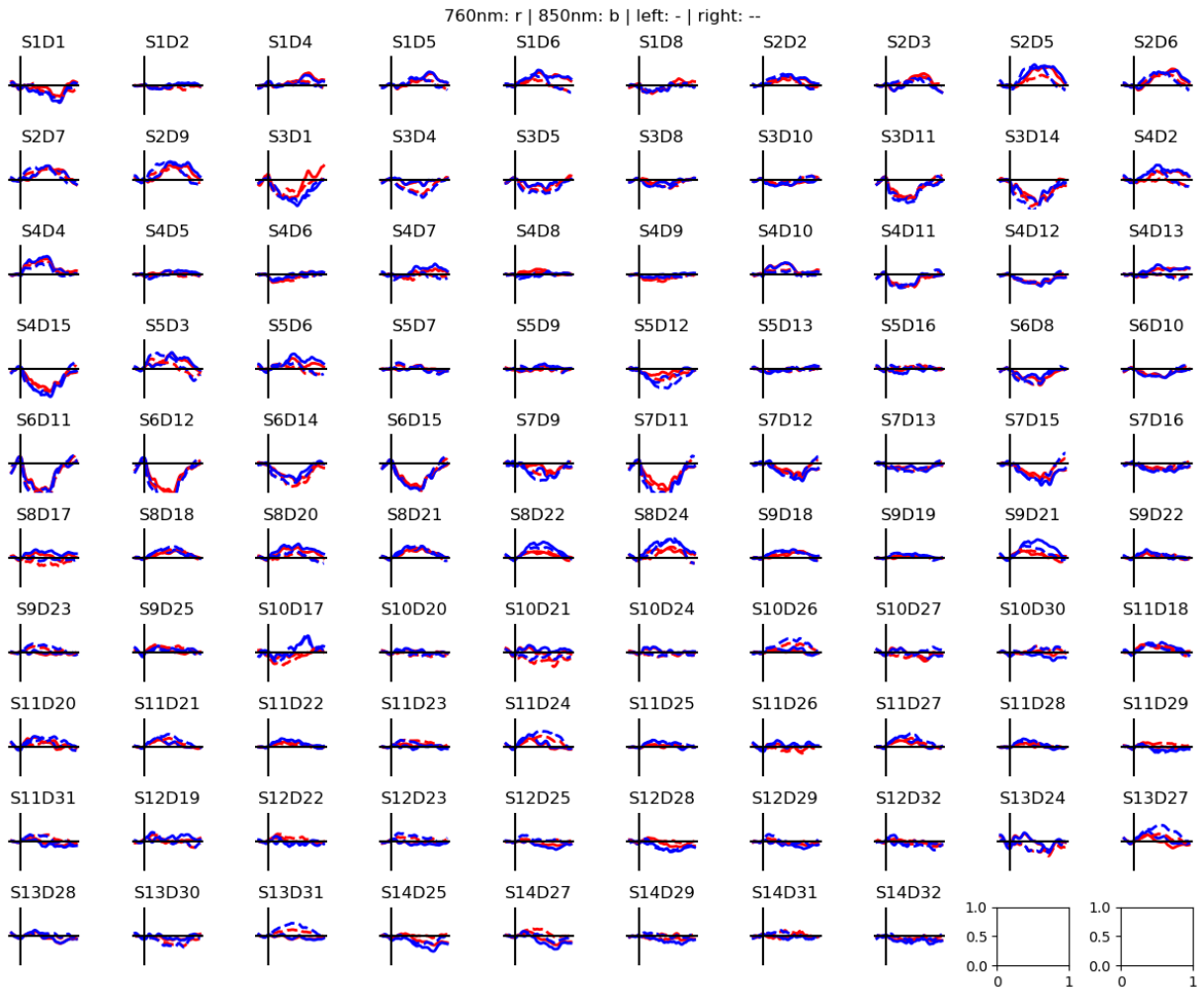

## Constructing the **TwoSurfaceHeadModel**

either factory or from segmentation masks notes about from_segmentation: The segmentation masks are in individual niftii files. The dict `mask_files` maps mask filenames relative to `SEG_DATADIR` to short labels. These labels describe the tissue type of the mask.

In principle the user is free to choose these labels. However, they are later used to lookup the tissue's optical properties. So they must be map to one of the tabulated tissue types (c.f. `cedalion.imagereco.tissue_properties.TISSUE_LABELS`).

The variable `landmarks_file` holds the path to a file containing landmark positions in scanner space (RAS). This file can be created with Slicer3D.

```
[11]: if HEAD_MODEL in ["colin27", "icbm152"]:
          # use a factory method for the standard head models
          head_ijk = cedalion.dot.get_standard_headmodel(HEAD_MODEL)

      elif HEAD_MODEL == "custom":
          # point these to a directory with segmentation masks
          segm_datadir = Path("/path/to/dir/with/segmentation_masks")
```

```python
    mask_files = {
        "csf": "mask_csf.nii",
        "gm": "mask_gray.nii",
        "scalp": "mask_skin.nii",
        "skull": "mask_bone.nii",
        "wm": "mask_white.nii",
    }
    # the landmarks must be in scanner (RAS) space.
    # For example Slicer3D can be used to pick the landmarks.
    landmarks_file = Path("/path/to/landmarks.mrk.json")

    # if available provide a mapping between vertices and labels
    parcel_file = None

    # Construct a head model from segmentation mask.
    head_ijk = cedalion.dot.TwoSurfaceHeadModel.from_segmentation(
        segmentation_dir=segm_datadir,
        mask_files=mask_files,
        landmarks_ras_file=landmarks_file,
        parcel_file=parcel_file,
        # adjust these to control mesh parameters
        brain_face_count=None,
        scalp_face_count=None,
        smoothing=0.0,
    )

    # Likely, better brain and scalp surfaces are achievable from
    # specialized segmentation tools.
    head_ijk = cedalion.dot.TwoSurfaceHeadModel.from_surfaces(
        segmentation_dir=segm_datadir,
        mask_files=mask_files,
        landmarks_ras_file=landmarks_file,
        parcel_file=parcel_file,
        brain_surface_file=segm_datadir / "mask_brain.obj",
        scalp_surface_file=segm_datadir / "mask_scalp.obj",
    )

else:
    raise ValueError("unknown head model")
```

## Optode Registration

The optode coordinates from the recording must be aligned with the scalp surface. Currently, cedaĺion offers a simple registration method, which finds an affine transformation (scaling, rotating, translating) that matches the landmark positions of the head model and their digitized counter parts. Afterwards, optodes are snapped to the nearest vertex on the scalp.

```python
[12]: geo3d_snapped_ijk = head_ijk.align_and_snap_to_scalp(rec.geo3d)
      display(geo3d_snapped_ijk)
```

```
<xarray.DataArray (label: 346, ijk: 3)> Size: 8kB
[] 15.1 140.9 100.2 30.75 144.6 130.6 … 172.5 137.6 42.28 172.7 126.0 43.07
Coordinates:
    type       (label) object 3kB PointType.SOURCE … PointType.LANDMARK
  * label      (label) <U6 8kB 'S1' 'S2' 'S3' 'S4' … 'FFT10h' 'FT10h' 'FTT10h'
Dimensions without coordinates: ijk
```

[13]:
```python
plt = pv.Plotter()
vbx.plot_surface(plt, head_ijk.brain, color="w")
vbx.plot_surface(plt, head_ijk.scalp, opacity=.1)
vbx.plot_labeled_points(plt, geo3d_snapped_ijk)
plt.show()
```

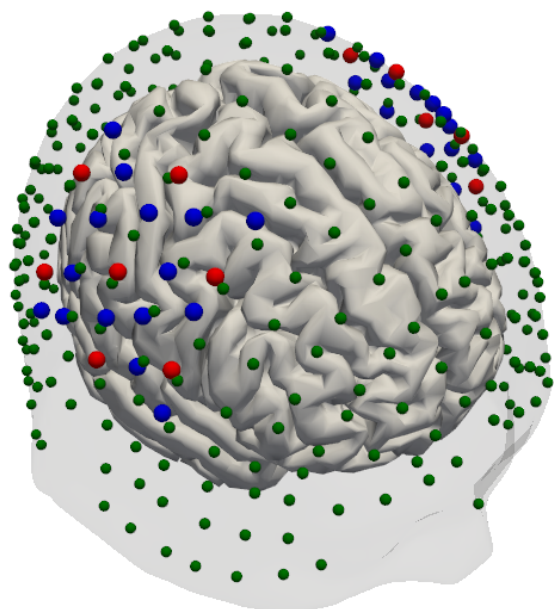

[14]:
```python
# remove landmarks from geo3d for subsequent plots
geo3d_plot = geo3d_snapped_ijk[geo3d_snapped_ijk.type != cdc.PointType.LANDMARK]
```

### Sensitivity matrix

[15]:
```python
Adot = cedalion.data.get_precomputed_sensitivity(DATASET, HEAD_MODEL)
```

**Plot Sensitivity Matrix**

```
[16]: plotter = cedalion.vis.anatomy.sensitivity_matrix.Main(
          sensitivity=Adot,
          brain_surface=head_ijk.brain,
          head_surface=head_ijk.scalp,
          labeled_points=geo3d_plot,
      )
      plotter.plot(high_th=0, low_th=-3)
      plotter.plt.show()
```

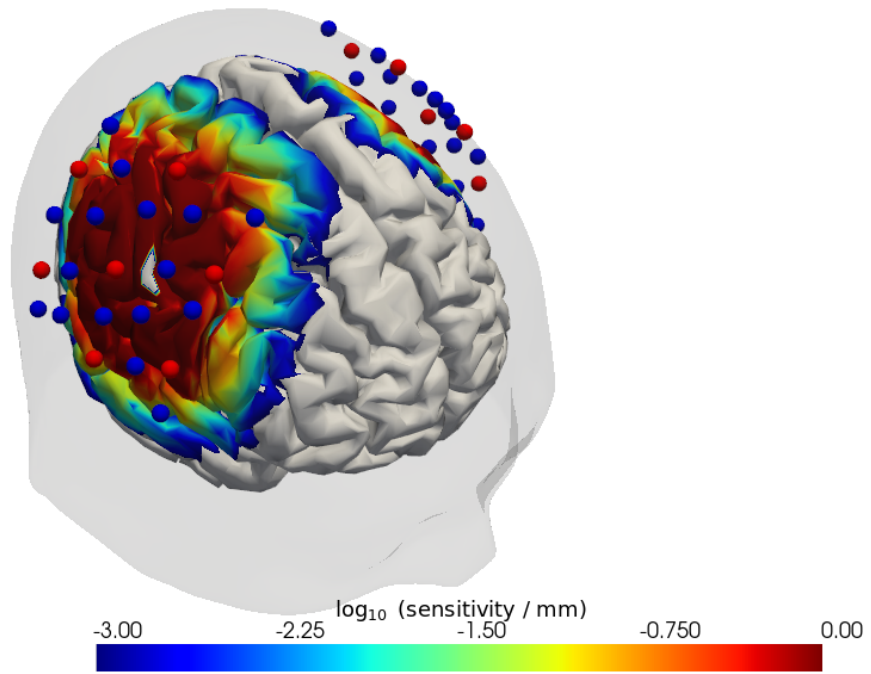

**Run the image reconstruction**

The class `cedalion.dot.ImageRecon` implements several methods to solve the inverse problem of recon-structing activations in brain and scalp tissue that would explain the measured optical density changes in channel space.

```
[17]: recon = cedalion.dot.ImageRecon(
          Adot,
          recon_mode="mua2conc",
          brain_only=False,
          alpha_meas=0.01,
          alpha_spatial=None,
```

```
        apply_c_meas=False,
        spatial_basis_functions=None,
)
```

In this example the channel-space block averages should be reconstructed into image space:

```
[18]: blockaverage
```

```
[18]: <xarray.DataArray (trial_type: 2, channel: 98, wavelength: 2, reltime: 154)> Size: 483kB
      0.001354 0.001405 0.001391 0.001314 … 0.0003461 0.0004625 0.0005884 0.0007154
      Coordinates:
        * reltime     (reltime) float64 1kB -5.038 -4.809 -4.58 … 29.54 29.77 30.0
        * channel     (channel) object 784B 'S1D1' 'S1D2' 'S1D4' … 'S14D31' 'S14D32'
          source      (channel) object 784B 'S1' 'S1' 'S1' 'S1' … 'S14' 'S14' 'S14'
          detector    (channel) object 784B 'D1' 'D2' 'D4' 'D5' … 'D29' 'D31' 'D32'
        * wavelength  (wavelength) float64 16B 760.0 850.0
        * trial_type  (trial_type) object 16B 'FTapping/Left' 'FTapping/Right'
```

Run the reconstruction and display results. Note how the OD time series with dimensions (`'channel'`, `'wavelength'`, ...) transformed into a concentration time series with dimensions (`'chromo'`, `'vertex'`, ...). The other dimensions (`'trial_type'`, `'reltime'`) did not change.

The `'vertex'` dimension has two coordinates: - `is_brain`: a boolean mask that discriminates between vertices on the brain and scalp surfaces - `parcel`: a string label attributed to each vertex according to a brain parcellation scheme (see section 'Parcel Space')

```
[19]: img = recon.reconstruct(blockaverage)
      img
```

```
[19]: <xarray.DataArray (chromo: 2, vertex: 25052, trial_type: 2, reltime: 154)> Size: 123MB
      [μM] -2.2e-11 -1.964e-11 -1.724e-11 … -1.489e-12 -1.522e-12 -1.547e-12
      Coordinates:
        * chromo      (chromo) <U3 24B 'HbO' 'HbR'
          parcel      (vertex) object 200kB 'VisCent_ExStr_8_LH' … 'scalp'
          is_brain    (vertex) bool 25kB True True True True … False False False
        * reltime     (reltime) float64 1kB -5.038 -4.809 -4.58 … 29.54 29.77 30.0
        * trial_type  (trial_type) object 16B 'FTapping/Left' 'FTapping/Right'
      Dimensions without coordinates: vertex
```

### Visualizing image reconstruction results

In the following, different cedalion plot functions will be showcased to visualize concentration changes on the brain and scalp surfaces.

### Channel Space

The function `cedalion.vis.anatomy.scalp_plot_gif` allows to create an animated gif of channel-space OD changes projected on the scalp.

Here, it is used to show the time course of the blockaveraged OD changes.

```
[20]: # configure the plot
      data_ts = blockaverage.sel(wavelength=850, trial_type="FTapping/Right")
      # scalp_plot_gif expects the time dimension to be named 'time'
```

```
data_ts = data_ts.rename({"reltime": "time"})
filename_scalp = "scalp_plot_ts"

# call plot function
cedalion.vis.anatomy.scalp_plot_gif(
    data_ts,
    rec.geo3d,
    filename=filename_scalp,
    time_range=(-5, 30, 0.5) * units.s,
    scl=(-0.01, 0.01),
    fps=6,
    optode_size=6,
    optode_labels=True,
    str_title="OD 850 nm",
)
```

[21]: 
```
display_image(f"{filename_scalp}.gif")
```

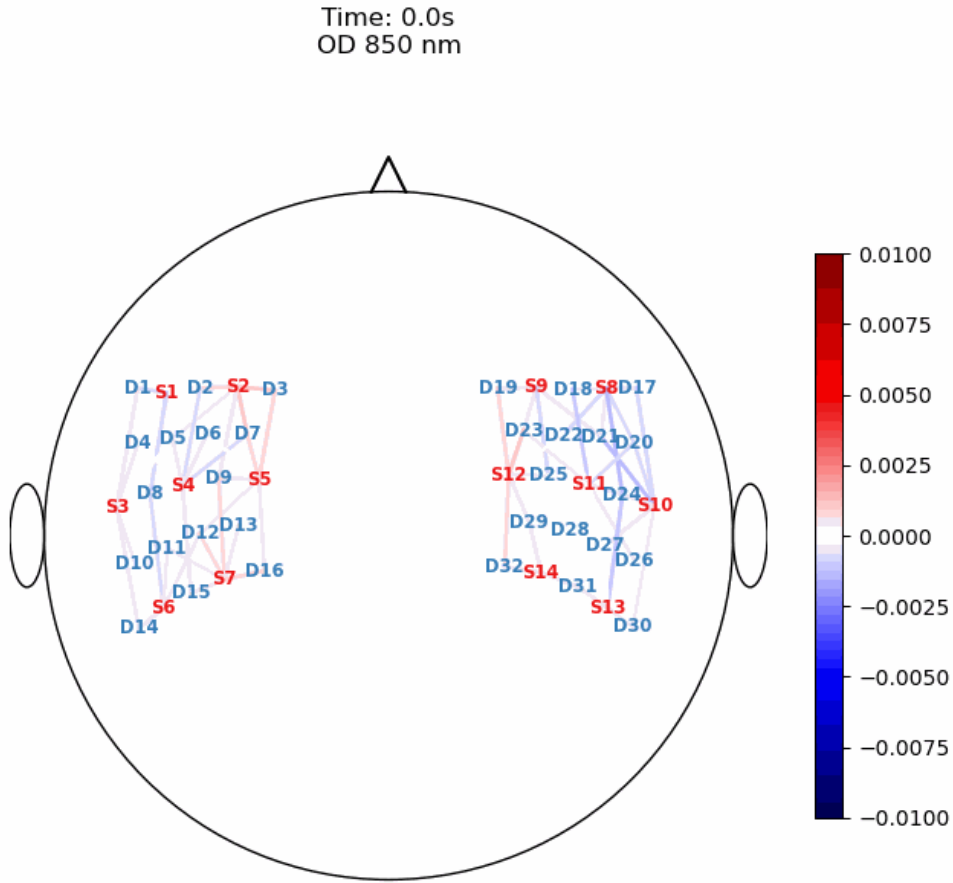

**Image Space**

**Single-View Animations of Activitations on the Brain**

The function `cedalion.vis.anatomy.image_recon_view` allows to create an animated gif of image-space concentration changes projected on the brain.

```
[22]: filename_view = 'image_recon_view'

      # image_recon_view expects input with dimensions
      # ("vertex", "chromo", "time") (in that order)
      X_ts = img.sel(trial_type="FTapping/Right").rename({"reltime": "time"})
      X_ts = X_ts.transpose("vertex", "chromo", "time")

      scl = np.percentile(np.abs(X_ts.sel(chromo='HbO')).pint.dequantify(), 99)
      clim = (-scl,scl)

      cedalion.vis.anatomy.image_recon_view(
          X_ts,   # time series data; can be 2D (static) or 3D (dynamic)
          head_ijk,
          cmap='seismic',
          clim=clim,
          view_type='hbo_brain',
          view_position='left',
          title_str='HbO / µM',
          filename=filename_view,
          SAVE=True,
          time_range=(-5,30,0.5)*units.s,
          fps=6,
          geo3d_plot = geo3d_plot,
          wdw_size = (1024, 768)
      )
```

```
[23]: display_image(f"{filename_view}.gif")
```

Time = 0.0 sec

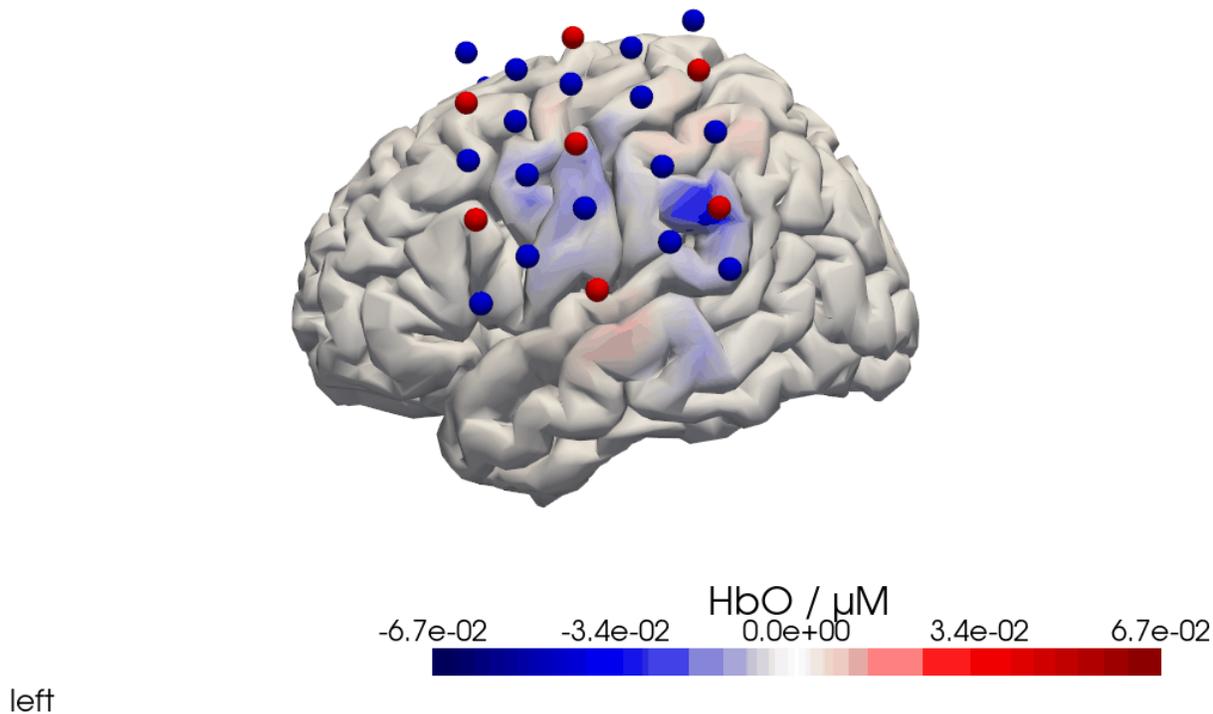

left

Alternatively, we can just select a single time point and plot activity as a still image at that time. Note the different file suffix (.png).

```
[24]:  # selects the nearest time sample at t=4s in X_ts
       # note: sel does not accept quantified units
       X_ts_plot = X_ts.sel(time=4, method="nearest")

       filename_view = 'image_recon_view_still'

       cedalion.vis.anatomy.image_recon_view(
           X_ts_plot,        # time series data; can be 2D (static) or 3D (dynamic)
           head_ijk,
           cmap='seismic',
           clim=clim,
           view_type='hbo_brain',
           view_position='left',
           title_str='HbO / µM',
           filename=filename_view,
           SAVE=True,
           time_range=(-5,30,0.5)*units.s,
           fps=6,
           geo3d_plot = geo3d_plot,
           wdw_size = (1024, 768)
```

)

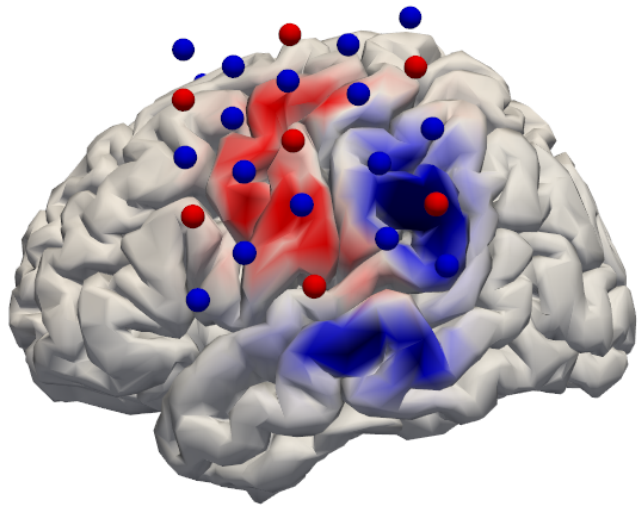

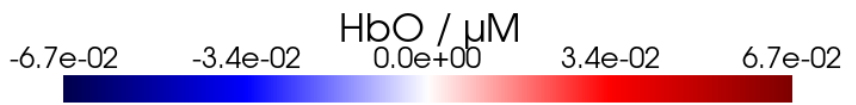

HbO / μM

left

[25]: `display_image(f"{filename_view}.png")`

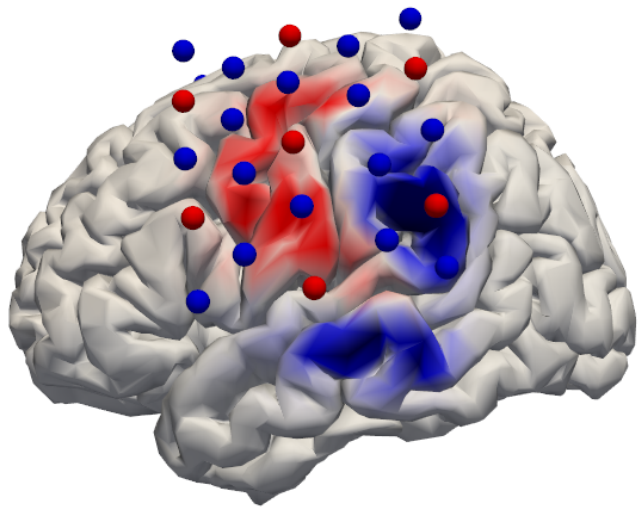

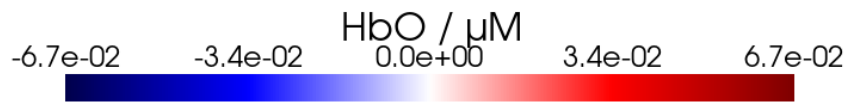

left

## Multi-View Animations of Activitations on the Brain

The function `cedalion.vis.anatomy.image_recon_multi_view` shows the activity on the brain from all angles as still image or animated across time:

```
[26]: filename_multiview = 'image_recon_multiview'

      # prepare data
      X_ts = img.sel(trial_type="FTapping/Right").rename({"reltime": "time"})
      X_ts = X_ts.transpose("vertex", "chromo", "time")

      scl = np.percentile(np.abs(X_ts.sel(chromo='HbO')).pint.dequantify(), 99)
      clim = (-scl,scl)

      cedalion.vis.anatomy.image_recon_multi_view(
          X_ts,  # time series data; can be 2D (static) or 3D (dynamic)
          head_ijk,
          cmap='seismic',
          clim=clim,
          view_type='hbo_brain',
          title_str='HbO / µM',
          filename=filename_multiview,
          SAVE=True,
```

```
        time_range=(-5,30,0.5)*units.s,
        fps=6,
        geo3d_plot = None, # geo3d_plot
        wdw_size = (1024, 768)
)
```

`[27]:` 
```
display_image(f"{filename_multiview}.gif")
```

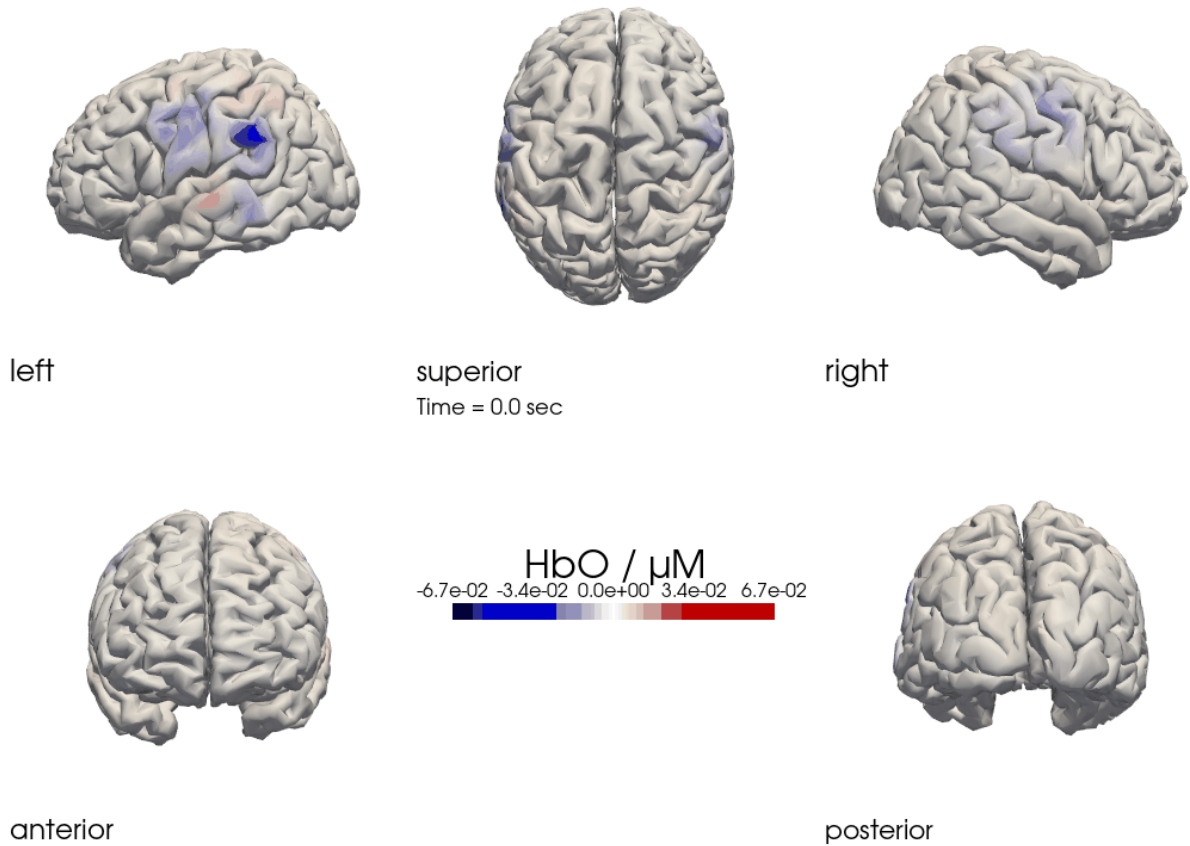

**Multi-View Animations of Activitations on the Scalp**

This gives us activity on the scalp after recon from all angles as still image or across time

`[28]:` 
```
filename_multiview_scalp = 'image_recon_multiview_scalp'

# prepare data
X_ts = img.sel(trial_type="FTapping/Right").rename({"reltime": "time"})
X_ts = X_ts.transpose("vertex", "chromo", "time")

scl = np.percentile(np.abs(X_ts.sel(chromo='HbO')).pint.dequantify(), 99)
clim = (-scl,scl)

cedalion.vis.anatomy.image_recon_multi_view(
    X_ts, # time series data; can be 2D (static) or 3D (dynamic)
```

```
        head_ijk,
        cmap='seismic',
        clim=clim,
        view_type='hbo_scalp',
        title_str='HbO / uM',
        filename=filename_multiview_scalp,
        SAVE=True,
        time_range=(-5,30,0.5)*units.s,
        fps=6,
        geo3d_plot = geo3d_plot,
        wdw_size = (1024, 768)
)
```

[29]: 
```
display_image(f"{filename_multiview_scalp}.gif")
```

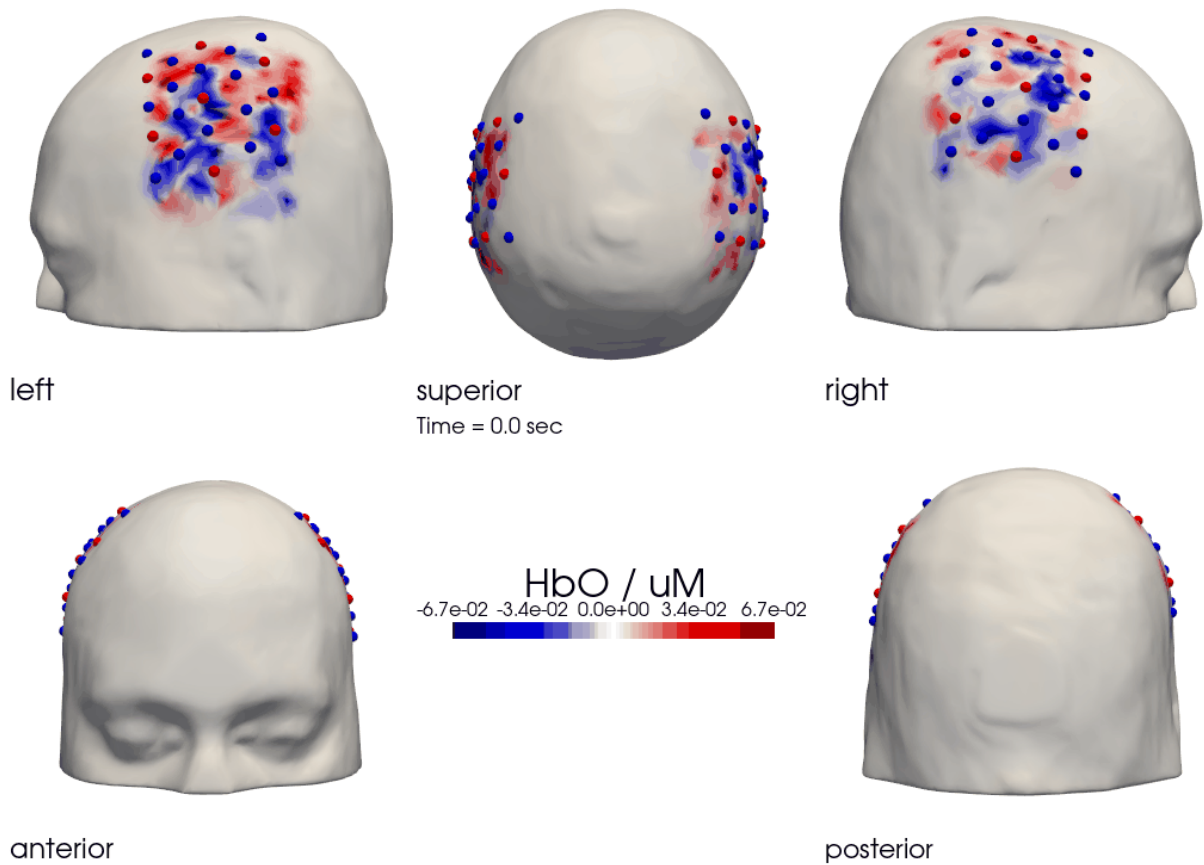

**Parcel Space**

The Schaefer Atlas [?] as implemented in Cedalion provides nearly 600 labels for different regions of the brain. Each vertex of the brain surface has its correspondng parcel label assigned as a coordinate.

[30]: 
```
head_ijk.brain.vertices
```

```
[30]: <xarray.DataArray (label: 15002, ijk: 3)> Size: 360kB
      []  77.71 20.23 74.63 83.52 19.84 69.31 … 128.0 152.7 100.5 97.28 105.3 87.73
      Coordinates:
        * label    (label) int64 120kB 0 1 2 3 4 5 … 14997 14998 14999 15000 15001
        * parcel   (label) <U44 3MB 'VisCent_ExStr_8_LH' … 'Background+FreeSurfer…
      Dimensions without coordinates: ijk
```

To obtain the average hemodynamic response in a parcel, the baseline-subtraced concentration changes of all vertices in a parcel are averaged. As baseline the first sample is used.

```
[31]: # only consider brain vertices in the following.
      img_brain = img.sel(vertex=img.is_brain)

      # use the first sample along the reltime dimension as baseline
      baseline = img_brain.isel(reltime=0)
      img_brain_blsubtracted = img_brain - baseline

      display(img_brain_blsubtracted.rename("img_brain_blsubtracted"))
```

```
<xarray.DataArray 'img_brain_blsubtracted' (chromo: 2, vertex: 15002,
                                            trial_type: 2, reltime: 154)> Size: 74MB
[µM] 0.0 2.36e-12 4.763e-12 7.323e-12 … 1.409e-11 1.376e-11 1.334e-11
Coordinates:
  * chromo      (chromo) <U3 24B 'HbO' 'HbR'
    parcel      (vertex) object 120kB 'VisCent_ExStr_8_LH' … 'Background+Fr…
    is_brain    (vertex) bool 15kB True True True True … True True True True
  * reltime     (reltime) float64 1kB -5.038 -4.809 -4.58 … 29.54 29.77 30.0
  * trial_type  (trial_type) object 16B 'FTapping/Left' 'FTapping/Right'
Dimensions without coordinates: vertex
```

Using parcel labels, vertices belonging to the same brain region can be easily grouped together with the `DataArray.groupby` function. On these vertex groups many reduction methods can be executed. He we use the function `mean()` to average over vertices.

Note how the time series dimension changed from `vertex` to `parcel`:

```
[32]: # average over parcels
      avg_HbO = img_brain_blsubtracted.sel(chromo="HbO").groupby('parcel').mean()
      avg_HbR = img_brain_blsubtracted.sel(chromo="HbR").groupby('parcel').mean()

      display(avg_HbO.rename("avg_HbO"))
```

```
<xarray.DataArray 'avg_HbO' (parcel: 602, trial_type: 2, reltime: 154)> Size: 1MB
[µM] 0.0 1.002e-09 2.205e-09 3.714e-09 … -8.357e-12 -8.706e-12 -8.967e-12
Coordinates:
    chromo      <U3 12B 'HbO'
  * reltime     (reltime) float64 1kB -5.038 -4.809 -4.58 … 29.54 29.77 30.0
  * trial_type  (trial_type) object 16B 'FTapping/Left' 'FTapping/Right'
  * parcel      (parcel) object 5kB 'Background+FreeSurfer_Defined_Medial_Wal…
```

The montage in this dataset covers only parts of the head. Consequently, many brain regions lack significant signal coverage due to the absence of optodes.

To focus on relevant regions, a subset of parcels from the somatosensory and motor regions in both hemispheres is selected.

```
[33]: selected_parcels = [
          "SomMotA_1_LH", "SomMotA_3_LH", "SomMotA_4_LH",
          "SomMotA_5_LH", "SomMotA_9_LH", "SomMotA_10_LH",

          "SomMotA_1_RH", "SomMotA_2_RH", "SomMotA_3_RH",
          "SomMotA_4_RH", "SomMotA_6_RH", "SomMotA_7_RH"
      ]
```

The following plot visualizes the montage and the selected parcels:

```
[34]: # map parcel labels to colors
      parcel_colors = {
          parcel: p.cm.jet(i / (len(selected_parcels) - 1))
          for i, parcel in enumerate(selected_parcels)
      }

      plt = pv.Plotter()

      vbx.plot_surface(
          plt,
          head_ijk.scalp,
          color="w",
          opacity=.1
      )

      vbx.plot_surface(
          plt,
          head_ijk.brain,
          color=np.asarray(
              [
                  parcel_colors.get(parcel, (0.8, 0.8, 0.8, 1.))
                  for parcel in head_ijk.brain.vertices.parcel.values
              ]
          ),
          silhouette=True,
      )

      vbx.plot_labeled_points(plt, geo3d_plot)

      plt.add_legend(labels=parcel_colors.items(), face="o", size=(0.3, 0.3))

      vbx.camera_at_cog(plt, head_ijk.brain, rpos=(0, 0, 1), up=(0, 1, 0), fit_scene=True)
      plt.show()
```

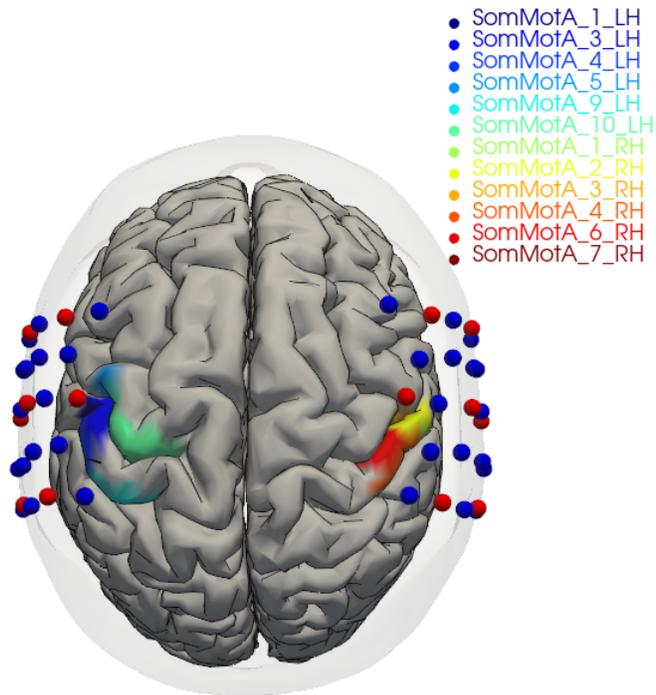

Plot averaged time traces in each parcel for the 'FTapping/Right' and 'FTapping/Left' conditions:

```
[35]: f,ax = p.subplots(2,6, figsize=(20,5))
      ylims = -0.02, 0.05
      ax = ax.flatten()
      for i_par, par in enumerate(selected_parcels):
          ax[i_par].plot(avg_HbO.sel(parcel = par, trial_type = "FTapping/Right").reltime,
          ↪avg_HbO.sel(parcel = par, trial_type = "FTapping/Right"), "r", lw=2, ls='-')
          ax[i_par].plot(avg_HbR.sel(parcel = par, trial_type = "FTapping/Right").reltime,
          ↪avg_HbR.sel(parcel = par, trial_type = "FTapping/Right"), "b", lw=2, ls='-')

          ax[i_par].grid(1)
          ax[i_par].set_title(par, color=parcel_colors[par])
          ax[i_par].set_ylim(*ylims)

          if i_par % 6 == 0:
              ax[i_par].set_ylabel("µM")
          if i_par >=6:
              ax[i_par].set_xlabel("$t_{rel} / s$")

      p.suptitle("Parcellations: HbO: r | HbR: b | FTapping/Right", y=1)
      p.tight_layout()
```

```
f,ax = p.subplots(2,6, figsize=(20,5))
ax = ax.flatten()
for i_par, par in enumerate(selected_parcels):
    ax[i_par].plot(avg_HbO.sel(parcel = par, trial_type = "FTapping/Left").reltime,⎵
↪avg_HbO.sel(parcel = par, trial_type = "FTapping/Left"), "r", lw=2, ls='-')
    ax[i_par].plot(avg_HbR.sel(parcel = par, trial_type = "FTapping/Left").reltime,⎵
↪avg_HbR.sel(parcel = par, trial_type = "FTapping/Left"), "b", lw=2, ls='-')

    ax[i_par].grid(1)
    ax[i_par].set_title(par, color=parcel_colors[par])
    ax[i_par].set_ylim(*ylims)

    if i_par % 6 == 0:
        ax[i_par].set_ylabel("µM")
    if i_par >=6:
        ax[i_par].set_xlabel("$t_{rel} / s$")

p.suptitle("Parcellations: HbO: r | HbR: b | FTapping/Left", y=1)
p.tight_layout()
```

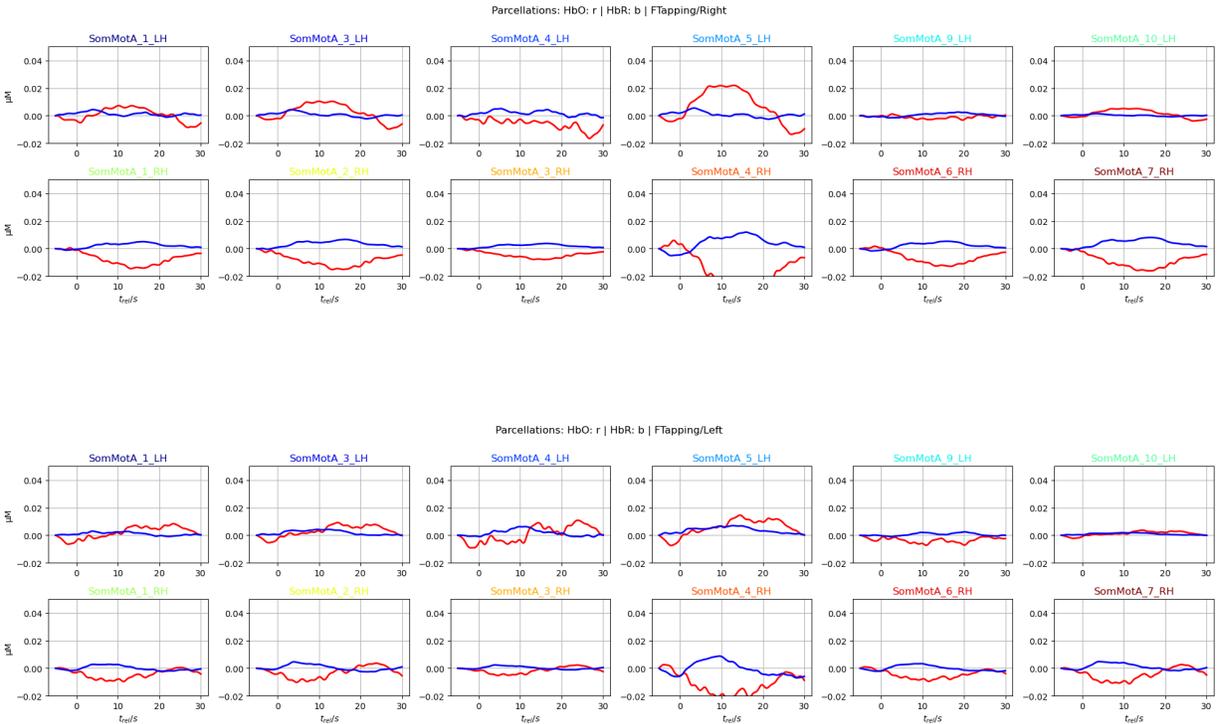

# S6: Data-Driven (ML) Analysis

This notebook has three sections:

- Multimodal-source-decomposition methods on simulated fNIRS-EEG data
- ICA Source Extraction
- Single Trial Classification

```python
[1]: # This cells setups the environment when executed in Google Colab.
     try:
         import google.colab
         !curl -s https://raw.githubusercontent.com/ibs-lab/cedalion/dev/scripts/
     ↪colab_setup.py -o colab_setup.py
         # Select branch with --branch "branch name" (default is "dev")
         %run colab_setup.py
     except ImportError:
         pass
```

```python
[2]: import cedalion
     import cedalion.data
     import cedalion.sigproc.quality as quality
     import matplotlib.pyplot as plt
     import numpy as np
     import scipy as sp
     import xarray as xr
     from cedalion import units
     import cedalion.sigproc.physio as physio
     import cedalion.sigproc.motion as motion
     from cedalion.sigdecomp.unimodal import ICA_ERBM
     from cedalion.sigproc.frequency import sampling_rate
     from cedalion.sim.datasets.synthetic_fnirs_eeg import (
         BimodalToyDataSimulation,
         standardize,
     )

     # Limit display to show 3 items at each edge, then "..."
     np.set_printoptions(threshold=20, edgeitems=2)
     xr.set_options(display_expand_data=False)

     # To reaload packages and modules automatically.
     %load_ext autoreload
     %autoreload 2
```

## Multimodal Source Decomposition Methods on Simulated fNIRS-EEG data

In this tutorial, we show how different multimodal source-decomposition methods can be used on an example toy fNIRS-EEG dataset to extract the underlying common (neural) sources. In particular, we cover Canonical Correlation Analysis (CCA), its regularized and temporally-embedded variants, and the multimodal Source Power Co-Modulation (mSPoC) algorithm from [Dähne, et.al. 2013]. All these methods can be found in the `cedalion.sigdecomp.multimodal` package.

```python
[3]: # Define helper plotting function for later use
     def plot_source_comparisson(
```

```
    sx_power, sy, sx_power_model, sy_model, corrx, corry, corrxy, title
):
    """Plot comparison between original and reconstructed sources."""

    fig, ax = plt.subplots(2, 1, figsize=(12, 4), sharex=True)

    ax[0].plot(sx_power.time, sx_power, label="sx_power", color="red")
    ax[0].plot(
        sx_power_model.time,
        sx_power_model,
        label="sx_power reconstructed",
        color="blue",
    )
    ax[0].set_title("Sx | Correlation: {:.3f}".format(corrx))
    ax[0].set_ylabel("Amplitude")
    ax[0].legend()
    ax[0].grid()

    ax[1].plot(sy.time, sy, label="sy", color="green")
    ax[1].plot(sy_model.time, sy_model, label="sy reconstructed", color="orange")
    ax[1].set_title("Sy | Correlation: {:.3f}".format(corry))
    ax[1].set_xlabel("Time")
    ax[1].set_ylabel("Amplitude")
    ax[1].legend()
    ax[1].grid()

    plt.suptitle(
        f"{title} Reconstructed Sources | Correlation: {corrxy:.3f}", fontsize=16
    )
    plt.tight_layout()
    plt.show()
```

**Simulated Dataset**

For the simulated data we follow a small extension of the approach presented in [Dähne et al. 2013] that can be found whitin Cedalion, inside the `sim.datasets.synthetic_fnirs_eeg` module. #### Brief toy data description: EEG ($x$) and fNIRS ($y$) recordings are generated from a pseudo-random linear mixing forward model. Sources $s_x, s_y$ are split into background (independent between modalities) and target (co-modulate between modalities). Each EEG background source is generated from random oscillatory signals in a given frequency band multiplied by a slow-varying random amplitude-modulation function. The latter is the envelope of $s_x$, and it serves as an estimate for the source bandpower timecourse. fNIRS background sources are generated from slow-varying random amplitude-modulation functions, using the same approach as for the envelope of $s_x$. The target sources are built using the same technique, but this time the same envelope used to modulate $s_x$ is used as the fNIRS source $s_y$. The $s_x$ target sources can be also time-lagged ($dT > 0$) with respect to the target source $s_y$ to simulate physiological delays between modalities. Upon simulation, fNIRS sources and recordings are downsampled to a new sampling interval, a.k.a. epochs, `T_epoch`.

Random background mixing matrices $A_x$ and $A_y$ are built using normal distributions, while we use Gaussian Radial Basis Functions (RBF) plus white noise to generate the ones for target sources. The latter approach gives the spatial patterns a more realistic local structure, by using the same center for the RBF for $A_x$ and $A_y$.

The SNR can be tuned with the `gamma` parameter, which regulates the relative strength between target source and background source contributions in channel space. The exact relationship between SNR [dB] and `gamma` is given by 20 $log_{10}(\gamma)$.

The EEG recordings and their channel-wise powerband timecourses can be accesed via the `x` and `x_power` attributes. The former has the full sampling rate used during simulation (i.e. `rate`), while the latter has been downsampled so to be in the same time basis as the fNIRS recordings, `y`, and therefore directly comparable. The simulated target sources are contained in `sx_t` and `sy_t`, each of them with their own sampling rate, and the powerband timecourse of the former source can be obtained from `sx_power`.

Finally, the simulated data can be fed into a small preprocessing step, where each dataset and source is standardized (zero mean and unit variance) and split into train and test sets.

```python
[4]:   # Configuration dictionary for the simulation
       # (equivalently, can be loaded from a YAML file)
       config_dict = {
           "Ny": 28,   # Number of channels
           "Nx": 32,
           "Ns_all": 100,   # Total number of sources
           "Ns_target": 1,   # Number of target sources
           "T": 300,   # Total simulation time (s)
           "T_epoch": 0.1,   # Length of epochs / Y sampling interval (s)
           "rate": 100,   # Sampling rate (Hz)
           "f_min": 8,   # Frequency band (Hz)
           "f_max": 12,
           "dT": 0,   # Time lag between target sources
           "invert_sy": False,   # Invert sy target source
           "gamma_e": 0.5,   # Noise strenght factor
           "gamma": 0.6,   # SNR parameter
           "ellx": 0.5,   # Width of RBF
           "elly": 0.2,
           "sigma_noise": 0.1,   # RBF white noise strength
       }

       # Create a simulation object with the configuration dictionary
       sim = BimodalToyDataSimulation(config_dict, seed=137, mixing_type="structured")
       SNR = 20 * np.log10(sim.args.gamma)   # Calculate SNR in dB
       print(f"SNR: {SNR:.2f}")

       # Plot target sources, channels, and mixing patterns (set xlim for better␣
       ↪visualization)
       sim.plot_targets(xlim=(0, 30))
       sim.plot_channels(N=2, xlim=(0, 30))
       sim.plot_mixing_patterns()

       # Run small preprocessing step to standardize and split data into train and test sets
       train_test_split = 0.8   # Proportion of data to use for training
       preprocess_data_dict = sim.preprocess_data(train_test_split)
       x_train, x_test = preprocess_data_dict["x_train"], preprocess_data_dict["x_test"]
       x_power_train, x_power_test = (
           preprocess_data_dict["x_power_train"],
           preprocess_data_dict["x_power_test"],
       )
       y_train, y_test = preprocess_data_dict["y_train"], preprocess_data_dict["y_test"]
```

```
sx, sx_power, sy = (
    preprocess_data_dict["sx"],
    preprocess_data_dict["sx_power"],
    preprocess_data_dict["sy"],
)
```

```
Random seed set as 137
Simulating sources…

Finished
SNR: -4.44
```

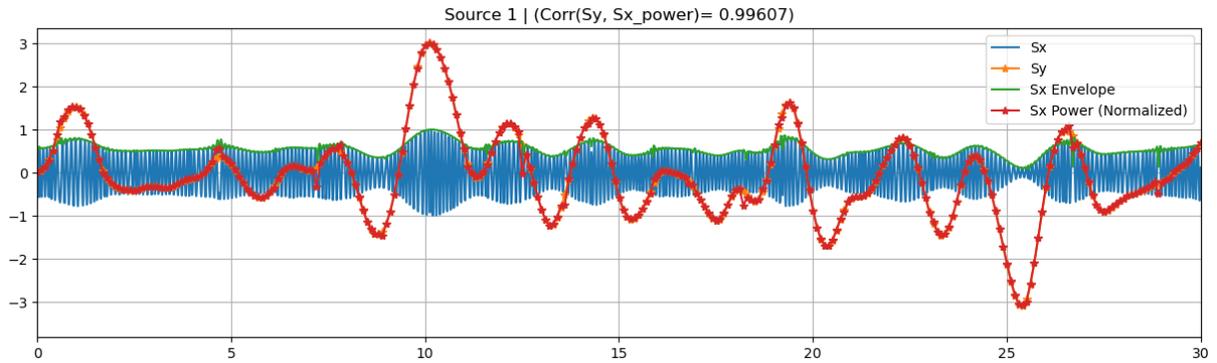

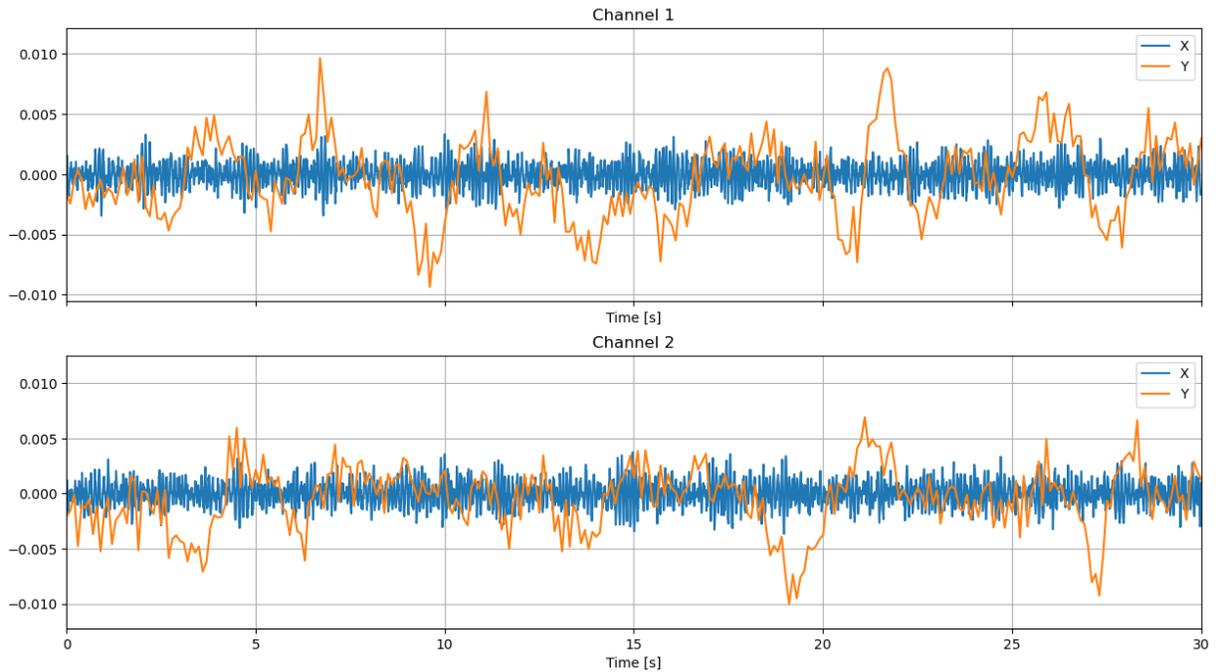

Target Source True Location

## Regularized CCA

Canonical Correlation Analysis (CCA) looks for linear projections that maximize the correlation between two datasets $X$ and $Y$. In certain scenarios, such as when dealing with high-dimensionality problems (more features than samples) or co-linearities between observations, or when looking for sparse or smooth solutions, regularized CCA is expected to achieve a better performance on the source reconstruction task and help with interpretability. Some of the most well-established regularization/penalty terms are the L1-norm, leading to Sparse CCA, the L2-norm, leading to Ridge CCA, and their combination, known as ElasticNet CCA.

All these methods can be found as classes in the `cca_models` module inside the multimodal package: `CCA`, `SparseCCA`, `RidgeCCA`, and `ElasticNetCCA`. The latter contains the first three models as particular cases, so each of them can be also instantiated by picking specific values of the regularization parameters in the ElasticNetCCA class. Its implementation is based on [Parkhomenki et al. 2009].

We now show example use cases of these classes utilizing the toy-model fNIRS-EEG data simulated above. To this end, we will use the bandpower timecourse of $x$ as one of the inputs, and the timecourse of $y$ as the second one, both sharing the same sampling rate. The reason behind this choice is that, from a neuroscientific perspective, we expect these quantities to co-modulate, rather than their directly timecourses.

```python
[5]: from cedalion.sigdecomp.multimodal.cca import (
         CCA,
         ElasticNetCCA,
         RidgeCCA,
         SparseCCA,
         StructuredSparseCCA,
     )
```

## CCA

We have four optional parameters when initializing `CCA`: `N_components` is the number of components to extract. If `None`, the number of components is set to the minimum number of features between modalities.

`max_iter` refers to the maximum number of iterations for the algorithm and `tol` regulates the tolerance for convergence by checking the norm of the difference between successive vector solutions during iteration. `scale` (bool) determines whether to scale the data during normalization to unit variance or just center it.

```
[6]:  # Initialize model
      cca = CCA(N_components=1, max_iter=1000, tol=1e-6, scale=True)
```

We can now fit the model to some (x_train, y_train). Each of the input datasets need to be an `xr.DataArray` object with exactly two dimensions: one for features and the other one for samples. Both x and y datasets need to share the same sample dimension name and shape, while features can be different. The dimension naming can be specified with optional parameters. By default, the sample dimension is expected to be named 'time' and the feature dimensions 'channel'.

```
[7]:  # Fit model
      cca.fit(
          x_power_train,
          y_train,
          sample_name="time",
          featureX_name="channel",
          featureY_name="channel",
      )

      display(cca.Wx)
```

```
<xarray.DataArray (channel: 32, CCA_X: 1)> Size: 256B
0.02794 0.1259 0.04475 0.174 0.03087 … 0.008922 0.02993 0.2225 -0.002854
Coordinates:
  * channel  (channel) <U3 384B 'X1' 'X2' 'X3' 'X4' … 'X29' 'X30' 'X31' 'X32'
  * CCA_X    (CCA_X) <U3 12B 'Sx1'
```

The learned filters, available via `cca.Wx`, and `cca.Wy`, can be used to transform a different pair (x_test, y_test). This test set must have the exact same feature dimension (name and shape) as the train data. The sample dimension should be the same between x_test and y_test, and its name must coincide with the one used during training. The number of samples, however, can be different between train and test sets.

```
[8]:  # Transform data
      sx_power_cca, sy_cca = cca.transform(x_power_test, y_test)

      print(f"Latent space dimensions: {cca.latent_featureX_name, cca.latent_featureY_name}")
      sx_power_cca
```

```
Latent space dimensions: ('CCA_X', 'CCA_Y')
```

```
[8]: <xarray.DataArray (time: 600, CCA_X: 1)> Size: 5kB
-0.2433 0.1107 0.5429 2.071 1.619 0.1281 … 2.217 1.67 2.113 1.802 1.129 0.6403
Coordinates:
  * time     (time) float64 5kB 240.1 240.2 240.3 240.4 … 299.8 299.9 300.0
  * CCA_X    (CCA_X) <U3 12B 'Sx1'
```

The `transform` method returns the "reconstructed sources", with the same sample dimension as the test input but with a new "latent_feature" dimension, generated automatically from the class/model name.

If the decomposition was performed successfully, the reconstructed sources should be close to the ground truth ones used in the linear forward model of the simulated data.

```
[9]:  # Normalize
      sx_power_cca = standardize(sx_power_cca).T
      sy_cca = standardize(sy_cca).T

      # Calculate correlations
      corrxy = np.corrcoef(sx_power_cca[0], sy_cca[0])[0, 1]
      corrx = np.corrcoef(sx_power_cca[0], sx_power[0])[0, 1]
      corry = np.corrcoef(sy_cca[0], sy[0])[0, 1]

      # Plot results
      plot_source_comparisson(
          sx_power[0], sy[0], sx_power_cca[0], sy_cca[0], corrx, corry, corrxy, title="CCA"
      )
```

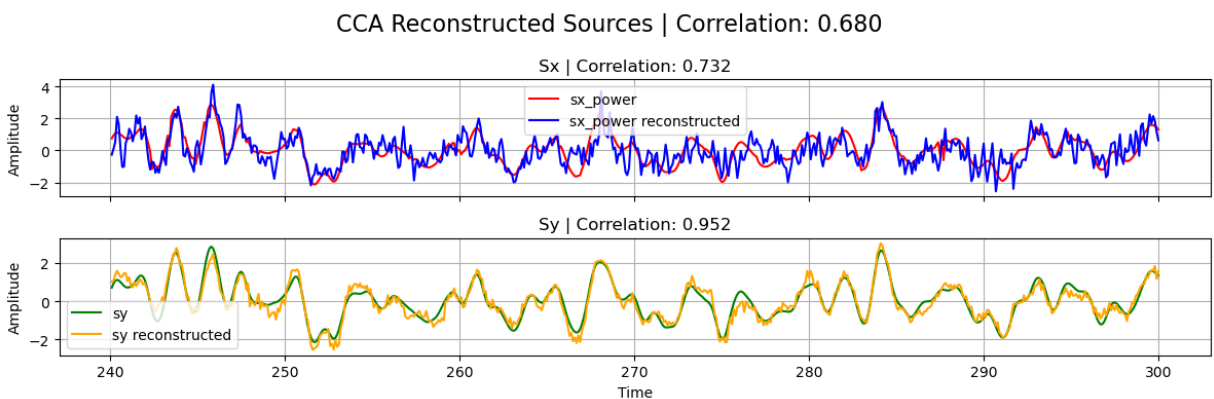

### Sparse CCA, Ridge CCA, and ElasticNet CCA

`ElasticNetCCA` admits the same optional parameters as `CCA` on top of `l1_reg` and `l2_reg` to encode the regularization parameteres for the L1 and L2 penalty terms, respectively. These parameters take a list of the form $[\lambda_x, \lambda_y]$, containing the regularization parameters for $W_x$, and $W_y$, respectively. If, instead, a float $\lambda$ is passed, then $\lambda_x = \lambda_y = \lambda$. L1 and L2 parameters need to be non-negative, with zero-values corresponding to no regularization. Additionally, `l1_reg` components must be smaller than 0.5, while `l2_reg` components are unbounded. `SparseCCA` is a sub-class of `ElasticNetCCA`, obtained by setting `l2_reg=0` on the latter. Analogously, `RidgeCCA` is equivalent to `ElasticNetCCA` with `l1_reg=0`. By choosing both regularizers to be zero, we recover the same `CCA` implementation we explored above.

With these regularized methods we can achieve better performance compared to standard CCA, even in lower SNR scenarios, as long as we make the right choice of hyperparamters. The latter gives us more flexibility, but requires a bit of extra work to achieve the optimal model's performance. For such a hyperparameter exploration, for instance, one can run a small random or grid search together with cross-validation in a validation set.

For the sake of simplicity, we won't follow that approach here, but rather show how the performance of `ElasticNetCCA` on the test set varies with different hyperparameter configurations. In particular, by looking at the `l1_reg=_l2_reg=0` in parameter space, we can see the increase in performance compared with CCA.

```
[10]:  # Define ranges for L1 and L2 parameters
       l1_reg_list = np.linspace(0, 0.1, 25)
       l2_reg_list = np.linspace(0, 2, 25)
```

```python
# Initialize a list to store correlations regularization parameters
correlations_list = np.zeros((len(l1_reg_list), len(l1_reg_list), 3))

# Loop through parameters
for i, l1 in enumerate(l1_reg_list):
    for j, l2 in enumerate(l2_reg_list):
        # Initialize (default names for sample and feature dimensions
        # is 'time' and 'channel')
        elastic_cca = ElasticNetCCA(
            N_components=1,
            l1_reg=[l1, l1],  # L1 regularization parameter (same for both datasets)
            l2_reg=l2,
        )  # L2 regularization parameter (same for both datasets)

        # Fit model
        elastic_cca.fit(x_power_train, y_train)

        # Transform data
        sx_power_elastic_cca, sy_elastic_cca = elastic_cca.transform(
            x_power_test, y_test
        )

        # Normalize
        sx_power_elastic_cca = standardize(sx_power_elastic_cca).T[0]
        sy_elastic_cca = standardize(sy_elastic_cca).T[0]

        # Calculate correlations
        corrxy_elastic = np.corrcoef(sx_power_elastic_cca, sy_elastic_cca)[0, 1]
        corrx_elastic = np.corrcoef(sx_power_elastic_cca, sx_power)[0, 1]
        corry_elastic = np.corrcoef(sy_elastic_cca, sy)[0, 1]

        correlations_list[i, j] = [corrxy_elastic, corrx_elastic, corry_elastic]

# Find (L1, L2) configuration with highest avarage correlation between corrx and corry
corr_avg = np.mean(correlations_list[:, :, 1:], axis=-1)
i, j = np.unravel_index(np.argmax(corr_avg), corr_avg.shape)
max_corrxy, max_corrx, max_corry = correlations_list[i, j]
best_l1_reg = l1_reg_list[i]
best_l2_reg = l2_reg_list[j]

# CCA correlation with L1=L2=0
cca_corr = correlations_list[0, 0, :]

print(
    f"Best L1 regularization: {best_l1_reg:.5f}, "
    f"Best L2 regularization: {best_l2_reg:.5f}"
)
print(
    f"Max correlation: {max_corrxy:.3f}, "
    f"Correlation X: {max_corrx:.3f}, "
    f"Correlation Y: {max_corry:.3f}"
```

```python
)
print(
    f"CCA correlation (L1=L2=0): {cca_corr[0]:.3f}, "
    f"Correlation X: {cca_corr[1]:.3f}, "
    f"Correlation Y: {cca_corr[2]:.3f}"
)

# Plot the correlation heatmap
plt.figure(figsize=(10, 8))
plt.imshow(
    corr_avg,
    cmap="magma",
    aspect="auto",
    origin="lower",
    extent=[l2_reg_list[0], l2_reg_list[-1], l1_reg_list[0], l1_reg_list[-1]],
    interpolation="nearest",
)
plt.colorbar(label="Average Correlation")
plt.xlabel("L2 Regularization")
plt.ylabel("L1 Regularization")
plt.title("Correlation Heatmap for ElasticNet CCA")
plt.scatter(best_l2_reg, best_l1_reg, color="red", label="Best Parameters")
plt.legend()
plt.show()
```

Best L1 regularization: 0.06250, Best L2 regularization: 0.58333
Max correlation: 0.688, Correlation X: 0.738, Correlation Y: 0.956
CCA correlation (L1=L2=0): 0.680, Correlation X: 0.732, Correlation Y: 0.952

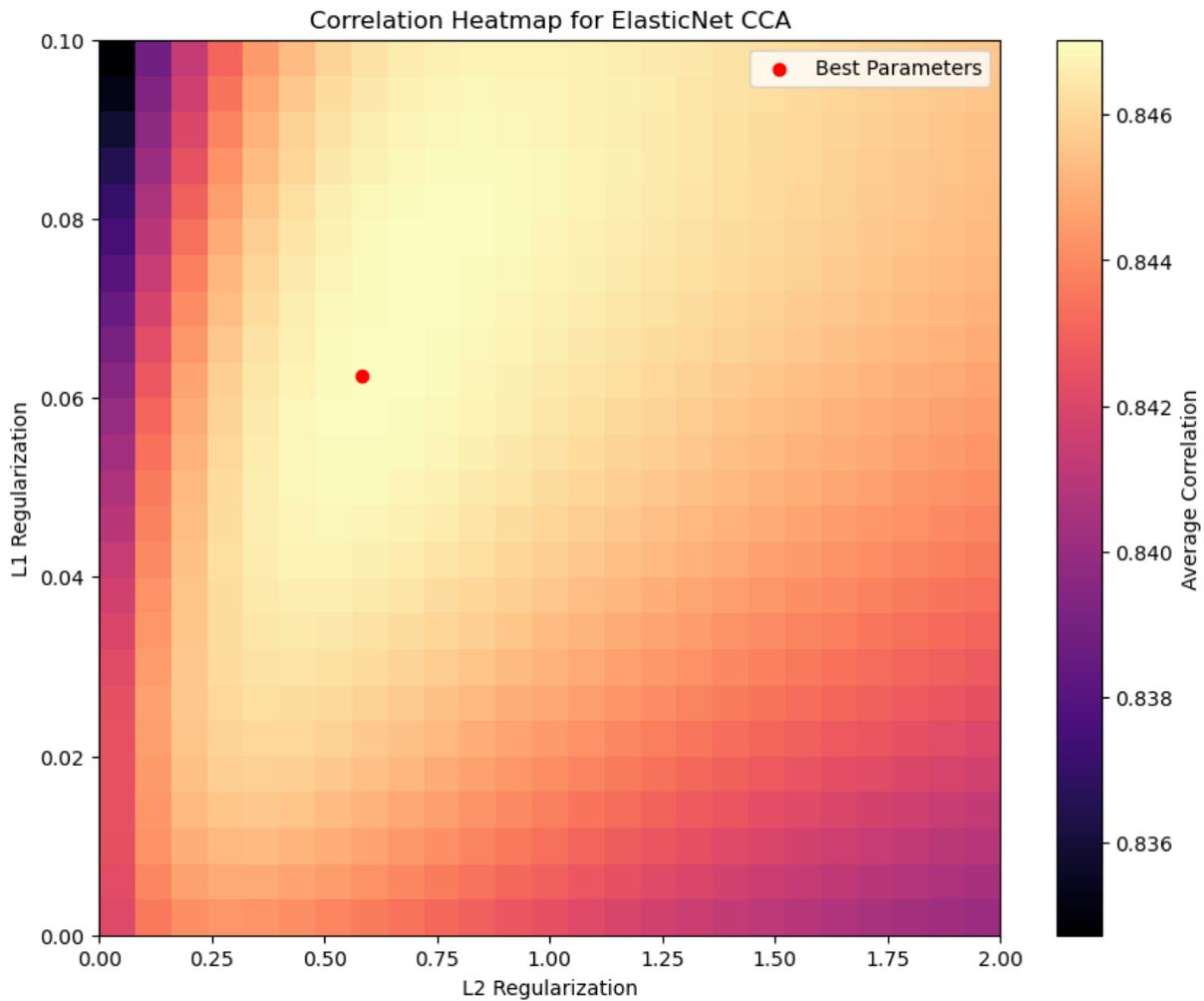

For the other regularization models, initializing, fitting and transforming work in the same way:

```
[11]: # Initialize models (use optimal parameters found above)
      elasticnet_cca = ElasticNetCCA(
          N_components=1,
          l1_reg=best_l1_reg,  # Use optimal L1, L2 found above
          l2_reg=best_l2_reg,
      )

      # Equals ElasticNetCCA with l1_reg=best_l1_reg and l2_reg=0
      sparse_cca = SparseCCA(N_components=1, l1_reg=best_l1_reg)
      # Equals ElasticNetCCA with l1_reg=0 and l2_reg=best_l2_reg
      ridge_cca = RidgeCCA(N_components=1, l2_reg=best_l2_reg)

      # Fit models
      elasticnet_cca.fit(x_power_train, y_train)
      sparse_cca.fit(x_power_train, y_train)
      ridge_cca.fit(x_power_train, y_train)

      # Transform data
```

```python
sx_power_elastic, sy_elastic = elasticnet_cca.transform(x_power_test, y_test)
sx_power_sparse, sy_sparse = sparse_cca.transform(x_power_test, y_test)
sx_power_ridge, sy_ridge = ridge_cca.transform(x_power_test, y_test)

# Normalize
sx_power_elastic = standardize(sx_power_elastic).T[0]
sy_elastic = standardize(sy_elastic).T[0]

sx_power_sparse = standardize(sx_power_sparse).T[0]
sy_sparse = standardize(sy_sparse).T[0]

sx_power_ridge = standardize(sx_power_ridge).T[0]
sy_ridge = standardize(sy_ridge).T[0]

# Calculate correlations
corrxy_elastic = np.corrcoef(sx_power_elastic, sy_elastic)[0, 1]
corrx_elastic = np.corrcoef(sx_power_elastic, sx_power)[0, 1]
corry_elastic = np.corrcoef(sy_elastic, sy)[0, 1]

corrxy_sparse = np.corrcoef(sx_power_sparse, sy_sparse)[0, 1]
corrx_sparse = np.corrcoef(sx_power_sparse, sx_power)[0, 1]
corry_sparse = np.corrcoef(sy_sparse, sy)[0, 1]

corrxy_ridge = np.corrcoef(sx_power_ridge, sy_ridge)[0, 1]
corrx_ridge = np.corrcoef(sx_power_ridge, sx_power)[0, 1]
corry_ridge = np.corrcoef(sy_ridge, sy)[0, 1]

# Plot
plot_source_comparisson(
    sx_power[0],
    sy[0],
    sx_power_elastic,
    sy_elastic,
    corrx_elastic,
    corry_elastic,
    corrxy_elastic,
    title="ElasticNet CCA",
)
plot_source_comparisson(
    sx_power[0],
    sy[0],
    sx_power_sparse,
    sy_sparse,
    corrx_sparse,
    corry_sparse,
    corrxy_sparse,
    title="Sparse CCA",
)
plot_source_comparisson(
    sx_power[0],
```

```
        sy[0],
        sx_power_ridge,
        sy_ridge,
        corrx_ridge,
        corry_ridge,
        corrxy_ridge,
        title="Ridge CCA",
)
```

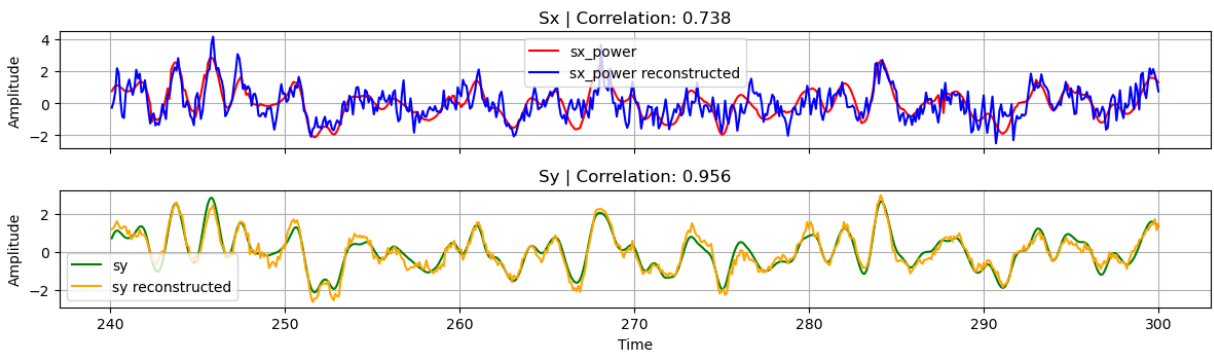

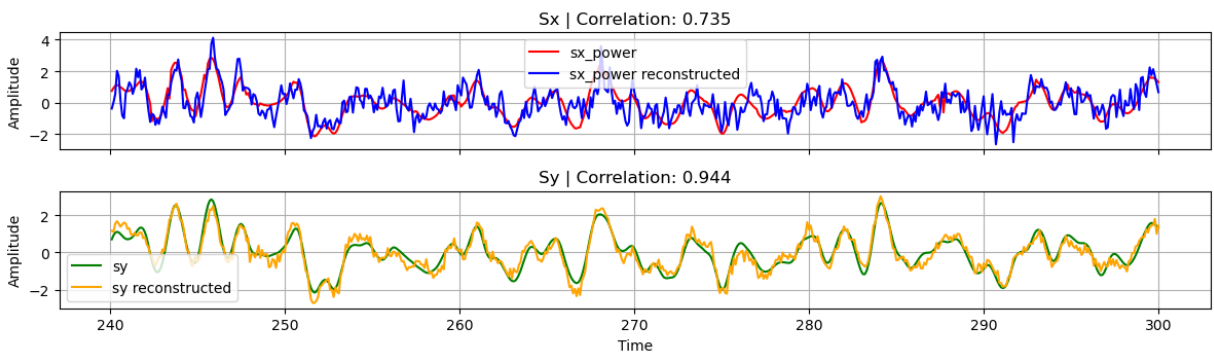

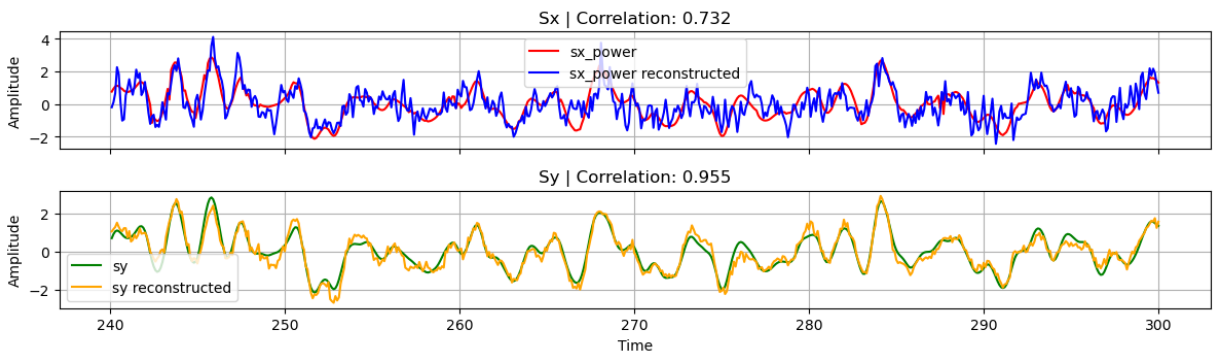

**Structured Sparse CCA (ssCCA)**

ElasticNet regularization can be modified to include contextual information about the local structure of the dataset, leading to structured CCA. This is particularly relevant in neuroimaging, where features (e.g. channels or brain regions) follow specific spatial distributions. One way of incorporating this prior information is to modify the L2-norm penalty term on the filters $W$ via $\|W\|_2^2 \to W^T L W$ for some matrix $L$ that captures the local dependencies among features. When this structure contraint is combined with an L1 penalization, the resulting method is called structured sparse CCA (ssCCA) and is implemented in Cedalion as the `StructuredSparseCCA` class, following [Chen et al. 2013].

This class admits the same parameters as `ElasticNetCCA` together with `Lx` and `Ly`, corresponding to structure matrices used for each dataset. The latter need to be square matrices of shapes $(Nx, Nx)$ and $(Ny, Ny)$, in terms of the number of features of each dataset. This time, the `l2_reg` parameter regulates the strength of the structured penalty term, and must be non-negative but otherwise unbounded.

One popular choice for such structure contraint is the Laplacian matrix, which is used as a graph representation of the dataset, in which features (channels) are the nodes, and (weighted) edges indicate which and how features are connected. We now show a simple example of a Laplacian matrix, build via an adjacency matrix that uses a binary nearest-neighbor strategy.

```python
[12]: # Helper function to build Laplacian matrix
      def build_laplace(nodes, eps):
          """Builds Laplacian matrix of a graph.

          The nodes are the components of the 1D vector nodes,
          by giving unit weight to connected nodes only if they are close enough.
          The latter condition is determined by comparing the 2-norm between nodes and eps.
          """

          N = len(nodes)
          Adj = np.zeros([N, N])  # Adjacency matrix
          D = np.eye(N)  # Degree matrix

          for i, xi in enumerate(nodes):
              for j, xj in enumerate(nodes):
                  if i == j:  # Skip diagonal entries
                      continue
                  are_close = np.linalg.norm(xi - xj) < eps
                  Adj[i, j] = 1 if are_close else 0

              D[i, i] = np.sum(Adj[i])

          L = D - Adj  # Laplace matrix

          return L, Adj
```

```python
[13]: # Read channel positions from simulation
      x_channels_pos = sim.x_montage.values
      y_channels_pos = sim.y_montage.values

      # Build Laplacian matrix from each montage
      # (eps thresholds chosen by small experimentation)
      Lx, Adjx = build_laplace(x_channels_pos, eps=sim.args.ellx)
```

```
Ly, Adjy = build_laplace(y_channels_pos, eps=sim.args.elly)

# Plot Adjacency matrices
plt.figure(figsize=(10, 5))
plt.subplot(1, 2, 1)
plt.imshow(Adjx, cmap="viridis", interpolation="nearest")
plt.title("Adjacency Matrix for X Montage")
plt.colorbar(label="Weight")
plt.subplot(1, 2, 2)
plt.imshow(Adjy, cmap="viridis", interpolation="nearest")
plt.title("Adjacency Matrix for Y Montage")
plt.colorbar(label="Weight")
plt.tight_layout()
```

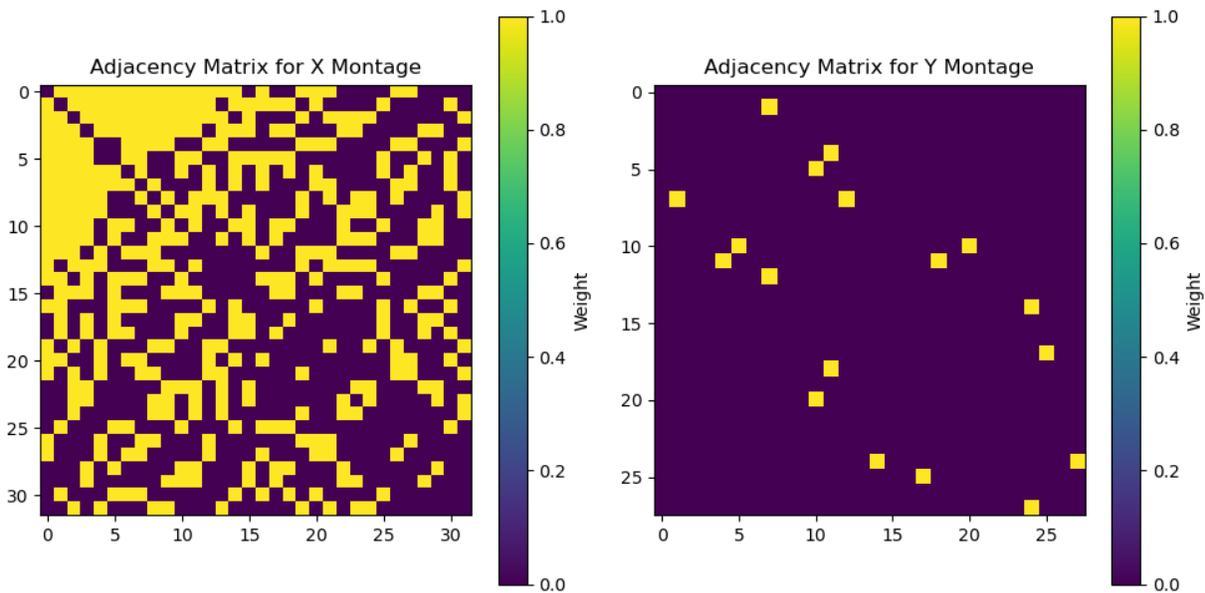

The choice of structure/Laplace matrix introduces yet another layer of complexity to the model, which needs further fine-tunning to achieve its optimal performance. Since making an exhaustive investigation is beyond the scope of this tutorial notebook, here we just chose the nearest-neighbor thresholds to coincide with the ground truth RBF width used in the simulated mixing matrices, and picked the same regularization parameters as for `ElasticNetCCA` above. Initialization, fitting, and transformation follows the same logic as before:

```
[14]: sscca = StructuredSparseCCA(
          N_components=1,
          l1_reg=best_l1_reg,   # Use optimal L1, L2 found above
          l2_reg=best_l2_reg,
          Lx=Lx,   # Laplacian matrix for X montage
          Ly=Ly,   # Laplacian matrix for Y montage
      )

      # Fit model
      sscca.fit(x_power_train, y_train)
```

```
# Transform data
sx_power_sscca, sy_sscca = sscca.transform(x_power_test, y_test)

# Normalize
sx_power_sscca = standardize(sx_power_sscca).T[0]
sy_sscca = standardize(sy_sscca).T[0]

# Calculate correlations
corrxy_sscca = np.corrcoef(sx_power_sscca, sy_sscca)[0, 1]
corrx_sscca = np.corrcoef(sx_power_sscca, sx_power)[0, 1]
corry_sscca = np.corrcoef(sy_sscca, sy)[0, 1]

# Plot
plot_source_comparisson(
    sx_power[0],
    sy[0],
    sx_power_sscca,
    sy_sscca,
    corrx_sscca,
    corry_sscca,
    corrxy_sscca,
    title="Structured Sparse CCA",
)
```

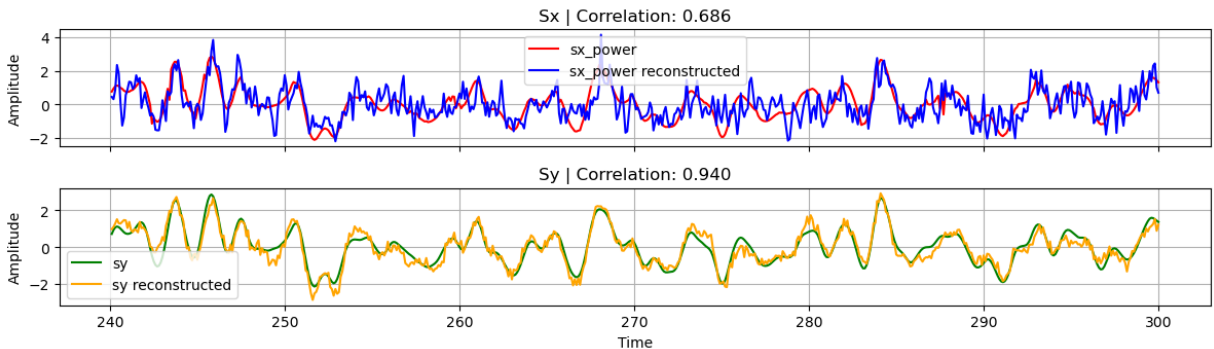

From the learned weights, `Wx` and `Wy`, one can estimate the spattial patterns `Ax` and `Ay` via a regression method [Haufe et al. 2014]. These quantities should be close to the original mixing matrices used in the linear forward model of the simulated data. Structured regularization is particularly promising for such a reconstruction task given the incorporation of spatial information during the decomposition:

```
[15]: # Helper function to estimate spatial patterns from
      # learned weights via regression approach
      def compute_spatial_pattern_from_weight(
          X_xr, W_xr, sample_name="time", feature_name="channel"
      ):
          # Bring to standard order
          X_xr = X_xr.transpose(sample_name, feature_name)
```

```python
    W_xr = W_xr.transpose(feature_name, ...)

    # Work with numpy arrays from now on
    N = len(X_xr[sample_name])
    X = X_xr.data
    W = W_xr.data
    # Covariance matrix for X
    C = (X.T @ X) / (N - 1)
    # Covariance matrix for reconstructed sources
    Cs = W.T @ C @ W
    # Estimated spattial pattern
    A = C @ W @ sp.linalg.pinv(Cs)

    # Bring to DataArray format\
    A = xr.DataArray(A, dims=W_xr.dims, coords=W_xr.coords)

    return A

# Compute spatial patterns from learned weights for ssCCA
Ax_sscca = compute_spatial_pattern_from_weight(x_test, sscca.Wx)
Ay_sscca = compute_spatial_pattern_from_weight(y_test, sscca.Wy)

# Normalize
Ax_sscca /= Ax_sscca.max()
Ay_sscca /= Ay_sscca.max()

# Plot original mixing patterns
sim.plot_mixing_patterns(title="Original Mixing Patterns")
# Plot estimated spatial patterns
sim.plot_mixing_patterns(
    Ax=Ax_sscca.values,
    Ay=Ay_sscca.values,
    title="Reconstructed Spatial Patterns (ssCCA))",
)
```

# Original Mixing Patterns

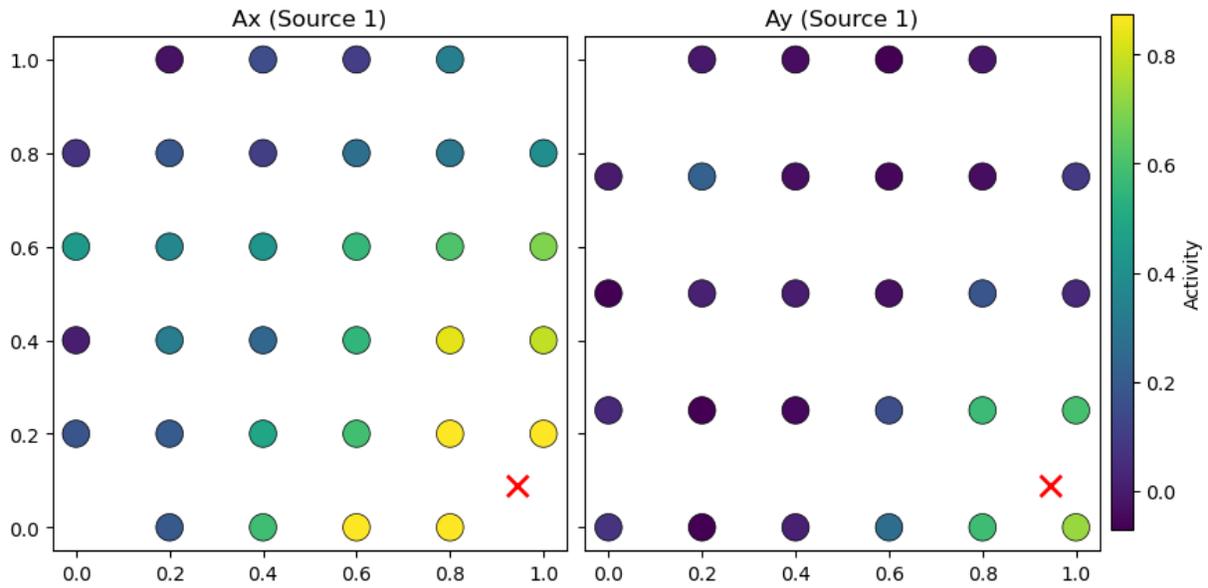

# Reconstructed Spatial Patterns (ssCCA)

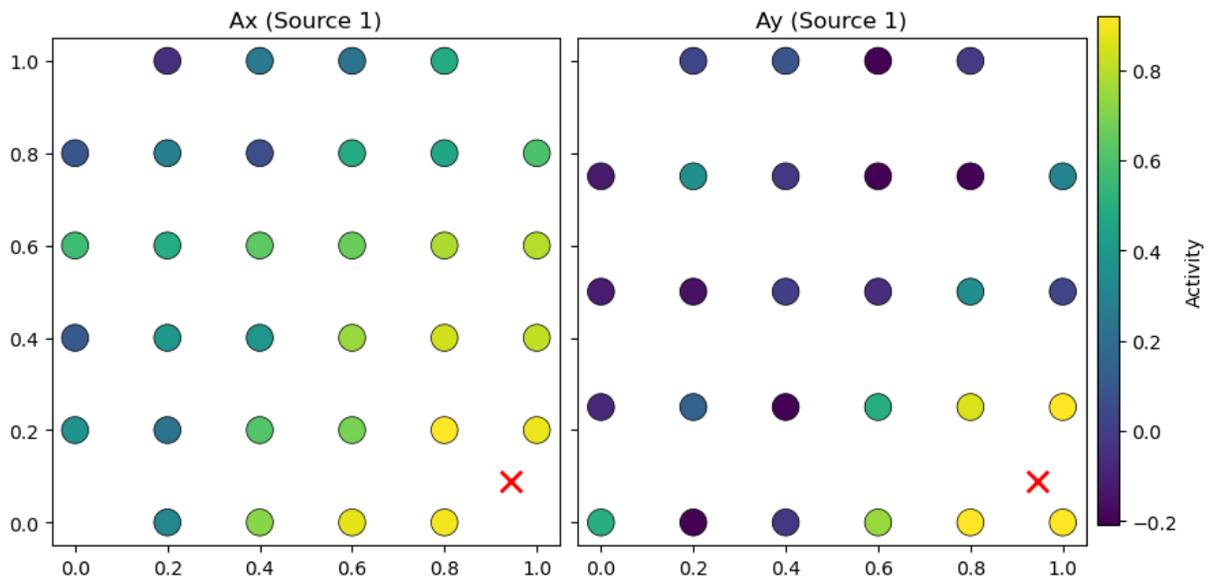

**Additional Functionalities**

Apart from the functionalities already covered for the regularized and standard CCA implementations, there are few addtional features that are worth mentioning. Given their usage are rather straightforward, we leave the exploration to the curious reader.

### *Multiple Components*

So far we just covered the one-unit algorithms, i.e. we extracted only one component/source. As mentioned above, all these methods can be used to extract several components by changing the value of the `N_component` parameter. That value cannot be bigger than the minimum between the number of X and Y features. The way these extra components are extracted is by following a deflation approach, where at each iteration the contribution from the previous component is removed from the input matrix entering the decomposition.

```
[16]:  # Example case of multiple components CCA
       cca_multiple_comp = CCA(N_components=10)
       cca_multiple_comp.fit(x_power_train, y_train)
       display(cca_multiple_comp.Wy)
```

```
<xarray.DataArray (channel: 28, CCA_Y: 10)> Size: 2kB
0.009316 -0.04114 0.3194 -0.1812 -0.2955 … -0.5022 -0.1405 -0.474 0.02501
Coordinates:
  * channel  (channel) <U3 336B 'Y1' 'Y2' 'Y3' 'Y4' … 'Y25' 'Y26' 'Y27' 'Y28'
  * CCA_Y    (CCA_Y) <U4 160B 'Sy1' 'Sy2' 'Sy3' 'Sy4' … 'Sy8' 'Sy9' 'Sy10'
```

### *Partial Least Squares (PLS)*

PLS is a multimodal source-decomposition algorithm very similar to CCA. The latter maximizes correlation between the projected views, while PLS maximizes covariance. From a practical point of view, PLS can be obtained from CCA by sending the covariance matrices $C_x$ and $C_y$ to the identity matrices in the cost function. Therefore, both algorithms coincide when the input datasets are whitened.

Within Cedalion, we can use these methods via the classes `PLS`, and `SparsePLS` [Witten et al. 2009]. The Ridge version of PLS simply does not exist, because PLS itself can be understood as an extreme case of RidgeCCA where $\lambda_2 \to \infty$. Internally, the classes are obtained from `ElasticNetCCA` with the extra parameter `pls=True`.

```
[17]:  from cedalion.sigdecomp.multimodal.cca import PLS, SparsePLS
```

```
[18]:  # Example initialization of PLS and SparsePLS models
       pls = PLS(N_components=1, max_iter=1000, tol=1e-6, scale=True)
       sparse_pls = SparsePLS(N_components=1, l1_reg=[0.1, 0.01])
```

### Temporally Embedded CCA (tCCA)

The methods covered so far assume modalities correlate instantaneously, which is certainly not the case in fNIRS-EEG fusion, for instance, where the common underlying sources are time-shifted due to hemodynamic delays. Temporally embedded CCA (tCCA) [Biessmann et al., 2010] captures such temporal offsets. While effective, the idea is rather simple: assuming $y$ is the "delayed" modality, $x$ is time-embedded by concatenating time-shifted copies along the feature dimension, leading to a time-embedded dataset $\tilde{x}$. The time lags must be preselected, introducing a new parameter to the model. Standard CCA (or any of the variants explored above) is then applied without further modifications on $(\tilde{x}, y)$, yielding temporal filters that capture delayed correlations between modalities.

All the CCA variants covered above admit a temporally embedded extension, and they are implemented in Cedalion in the `tcca_models` module, which contains the classes `tCCA`, `ElasticNetTCCA`, and `StructuredSparseTCCA`. Analogously to the simultaneuously-coupled models, Sparse tCCA and Ridge tCCA can be obtained from `ElasticNetTCCA` by the right choice of `l1_reg` and `l2_reg` parameters, but in this case they do not have their own classes.

```
[19]: from cedalion.sigdecomp.multimodal.tcca import (
           tCCA,
           # ElasticNetTCCA,
           # StructuredSparseTCCA,
       )
```

In order to show how these models are used, let's build a time-shifted version of the simulated data from above:

```
[20]: # Use the same configuration dictionary as before but with a non-zero time lag
       config_dict["dT"] = 2  # Time lag between target sources (s)

       # Simulate
       sim = BimodalToyDataSimulation(config_dict, seed=137, mixing_type="structured")
       SNR = 20 * np.log10(sim.args.gamma)  # Calculate SNR in dB
       print(f"SNR: {SNR:.2f}")
       print("Time lag between target sources:", sim.args.dT, "s")

       # Plot target sourcers, channels, and mixing patterns
       # (add xlim now to see the effect of time lag)
       sim.plot_targets(xlim=(0, 20))
       # sim.plot_channels(N=2)
       # sim.plot_mixing_patterns()

       # Run small preprocessing step to standardize and
       # split the data into train and test sets
       train_test_split = 0.8  # Proportion of data to use for training
       preprocess_data_dict = sim.preprocess_data(train_test_split)

       x_train, x_test = preprocess_data_dict["x_train"], preprocess_data_dict["x_test"]

       x_power_train, x_power_test = (
           preprocess_data_dict["x_power_train"],
           preprocess_data_dict["x_power_test"],
       )

       y_train, y_test = preprocess_data_dict["y_train"], preprocess_data_dict["y_test"]

       sx, sx_power, sy = (
           preprocess_data_dict["sx"],
           preprocess_data_dict["sx_power"],
           preprocess_data_dict["sy"],
       )
```

```
Random seed set as 137
Simulating sources…

Finished
SNR: -4.44
Time lag between target sources: 2 s
```

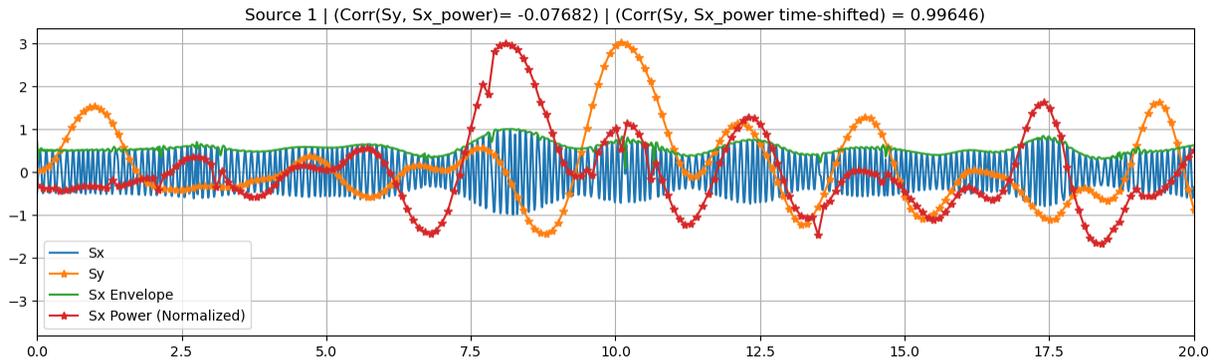

These time-embedded classes are built on top of their simultaneuously-coupled counterparts, sharing several attributes and methods. In particular, they accept the same parameters as before, in addition to `time_shifts` and `shift_source`.

The former must be a `Numpy` array, containing the time-lags to be applied to the $x$ modality. These lags should be non-negative float numbers within the time domain of the data. Because of the latter, a validation/consistency check of this array is carried only after fitting the model. During that validation, a time-lag = 0 component would be added to `time_shift` if not present already (zero-lag $x$ copy), and the resulting array is sorted in ascending order.

If the `shift_source` parameter is set to `True`, the reconstructed source $s_x$, obtained after transforming the data, is shifted by an `optimal_shift`, so it is temporally aligned with the reconstructed source $s_y$. This optimal shift attribute is estimated right after training by looking for the time-shifted x that produces the biggest correlation between reconstructed sources $s_x$ and $s_y$. Note that the `optimal_shift` attribute is available even when `shift_source=False`, so one can apply the shift a posteri if desired. It is worth clarifying that this shift is only estimated using the (train) data used during fitting.

```
[21]: # Temporal embedding parameters
      dt = 1
      N_lags = 5
      time_shifts = np.arange(0, dt * N_lags, dt)
      print(f"Time shifts: {time_shifts}")
      print("True time lag between target sources:", sim.args.dT, "s")

      # Initialize tCCA model
      tcca = tCCA(
          N_components=1,
          max_iter=1000,
          tol=1e-6,
          scale=True,
          time_shifts=time_shifts,
          shift_source=True,
      )
```

```
Time shifts: [0 1 2 3 4]
True time lag between target sources: 2 s
```

Fitting works in the same way as before, except that `sample_name` is assumed always to be `time`.

```python
[22]: # Fit model
      tcca.fit(x_power_train, y_train, featureX_name="channel", featureY_name="channel")

      # At this point we have an estimate for the time lag between target sources
      print(
          f"Estimated time lag between target sources "
          f"during training: {tcca.optimal_shift[0]} s"
      )
      display(tcca.Wx)
```

Estimated time lag between target sources during training: 2.0 s

```
<xarray.DataArray (time_shift: 5, channel: 32, tCCA_X: 1)> Size: 1kB
0.00386 0.002902 0.01951 -0.01833 -0.0143 … 0.03439 0.01409 0.003379 0.05622
Coordinates:
  * time_shift  (time_shift) int64 40B 0 1 2 3 4
  * channel     (channel) <U3 384B 'X1' 'X2' 'X3' 'X4' … 'X30' 'X31' 'X32'
  * tCCA_X      (tCCA_X) <U3 12B 'Sx1'
```

The learned weights `Wx` now have an extra `time_shift` dimension, where each component contains the filters applied to the copy of $x$ shifted by that particular time-lag. `Wy` remains with the same shape as previous models.

Data transformation is also applied in the same manner, but in this case, if `shift_sources=True`, the reconstructed sources are truncated by removing the last `optimal shift` seconds, containing null values introduced by the zero-padding used during shifting.

```python
[23]: # Transform data
      sx_power_tcca, sy_tcca = tcca.transform(x_power_test, y_test)

      display(sx_power_tcca, sx_power)
```

```
<xarray.DataArray (time: 580, tCCA_X: 1)> Size: 5kB
2.014 0.707 0.3418 0.8845 -0.2528 … -0.06712 -0.4116 0.5565 0.4488 0.2566
Coordinates:
  * time    (time) float64 5kB 240.1 240.2 240.3 240.4 … 297.8 297.9 298.0
  * tCCA_X  (tCCA_X) <U3 12B 'Sx1'

<xarray.DataArray (source: 1, time: 600)> Size: 5kB
0.8019 0.3854 -0.06178 -0.4672 -0.7737 -0.9842 … 1.732 2.067 2.226 2.188 1.977
Coordinates:
  * time    (time) float64 5kB 240.1 240.2 240.3 240.4 … 299.8 299.9 300.0
  * source  (source) <U2 8B 'S1'
```

Because of this sample number mismatch with the ground truth sources, we need to truncate the latter before any comparisson:

```python
[24]: # Normalize
      sx_power_tcca = standardize(sx_power_tcca).T[0]
      sy_tcca = standardize(sy_tcca).T[0]

      # Truncate ground truth sources to match the reconstructed sources
      if len(sx_power_tcca) < len(sx_power.time):
          sx_power_trunc = sx_power[0, : len(sx_power_tcca)]
      if len(sy_tcca) < len(sy.time):
```

```
    sy_trunc = sy[0, : len(sy_tcca)]

    # Calculate correlations. Should be low because sx is shifted!
    corrxy_tcca = np.corrcoef(sx_power_tcca, sy_tcca)[0, 1]
    corrx_tcca = np.corrcoef(sx_power_tcca, sx_power_trunc)[0, 1]
    corry_tcca = np.corrcoef(sy_tcca, sy_trunc)[0, 1]
```

We also train a standard CCA method so we can compare and see the importance of capturing the time lags in the data:

```
[25]:  # CCA
       cca = CCA(N_components=1)
       cca.fit(x_power_train, y_train)
       sx_cca, sy_cca = cca.transform(x_power_test, y_test)
       # Normalize
       sx_cca = standardize(sx_cca).T[0]
       sy_cca = standardize(sy_cca).T[0]
       # Calculate correlations for CCA
       corrxy_cca = np.corrcoef(sx_cca, sy_cca)[0, 1]
       corrx_cca = np.corrcoef(sx_cca, sx_power)[0, 1]
       corry_cca = np.corrcoef(sy_cca, sy)[0, 1]
```

```
[26]:  # Plot results
       plot_source_comparisson(
           sx_power_trunc,
           sy_trunc,
           sx_power_tcca,
           sy_tcca,
           corrx_tcca,
           corry_tcca,
           corrxy_tcca,
           title="tCCA",
       )
       plot_source_comparisson(
           sx_power[0], sy[0], sx_cca, sy_cca, corrx_cca, corry_cca, corrxy_cca, title="CCA"
       )
```

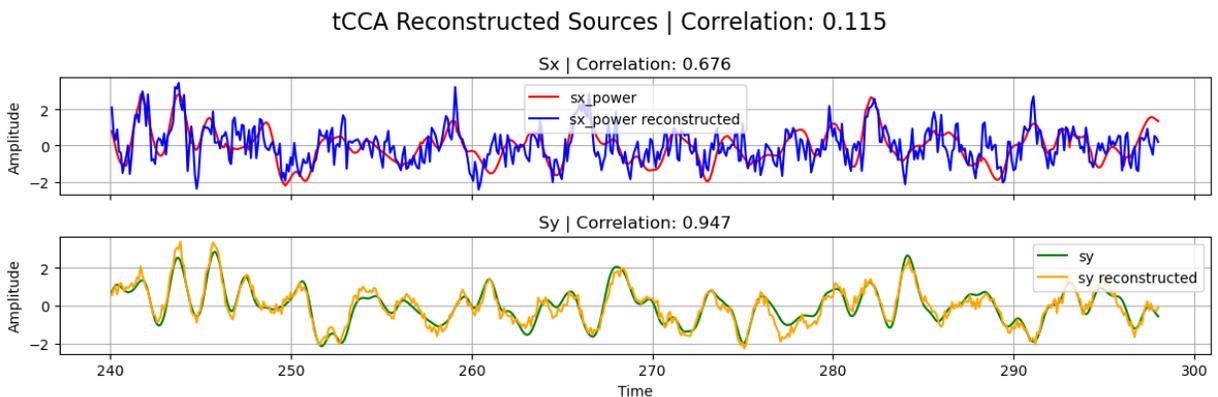

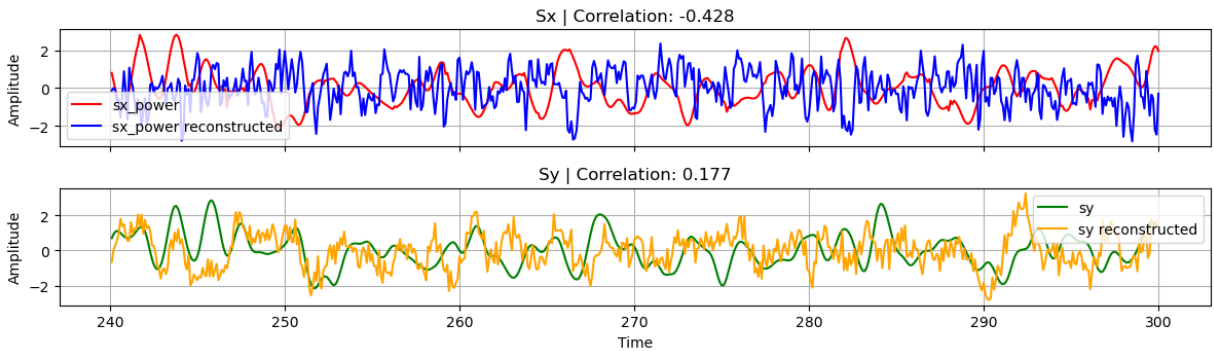

Note that, in the previous plot, the correlation between the reconstructed sources from tCCA is naturally low since $s_x$ has been shifted with the estimated `optimal_shift`.

**Additional Functionalities**

As for the regularized CCA classes, the temporally-embedded extensions can also extract multiple components (using the same deflation algorithm), and the PLS variant can be used by settting `pls=True` during initialization.

**Multimodal Source Power Co-modulation (mSPoC)**

EEG bandpower and hemodynamic responses captured by fNIRS are proven to comodulate during cognitive tasks. Conventional methods, such as CCA and its variants, however, present some limitations to integrate bandpower signals. More precisely, computing bandpower at channel level and then applying a backward linear method, as we have been doing so far, is not in line with the assumed linear generative model of EEG. mSPoC [Dähne, et.al. 2013] arises as a method that avoids these pitfalls by inverting the generative model prior to computing bandpower, and is implemented in Cedalion as the `mSPoC` class inside the `mspoc` module.

The algorithm takes two vector-valued time series $x(t)$, and $y(e)$, which are time-aligned but have different sample dimensions `Ntx` > `Nty`. It is expected that $x(t)$ has been previously bandpassed filtered to the band of interest. It finds components $s_x(t)$ and $s_y(e)$ via linear projection of the observations, such that the covariance between the temporally-embedded,bandpower of $s_x(t)$ and the time course of $s_y(e)$ is maximized. The bandpower of $s_x(t)$ is estimated by dividing the source into non-overlapping time windows (or epochs), such that there is one window per data point $y(e)$, and computing the variance within epochs. After such operation, both signals share the same sampling rate and their cross-covariance can be calculated. The solution to the optimization problem is captured by the spatial (`Wx`, `Wy`), and temporal (`Wt`) filters.

```
[27]: from cedalion.sigdecomp.multimodal.mspoc import mSPoC
```

The `mSPoC` class accepts the very same parameters as the `tCCA` class, namely `N_components`, `max_iter`, `tol`, `scale`, `time_shifts`, and `shift_source`, with the addition of `N_restarts`. The latter determines the number of times the algorithm is repeated, which may find better extrema in the optimization problem, due to its stochastic nature. Typical values for `N_restarts` lie between 2 and 10, and it is recommended to choose smaller `max_iter` and bigger `tol` values, compared to simpler CCA methods, due the increased computational time required in mSPoC. As all previous CCA-based methods, mSPoC can also extract multiple components.

```
[28]: # We run the same simulation as for tCCA above
config_dict["dT"] = 2
sim = BimodalToyDataSimulation(config_dict, seed=137, mixing_type="structured")
```

```python
SNR = 20 * np.log10(sim.args.gamma)  # Calculate SNR in dB
print(f"SNR: {SNR:.2f}")
print("Time lag between target sources:", sim.args.dT, "s")
sim.plot_targets(xlim=(0, 20))

# Run small preprocessing step to standardize and
# split the data into train and test sets
train_test_split = 0.8
preprocess_data_dict = sim.preprocess_data(train_test_split)

x_train, x_test = preprocess_data_dict["x_train"], preprocess_data_dict["x_test"]
x_power_train, x_power_test = (
    preprocess_data_dict["x_power_train"],
    preprocess_data_dict["x_power_test"],
)

y_train, y_test = preprocess_data_dict["y_train"], preprocess_data_dict["y_test"]
sx, sx_power, sy = (
    preprocess_data_dict["sx"],
    preprocess_data_dict["sx_power"],
    preprocess_data_dict["sy"],
)

# Temporal embedding parameters (same as for tCCA)
dt = 1
N_lags = 5
time_shifts = np.arange(0, dt * N_lags, dt)
print(f"Time shifts: {time_shifts}")
print("True time lag between target sources:", sim.args.dT, "s")

# Initialize mSPoC model
mspoc = mSPoC(
    N_components=1,
    N_restarts=2,
    max_iter=100,  # Note we use much smaller values than before
    tol=1e-4,
    scale=True,
    time_shifts=time_shifts,
    shift_source=True,
)
```

```
Random seed set as 137
Simulating sources…

Finished
SNR: -4.44
Time lag between target sources: 2 s
```

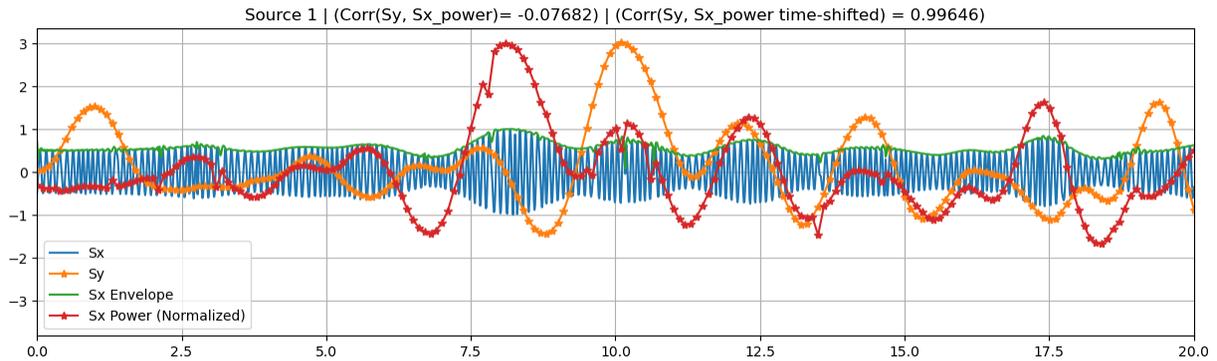

Time shifts: [0 1 2 3 4]
True time lag between target sources: 2 s

Fit and transform work analogously to the `tCCA` method, except that this time the number of samples in the $x$ input should be bigger than the number of $y$ samples.

```
[29]: # Fit model (sample_name fixed to be 'time' in mSPoC)
mspoc.fit(
    x_train,
    y_train,  # Note that we use x_train here, not x_power_train
    featureX_name="channel",
    featureY_name="channel",
)

# At this point we have an estimate for the time lag between target sources
print(
    "Estimated time lag between target sources "
    f"during training: {mspoc.optimal_shift[0]} s"
)

# Learned filters
display(mspoc.Wx)
display(mspoc.Wt)
```

Estimated time lag between target sources during training: 2 s

```
<xarray.DataArray (channel: 32, mSPoC_X: 1)> Size: 256B
0.009032 -0.02669 0.05469 -0.04415 -0.02208 … -0.1543 -0.05827 -0.1278 -0.0265
Coordinates:
  * channel  (channel) <U3 384B 'X1' 'X2' 'X3' 'X4' … 'X29' 'X30' 'X31' 'X32'
  * mSPoC_X  (mSPoC_X) <U3 12B 'Sx1'
```

```
<xarray.DataArray (time_embedding: 5, mSPoC_T: 1)> Size: 40B
-0.06374 -0.0625 0.991 -0.08447 -0.05272
Coordinates:
  * time_embedding  (time_embedding) int64 40B 0 1 2 3 4
  * mSPoC_T         (mSPoC_T) <U3 12B 'St1'
```

The reconstructed sources returned by `transform` correspond to the temporally-embedded bandpower of the source $s_x$, and the time course of $s_y$. Because of this choice, the outputs are directly comparable, since they have the same sampling time. As before, whenever `shift_source=True`, the time dimensions of the sources are truncated, removing the last `optimal_shift` seconds.

```
[30]: # Transform data
      sx_power_mspoc, sy_mspoc = mspoc.transform(x_test, y_test)
      display(sx_power_mspoc, sx_power)
      display(sy_mspoc, sy)

      # Normalize
      sx_power_mspoc = standardize(sx_power_mspoc).T[0]
      sy_mspoc = standardize(sy_mspoc).T[0]
```

```
<xarray.DataArray (time: 580, mSPoC_X: 1)> Size: 5kB
0.9518 0.8879 0.838 1.385 1.158 1.05 … 0.8189 0.9737 0.626 1.605 2.467 1.912
Coordinates:
  * time       (time) float64 5kB 240.1 240.2 240.3 240.4 … 297.8 297.9 298.0
  * mSPoC_X    (mSPoC_X) <U3 12B 'Sx1'

<xarray.DataArray (source: 1, time: 600)> Size: 5kB
0.8019 0.3854 −0.06178 −0.4672 −0.7737 −0.9842 … 1.732 2.067 2.226 2.188 1.977
Coordinates:
  * time       (time) float64 5kB 240.1 240.2 240.3 240.4 … 299.8 299.9 300.0
  * source     (source) <U2 8B 'S1'

<xarray.DataArray (time: 580, mSPoC_Y: 1)> Size: 5kB
0.6547 1.091 1.502 0.9325 1.586 … −0.07052 −0.5063 −0.2778 −0.4498 −0.1842
Coordinates:
  * time       (time) float64 5kB 240.1 240.2 240.3 240.4 … 297.8 297.9 298.0
  * mSPoC_Y    (mSPoC_Y) <U3 12B 'Sy1'

<xarray.DataArray (source: 1, time: 600)> Size: 5kB
0.7055 0.9216 1.062 1.11 1.077 0.9927 … 1.532 1.584 1.563 1.5 1.425 1.356
Coordinates:
  * source     (source) <U2 8B 'S1'
  * time       (time) float64 5kB 240.1 240.2 240.3 240.4 … 299.8 299.9 300.0
```

As before, we truncate the ground truth sources for a direct comparison

```
[31]: # Truncate ground truth sources to match the reconstructed sources
      if len(sx_power_mspoc) < len(sx_power.time):
          sx_power_trunc = sx_power[0, : len(sx_power_mspoc)]
      else:
          sx_power_trunc = sx_power[0]
      if len(sy_mspoc) < len(sy.time):
          sy_trunc = sy[0, : len(sy_mspoc)]
      else:
          sy_trunc = sy[0]

      # Calculate correlations. corrxy_mspoc should be low because sx is shifted!
      corrxy_mspoc = np.corrcoef(sx_power_mspoc, sy_mspoc)[0, 1]
      corrx_mspoc = np.corrcoef(sx_power_mspoc, sx_power_trunc)[0, 1]
      corry_mspoc = np.corrcoef(sy_mspoc, sy_trunc)[0, 1]

      # Plot results
```

```
plot_source_comparisson(
    sx_power_trunc,
    sy_trunc,
    sx_power_tcca,
    sy_tcca,
    corrx_mspoc,
    corry_mspoc,
    corrxy_mspoc,
    title="mSPoC",
)
```

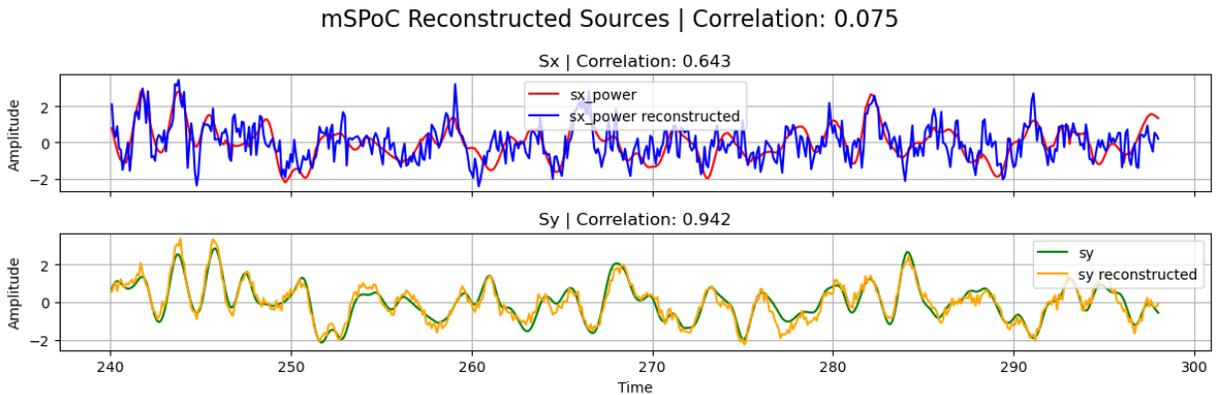

## ICA Source Extraction

In this notebook we will investigate an example on how Independent Component Analysis by Entropy Rate Bound Minimization (ICA-ERBM) can be used to extract physiological sources from fNIRS data. For this example, we will use a finger tapping dataset.

Let $X \in \mathbb{R}^{N \times T}$ denote the finger tapping data with $N$ channels and $T$ time points. We assume that the data $X$ consists of unknown independent sources $S \in \mathbb{R}^{N \times T}$ that were mixed through an unknown mixing matrix $A \in \mathbb{R}^{N \times N}$, such that

$$X = A \cdot S.$$

Source reconstruction in ICA-ERBM is done by minimizing the mutual information rate of the estimated sources. A demixing matrix $W$ is determined and the estimated sources $\hat{S} \in \mathbb{R}^{N \times T}$ can be computed as

$$\hat{S} = W \cdot X.$$

Among the extracted sources, we will identify the ones that correspond to the PPG and Mayer Wave signals.

### Loading Raw Finger Tapping Data

```
[32]: # Load finger tapping data set
      finger_tapping_data = cedalion.data.get_fingertappingDOT()

      # Extract the fnirs recording
      fnirs_data = finger_tapping_data['amp']
```

```
# Plot three channels of the fnirs data
fig, ax = plt.subplots(3, 1, sharex=True, figsize=(10, 5))
for i, ch in enumerate(["S1D1", "S1D2", "S7D9"]):
    ax[i].plot(fnirs_data.time, fnirs_data.sel(channel=ch, wavelength="760"), "r-",␣
    ↪label="760nm")
    ax[i].plot(fnirs_data.time, fnirs_data.sel(channel=ch, wavelength="850"), "b-",␣
    ↪label="850nm")
    ax[i].set_title(f"Channel {ch}")

ax[0].legend()
ax[2].set_xlim(0,60)
ax[2].set_xlabel("time / s")
plt.tight_layout()
```

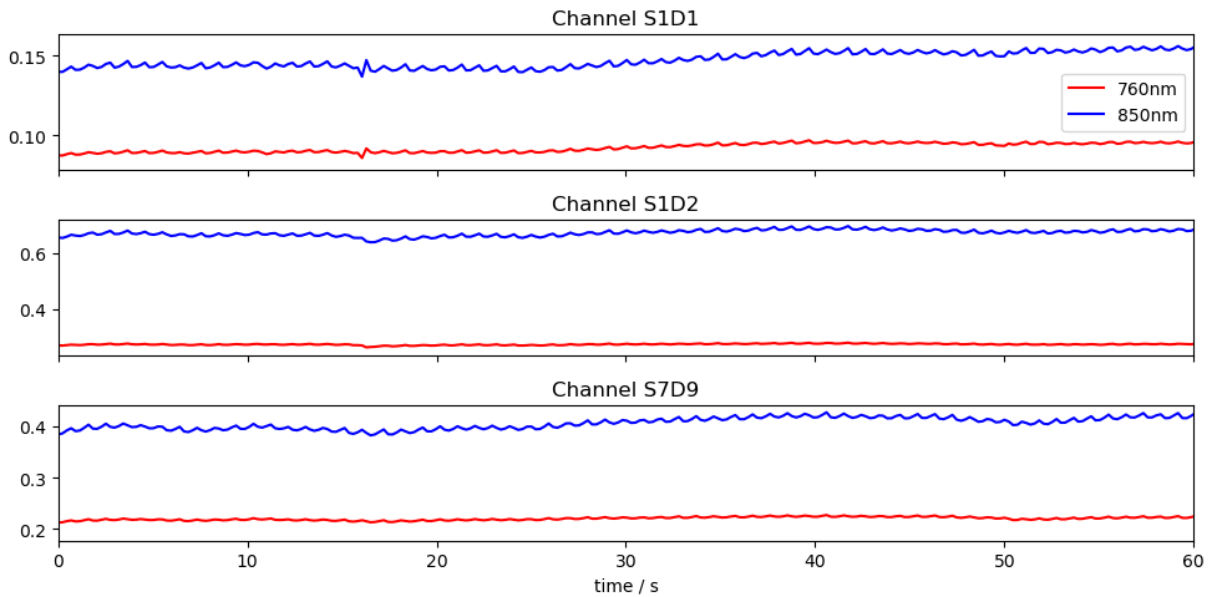

**Conversion to Optical Density**

```
[33]: # Convert to Optical Density (OD)
      fnirs_data_od = cedalion.nirs.cw.int2od(fnirs_data)
```

**Channel Quality Assessment and Pruning**

The Scalp Coupling Index (SCI) and Peak Spectral Power (PSP) are used for quality assessment. We compute SCI and PSP for each channel, and remove channels with less than 75% of clean time.

```
[34]: # Calculate masks for SCI and PSP quality metrics
      window_length = 5 * units.s
      sci_thresh = 0.75
      psp_thresh = 0.1
      sci_psp_percentage_thresh = 0.75
```

```python
sci, sci_mask = quality.sci(fnirs_data_od, window_length, sci_thresh)
psp, psp_mask = quality.psp(fnirs_data_od, window_length, psp_thresh)
sci_x_psp_mask = sci_mask & psp_mask
perc_time_clean = sci_x_psp_mask.sum(dim="time") / len(sci.time)
sci_psp_mask = [perc_time_clean >= sci_psp_percentage_thresh]

# Prune channels that do not pass the quality test
fnirs_data_pruned, drop_list = quality.prune_ch(fnirs_data_od, sci_psp_mask, "all")

# Display pruned channels
print(f"List of pruned channels: {drop_list}  ({len(drop_list)})")
```

List of pruned channels: ['S13D26']  (1)

**High-pass filter**

```python
[35]:  # Filter the data
       # fmax = 0 is used to indicate high-pass filtering
       fnirs_data_filtered = fnirs_data_pruned.cd.freq_filter(fmin= 0.01, fmax= 0,
        ↪butter_order=4)

       # Store sampling rate
       fnirs_data_samplingrate = sampling_rate(fnirs_data_pruned.time).magnitude

       # Plot the filtered data
       fig, ax = plt.subplots(3, 1, sharex=True, figsize=(10, 5))
       for i, ch in enumerate(["S1D1", "S1D2", "S7D9"]):
           ax[i].plot(fnirs_data_filtered.time, fnirs_data_filtered.sel(channel=ch,
        ↪wavelength="760"), "r-", label="760nm")
           ax[i].plot(fnirs_data_filtered.time, fnirs_data_filtered.sel(channel=ch,
        ↪wavelength="850"), "b-", label="850nm")
           ax[i].set_title(f"Channel {ch}")

       ax[0].legend()
       ax[2].set_xlim(0,60)
       ax[2].set_label("time / s")
       plt.tight_layout()
```

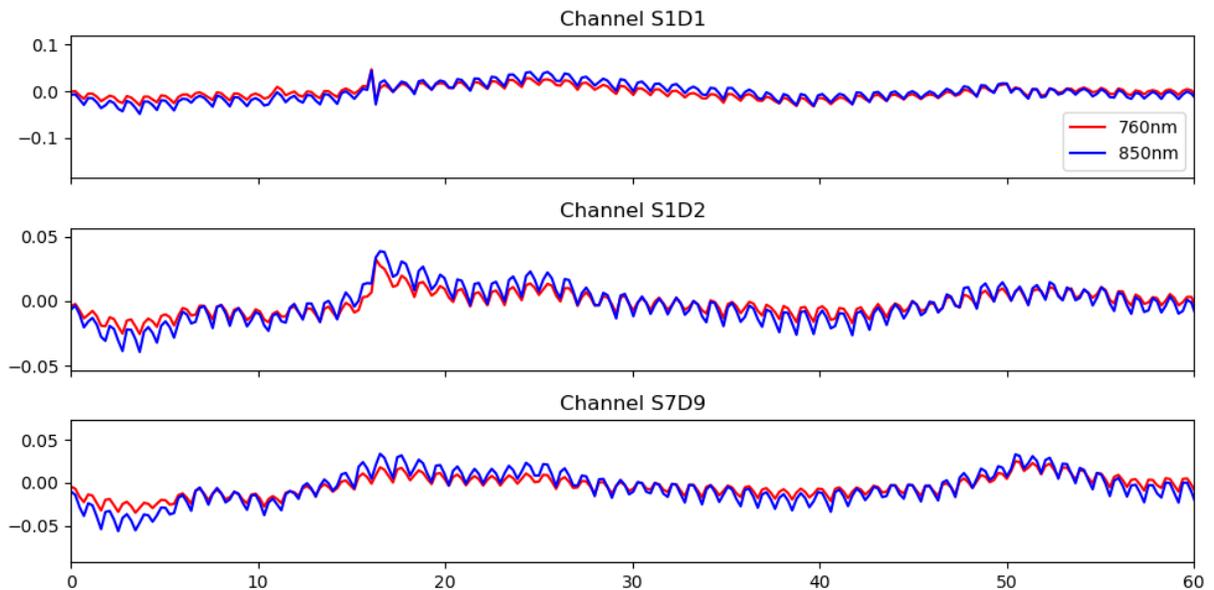

**Select Channels and Time Slice for ICA**

The entire finger tapping dataset was recorded over 30 minutes and contains 99 channels after pruning. Unfortunately, these dimensions result in a long runtime for ICA-ERBM. For this reason, we will use only a subset of the channels and a 10-minute slice of the selected channels. Despite the longer runtime, this example is also applicable to the full dataset.

```
[36]: # Choose the best 30 channels based on the percentage of time clean
      id_best_channels = np.argsort(perc_time_clean).values[-30:]
      best_channels = fnirs_data['channel'][id_best_channels]

      # Extract the best channels from the filtered data
      fnirs_best_channels = fnirs_data_filtered.sel(channel = best_channels)

      # Select a 10 min interval
      duration = 10 * 60
      buffer = 60
      fnirs_best_channels = fnirs_best_channels.sel(time=slice(buffer, buffer + duration))

      # Select the first wavelength
      X = fnirs_best_channels.values[:, 0, :]
      print(f"Shape of data for ICA-ERBM: {X.shape}")
```

Shape of data for ICA-ERBM: (30, 2616)

/opt/miniconda3/envs/cedalion_250922/lib/python3.11/site-
packages/xarray/core/variable.py:315: UnitStrippedWarning: The unit of the quantity is
stripped when downcasting to ndarray.
  data = np.asarray(data)

**Apply ICA-ERBM**

ICA-ERBM is applied to the selected channels. For the autoregressive filter used in ICA-ERBM, we use the default parameter $p = 11$. The source estimates are then computed as $\hat{S} = W \cdot X$.

```
[37]:  # Set filter length
       p = 11

       # Apply ICA-ERBM to the data
       W = ICA_ERBM(X, p)

       # Compute separated source as S = W * X
       sources = W.dot(X)
```

```
[38]:  # Apply z-score normalization to the sources
       sources_zscore = sp.stats.zscore(sources, axis=0)
```

**Selection of PPG Source**

From the reconstructed sources, we now want to identify those that are most similar to a PPG signal. To this end, we compare the frequency band in which the PPG signal is expected to have large amplitudes with the surrounding frequency bands. The sources with the highest contrast are selected. The PPG signal is expected to exhibit high amplitudes in a frequency band around 1 Hz.

```
[39]:  # Compute the frequency spectrum for each source
       psd_sources = np.abs(np.fft.fft(sources, axis = 1))

       # The frequencies corresponding to the spectrum
       freqs = np.fft.fftfreq(sources.shape[1], 1/fnirs_data_samplingrate)

       # Choose the indices of frequencies that are in the ppg band (0.75 - 1.25 Hz)
       ppg_band_ind = np.logical_and(freqs >= 0.75, freqs <= 1.25)

       # Choose the indices of frequencies that are in the band (0 - 0.75 Hz and 1.25 - 3.0⌴
       ↪Hz)
       comp_band =  np.logical_and(freqs >= 0, freqs < 0.75) +  np.logical_and(freqs > 1.25 ,⌴
       ↪freqs <= 3.0)

       # Compute the quotient of the  ppg band and the contrast band
       psd_quotient = np.sum(psd_sources[:, ppg_band_ind], axis = 1 ) / np.sum(psd_sources[:,⌴
       ↪comp_band], axis = 1 )

       # Choose the indices of the sources with the highest contrast
       max_contrast_index = np.argsort(psd_quotient, axis = 0 )[-5:]

       # Reverse the order of the indices to have the highest contrast first
       max_contrast_index = max_contrast_index[::-1]

       # Choose the sources with the highest contrast
       ppg_sources = sources_zscore[max_contrast_index, :]
```

```
[40]:  # Plot the sources with the highest contrast and their frequency spectrum
        fig, ax = plt.subplots(ppg_sources.shape[0], 2, figsize=(12, 2 * ppg_sources.shape[0]))

        for i in range(ppg_sources.shape[0]):
            # Plot the source for 60 seconds
            samples = int(fnirs_data_samplingrate * 60 * 1)
            ax[i, 0].plot( 1/fnirs_data_samplingrate * np.arange(0,samples), ppg_sources[i, :
        ↪samples], label=f"Source {max_contrast_index[i]+1}")
            ax[i, 0].set_title(f"Source {max_contrast_index[i] + 1}", fontsize=10)

            # Plot frequency spectrum of the source
            psd = np.abs(np.fft.rfft(ppg_sources[i, :])) ** 2
            x_freqs = np.fft.rfftfreq(ppg_sources.shape[1], 1 / fnirs_data_samplingrate)
            ax[i, 1].plot(x_freqs, psd, label="Contrast Band")
            ax[i, 1].set_title(f"Frequency Spectrum of Source {max_contrast_index[i]+1},␣
        ↪Contrast Quotient: {psd_quotient[max_contrast_index[i]]:.2f}",  fontsize=10)

            # Highlight the PPG band in the frequency spectrum
            highlight_ppg_band = np.logical_and(x_freqs >= 0.75, x_freqs <= 1.25)
            ax[i, 1].plot(x_freqs[highlight_ppg_band], psd[highlight_ppg_band],␣
        ↪color='orange', label='PPG Band')

        ax[0, 1].legend()
        ax[i, 0].set_xlabel("Time / s")
        ax[i, 1].set_xlabel("Frequency / Hz")
        fig.suptitle("PPG Sources", fontsize=16)
        plt.tight_layout()
```

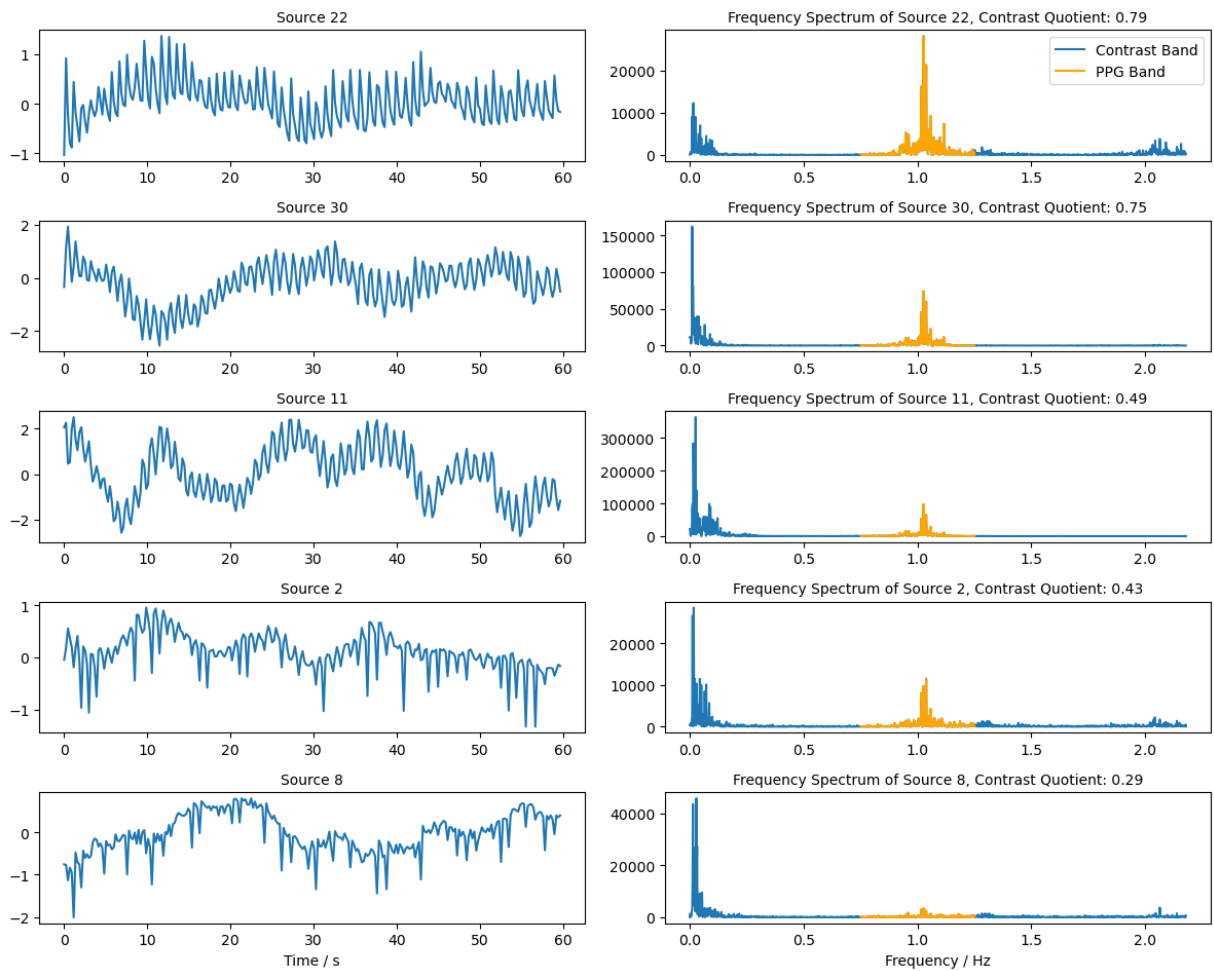

## Selection of Mayer Wave Source

Mayer waves are expected to have a frequency around 0.1 Hz. Similar to the PPG sources above, we will use the contrast between the frequency band around 0.1 Hz and the surrounding bands to rank the sources and identify those that are most similar to the Mayer wave.

```python
[41]: # Choose the indices of frequencies that are in the Mayer Wave band (0.05 - 0.15 Hz)
mw_band_ind = np.logical_and(freqs >= 0.05, freqs <= 0.15)

# Choose the indices of frequencies that are in the band (0 - 0.05 Hz and 0.15 - 3.0
↪Hz)
comp_band =  np.logical_and(freqs >= 0, freqs < 0.05) +  np.logical_and(freqs > 0.15 ,
↪freqs <= 3.0)

# Compute the quotient of the Mayer Wave band and the contrast band
psd_quotient = np.sum(psd_sources[:, mw_band_ind], axis = 1 ) / np.sum(psd_sources[:,
↪comp_band], axis = 1 )

# Choose the indices of the sources with the highest contrast
max_contrast_index = np.argsort(psd_quotient, axis = 0 )[-5:]
```

```python
# Reverse the order of the indices to have the highest contrast first
max_contrast_index = max_contrast_index[::-1]

# Extract the sources with the highest contrast
mw_sources = sources_zscore[max_contrast_index, :]
```

```python
[42]: # Plot the sources with the highest contrast and their frequency spectrum
fig, ax = plt.subplots(mw_sources.shape[0], 2, figsize=(12, 2 * mw_sources.shape[0]))

for i in range(mw_sources.shape[0]):
    # Plot the source for 60 seconds
    samples = int(fnirs_data_samplingrate * 60 * 1)
    ax[i, 0].plot( 1/fnirs_data_samplingrate * np.arange(0,samples), mw_sources[i, :
     samples], label=f"Source {max_contrast_index[i]+1}")
    ax[i, 0].set_title(f"Source {max_contrast_index[i] + 1}", fontsize=10)

    # Plot frequency spectrum of the source
    psd = np.abs(np.fft.rfft(mw_sources[i, :])) ** 2
    x_freqs = np.fft.rfftfreq(mw_sources.shape[1], 1 / fnirs_data_samplingrate)
    ax[i, 1].plot(x_freqs, psd, label="Contrast Band")
    ax[i, 1].set_title(f"Frequency Spectrum of Source {max_contrast_index[i]+1},
     Contrast Quotient: {psd_quotient[max_contrast_index[i]]:.2f}",  fontsize=10)

    # Highlight the Mayer Wave band in the frequency spectrum
    highlight_mw_band = np.logical_and(x_freqs >= 0.05, x_freqs <= 0.15)
    ax[i, 1].plot(x_freqs[highlight_mw_band], psd[highlight_mw_band], color='orange',
     label='Mayer Wave Band')

ax[0, 1].legend()
ax[i, 0].set_xlabel("Time / s")
ax[i, 1].set_xlabel("Frequency / Hz")
fig.suptitle("Mayer Wave Sources", fontsize=16)
plt.tight_layout()
```

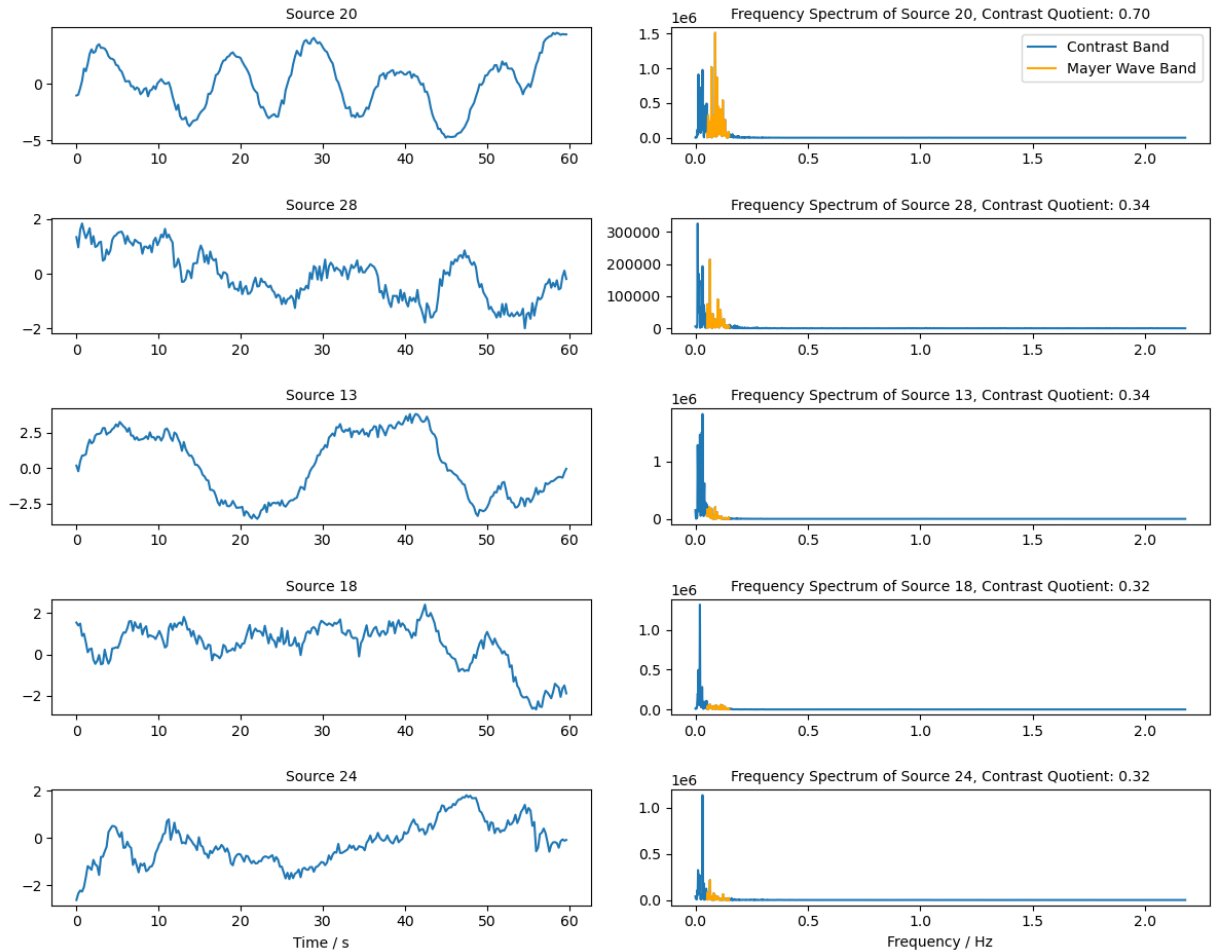

## Single Trial Classification

The last section of this notebook demonstrates how Cedalion interfaces with scikit-learn to train a simple single-subject, single-trial classifier. The focus is on data flow: performing preprocessing with short-channel regression within a cross-validation scheme, extracting features from epochs, and passing these features to scikit-learn for training and evaluation while preserving feature metadata to allow tracing each feature back to its origin.

```
[43]: from sklearn.discriminant_analysis import LinearDiscriminantAnalysis
      from sklearn.preprocessing import LabelEncoder
      from sklearn.inspection import permutation_importance

      import cedalion.models.glm as glm
      import cedalion.mlutils as mlutils
```

```
[44]: rec = cedalion.data.get_fingertappingDOT()

      # assign string labels to events and pool finger-tapping and ball-squeezing trials
      rec.stim.cd.rename_events(
          {
```

```
            "1": "Control",
            "2": "Motor/Left",   # "FTapping/Left",
            "3": "Motor/Right",  # "FTapping/Right",
            "4": "Motor/Left",   # "BallSqueezing/Left",
            "5": "Motor/Right",  # "BallSqueezing/Right",
        }
)

# Keep only motor trials. Also remove the last trial so that there
# are equal numbers of trials for Motor/Left and Motor/Right
rec.stim = (
        rec.stim[rec.stim.trial_type.str.startswith("Motor")]
        .sort_values("onset")
        .reset_index(drop=True)
        .iloc[:-1]
)

display(rec.stim.groupby("trial_type").count())
display(rec.stim)
```

|            | onset | duration | value |
|------------|-------|----------|-------|
| trial_type |       |          |       |
| Motor/Left | 32    | 32       | 32    |
| Motor/Right| 32    | 32       | 32    |

|    | onset       | duration | value | trial_type  |
|----|-------------|----------|-------|-------------|
| 0  | 8.486912    | 10.0     | 1.0   | Motor/Left  |
| 1  | 38.764544   | 10.0     | 1.0   | Motor/Right |
| 2  | 69.042176   | 10.0     | 1.0   | Motor/Right |
| 3  | 99.549184   | 10.0     | 1.0   | Motor/Left  |
| 4  | 129.597440  | 10.0     | 1.0   | Motor/Left  |
| .. | ...         | ...      | ...   | ...         |
| 59 | 1837.301760 | 10.0     | 1.0   | Motor/Left  |
| 60 | 1868.496896 | 10.0     | 1.0   | Motor/Left  |
| 61 | 1899.003904 | 10.0     | 1.0   | Motor/Right |
| 62 | 1931.116544 | 10.0     | 1.0   | Motor/Right |
| 63 | 1962.541056 | 10.0     | 1.0   | Motor/Left  |

[64 rows x 4 columns]

**Preprocessing the dataset**

As in previous notebooks, motion artifacts are corrected with the TDDR and wavelet algorithms. The data is transformed into optical densities and a global component is subtracted. After bandpass filtering, concentration changes are calculated.

```
[45]: rec["od"] = cedalion.nirs.cw.int2od(rec["amp"])
      rec["od_tddr"] = motion.tddr(rec["od"])
      rec["od_wavelet"] = motion.wavelet(rec["od_tddr"])

      # see 2_tutorial_preprocessing.ipynb for channel selection
      bad_channels = ['S13D26', 'S14D28']
```

```
rec["od_clean"] = rec["od_wavelet"].sel(channel=~rec["od"].channel.isin(['S13D26',
    'S14D28']))

od_var = quality.measurement_variance(rec["od_clean"], calc_covariance=False)

rec["od_mean_subtracted"], global_comp = physio.global_component_subtract(
    rec["od_clean"], ts_weights=1 / od_var, k=0
)

rec["od_freqfiltered"] = rec["od_mean_subtracted"].cd.freq_filter(
#rec["od_freqfiltered"] = rec["od_clean"].cd.freq_filter(
    fmin=0.01, fmax=0.5, butter_order=4
)

dpf = xr.DataArray(
    [6, 6],
    dims="wavelength",
    coords={"wavelength": rec["amp"].wavelength},
)

rec["conc"] = cedalion.nirs.cw.od2conc(rec["od_freqfiltered"], rec.geo3d, dpf)
```

**Short-channel regression**

Following the approach of von Lühmann et al. (2020), we apply a GLM to regress out physiological noise, thereby improving single-trial classification performance. To incorporate the GLM into a cross-validaton scheme, the design matrix will be masked for each cross-validation fold.

```
[46]:  # separate long and short channels
       rec["conc_long"], rec["conc_short"] = cedalion.nirs.split_long_short_channels(
           rec["conc"], rec.geo3d, distance_threshold=22.5 * units.mm
       )

       # define the design matrix
       dms = (
           glm.design_matrix.hrf_regressors(
               rec["conc_long"],
               rec.stim,
               glm.Gamma(tau=0 * units.s, sigma=3 * units.s),
           )
           & glm.design_matrix.drift_regressors(rec["conc_long"], drift_order=1)
           & glm.design_matrix.average_short_channel_regressor(rec["conc_short"])
       )
       dms
```

```
[46]:  DesignMatrix(common=['HRF Motor/Left','HRF Motor/Right','Drift 0','Drift 1','short'],
       channel_wise=[])
```

The dataset has 64 trials, with equal numbers of "Motor/Left" and "Motor/Right" conditions. In the following this dataset is split into 4 folds using the function **mlutils.cv.create_cv_splits**.

For each fold, a masked design matrix is created by setting the time segment used for testing to zero.

The GLM parameters are estimated and the signal components explained by nuisance regressors are subtracted.

Finally, the time series is segmented into epochs, yielding for each cross-validation fold an array of samples, to which the trial information required for training is appended as additional coordinates.

```
[47]: n_splits = 4
      before = 5 * cedalion.units.s
      after = 25 * cedalion.units.s

      cv_folds = []

      for i_split, (df_stim_train, df_stim_test) in enumerate(
          mlutils.cv.create_cv_splits(rec.stim, n_splits)
      ):
          # short-channel regression (SCR):

          # zero-out design matrix in test segment
          dms_masked = mlutils.cv.mask_design_matrix(
              dms,
              df_stim_test,
              before=before,
              after=after,
          )

          # fit long channels with masked design matrix
          result = glm.fit(rec["conc_long"], dms_masked, noise_model="ols")

          # compute component explained by short-channel regressor
          short_component = glm.predict(
              rec["conc_long"],
              result.sm.params.sel(regressor=["short", "Drift 0", "Drift 1"]),
              dms_masked,
          )
          short_component = short_component.pint.quantify(rec["conc_long"].pint.units)

          # subtract short component
          conc_long_scr = rec["conc_long"] - short_component

          # split time series into epochs
          epochs = conc_long_scr.cd.to_epochs(
              rec.stim,
              ["Motor/Left", "Motor/Right"],
              before=before,
              after=after,
          )

          # baseline correction
          baseline = epochs.sel(reltime=(epochs.reltime < 0)).mean("reltime")
          epochs = epochs - baseline
```

```python
        # assign train-/test-set membership...
        is_train = np.zeros(epochs.sizes["epoch"], dtype=bool)
        is_test = np.zeros(epochs.sizes["epoch"], dtype=bool)
        is_train[df_stim_train.index.values] = True
        is_test[df_stim_test.index.values] = True

        # ... and digitized trial labels ...
        label_encoder = LabelEncoder()
        y = label_encoder.fit_transform(epochs.trial_type.values)

        # ... as coordinates to the DataArray
        epochs = epochs.assign_coords(
            {
                "is_train": ("epoch", is_train),
                "is_test": ("epoch", is_test),
                "y" : ("epoch", y)
            }
        )

        cv_folds.append(epochs)
```

For each cross-validation fold the epoched time series looks like this:

```
[48]: cv_folds[0]
```

```
[48]: <xarray.DataArray (epoch: 64, chromo: 2, channel: 44, reltime: 132)> Size: 6MB
      [µM] 0.06823 0.04856 0.04232 0.03702 0.03248 … 0.04141 0.0395 0.03601 0.03161
      Coordinates:
        * reltime      (reltime) float64 1kB -5.038 -4.809 -4.58 … 24.5 24.73 24.96
          trial_type   (epoch) <U11 3kB 'Motor/Left' 'Motor/Right' … 'Motor/Left'
        * chromo       (chromo) <U3 24B 'HbO' 'HbR'
        * channel      (channel) object 352B 'S1D6' 'S1D8' 'S2D5' … 'S14D25' 'S14D27'
          source       (channel) object 352B 'S1' 'S1' 'S2' 'S2' … 'S13' 'S14' 'S14'
          detector     (channel) object 352B 'D6' 'D8' 'D5' 'D9' … 'D28' 'D25' 'D27'
          is_train     (epoch) bool 64B False False False False … True True True True
          is_test      (epoch) bool 64B True True True True … False False False False
          y            (epoch) int64 512B 1 1 0 0 1 1 1 0 1 … 0 1 1 0 1 0 0 1 0
      Dimensions without coordinates: epoch
```

The hemodynamic response to the two motor tasks can be visualized by block-averaging these epochs.

```python
[49]: blockaverage = cv_folds[0].groupby("trial_type").mean("epoch")

      # Plot block averages. Please ignore errors if the plot is too small in the HD case

      noPlts2 = int(np.ceil(np.sqrt(len(blockaverage.channel))))
      f,ax = plt.subplots(noPlts2,noPlts2, figsize=(12,10))
      ax = ax.flatten()
      for i_ch, ch in enumerate(blockaverage.channel):
          for ls, trial_type in zip(["-", "--"], blockaverage.trial_type):
              ax[i_ch].plot(blockaverage.reltime, blockaverage.sel(chromo="HbO",␣
      ↪trial_type=trial_type, channel=ch), "r", lw=2, ls=ls)
```

```
        ax[i_ch].plot(blockaverage.reltime, blockaverage.sel(chromo="HbR",␣
↪trial_type=trial_type, channel=ch), "b", lw=2, ls=ls)

    ax[i_ch].grid(1)
    ax[i_ch].set_title(ch.values)
    ax[i_ch].set_ylim(-.3, .3)
    ax[i_ch].set_axis_off()
    ax[i_ch].axhline(0, c="k")
    ax[i_ch].axvline(0, c="k")

for i in range(len(blockaverage.channel), len(ax)):
    ax[i].set_axis_off()

plt.suptitle("HbO: r | HbR: b | left: - | right: --")
plt.tight_layout()
```

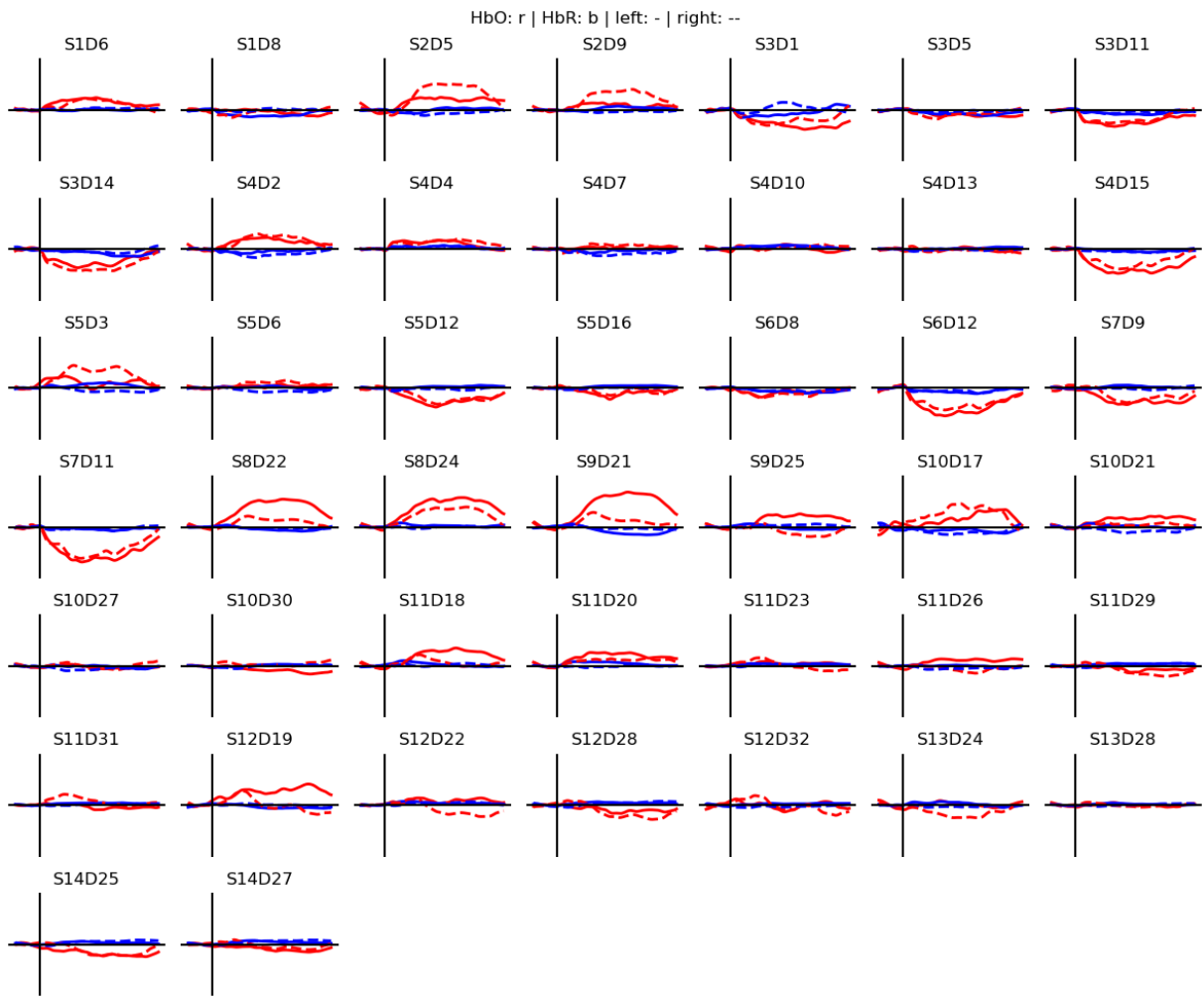

For each of the 64 epochs there are $N_{chromo} \times N_{channel}$ time traces. From these time traces, features must get extracted.

The function `mlutils.features.epoch_features` calculates common features of the hemodynamic response, such as slope, mean, maximum, minimum and area under the curve. For each feature type, a time range can be specified, over which the feature is calculated. In the present case, this yields features for each channel and chromophore, which are then stacked into a single feature dimensions. The resulting array has the shape expected by scikit-learn, $(N_{samples}, N_{features})$.

```
[50]: X = mlutils.features.epoch_features(
          cv_folds[0],
          feature_types=["slope", "mean", "max", "min", "auc"],
          reltime_slices={
              "slope": slice(0, 9),
              "mean": slice(3, 10),
              "max": slice(2, 8),
              "min": slice(2, 8),
          },
      )
      X
```

```
[50]: <xarray.DataArray (epoch: 64, feature: 440)> Size: 225kB
      0.02127 -0.01976 0.009303 -0.007574 -0.01678 … 0.07023 -0.04713 0.1401 0.214
      Coordinates:
          trial_type   (epoch)   <U11 3kB 'Motor/Left' 'Motor/Right' … 'Motor/Left'
          source       (feature) object 4kB 'S1' 'S1' 'S2' 'S2' … 'S13' 'S14' 'S14'
          detector     (feature) object 4kB 'D6' 'D8' 'D5' 'D9' … 'D28' 'D25' 'D27'
          is_train     (epoch)   bool 64B False False False False … True True True
          is_test      (epoch)   bool 64B True True True True … False False False
          y            (epoch)   int64 512B 0 1 1 0 0 0 1 1 1 0 … 0 1 1 0 1 0 0 1 1 0
        * feature      (feature) object 4kB MultiIndex
        * feature_type (feature) object 4kB 'slope' 'slope' 'slope' … 'auc' 'auc'
        * chromo       (feature) <U3 5kB 'HbO' 'HbO' 'HbO' … 'HbR' 'HbR' 'HbR'
        * channel      (feature) object 4kB 'S1D6' 'S1D8' … 'S14D25' 'S14D27'
      Dimensions without coordinates: epoch
```

During stacking, the coordinates of the `'chromo'` and `'channel'` dimensions don't get lost. They are combined and reassigned to the new `'feature'` dimension. This means that for every feature in `X`, you can trace back which channel and chromophore it originated from.

```
[51]: X.feature
```

```
[51]: <xarray.DataArray 'feature' (feature: 440)> Size: 4kB
      MultiIndex
      Coordinates:
          source       (feature) object 4kB 'S1' 'S1' 'S2' 'S2' … 'S13' 'S14' 'S14'
          detector     (feature) object 4kB 'D6' 'D8' 'D5' 'D9' … 'D28' 'D25' 'D27'
        * feature      (feature) object 4kB MultiIndex
        * feature_type (feature) object 4kB 'slope' 'slope' 'slope' … 'auc' 'auc'
        * chromo       (feature) <U3 5kB 'HbO' 'HbO' 'HbO' … 'HbR' 'HbR' 'HbR'
        * channel      (feature) object 4kB 'S1D6' 'S1D8' … 'S14D25' 'S14D27'
```

**Training and Evaluating a LDA classifier**

For each cross-validation fold, extract features, train the classifier and estimate classification accuracy.

```
[52]: accuracies = []
```

```python
for epochs in cv_folds:
    # extract features
    X = mlutils.features.epoch_features(
        epochs.sel(chromo="HbO"), # HbO only
        feature_types=["slope", "max", "mean"],
        reltime_slices={
            "slope": slice(0, 9),
            "mean": slice(3, 10),
            "max": slice(2, 8),
        },
    )

    # separate train and test sets
    X_train = X[X.is_train]
    y_train = X_train.y
    X_test = X[X.is_test]
    y_test = X_test.y

    # train a LDA classifier
    clf = LinearDiscriminantAnalysis(n_components=1, solver='lsqr', shrinkage="auto")
    clf.fit(X_train, y_train)

    # evaluate perfomance
    accuracy = clf.score(X_test, y_test)
    accuracies.append(accuracy)
    print(f"#train: {len(y_train)} #test: {len(y_test)} #features: {X_train.shape[1]}
    ↪accuracy: {accuracy}")

print()
print(rf"average accuracy over cross-validation splits: {np.mean(accuracies):.3f} ±
↪{np.std(accuracies):.3f}")
```

```
#train: 48 #test: 16 #features: 132 accuracy: 0.8125
#train: 48 #test: 16 #features: 132 accuracy: 0.8125
#train: 48 #test: 16 #features: 132 accuracy: 0.6875
#train: 48 #test: 16 #features: 132 accuracy: 0.8125

average accuracy over cross-validation splits: 0.781 ± 0.054
```

**What did it learn?**

One way of estimating feature importance is the sklearn function `permuatation_importance`. It permutes feature columns and assesses the effect on the classification accuracy.

```python
[53]: result = permutation_importance(
          clf, X_test, y_test, scoring="accuracy", n_repeats=10, random_state=0
      )

importances = result.importances_mean
importance_normalized = importances / importances.sum()
```

The returned metric highlights only a few important features:

```
[54]: plt.figure(figsize=(10,4))
      plt.plot(importance_normalized)
      plt.xlabel("feature")
      plt.ylabel("normalized permuation importance");
```

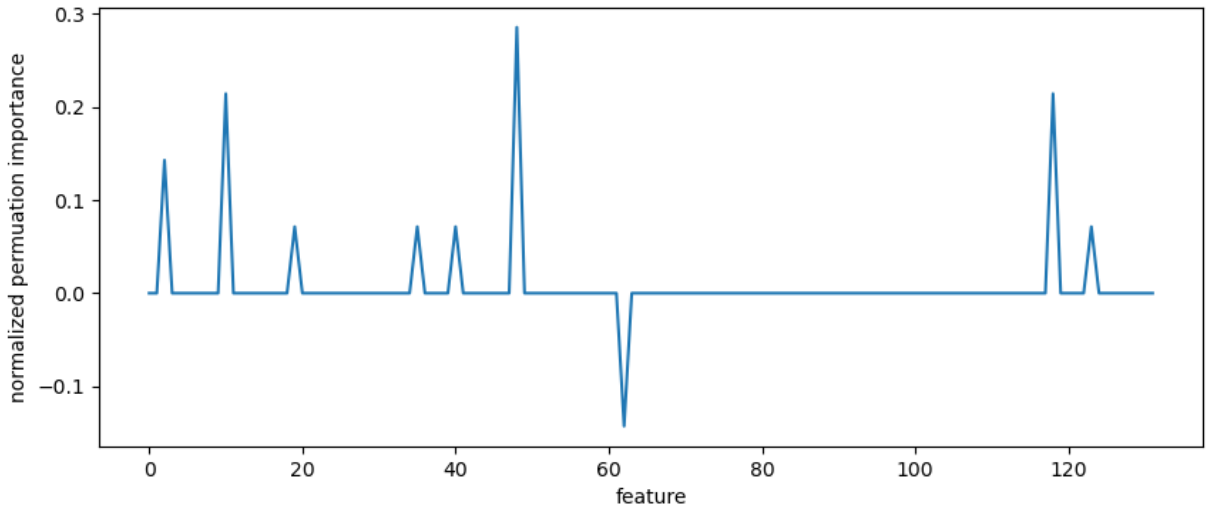

Use the preserved coordinates to identify these features:

```
[55]: X_train.feature[importance_normalized != 0]
```

```
[55]: <xarray.DataArray 'feature' (feature: 9)> Size: 72B
      MultiIndex
      Coordinates:
          chromo          <U3 12B 'HbO'
          source          (feature) object 72B 'S2' 'S4' 'S6' 'S11' … 'S6' 'S11' 'S11'
          detector        (feature) object 72B 'D5' 'D7' 'D12' … 'D8' 'D18' 'D31'
        * feature         (feature) object 72B MultiIndex
        * feature_type    (feature) object 72B 'slope' 'slope' 'slope' … 'mean' 'mean'
        * channel         (feature) object 72B 'S2D5' 'S4D7' … 'S11D18' 'S11D31'
```

By visualizing the feature distributions and the block-averaged epochs of the corresponding channels, we can observe that the HRFs in these channels, as well as the derived features, indeed have discriminative power.

```
[56]: important_features = X_train.feature[importance_normalized != 0].values

      f, ax = plt.subplots(2, len(important_features), figsize=(16,6))

      for i, ftr in enumerate(important_features):
          xx_train = X_train.loc[:, ftr]
          xx_test = X_test.loc[:, ftr]
          bins = np.linspace(xx_train.min().item(),xx_train.max().item(), 11)

          ax[0,i].hist(xx_train[xx_train.trial_type == "Motor/Left"], bins,fc="r",␣
      ↪label="Motor/Left", alpha=.5)
```

```
    ax[0,i].hist(xx_train[xx_train.trial_type == "Motor/Right"], bins, fc="g",␣
↪label="Motor/Right", alpha=.5)

    ax[1,i].hist(xx_test[xx_test.trial_type == "Motor/Left"], bins,fc="r",␣
↪label="Motor/Left", alpha=.5)
    ax[1,i].hist(xx_test[xx_test.trial_type == "Motor/Right"], bins, fc="g",␣
↪label="Motor/Right", alpha=.5)

    ax[1,i].set_xlabel(str(ftr))

ax[0,0].set_ylabel("# trials")
ax[1,0].set_ylabel("# trials")
f.suptitle("Top: Train | Bottom: Test | Red: 'Motor/Left' | Green: 'Motor/Right'");
plt.tight_layout()

f, ax = plt.subplots(1, len(important_features), figsize=(16,3))
for i, ftr in enumerate(important_features):
    ax[i].plot(blockaverage.reltime, blockaverage.sel(channel=ftr[1], chromo="HbO",␣
↪trial_type="Motor/Left"), "r-")
    ax[i].plot(blockaverage.reltime, blockaverage.sel(channel=ftr[1], chromo="HbR",␣
↪trial_type="Motor/Left"), "b-")
    ax[i].plot(blockaverage.reltime, blockaverage.sel(channel=ftr[1], chromo="HbO",␣
↪trial_type="Motor/Right"), "r--")
    ax[i].plot(blockaverage.reltime, blockaverage.sel(channel=ftr[1], chromo="HbR",␣
↪trial_type="Motor/Right"), "b--")
    ax[i].set_ylim(-.2,.2)
    ax[i].grid()
    ax[i].set_title(ftr)
    ax[i].set_xlabel("$t_{rel}$ / s")

ax[0].set_ylabel(r"$\Delta$ concentration / µM")
f.suptitle("Red: HbO | Blue: HbR | - 'Motor/Left' | -- 'Motor/Right'");
plt.tight_layout()
```

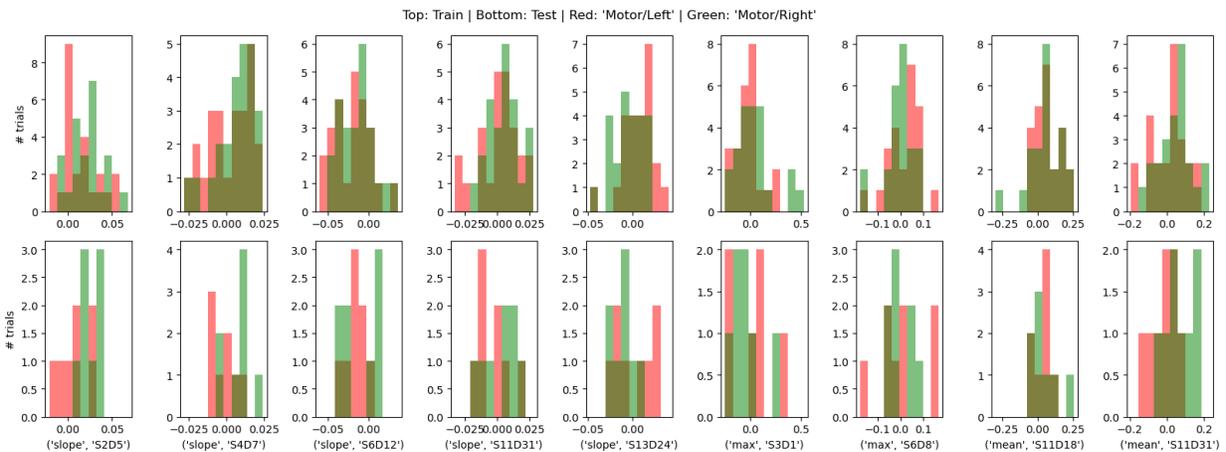

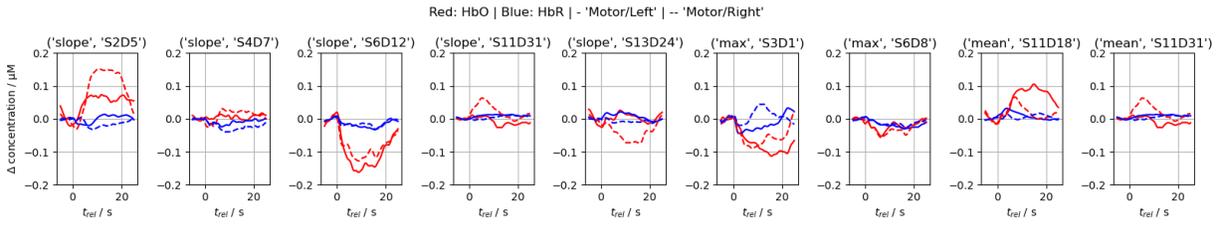

# S7: Data Augmentation

This example notebook illustrates the functionality in `cedalion.sim.synthetic_hrf` and `cedalion.sim.synthetic_artifact` to create simulated datasets with added activations and motion artifacts.

It has two parts:

1. Adding synthetic activations
2. Adding synthetic motion artifacts

```python
[1]: # This cells setups the environment when executed in Google Colab.
     try:
         import google.colab
         !curl -s https://raw.githubusercontent.com/ibs-lab/cedalion/dev/scripts/
     ↪colab_setup.py -o colab_setup.py
         # Select branch with --branch "branch name" (default is "dev")
         %run colab_setup.py
     except ImportError:
         pass
```

```python
[2]: # set this flag to True to enable interactive 3D plots
     INTERACTIVE_PLOTS = False
```

```python
[3]: import matplotlib.pyplot as plt
     import numpy as np
     import pyvista as pv
     import xarray as xr

     import cedalion
     import cedalion.dataclasses as cdc
     import cedalion.data
     import cedalion.geometry.landmarks as cd_landmarks
     import cedalion.dot as dot
     import cedalion.models.glm as glm
     import cedalion.nirs

     import cedalion.vis.anatomy
     import cedalion.vis.blocks as vbx

     import cedalion.sigproc.quality as quality
     import cedalion.sim.synthetic_hrf as synhrf
     import cedalion.sim.synthetic_artifact as synart
     import cedalion.xrutils as xrutils
     from cedalion import units
     from IPython.display import Image

     xr.set_options(display_expand_data=False)

     if INTERACTIVE_PLOTS:
         pv.set_jupyter_backend('server')
     else:
         pv.set_jupyter_backend('static')
```

```
[4]: # helper function to display gifs in rendered notbooks
     def display_image(fname : str):
         display(Image(data=open(fname,'rb').read(), format='png'))
```

## Adding Synthetic Activations

### Loading and preprocessing the dataset

This notebook uses a high-density, whole head resting state dataset recorded with a NinjaNIRS 22.

```
[5]: rec = cedalion.data.get_nn22_resting_state()

     geo3d = rec.geo3d
     meas_list = rec._measurement_lists["amp"]

     amp = rec["amp"]
     amp = amp.pint.dequantify().pint.quantify("V")

     display(amp)
```

```
<xarray.DataArray (channel: 567, wavelength: 2, time: 3309)> Size: 30MB
[V] 0.0238 0.02385 0.02375 0.02361 0.02364 … 0.2887 0.2892 0.2897 0.291 0.2923
Coordinates:
  * time        (time) float64 26kB 0.0 0.1112 0.2225 … 367.8 367.9 368.0
    samples     (time) int64 26kB 0 1 2 3 4 5 … 3303 3304 3305 3306 3307 3308
  * channel     (channel) object 5kB 'S10D87' 'S10D94' … 'S47D29' 'S5D137'
    source      (channel) object 5kB 'S10' 'S10' 'S10' … 'S56' 'S47' 'S5'
    detector    (channel) object 5kB 'D87' 'D94' 'D89' … 'D129' 'D29' 'D137'
  * wavelength  (wavelength) float64 16B 760.0 850.0
Attributes:
    data_type_group:  unprocessed raw
```

```
[6]: cedalion.vis.anatomy.plot_montage3D(rec["amp"], geo3d)
```

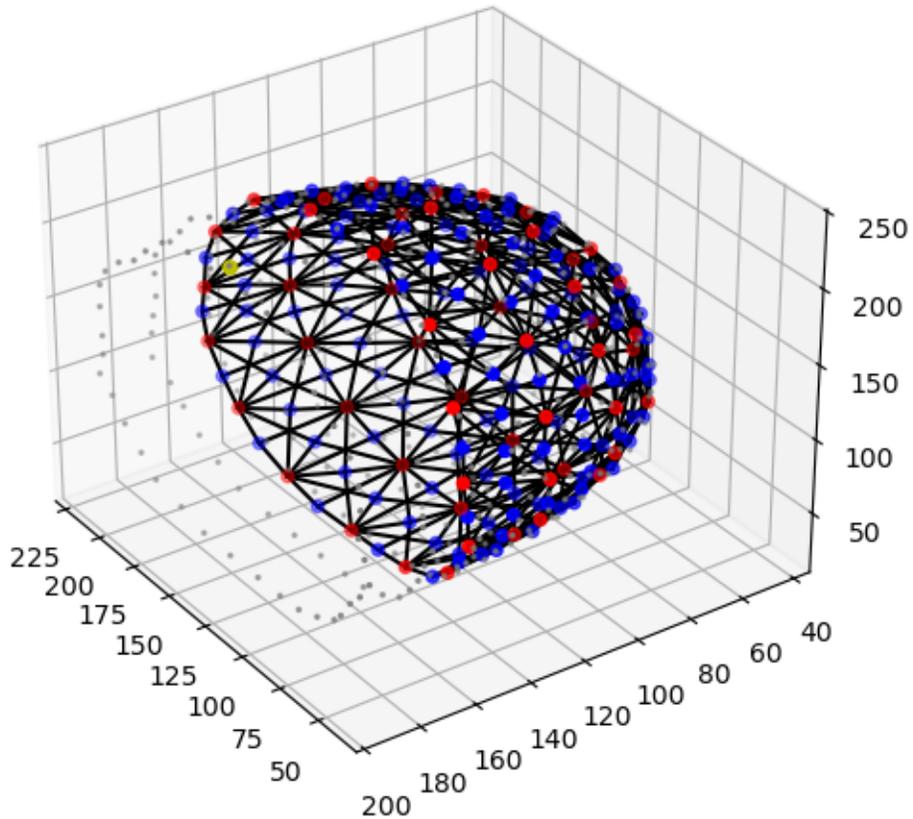

```
[7]:  # Select channels which have at least a signal-to-noise ratio of 10
      snr_thresh = 10   # the SNR (std/mean) of a channel.
      snr, snr_mask = quality.snr(rec["amp"], snr_thresh)
      amp_selected, masked_channels = xrutils.apply_mask(
          rec["amp"], snr_mask, "drop", "channel"
      )

      print(f"Removed {len(masked_channels)} channels because of low SNR.")
```

```
Removed 22 channels because of low SNR.
```

```
[8]:  # Calculate optical density
      od = cedalion.nirs.cw.int2od(amp_selected)
```

**Construct headmodel**

We load the the Colin27 headmodel, since we need the geometry for image reconstruction.

```
[9]:  head_ijk = dot.get_standard_headmodel("colin27")

      # transform coordinates to a RAS coordinate system
      head_ras = head_ijk.apply_transform(head_ijk.t_ijk2ras)
```

```
[10]:  display(head_ijk.brain)
       display(head_ras.brain)
```

TrimeshSurface(faces: 29988 vertices: 15002 crs: ijk units: dimensionless vertex_coords:↵
↳['parcel'])

TrimeshSurface(faces: 29988 vertices: 15002 crs: mni units: millimeter vertex_coords:↵
↳['parcel'])

```
[11]:  head_ras.landmarks
```

```
[11]:  <xarray.DataArray (label: 73, mni: 3)> Size: 2kB
       [mm] 0.0045 -118.6 -23.08 -86.08 -19.99 … -100.4 37.02 49.81 -99.34 21.47
       Coordinates:
         * label    (label) <U3 876B 'Iz' 'LPA' 'RPA' 'Nz' … 'PO1' 'PO2' 'PO4' 'PO6'
           type     (label) object 584B PointType.LANDMARK … PointType.LANDMARK
       Dimensions without coordinates: mni
```

We want to build the synthetic HRFs at C3 and C4 (green dots in the image below)

```
[12]:  plt_pv = pv.Plotter()
       vbx.plot_surface(plt_pv, head_ras.brain, color="#d3a6a1")
       vbx.plot_surface(plt_pv, head_ras.scalp, opacity=0.1)
       vbx.plot_labeled_points(
           plt_pv, head_ras.landmarks.sel(label=["C3", "C4"]), show_labels=True
       )
       vbx.camera_at_cog(plt_pv, head_ras.brain, (-400,500,400), fit_scene=True)

       plt_pv.show()
```

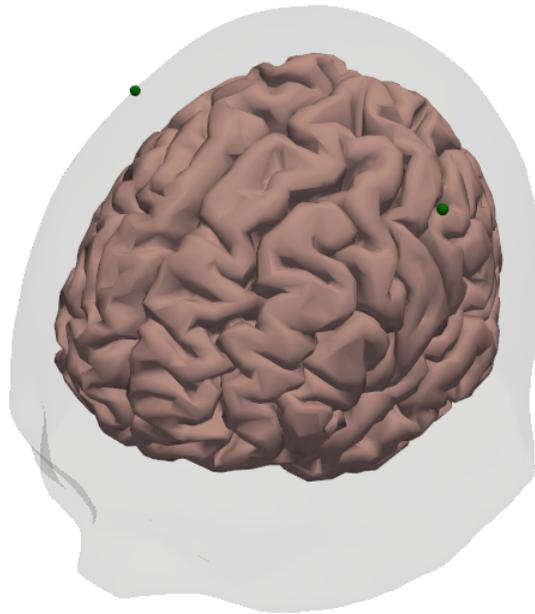

**Build spatial activation pattern on brain surface for landmarks C3 and C4**

The function `build_spatial_activation` is used to place a spatial activation pattern on the brain surface. The activation pattern is a Gaussian function of the geodesic distance to a seed vertex. Hence, the size of the activation is determined by the standard deviation of this Gaussian, specified by the parameter `spatial_scale`. The peak intensity in HbO is determined by the parameter `intensity_scale`. The intensity of HbR activation is specified relative to the HbO peak intensity. So if the HbO pattern describes an increase in Hbo then providing a negative factor smaller than 1 yields a decrease in HbR with smaller amplitude. The seed vertex (integer) can be selected as the closest vertex to a given landmark:

```
[13]:  # obtain the closest vertices to C3 and C4
       _, c3_seed = head_ras.brain.mesh.kdtree.query(
           head_ras.landmarks.sel(label="C3").pint.magnitude
       )
       _, c4_seed = head_ras.brain.mesh.kdtree.query(
           head_ras.landmarks.sel(label="C4").pint.magnitude
       )
```

```
[14]:  # create the spatial activation
       spatial_act = synhrf.build_spatial_activation(
           head_ras.brain,
           c3_seed,
           spatial_scale=2 * cedalion.units.cm,
```

```
        intensity_scale=1 * units.micromolar,
        hbr_scale=-0.4,
)
```

The resulting `DataArray` contains an activation value for each vertex and chromophore on the brain surface.

```
[15]: display(spatial_act)
```

```
<xarray.DataArray (vertex: 15002, chromo: 2)> Size: 240kB
[M] 1.115e-28 -4.461e-29 2.003e-29 -8.013e-30 5.23e-30 … 0.0 -0.0 0.0 -0.0
Coordinates:
  * chromo     (chromo) <U3 24B 'HbO' 'HbR'
Dimensions without coordinates: vertex
```

```
[16]: f,ax = plt.subplots(1,2,figsize=(10,5))
cedalion.vis.anatomy.plot_brain_in_axes(
        od,
        head_ras.landmarks,
        spatial_act.sel(chromo="HbO").pint.to("uM"),
        head_ras.brain,
        ax[0],
        camera_pos="C3",
        cmap="RdBu_r",
        vmin=-1,
        vmax=+1,
        cb_label=r"$\Delta$ HbO / µM",
        title=None,
    )
ax[0].set_title("HbO")
cedalion.vis.anatomy.plot_brain_in_axes(
        od,
        head_ras.landmarks,
        spatial_act.sel(chromo="HbR").pint.to("uM"),
        head_ras.brain,
        ax[1],
        camera_pos="C3",
        cmap="RdBu_r",
        vmin=-1,
        vmax=+1,
        cb_label=r"$\Delta$ HbR / µM",
        title=None,
    )
ax[1].set_title("HbR");
```

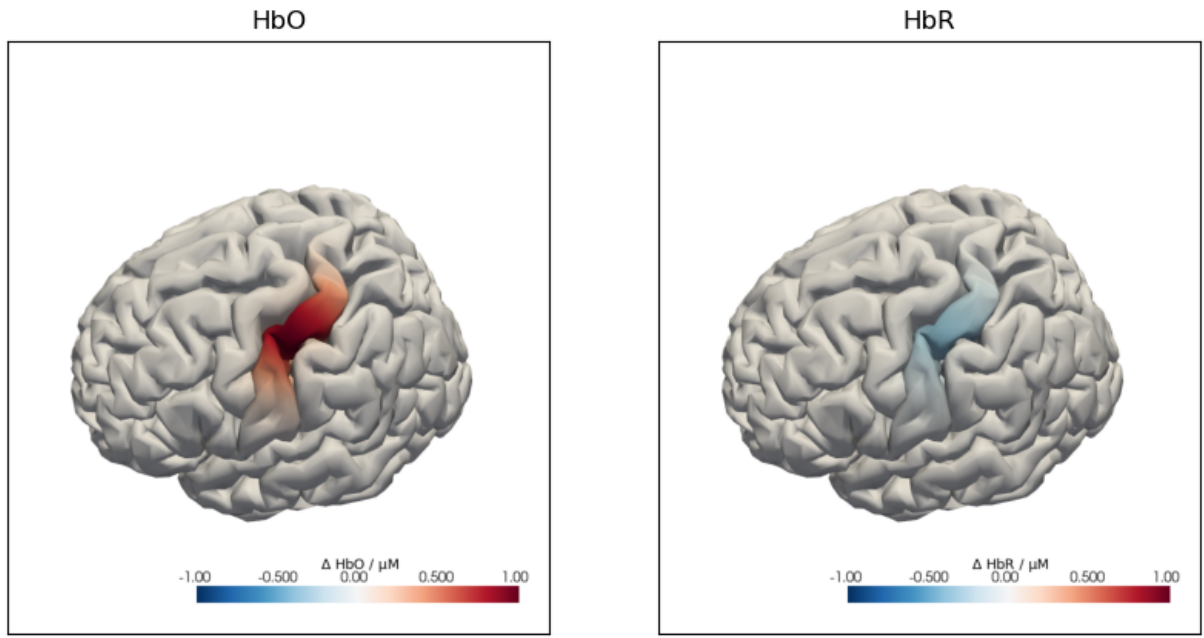

The following plot illustrates the effects of the `spatial_scale` and `intensity_scale` parameters:

```
[17]: f, ax = plt.subplots(2, 3, figsize=(9, 6))
      for i, spatial_scale in enumerate([0.5 * units.cm, 2 * units.cm, 3 * units.cm]):
          spatial_act = synhrf.build_spatial_activation(
              head_ras.brain,
              c3_seed,
              spatial_scale=spatial_scale,
              intensity_scale=1 * units.micromolar,
              hbr_scale=-0.4,
          )

          cedalion.vis.anatomy.plot_brain_in_axes(
              od,
              head_ras.landmarks,
              spatial_act.sel(chromo="HbO").pint.to("uM"),
              head_ras.brain,
              ax[0, i],
              camera_pos="C3",
              cmap="RdBu_r",
              vmin=-1,
              vmax=+1,
              cb_label=r"$\Delta$ HbO / µM",
              title=None,
          )
          ax[0, i].set_title(f"spatial_scale: {spatial_scale.magnitude} cm")

      for i, intensity_scale in enumerate(
          [
              0.5 * units.micromolar,
```

```python
        1.0 * units.micromolar,
        2.0 * units.micromolar,
    ]
):
    spatial_act = synhrf.build_spatial_activation(
        head_ras.brain,
        c3_seed,
        spatial_scale=2 * units.cm,
        intensity_scale=intensity_scale,
        hbr_scale=-0.4,
    )

    cedalion.vis.anatomy.plot_brain_in_axes(
        od,
        head_ras.landmarks,
        spatial_act.sel(chromo="HbO").pint.to("uM"),
        head_ras.brain,
        ax[1, i],
        camera_pos="C3",
        cmap="RdBu_r",
        vmin=-2,
        vmax=+2,
        cb_label=r"$\Delta$ HbO / µM",
        title=None,
    )
    ax[1, i].set_title(f"intensity_scale: {intensity_scale.magnitude} µM")

f.tight_layout()
```

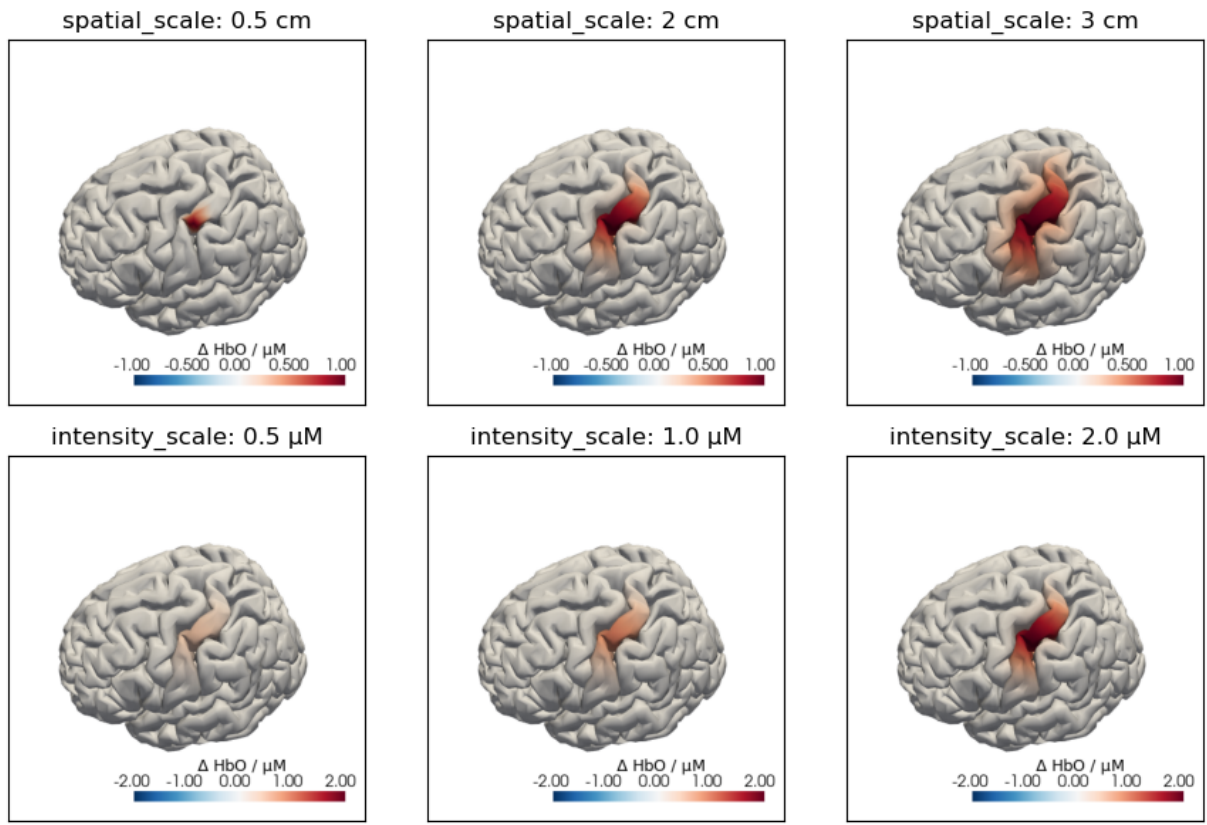

For this example notebook two activations are placed below C3 and C4:

```
[18]: spatial_act_c3 = synhrf.build_spatial_activation(
          head_ras.brain,
          c3_seed,
          spatial_scale=2 * cedalion.units.cm,
          intensity_scale=1 * units.micromolar,
          hbr_scale=-0.4,
      )
      spatial_act_c4 = synhrf.build_spatial_activation(
          head_ras.brain,
          c4_seed,
          spatial_scale=2 * cedalion.units.cm,
          intensity_scale=1 * units.micromolar,
          hbr_scale=-0.4,
      )
```

We concatenate the two images for C3 and C4 along dimension `trial_type` to get a single `DataArray` with the spatial information for both landmarks.

```
[19]: # concatenate the two spatial activations along a new dimension
      spatial_imgs = xr.concat(
          [spatial_act_c3, spatial_act_c4], dim="trial_type"
      ).assign_coords(trial_type=["Stim C3", "Stim C4"])
      spatial_imgs
```

```
[19]: <xarray.DataArray (trial_type: 2, vertex: 15002, chromo: 2)> Size: 480kB
      [M] 1.115e-28 -4.461e-29 2.003e-29 … -2.887e-14 1.427e-21 -5.706e-22
      Coordinates:
        * chromo      (chromo) <U3 24B 'HbO' 'HbR'
        * trial_type  (trial_type) <U7 56B 'Stim C3' 'Stim C4'
      Dimensions without coordinates: vertex
```

**Plots of spatial patterns**

Using the helper function `cedalion.vis.anatomy.plot_brain_in_axes`, the created activations on the brain surface below C3 and C4 are plotted:

```python
[20]: f, ax = plt.subplots(1,2, figsize=(10,5))

      cedalion.vis.anatomy.plot_brain_in_axes(
          od,
          head_ras.landmarks,
          spatial_imgs.sel(trial_type="Stim C3", chromo="HbO").pint.to("uM"),
          head_ras.brain,
          ax[0],
          camera_pos="C3",
          cmap="RdBu_r",
          vmin=-1,
          vmax=+1,
          cb_label=r"$\Delta$ HbO / µM",
          title="C3 Activation",
      )

      cedalion.vis.anatomy.plot_brain_in_axes(
          od,
          head_ras.landmarks,
          spatial_imgs.sel(trial_type="Stim C4", chromo="HbO").pint.to("uM"),
          head_ras.brain,
          ax[1],
          camera_pos="C4",
          cmap="RdBu_r",
          vmin=-1,
          vmax=+1,
          cb_label=r"$\Delta$ HbO / µM",
          title="C4 Activation",
      )

      f.tight_layout()
```

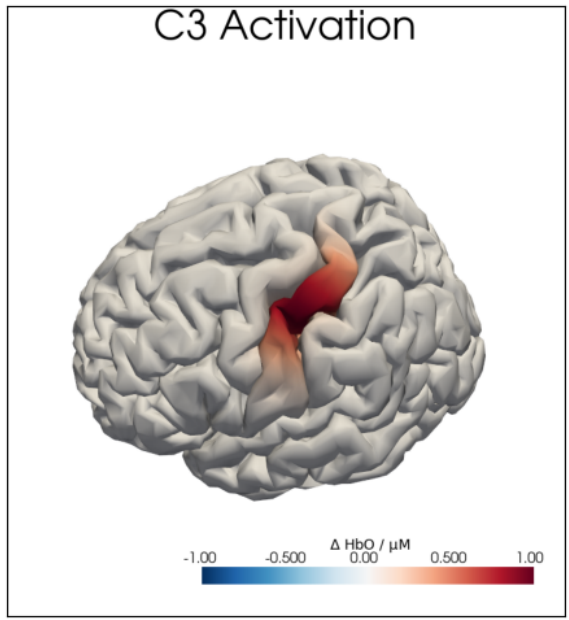
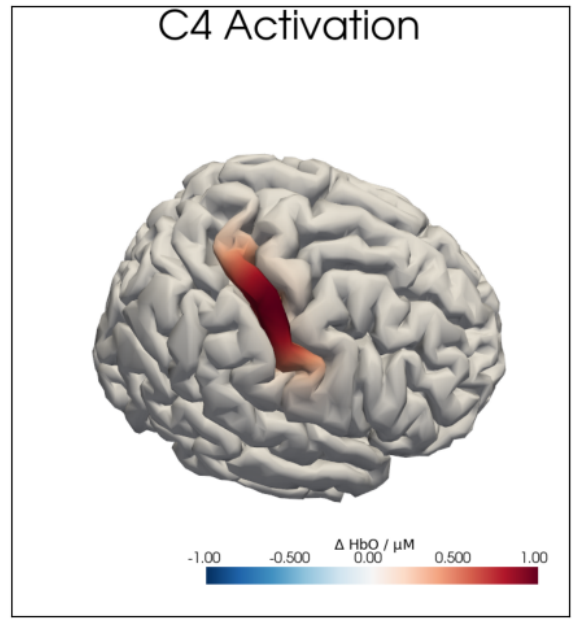

**Image Reconstruction**

We load the precomputed Adot matrix to be able to map from image to channel space. (For details see S5 image reconstruction tutorial).

```
[21]: Adot = cedalion.data.get_precomputed_sensitivity("nn22_resting", "colin27")
```

```
[22]: # we only consider brain vertices, not scalp and drop pruned channels
      Adot_brain = Adot.sel(vertex=Adot.is_brain, channel=od.channel)
      Adot_brain
```

```
[22]: <xarray.DataArray (channel: 545, vertex: 15002, wavelength: 2)> Size: 131MB
      4.881e-18 4.881e-18 7.405e-18 7.405e-18 … 0.0001423 1.332e-12 1.332e-12
      Coordinates:
          parcel      (vertex) object 120kB 'VisCent_ExStr_8_LH' … 'Background+Fr…
          is_brain    (vertex) bool 15kB True True True True … True True True True
        * channel     (channel) object 4kB 'S10D87' 'S10D94' … 'S47D29' 'S5D137'
          detector    (channel) object 4kB 'D87' 'D94' 'D89' … 'D129' 'D29' 'D137'
          source      (channel) object 4kB 'S10' 'S10' 'S10' … 'S56' 'S47' 'S5'
        * wavelength  (wavelength) float64 16B 760.0 850.0
      Dimensions without coordinates: vertex
      Attributes:
          units:    mm
```

The forward model and image reconstruction transform time series in image space into channel space, and vice versa. In the present case, we need to translate between optical-density time series of different wavelengths in channel space and concentration time series of different chromophores in image space. This is facilitated by the function **dot.forward_model.image_to_channel_space**:

```
[23]: # propagate concentration changes in the brain to optical density
      # changes in channel space
```

```
spatial_chan = dot.forward_model.image_to_channel_space(
    Adot_brain, spatial_imgs, spectrum="prahl"
)
spatial_chan
```

[23]: `<xarray.DataArray (trial_type: 2, wavelength: 2, channel: 545)> Size: 17kB`
`[] -1.583e-12 -6.242e-11 -3.406e-11 -2.385e-13 … 4.92e-06 7.374e-08 5.261e-08`
`Coordinates:`
`  * wavelength    (wavelength) float64 16B 760.0 850.0`
`  * channel       (channel) object 4kB 'S10D112' 'S10D133' … 'S9D94' 'S9D96'`
`  * trial_type    (trial_type) <U7 56B 'Stim C3' 'Stim C4'`
`    source        (channel) object 4kB 'S10' 'S10' 'S10' 'S10' … 'S9' 'S9' 'S9'`
`    detector      (channel) object 4kB 'D112' 'D133' 'D135' … 'D92' 'D94' 'D96'`

Show the spatial activation in channel space with a scalp plot.

[24]:
```
fig, ax = plt.subplots(1, 2)
# adjust plot size
fig.set_size_inches(12, 6)
cedalion.vis.anatomy.scalp_plot(
    od,
    rec.geo3d,
    spatial_chan.sel(trial_type="Stim C3", wavelength=850),
    ax[0],
    cmap="YlOrRd",
    title="850nm, activation under C3",
    vmin=spatial_chan.min().item(),
    vmax=spatial_chan.max().item(),
    cb_label="max peak amplitude",
)
cedalion.vis.anatomy.scalp_plot(
    od,
    rec.geo3d,
    spatial_chan.sel(trial_type="Stim C4", wavelength=850),
    ax[1],
    cmap="YlOrRd",
    title="850nm, activation under C4",
    vmin=spatial_chan.min().item(),
    vmax=spatial_chan.max().item(),
    cb_label="Max peak amplitude",
)
plt.show()
```

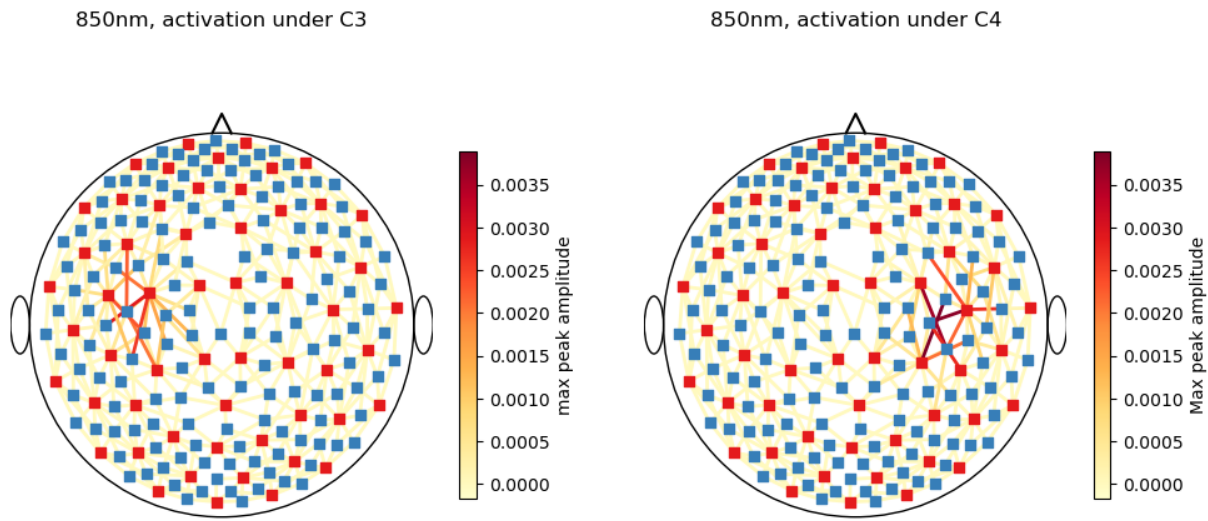

850nm, activation under C3 · 850nm, activation under C4

Get top 5 channels for each trial type where synthetic activation is highest

```
[25]: roi_chans_c3 = spatial_chan.channel[
          spatial_chan.sel(trial_type="Stim C3").max("wavelength").argsort()[-5:].values
      ].values
      roi_chans_c4 = spatial_chan.channel[
          spatial_chan.sel(trial_type="Stim C4").max("wavelength").argsort()[-5:].values
      ].values
```

**Concentration Scale**

The activations were simulataed in image space with a peak concentration change of 1 µM in one vertex. The change in optical density in one channel reflects concentration changes in the ensemble of vertices that this channel is sensitive to.

When applying the Beer-Lambert-transformation in channel space, a change in concentration is calculated for each channel. However, the scales of these concentration changes and the concentration changes in single vertices are not the same.

Here, a correction factor is calculated to scale the activation in channel space to 1uM.

```
[26]: dpf = xr.DataArray(
          [6, 6],
          dims="wavelength",
          coords={"wavelength": rec["amp"].wavelength},
      )

      # add time axis with one time point so we can convert to conc
      spatial_chan_w_time = spatial_chan.expand_dims("time")
      spatial_chan_w_time = spatial_chan_w_time.assign_coords(time=[0])
      spatial_chan_w_time.time.attrs["units"] = "second"
      display(spatial_chan_w_time)
```

```
spatial_chan_conc = cedalion.nirs.cw.od2conc(
        spatial_chan_w_time, geo3d, dpf, spectrum="prahl"
)
```

```
<xarray.DataArray (time: 1, trial_type: 2, wavelength: 2, channel: 545)> Size: 17kB
[] -1.583e-12 -6.242e-11 -3.406e-11 -2.385e-13 … 4.92e-06 7.374e-08 5.261e-08
Coordinates:
  * wavelength   (wavelength) float64 16B 760.0 850.0
  * channel      (channel) object 4kB 'S10D112' 'S10D133' … 'S9D94' 'S9D96'
  * trial_type   (trial_type) <U7 56B 'Stim C3' 'Stim C4'
    source       (channel) object 4kB 'S10' 'S10' 'S10' 'S10' … 'S9' 'S9' 'S9'
    detector     (channel) object 4kB 'D112' 'D133' 'D135' … 'D92' 'D94' 'D96'
  * time         (time) int64 8B 0
```

```
[27]: # rescale so that synthetic hrfs add 1 micromolar at peak.
rescale_factor = (1* units.micromolar / spatial_chan_conc.max()).item()

display(rescale_factor)
spatial_chan *= rescale_factor
```

) (dimensionless

**HRFs in channel space**

So far the notebook focused on the spatial extent of the activation. To build the temporal HRF model we use the same functionality that generates hrf regressors for the GLM.

First we select a basis function, which defines the temporal shape of the HRF.

```
[28]: basis_fct = glm.Gamma(tau=0 * units.s, sigma=3 * units.s, T=3 * units.s)
```

```
[29]: od.time
```

```
[29]: <xarray.DataArray 'time' (time: 3309)> Size: 26kB
      0.0 0.1112 0.2225 0.3337 0.445 0.5562 … 367.5 367.6 367.7 367.8 367.9 368.0
      Coordinates:
        * time       (time) float64 26kB 0.0 0.1112 0.2225 0.3337 … 367.8 367.9 368.0
          samples    (time) int64 26kB 0 1 2 3 4 5 6 … 3303 3304 3305 3306 3307 3308
      Attributes:
          units:      second
```

A Stim DataFrame, which contains the onset, duration and amplitude of the synthetic HRFs, is created.

```
[30]: stim_df = synhrf.build_stim_df(
        max_time=od.time.values[-1] * units.seconds,
        trial_types=["Stim C3", "Stim C4"],
        min_interval=10 * units.seconds,
        max_interval=20 * units.seconds,
        min_stim_dur = 10 * units.seconds,
        max_stim_dur = 10 * units.seconds,
        min_stim_value = 1.0,
        max_stim_value = 1.0,
        order="random",
)
```

```
[31]: stim_df.head()
```

```
[31]:     onset  duration  value  trial_type
      0   14.48      10.0    1.0    Stim C4
      1   43.62      10.0    1.0    Stim C4
      2   73.32      10.0    1.0    Stim C4
      3  101.25      10.0    1.0    Stim C4
      4  121.93      10.0    1.0    Stim C3
```

We can now use our stim dataframe, basis function, and spatial information to create the synthetic HRF timeseries

```
[32]: syn_ts = synhrf.build_synthetic_hrf_timeseries(od, stim_df, basis_fct, spatial_chan)
```

We get a synthetic HRF timeseries for each channel, trial_type and chromo / wavelength

```
[33]: syn_ts
```

```
[33]: <xarray.DataArray (trial_type: 2, wavelength: 2, channel: 545, time: 3309)> Size: 58MB
      [] -0.0 -0.0 -0.0 -0.0 -0.0 -0.0 -0.0 -0.0 … 0.0 0.0 0.0 0.0 0.0 0.0 0.0 0.0
      Coordinates:
        * wavelength  (wavelength) float64 16B 760.0 850.0
        * channel     (channel) object 4kB 'S10D112' 'S10D133' … 'S9D94' 'S9D96'
        * trial_type  (trial_type) <U7 56B 'Stim C3' 'Stim C4'
          source      (channel) object 4kB 'S10' 'S10' 'S10' 'S10' … 'S9' 'S9' 'S9'
          detector    (channel) object 4kB 'D112' 'D133' 'D135' … 'D92' 'D94' 'D96'
        * time        (time) float64 26kB 0.0 0.1112 0.2225 … 367.8 367.9 368.0
```

Sum the synthetic timeseries over trial_type dimension, so it has the same shape as the resting state data

```
[34]: syn_ts_sum = syn_ts.sum(dim='trial_type')
      syn_ts_sum
```

```
[34]: <xarray.DataArray (wavelength: 2, channel: 545, time: 3309)> Size: 29MB
      [] 0.0 0.0 0.0 0.0 0.0 0.0 0.0 0.0 0.0 0.0 … 0.0 0.0 0.0 0.0 0.0 0.0 0.0 0.0 0.0
      Coordinates:
        * wavelength  (wavelength) float64 16B 760.0 850.0
        * channel     (channel) object 4kB 'S10D112' 'S10D133' … 'S9D94' 'S9D96'
          source      (channel) object 4kB 'S10' 'S10' 'S10' 'S10' … 'S9' 'S9' 'S9'
          detector    (channel) object 4kB 'D112' 'D133' 'D135' … 'D92' 'D94' 'D96'
        * time        (time) float64 26kB 0.0 0.1112 0.2225 … 367.8 367.9 368.0
```

**Adding HRFs to measured data**

Here, the simulated activations are combined with physiological noise by adding the synthetic HRFs to the resting state dataset:

```
[35]: # set to false to process only the synth. HRF in the following
      ADD_RESTING_STATE = True

      if ADD_RESTING_STATE:
          od_w_hrf = od + syn_ts_sum
      else:
          od_w_hrf = syn_ts_sum
```

```python
        od_w_hrf = od_w_hrf.assign_coords(
            {
                "samples": ("time", od.samples.values),
                "time": ("time", od.time.data),
            }
        )
        od_w_hrf = od_w_hrf.pint.quantify({"time": "s"})
```

**Recover the HRFs again**

In the following, the added activations should be extracted from the simulated dataset again. To this end, the data is frequency filtered and block averages are calculated.

```
[36]: od_w_hrf_filtered = od_w_hrf.cd.freq_filter(fmin=0.02, fmax=0.5, butter_order=4)
```

```
[37]: od_w_hrf_filtered
```

```
[37]: <xarray.DataArray (channel: 545, wavelength: 2, time: 3309)> Size: 29MB
      [] 0.002073 0.002634 0.00317 0.003663 … 0.006999 0.006033 0.004973 0.00387
      Coordinates:
        * time        (time) float64 26kB 0.0 0.1112 0.2225 … 367.8 367.9 368.0
        * channel     (channel) object 4kB 'S10D87' 'S10D94' … 'S47D29' 'S5D137'
        * wavelength  (wavelength) float64 16B 760.0 850.0
          samples     (time) int64 26kB 0 1 2 3 4 5 … 3303 3304 3305 3306 3307 3308
          source      (channel) object 4kB 'S10' 'S10' 'S10' … 'S56' 'S47' 'S5'
          detector    (channel) object 4kB 'D87' 'D94' 'D89' … 'D129' 'D29' 'D137'
```

```python
[38]: epochs = od_w_hrf_filtered.cd.to_epochs(
          stim_df,  # stimulus dataframe
          ["Stim C3", "Stim C4"],  # select events
          before=5 * units.seconds,  # seconds before stimulus
          after=20 * units.seconds,  # seconds after stimulus
      )

      # calculate baseline
      baseline = epochs.sel(reltime=(epochs.reltime < 0)).mean("reltime")
      # subtract baseline
      epochs_blcorrected = epochs - baseline

      # group trials by trial_type. For each group individually average the epoch dimension
      blockaverage = epochs_blcorrected.groupby("trial_type").mean("epoch")

      n_roi = roi_chans_c3.size
      # show results
      f, ax = plt.subplots(2, n_roi, figsize=(16, 8))
      ax = ax.flatten()
      for i_ch, ch in enumerate(roi_chans_c3):
          for ls, trial_type in zip(["-", "--"], blockaverage.trial_type):
              ax[i_ch].plot(
                  blockaverage.reltime,
                  blockaverage.sel(wavelength=760, trial_type=trial_type, channel=ch),
                  "r",
```

```python
                lw=2,
                ls=ls,
            )
            ax[i_ch].plot(
                blockaverage.reltime,
                blockaverage.sel(wavelength=850, trial_type=trial_type, channel=ch),
                "b",
                lw=2,
                ls=ls,
            )
        ax[i_ch].grid(1)
        ax[i_ch].set_title(ch)
        ax[i_ch].set_ylim(-0.05, 0.05)

for i_ch, ch in enumerate(roi_chans_c4):
    for ls, trial_type in zip(["-", "--"], blockaverage.trial_type):
        ax[i_ch + n_roi].plot(
            blockaverage.reltime,
            blockaverage.sel(wavelength=760, trial_type=trial_type, channel=ch),
            "r",
            lw=2,
            ls=ls,
        )
        ax[i_ch + n_roi].plot(
            blockaverage.reltime,
            blockaverage.sel(wavelength=850, trial_type=trial_type, channel=ch),
            "b",
            lw=2,
            ls=ls,
        )
    ax[i_ch + n_roi].grid(1)
    ax[i_ch + n_roi].set_title(ch)
    ax[i_ch + n_roi].set_ylim(-0.05, 0.05)

plt.suptitle(
    "Blockaverage for channels most sensitive to C3 (top) and C4 (bottom): 760nm: r |␣
    ↪850nm: b | C3: - | C4: --"
)
plt.tight_layout()
plt.show()
```

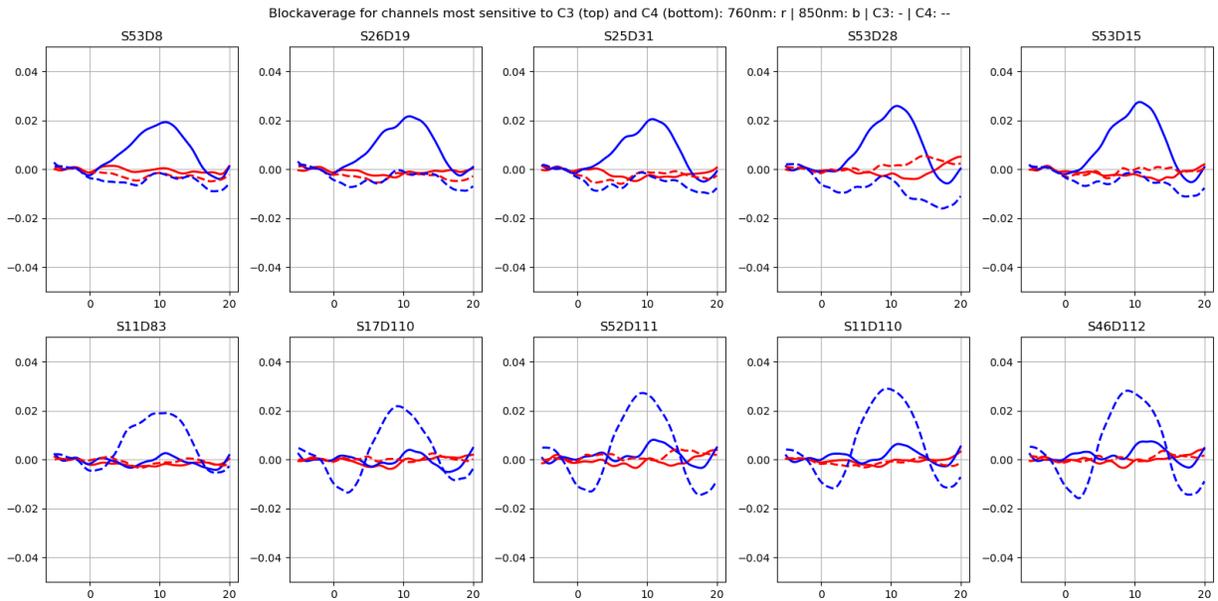

**Map block average back to brain surface**

We map our extracted block averages back to the brain surface to visualize the recovered HRFs activation for Stim C3. We can compare it to the synthetic HRF image we created earlier.

```
[39]: sbf = dot.GaussianSpatialBasisFunctions(head_ras, Adot, **dot.SBF_GAUSSIANS_DENSE,
      verbose=False)
```

```
[40]: recon = dot.ImageRecon(
          Adot,
          recon_mode="mua2conc",
          alpha_meas=0.01,
          brain_only=True,
          spatial_basis_functions=sbf
      )

      blockaverage_img = recon.reconstruct(blockaverage)
      blockaverage_img
```

```
[40]: <xarray.DataArray (chromo: 2, vertex: 15002, trial_type: 2, reltime: 226)> Size: 108MB
      [µM] -0.03861 -0.02582 -0.01396 -0.003175 0.006414 … 0.0 0.0 0.0 0.0 0.0
      Coordinates:
        * chromo      (chromo) <U3 24B 'HbO' 'HbR'
        * reltime     (reltime) float64 2kB -4.995 -4.884 -4.773 … 19.76 19.87 19.98
        * trial_type  (trial_type) object 16B 'Stim C3' 'Stim C4'
          is_brain    (vertex) bool 15kB True True True True … True True True True
          parcel      (vertex) object 120kB 'VisCent_ExStr_8_LH' … 'Background+Fr…
      Dimensions without coordinates: vertex
```

```
[41]: # plot HbO time trace of left and right brain hemisphere during FTapping/Right

      for view in ["left_hemi", "right_hemi"]:
```

```python
trial_type = "Stim C3"
gif_fname = "Ftapping-right" + "_HbO_" + view + ".gif"

hbo = blockaverage_img.sel(chromo="HbO", trial_type=trial_type).pint.dequantify()
hbo_brain = hbo.transpose("vertex", "reltime")

ntimes = hbo.sizes["reltime"]

b = cdc.VTKSurface.from_trimeshsurface(head_ras.brain)
b = pv.wrap(b.mesh)
b["reco_hbo"] = hbo_brain[:, 0] - hbo_brain[:, 0]

p = pv.Plotter()

p.add_mesh(
    b,
    scalars="reco_hbo",
    cmap="seismic",  # 'gist_earth_r',
    clim=(-2.5, 2.5),
    scalar_bar_args={"title": "HbO / µM"},
    smooth_shading=True,
)

tl = lambda tt: f"{trial_type} HbO rel. time: {tt:.3f} s"
time_label = p.add_text(tl(0))

cog = head_ras.brain.vertices.mean("label")
cog = cog.pint.dequantify().values
if view == "left_hemi":
    p.camera.position = cog + [-400, 0, 0]
else:
    p.camera.position = cog + [400, 0, 0]
p.camera.focal_point = cog
p.camera.up = [0, 0, 1]
p.reset_camera()

p.open_gif(gif_fname)

for i in range(0, ntimes, 3):
    b["reco_hbo"] = hbo_brain[:, i] - hbo_brain[:, 0]
    time_label.set_text("upper_left", tl(hbo_brain.reltime[i]))

    p.write_frame()

p.close()
```

```python
[42]: display_image("Ftapping-right_HbO_left_hemi.gif")
      display_image("Ftapping-right_HbO_right_hemi.gif")
```

# Stim C3 HbO rel. time: -1.665 s

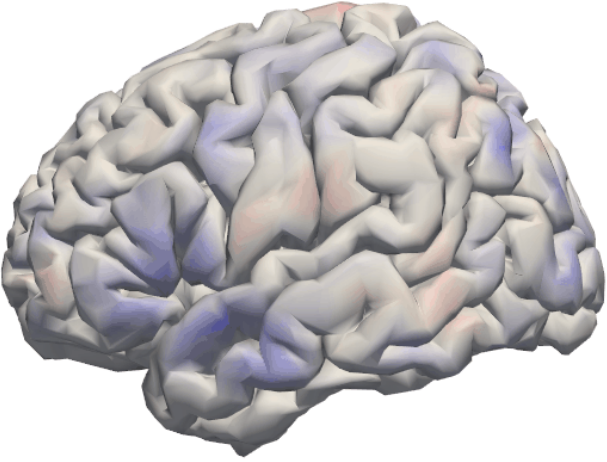

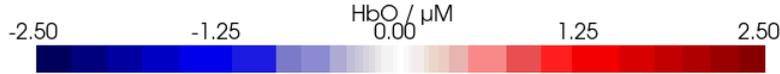

HbO / µM

-2.50   -1.25   0.00   1.25   2.50

## Stim C3 HbO rel. time: -1.665 s

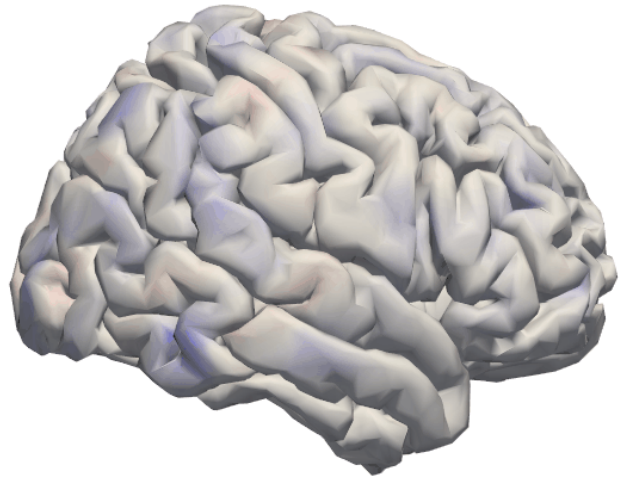

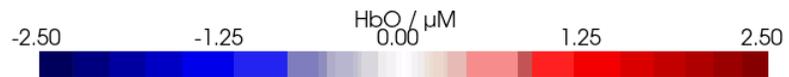

HbO / µM

-2.50    -1.25    0.00    1.25    2.50

```
[43]: for c, channel in [("C3", roi_chans_c3[-1]), ("C4", roi_chans_c4[-1])]:
          trial_type = f"Stim {c}"

          f,ax = plt.subplots(2,4, figsize=(14,6))
          vmin,vmax = -2.5,2.5

          cedalion.vis.anatomy.plot_brain_in_axes(
              od,
              head_ras.landmarks,
              spatial_imgs.sel(trial_type=trial_type, chromo="HbO").pint.to("uM"),
              head_ras.brain,
              ax[0,0],
              camera_pos=c,
              cmap="RdBu_r",
              vmin=vmin,
              vmax=vmax,
              cb_label=r"$\Delta$ HbO / µM",
              #title=f"{c} Activation",
          )
          ax[0,0].set_title(f"true activation at {c}")

          trels = [0., 10., 15.]
```

```python
for i_wl, wl in enumerate(blockaverage.wavelength.values):
    tmp = blockaverage.sel(
        trial_type=trial_type, channel=channel, wavelength=wl
    )
    ax[1,0].plot(tmp.reltime, tmp, label=f"{wl} nm")
for trel in trels:
    ax[1,0].axvline(trel, c="k", ls=":")
ax[1,0].legend(loc="upper left")
ax[1,0].set_xlabel("$t_{rel}$")
ax[1,0].set_ylabel("OD")
ax[1,0].set_title(channel)

for i_col, trel in enumerate(trels, start=1):
    for i_row, cam_pos in enumerate(["C3", "C4"]):
        cedalion.vis.anatomy.plot_brain_in_axes(
            od,
            head_ras.landmarks,
            blockaverage_img.sel(chromo="HbO", trial_type=trial_type)
            .sel(reltime=trel, method="nearest"),
            head_ras.brain,
            ax[i_row, i_col],
            camera_pos=cam_pos,
            cmap="RdBu_r",
            vmin=vmin,
            vmax=vmax,
            cb_label=r"$\Delta$ HbO / µM",
            #title="$t_{rel}=" + f"{trel:.1f} s$",
        )
        ax[i_row, i_col].set_title(f"view at {cam_pos} | " + "$t_{rel}=" + f"{trel:
↪.1f} s$")
```

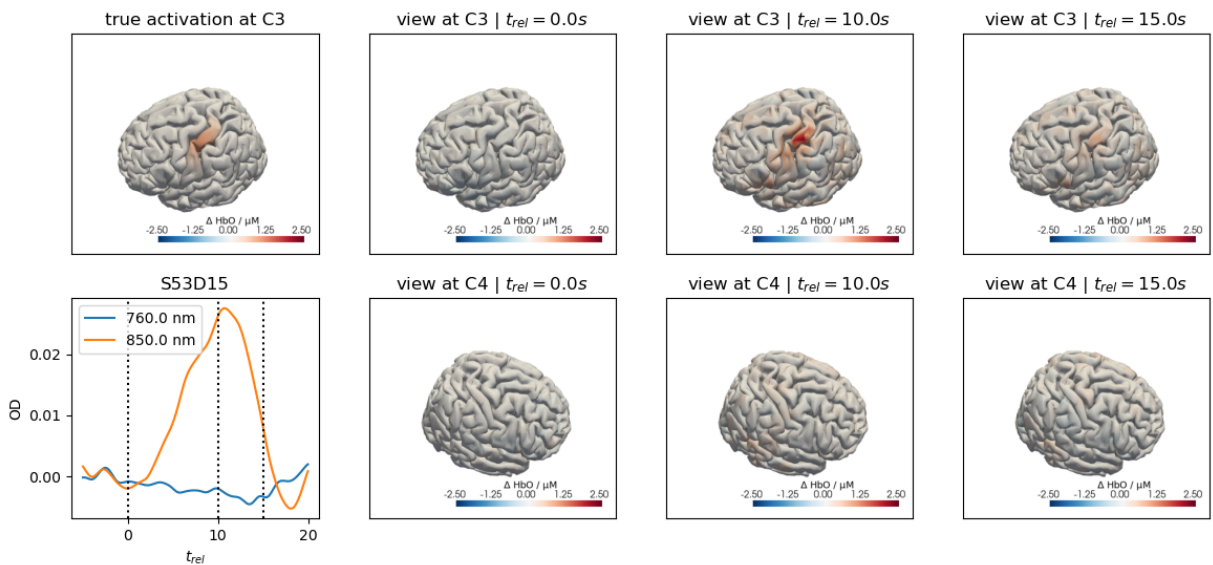

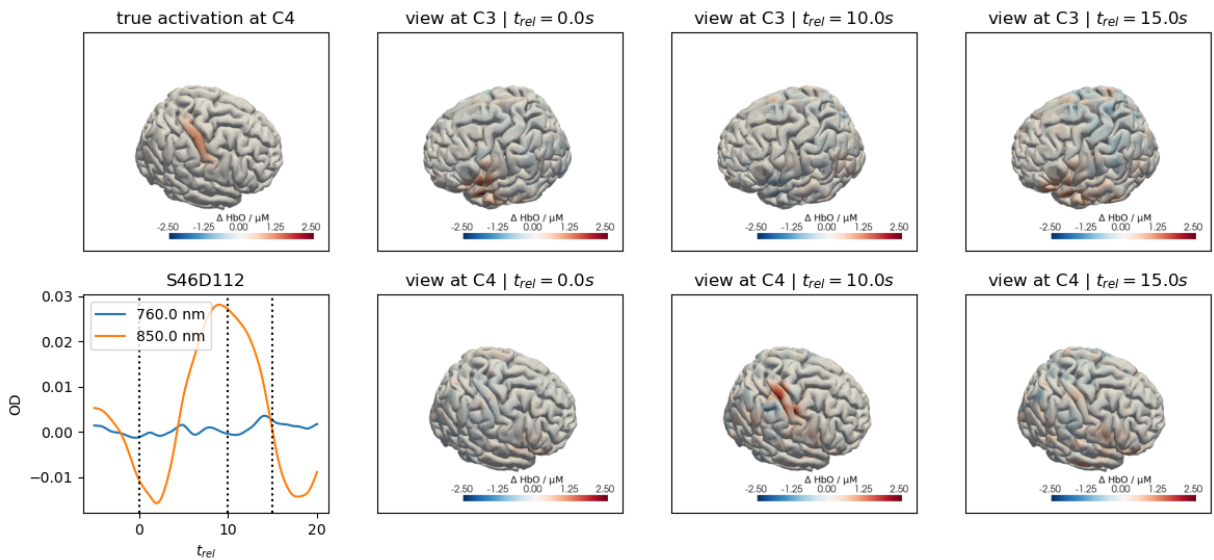

## Adding Motion Artifacts

The second part of this notebook demonstrates how to add spike and baseline-shift artifacts to existing time series.

First, we'll load some example data.

```
[44]: rec = cedalion.data.get_fingertappingDOT()
rec["od"] = cedalion.nirs.cw.int2od(rec["amp"])

f, ax = plt.subplots(1, 1, figsize=(12, 4))
ax.plot(
    rec["amp"].time,
    rec["amp"].sel(channel="S1D2", wavelength="760"),
    "r-",
    label="760nm",
)
ax.plot(
    rec["amp"].time,
    rec["amp"].sel(channel="S1D2", wavelength="850"),
    "g-",
    label="850nm",
)

plt.legend()
ax.set_xlabel("time / s")
ax.set_ylabel("intensity / V")

display(rec["od"])
```

```
<xarray.DataArray (channel: 100, wavelength: 2, time: 8794)> Size: 14MB
[] 0.02263 0.02322 0.01368 0.005969 0.01692 … 0.01558 0.02343 0.02274 0.02761
Coordinates:
  * time          (time) float64 70kB 0.0 0.2294 0.4588 … 2.017e+03 2.017e+03
```

```
samples        (time) int64 70kB 0 1 2 3 4 5 … 8788 8789 8790 8791 8792 8793
*  channel     (channel) object 800B 'S1D1' 'S1D2' 'S1D4' … 'S14D31' 'S14D32'
   source      (channel) object 800B 'S1' 'S1' 'S1' 'S1' … 'S14' 'S14' 'S14'
   detector    (channel) object 800B 'D1' 'D2' 'D4' 'D5' … 'D29' 'D31' 'D32'
*  wavelength  (wavelength) float64 16B 760.0 850.0
```

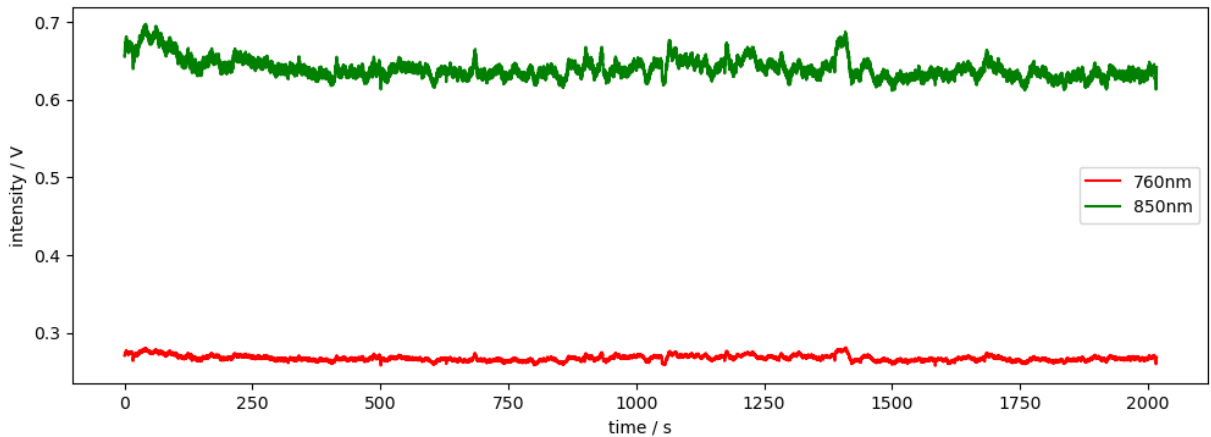

**Artifact Generation**

Artifacts are generated by the functions `gen_bl_shift` and `gen_spike`.

These function take as arguments: - the time axis of time series - onset time - duration

The amplitude of the generated artifacts is set to 1.

```
[45]: time = rec["amp"].time

      sample_bl_shift = synart.gen_bl_shift(time, 1000) # baseline shift a t = 1000s
      sample_spike = synart.gen_spike(time, 1500, 3) # spike artifact at t=1500s for 3s

      display(sample_bl_shift)

      fig, ax = plt.subplots(1, 1, figsize=(12,2))
      ax.plot(time, sample_bl_shift, "r-", label="bl_shift")
      ax.plot(time, sample_spike, "g:", label="spike")
      ax.set_xlabel('Time / s')
      ax.set_ylabel('Amp')
      ax.legend()

      plt.tight_layout()
      plt.show()
```

```
<xarray.DataArray 'time' (time: 8794)> Size: 70kB
0.0 0.0 0.0 0.0 0.0 0.0 0.0 0.0 0.0 0.0 0.0 0.0 … 1.0 1.0 1.0 1.0 1.0 1.0 1.0 1.0
Coordinates:
  * time     (time) float64 70kB 0.0 0.2294 0.4588 … 2.017e+03 2.017e+03
```

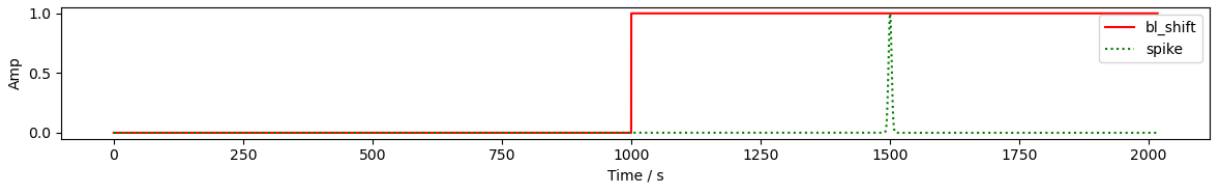

## Controlling Artifact Timing

Artifacts can be placed using a timing data frame with columns 'onset', 'duration', 'trial_type', 'value', and 'channel', i.e. it resembles the stimulus data frame with the added 'channel' column.

We can use the function `add_event_timing` to create such timing data frames. The function allows precise control over each event.

The function `sel_chans_by_opt` allows us to select a list of channels by way of a list of optode labels. This reflects the fact that motion artifacts usually stem from the motion of a specific optode or set of optodes, which in turn affects all related channels.

We can also use the functions `random_events_num` and `random_events_perc` to add random events to the data frame, specifying either the number of events or the percentage of the time series duration, respectively.

```
[46]:   # Create a list of events in the format (onset, duration)
        events = [(1000, 1), (2000, 1)]

        # Creates a new timing dataframe with the specified events.
        # Setting channel to None indicates that the artifact applies to all channels.
        timing_amp = synart.add_event_timing(events, 'bl_shift', None)

        # Select channels by optode
        chans = synart.sel_chans_by_opt(["S1"], rec["od"])

        # Add random events to the timing dataframe
        timing_od = synart.random_events_perc(time, 0.01, ["spike"], chans)

        display(timing_amp)
        display(timing_od)
```

|   | onset | duration | trial_type | value | channel |
|---|-------|----------|------------|-------|---------|
| 0 | 1000  | 1        | bl_shift   | 1     | None    |
| 1 | 2000  | 1        | bl_shift   | 1     | None    |

|    | onset       | duration | trial_type | value | \ |
|----|-------------|----------|------------|-------|---|
| 0  | 1190.982338 | 0.388722 | spike      | 1     |   |
| 1  | 1167.575165 | 0.300457 | spike      | 1     |   |
| 2  | 688.302535  | 0.115636 | spike      | 1     |   |
| 3  | 315.660854  | 0.373323 | spike      | 1     |   |
| 4  | 1491.663332 | 0.151644 | spike      | 1     |   |
| .. | ...         | ...      | ...        | ...   |   |
| 76 | 454.468384  | 0.393736 | spike      | 1     |   |
| 77 | 515.875056  | 0.346139 | spike      | 1     |   |
| 78 | 474.319191  | 0.216854 | spike      | 1     |   |
| 79 | 1489.173013 | 0.120179 | spike      | 1     |   |

```
80   1428.407956   0.373444        spike        1
```

```
                               channel
0    [S1D1, S1D2, S1D4, S1D5, S1D6, S1D8]
1    [S1D1, S1D2, S1D4, S1D5, S1D6, S1D8]
2    [S1D1, S1D2, S1D4, S1D5, S1D6, S1D8]
3    [S1D1, S1D2, S1D4, S1D5, S1D6, S1D8]
4    [S1D1, S1D2, S1D4, S1D5, S1D6, S1D8]
..                                    ...
76   [S1D1, S1D2, S1D4, S1D5, S1D6, S1D8]
77   [S1D1, S1D2, S1D4, S1D5, S1D6, S1D8]
78   [S1D1, S1D2, S1D4, S1D5, S1D6, S1D8]
79   [S1D1, S1D2, S1D4, S1D5, S1D6, S1D8]
80   [S1D1, S1D2, S1D4, S1D5, S1D6, S1D8]

[81 rows x 5 columns]
```

**Adding Artifacts to Data**

The function `add_artifacts` scales artifacts and adds them to time-series data.

The artifact generation functions (see above) are passed as a dictionary. Keys correspond to entries in the column 'trial_type' of the timing data frame, i.e. each event specified in the timing data frame is generated using the function `artifacts[trial_type]`. If mode is 'manual', artifacts are scaled directly by the `scale` parameter, otherwise artifacts are automatically scaled by a parameter `alpha` which is calculated using a sliding window approach.

Finally, artifacts to be added to optical-density time series can be automatically scaled to yield specific concentration amplitudes with the function `add_chromo_artifacts_2_od`.

```python
[47]: artifacts = {"spike": synart.gen_spike, "bl_shift": synart.gen_bl_shift}

      # Add baseline shifts to the amp data
      rec["amp2"] = synart.add_artifacts(rec["amp"], timing_amp, artifacts)

      # Convert the amp data to optical density
      rec["od2"] = cedalion.nirs.cw.int2od(rec["amp2"])

      dpf = xr.DataArray(
          [6, 6],
          dims="wavelength",
          coords={"wavelength": rec["amp"].wavelength},
      )

      # add spikes to od based on conc amplitudes
      rec["od2"] = synart.add_chromo_artifacts_2_od(
          rec["od2"], timing_od, artifacts, rec.geo3d, dpf, 1.5
      )

      # Plot the OD data
      channels = rec["od"].channel.values[0:6]
      fig, axes = plt.subplots(len(channels), 1, figsize=(12, len(channels) * 2))
      if len(channels) == 1:
```

```python
    axes = [axes]
for i, channel in enumerate(channels):
    ax = axes[i]
    ax.plot(
        rec["od2"].time,
        rec["od2"].sel(channel=channel, wavelength="850"),
        "g-",
        label="850nm + artifacts",
    )
    ax.plot(
        rec["od"].time,
        rec["od"].sel(channel=channel, wavelength="850"),
        "r-",
        label="850nm - od",
    )
    ax.set_title(f"Channel: {channel}")
    ax.set_xlabel("Time / s")
    ax.set_ylabel("OD")
    ax.legend()
plt.tight_layout()
```

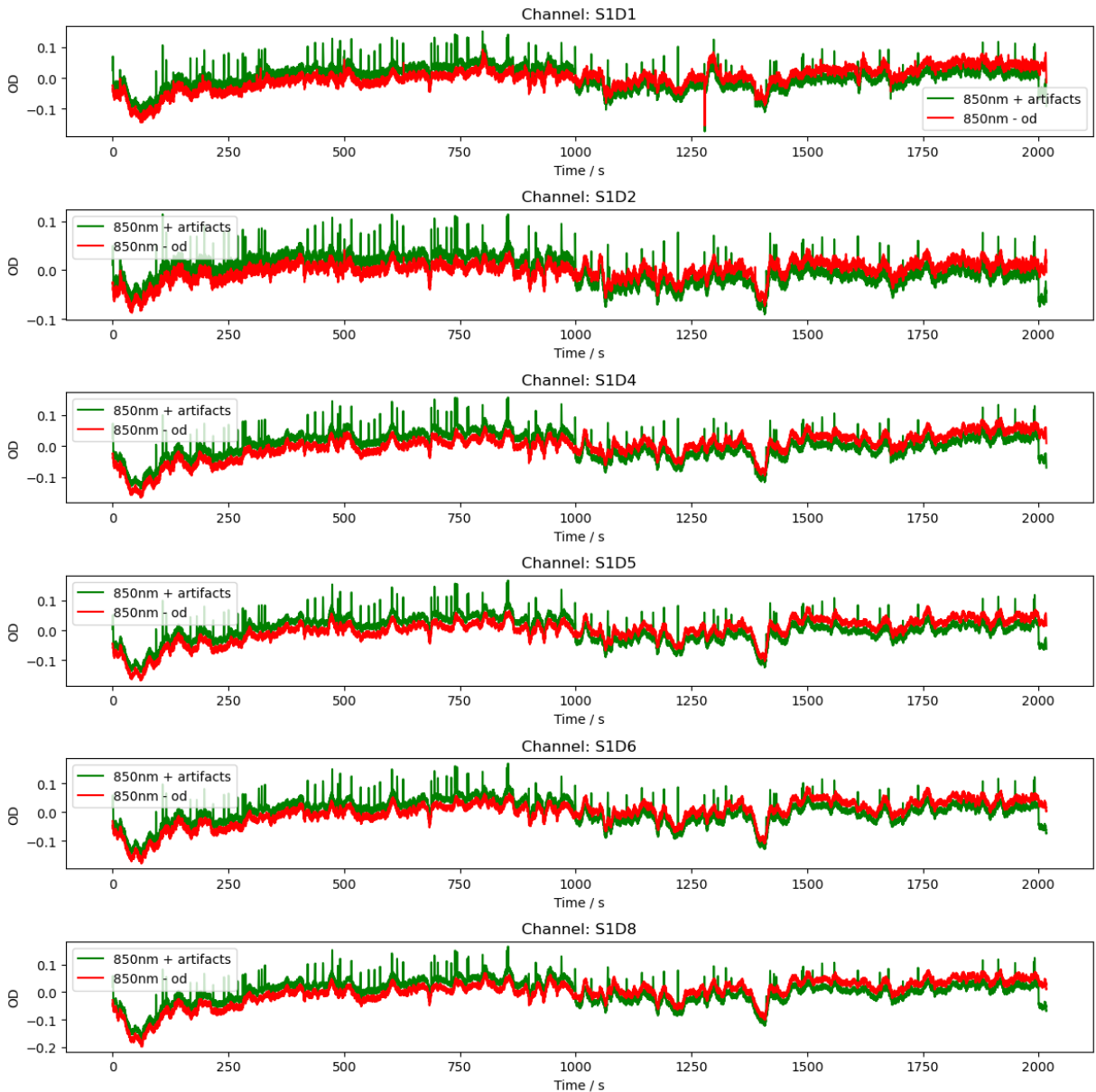

```
[48]:  # Plot the data in conc

rec["conc"] = cedalion.nirs.cw.od2conc(rec["od"], rec.geo3d, dpf)
rec["conc2"] = cedalion.nirs.cw.od2conc(rec["od2"], rec.geo3d, dpf)
channels = rec["od"].channel.values[0:6]
fig, axes = plt.subplots(len(channels), 1, figsize=(12, len(channels) * 2))
if len(channels) == 1:
    axes = [axes]
for i, channel in enumerate(channels):
    ax = axes[i]
    ax.plot(
        rec["conc2"].time,
        rec["conc2"].sel(channel=channel, chromo="HbR"),
        "g-",
```

```
            label="HbR + artifacts",
        )
    ax.plot(
        rec["conc"].time,
        rec["conc"].sel(channel=channel, chromo="HbR"),
        "b-",
        label="HbR",
    )
    ax.set_title(f"Channel: {channel}")
    ax.set_xlabel("Time / s")
    ax.set_ylabel("conc")
    ax.legend()
plt.tight_layout()
plt.show()
```

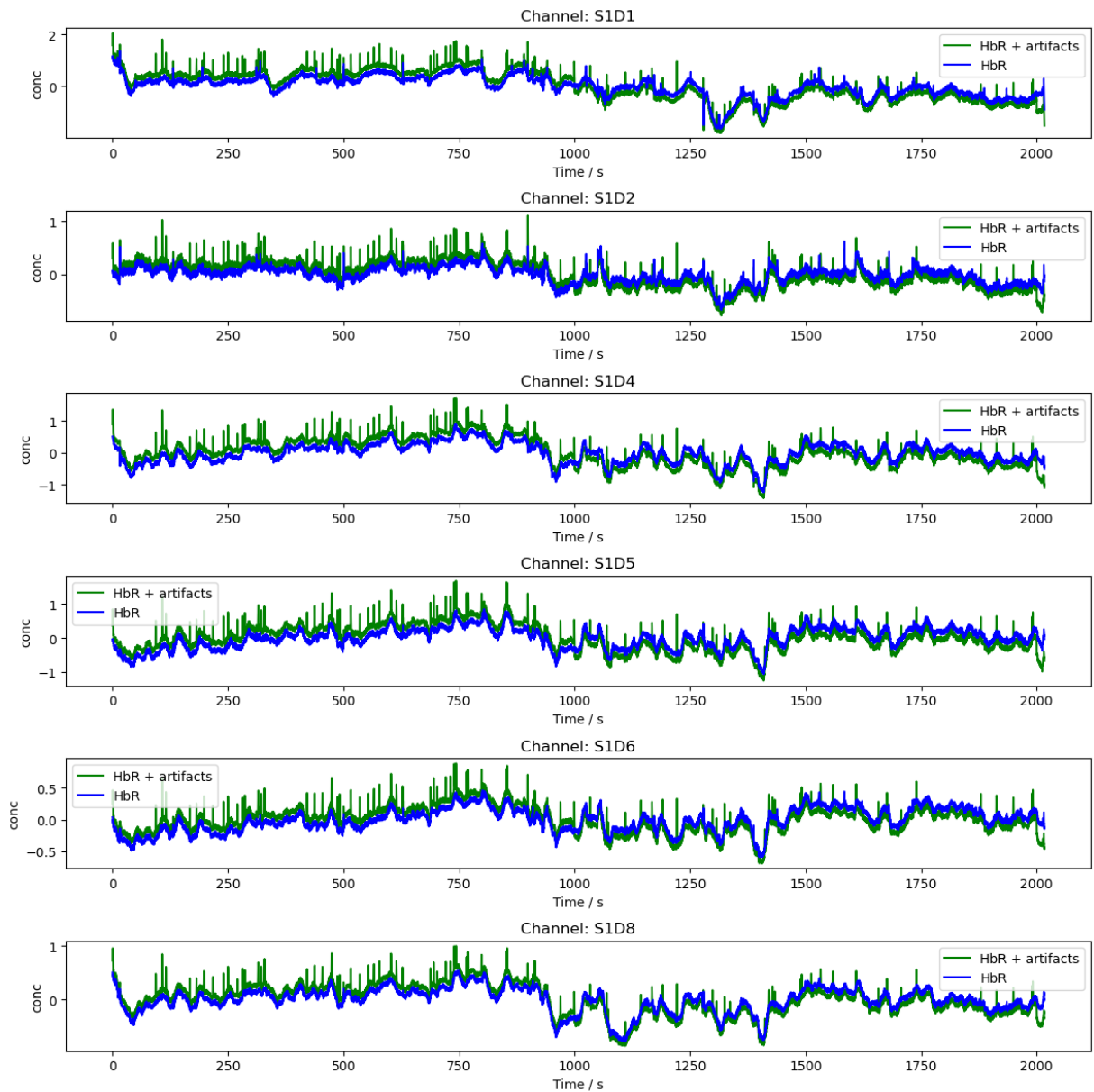